\documentclass[journal,twocolumn]{IEEEtran}

\usepackage{amsmath}
\usepackage{amssymb}
\usepackage{times}
\usepackage{graphicx}
\usepackage{xspace}
\usepackage{paralist} 
\usepackage{xypic}
\xyoption{curve}
\usepackage{latexsym}
\usepackage{theorem}
\usepackage{ifthen}
\usepackage{subfigure}

\usepackage{booktabs}
\newcommand{\ra}[1]{\renewcommand{\arraystretch}{#1}}

\newboolean{draft} 
\newcommand{\isdraft}[2]{\ifthenelse{\boolean{draft}}{#1}{#2}}

\newcommand{\DFT}[0]{\operatorname{DFT}}
\newcommand{\DFTt}[1]{{\DFT\text{\rm \!-#1}}}
\newcommand{\RDFT}[0]{\operatorname{RDFT}}
\newcommand{\RDFTt}[1]{{\RDFT\text{\rm \!-#1}}}
\newcommand{\QDFT}[0]{\operatorname{QDFT}}
\newcommand{\DHT}{\operatorname{DHT}}
\newcommand{\DHTt}[1]{{\DHT\text{\rm \!-#1}}}
\newcommand{\DWT}{\operatorname{DWT}}
\newcommand{\DWTt}[1]{{\DWT\text{\rm \!-#1}}}

\newcommand{\DCT}{\operatorname{DCT}}
\newcommand{\iDCT}{\operatorname{iDCT}}
\newcommand{\iDST}{\operatorname{iDST}}
\newcommand{\iDTT}{\operatorname{iDTT}}
\newcommand{\DST}{\operatorname{DST}}
\newcommand{\DTT}{\operatorname{DTT}}
\newcommand{\pDTT}{\operatorname{\overline{DTT}}}

\newcommand{\DCTt}[1]{{\DCT\text{\rm \!-#1}}}
\newcommand{\DSTt}[1]{{\DST\text{\rm \!-#1}}}
\newcommand{\pDCTt}[1]{{\overline{\DCT\text{\rm \!-#1}}}}
\newcommand{\pDSTt}[1]{{\overline{\DST\text{\rm \!-#1}}}}
\newcommand{\iDCTt}[1]{{\iDCT\text{\rm \!-#1}}}

\newcommand{\cas}[0]{\operatorname{cas}}
\newcommand{\prob}[0]{\operatorname{prob}}

\newcommand{\tensor}[0]{\otimes}
\newcommand{\dirsum}[0]{\oplus}
\newcommand{\bigdirsum}[0]{\bigoplus}
\newcommand{\diag}[0]{\operatorname{diag}}

\newcommand{\C}[0]{{\mathbb{C}}}
\newcommand{\R}[0]{{\mathbb{R}}}
\newcommand{\Z}[0]{{\mathbb{Z}}}
\newcommand{\Q}[0]{{\mathbb{Q}}}
\newcommand{\N}[0]{{\mathbb{N}}}

\newcommand{\ts}{\textstyle}
\newcommand{\dps}[0]{\displaystyle}
\newcommand{\fh}[0]{{\frac 12}}
\newcommand{\alg}[0]{{\cal A}}
\newcommand{\md}[0]{{\cal M}}

\newcommand{\four}[0]{{\cal F}}
\newcommand{\poly}[0]{{\cal P}}
\newcommand{\bc}[0]{b.c.\xspace}

\newcommand{\lifta}[1]{\raisebox{1.7mm}[-1.7mm]{$#1$}}
\newcommand{\markit}[0]{\underline}
\newcommand{\mypar}[1]{{\bf #1.}}

\newcommand{\myleft}[0]{\Bigl}
\newcommand{\myright}[0]{\Bigr}
\newcommand{\oddots}[0]%
  {\rule{0pt}{7pt}\raisebox{1pt}{.}\,\raisebox{4pt}{.}\,
     \raisebox{7pt}{.}}

\newcommand{\coorda}[1]{\text{\boldmath$#1$\unboldmath}}
\newcommand{\coord}[1]{\text{\bf #1}}
\newcommand{\multvar}[1]{\overline{#1}}

\newcommand{\one}[0]{\operatorname{I}}
\newcommand{\oppone}[0]{\operatorname{J}}

\newcommand{\sst}[0]{\scriptstyle}

\newcommand{\da}[0]{\raisebox{0.5mm}[-0.5mm]{---}}

\newsavebox{\upline}
\savebox{\upline}(0,0)[lb]{
  \put(0,0){\makebox(0,0){$\bullet$}}
  \put(0,0){\line(1,1){20}}
  \put(20,20){\makebox(0,0){$\bullet$}}
}
\newsavebox{\bupline}
\savebox{\bupline}(0,0)[lb]{
  \put(0,0){\makebox(0,0){$\bullet$}}
  \put(0,0){\line(1,1){30}}
  \put(30,30){\makebox(0,0){$\bullet$}}
}
\newsavebox{\upbline}
\savebox{\upbline}(0,0)[lb]{
  \put(0,0){\makebox(0,0){$\bullet$}}
  \thicklines
  \put(0,0){\line(1,1){20}}
  \thinlines
  \put(20,20){\makebox(0,0){$\bullet$}}
}
\newsavebox{\bupbline}
\savebox{\bupbline}(0,0)[lb]{
  \put(0,0){\makebox(0,0){$\bullet$}}
  \thicklines
  \put(0,0){\line(1,1){30}}
  \put(30,30){\makebox(0,0){$\bullet$}}
}
\newsavebox{\updline}
\savebox{\updline}(0,0)[lb]{
  \put(0,0){\makebox(0,0){$\circ$}}
  \multiput(2,2)(1.45,1.45){12}{\makebox(0,0){$\cdot$}}
  \put(20,20){\makebox(0,0){$\circ$}}
}
\newsavebox{\downline}
\savebox{\downline}(0,0)[lb]{
  \put(0,20){\makebox(0,0){$\bullet$}}
  \put(0,20){\line(1,-1){20}}
  \put(20,0){\makebox(0,0){$\bullet$}}
}
\newsavebox{\bdownline}
\savebox{\bdownline}(0,0)[lb]{
  \put(0,30){\makebox(0,0){$\bullet$}}
  \put(0,30){\line(1,-1){30}}
  \put(30,0){\makebox(0,0){$\bullet$}}
}
\newsavebox{\downdline}
\savebox{\downdline}(0,0)[lb]{
  \put(0,20){\makebox(0,0){$\circ$}}
  \multiput(2,18)(1.45,-1.45){12}{\makebox(0,0){$\cdot$}}
  \put(20,0){\makebox(0,0){$\circ$}}
}

\newcommand{\algogen}[8]{%
\xymatrix{
#1 \ar[rr]^{\dps #2} \ar[dd]^{\dps #4} && #3 \ar[dd]^{\dps #5} \\ \\
#6 \ar[rr]^{\dps #7} && #8
}
}

{\theoremheaderfont{\it}
\theorembodyfont{\rmfamily}
\newtheorem{theorem}{Theorem}
\newtheorem{lemma}[theorem]{Lemma}
\newtheorem{definition}[theorem]{Definition}

}

\title{Algebraic Signal Processing Theory}
\author{Markus P\"uschel and Jos\'e M.~F.~Moura\thanks{This work
was supported by NSF through awards 9988296 and 0310941.}
\thanks{Markus P\"uschel and Jos\'e M.~F.~Moura
are with the Department of Electrical and Compu{\-}ter Engineering,
Carnegie Mellon University, Pittsburgh. 
E-mail: \{pueschel,moura\}@ece.cmu.edu .}}

\begin{document}

\maketitle

\begin{abstract}
This paper presents an algebraic theory of linear signal processing.
At the core of algebraic signal processing is the concept of a linear
signal model defined as a triple
\isdraft{}{\boldmath}$(\alg,\md,\Phi)$, where familiar concepts like
the filter space and the signal space are cast as an algebra~$\alg$
and a module~$\md$, respectively, and $\Phi$ generalizes the concept
of the \isdraft{}{\boldmath}$z$-transform to bijective linear mappings
from a vector space of, e.g., signal samples, into the module~$\md$.
A signal model provides the structure for a particular linear signal
processing application, such as infinite and finite discrete time, or
infinite or finite discrete space, or the various forms of
multidimensional linear signal processing.  As soon as a signal
model is chosen, basic ingredients follow, including the associated
notions of filtering, spectrum, and Fourier transform. 

The shift operator~$q$, which is at the heart of ergodic theory and
dynamical systems, is a key concept in the algebraic theory: it is the
generator of the algebra of filters~$\alg$.  Once the shift is chosen,
a well-defined methodology leads to the associated signal
model. Different shifts correspond to infinite and finite time models
with associated infinite and finite
\isdraft{}{\boldmath}$z$-transforms, and to infinite and finite space
models with associated infinite and finite $C$-transforms (that we
introduce). In particular, we show that the 16 discrete cosine and
sine transforms are Fourier transforms for the finite space
models. Other definitions of the shift naturally lead to new signal
models and to new transforms as associated Fourier transforms in one
and higher dimensions, separable and non-separable. 

We explain in algebraic terms shift-invariance (the algebra of
filters~$\alg$ is commutative), the role of boundary conditions and
signal extensions, the connections between linear transforms and
linear finite Gauss-Markov fields, and several other concepts and
connections. Finally, the algebraic theory is a means to discover,
concisely derive, explain, and classify fast transform algorithms,
which is the subject of a future paper.
\end{abstract}


\tableofcontents

\ \bigskip\ 

\section{Introduction}\label{intro}

The paper presents an algebraic theory of signal processing that
provides a new interpretation to linear signal processing, extending
the existing theory in several directions. Linear signal processing is
built around signals, filters, $z$-transform, spectrum, Fourier
transform, as well as several other fundamental concepts; it is a
well-developed theory for continuous and discrete time. In linear
signal processing, signals are modeled as elements of vector spaces
over some basefield, e.g, the real or complex field, and filters
operate as linear mappings on the vector spaces of signals.

The assumption of linearity has made the theory of vector spaces, or
linear algebra, the predominant mathematical discipline in linear
signal processing. This paper proposes that the basic structure in
linear signal processing actually goes beyond vector spaces and linear
algebra.  The algebraic theory that we describe will show that this
structure is better exploited by the {\em representation theory of
algebras}, which is a well established branch of {\em algebra}, the
theory of groups, rings, and fields.\footnote{The word algebra
describes the discipline as well as an algebraic structure (namely a
vector space that is also a ring, to be defined later).}  
We will show that by appropriate choices for the space of signals, the
space of filters, and the filtering operation---what we call the
\emph{signal model}---the algebraic theory captures within the same
general framework many important instantiations of linear signal
processing, namely: linear signal processing for infinite and finite
discrete time; linear signal processing for infinite or finite
discrete ``space;'' and linear signal processing for higher order
linear models, e.g., separable and non-separable linear signal
processing on infinite and finite lattices in two or more dimensions.
To get a better understanding and appreciation for what we mean, we
expand on some of these examples in the next subsection.

\mypar{Remark} This paper focuses on \emph{discrete} parameter (time
or space) finite or infinite \emph{linear} signal processing, which for
the sake of brevity will simply be referred to as signal processing
and abbreviated by~SP. Much of the paper extends to continuous
parameter~SP, but this will not be considered here.

\begin{table*}\centering
\ra{1.5}
\caption{1-D discrete infinite and finite time and space signal
processing as four instantiations of the general algebraic theory. The
bolded concepts are supplied by the algebraic theory.\label{fourmodels}}
\begin{tabular}{@{}l@{\ }l@{\hspace*{10mm}}llll@{}}\toprule
\multicolumn{2}{@{}l}{generic theory}
  & infinite time & finite time & infinite space &
  finite space \\ \midrule
$\Phi$ & = ``$z$-transform'' & $z$-transform & 
  \bf finite \boldmath$z$-transform(s) &
  \bf \boldmath$C$-transform(s) & \bf finite \boldmath$C$-transform(s) \\
$\alg$ & = algebra of filters & 
  series in $z^{-n}$ & polynomials in $z^{-n}$ &
  \bf series in \boldmath$T_n$ & \bf polynomials in \boldmath$T_n$ \\
$\md$ & = $\alg$-module of signals & series in $z^{-n}$ & 
  polynomials in $z^{-n}$ &
  \bf series in \boldmath$C_n$ & \bf polynomials in \boldmath$C_n$ \\
$\four$ & = Fourier transform & DTFT & DFTs & \bf DSFTs & DCTs/DSTs\\ 
\bottomrule
\end{tabular}
\end{table*}

\subsection{Overview}

\mypar{The basic idea: Signal models} Consider Table~\ref{fourmodels},
ignoring for the time being the bold-faced entries and focusing first
on the second and third columns labeled infinite time and finite time.
Rows~2 to~5 represent the four basic concepts in~SP: the
$z$-transform, filters, signals, and the Fourier transform. Column~2
recalls that, for infinite discrete time~SP, we have the well defined
concept of $z$-transform and that signals and filters (in their
$z$-transform representation) are described by power series in the
variable~$z^{-1}$.  Filtering becomes \emph{multiplication} of series,
and the associated Fourier transform is the well known DTFT (discrete
time Fourier transform).

When only a finite number~$N$ of samples is available, we are in the
domain of discrete \emph{finite} time~SP, which is considered in the
third column in the table.  The Fourier transform is the well known
$\DFT_N$ (discrete Fourier transform).  However, attempting to extend
the infinite time case, column~2, to the finite time case, column~3,
by simply truncating the $z$-transform to obtain polynomials as
signals and filters leads to problems. Namely, if signals and filters
are polynomials $S(z^{-1})$ and $H(z^{-1})$ of degree $N-1$, then
their product is in general of higher degree. In other words, the
space of signals (polynomials of degree $N-1$) is not closed under
this notion of filtering. The solution, which is well-known (e.g.,
\cite{Nussbaumer:82}), casts filtering, as multiplication modulo
$z^{-N}-1$,
\begin{equation}\label{circconv}
H(z^{-1})S(z^{-1})\text{ mod }(z^{-N}-1).
\end{equation}
Correspondingly, signals and filters are now in the space of
``polynomials in $z^{-1}$ modulo $(z^{-N}-1)$,'' which is denoted by
$\C[z^{-1}]/(z^{-N}-1)$ and is called a {\em polynomial algebra}.  The
definition of a finite $z$-transform, not found in the literature, is
now straightforward; it is bold-faced in Table~\ref{fourmodels} and is
provided by the algebraic theory. Filtering as described in
\eqref{circconv} is equivalent to the well known circular convolution.

Besides the DFT, there are numerous other transforms available for
finite signals, for example, the discrete cosine and sine transforms
(DCTs and DSTs) considered in the fifth column in
Table~\ref{fourmodels}. They have been successfully used in image
processing, so, intuitively, we refer to them as associated to
``space'' signals, in contradistinction to time signals.  Note that
``space'' here is one-dimensional~(1-D), not necessarily 2-D, since
the DCTs/DSTs are 1-D transforms. We will discuss in more detail below
what we mean by space. More importantly, we go back to
Table~\ref{fourmodels} and ask for the DCTs and DSTs: What are their
analogues of the finite $z$-transform, signals, and filters?
Likewise, since the DCTs/DSTs are {\em finite} transforms, we consider
the corresponding {\em infinite} ``space'' analogues in column~4,
again asking what are the appropriate notions of $z$-transform,
signals, filters, and Fourier transform for infinite space signals.

The algebraic theory presented in the paper identifies the basic
structure to answer these questions as hinted at in
Table~\ref{fourmodels} and provides the bold-faced entries, which will
be defined in subsequent sections. In other words, the algebraic
approach leads to a single theory that instantiates itself to the four
right columns in Table~\ref{fourmodels} {\em and}, furthermore, to
several other columns corresponding to existing or new ways of doing
time and space SP.

\begin{figure}\centering
\isdraft{\renewcommand{\baselinestretch}{1}\small}{}
\begin{picture}(230,90)
\put(30,0){\framebox(40,30){\begin{tabular}{c}vector\\space\end{tabular}}}
\thicklines
\put(70,15){\makebox(70,0){%
  \begin{tabular}{c}\bf bijective\\\bf mapping\end{tabular}}}
\put(70,15){\vector(1,0){70}}
\put(140,0){\framebox(40,30){\bf module}}
\put(160,60){\vector(0,-1){30}}
\put(140,60){\framebox(40,30){\bf algebra}}
\thinlines
\put(165,30){\makebox(0,30)[l]{operates on}}
\put(220,0){\makebox(0,30){\it signals}}
\put(220,60){\makebox(0,30){\it filters}}
%
\end{picture}
\isdraft{\renewcommand{\baselinestretch}{1.5}}{}
\caption{The central concept in the algebraic theory of signal
processing is the {\em signal model}, which is a triple of an algebra,
an associated module, and a bijective linear mapping (all bolded) from
a vector space of signal samples into the module. Defining this
mapping fixes the notion of filtering.\label{modelfig}}
\end{figure}
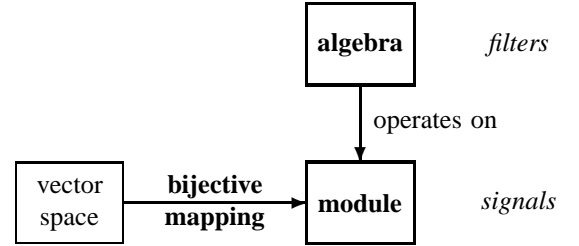

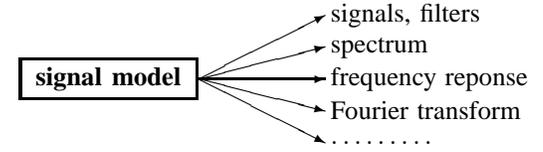
\begin{figure}\centering
\begin{picture}(170,60)
\thicklines \put(0,17){\framebox(66,14){\bf signal model}} \thinlines
\put(67,24){\vector(2,1){48}}
\put(67,24){\vector(4,1){48}}
\put(67,24){\vector(4,0){48}}
\put(67,24){\vector(4,-1){48}}
\put(67,24){\vector(2,-1){48}}
\put(117,0){\makebox(0,0)[l]{$\ldots\ldots\ldots$}}
\put(117,12){\makebox(0,0)[l]{Fourier transform}}
\put(117,24){\makebox(0,0)[l]{frequency reponse}}
\put(117,36){\makebox(0,0)[l]{spectrum}}
\put(117,48){\makebox(0,0)[l]{signals, filters}}
\end{picture}
\caption{Defining the signal model fixes
well-defined notions of signals, filters, spectrum, frequency response,
Fourier transform, and the other ingredients for~SP.\label{smdefine}}
\end{figure}

Central in the algebraic theory of~SP is the concept of the {\em signal
model}. It is defined as a triple $(\alg, \md, \Phi)$ (see the first
column in Table~\ref{fourmodels}): 
\begin{itemize}
\item $\alg$ is the chosen {\em algebra}
of filters, i.e., a vector space where multiplication (of filters) 
is also defined.
\item $\md$ is an {\em
$\alg$-module} of signals, i.e., a vector space whose elements can be
multiplied by elements of~$\alg$. We say that~$\alg$ operates
on~$\md$ through filtering.
\item $\Phi$ generalizes the $z$-transform. It is defined as a
bijective mapping from a vector space~$V$ of signal samples
into the module $\md$ of signals (see Figure~\ref{modelfig}).
\end{itemize}
The vector space~$V$ is a product space of countable or finite many
copies of, say, the real numbers~$\R$ or the complex numbers~$\C$, so
that elements of~$V$ are series or vectors of signal samples. The
purpose of the bijective mapping $\Phi$ is to assign a choice of
filtering, which is given by the operation (multiplication) of $\alg$
on $\md$.  We will show that once we fix a signal model, application
of the well-developed representation theory of algebras provides a
systematic methodology to derive the main ingredients in SP, such as
the notions of spectrum, Fourier transform, and frequency response
(see Figure~\ref{smdefine}).

To be specific, we consider briefly two illustrations of the signal
model for finite SP. 

For finite time, the signal model is given by (we
set $x = z^{-1}$ for simplicity) $\alg = \md = \C[x]/(x^{N}-1)$ and
$$
\Phi:\ \coord{s}\mapsto\sum_{0\leq n<N}s_nx^{n}\in\md,
$$ for $\coord{s} = (s_0,\dots,s_{N-1})\in V=\C^n$ (assuming complex
valued signals). This $\Phi$ is the finite $z$-transform indicated in
Table~\ref{fourmodels}. As we saw in this table, the Fourier transform
associated with this signal model is the DFT.

We now consider a second signal model, a finite space model, given by
a different polynomial algebra: $\alg = \md = \C[x]/T_N(x)$, and
$$
\Phi:\ \coord{s}\mapsto\sum_{0\leq n<N}s_nT_{n}(x)\in\md,
$$ 
where $T_n$ are Chebyshev polynomials of the first kind. We will
show that the corresponding Fourier transform for this model is the
DCT, type~3.

\mypar{Derivation of signal models: Shift and boundary conditions}
These two examples illustrate how we can fill Table~\ref{fourmodels}
to obtain a consistent set of SP~concepts based on the concept of a
signal model. We address now the question of which algebras and
modules are implicitly assumed in common instantiations of~SP and
why. This question is important because it will lead to a method to
develop signal models beyond the ones shown in Table~\ref{fourmodels}.

A first high-level answer to this question is provided by the
algebraic theory: Every signal model $(\alg,\md,\Phi)$ that has the
\emph{shift invariance} property has necessarily a commutative
$\alg$. If the model is in addition finite, then $\alg$ has to be a
polynomial algebra. As examples we saw above the models for the DFT
and DCT, type~3.  To obtain a more detailed answer to the question
which signal models occur in SP, we explain how to derive signal
models from basic principles, namely from a chosen definition of the
{\em shift operator}. The shift plays a fundamental role in many areas
including ergodic theory, random processes and statistics, dynamical
systems, and information theory. As an abstract concept, once there is
a group structure, e.g., \cite{Gray:88}, shifts can be
defined for time or space, or in multiple dimensions. In the algebraic
theory, the shift has a particularly simple interpretation: it is the
{\em generator of the filter algebra}. Specifically, we describe a
procedure that, starting from the definition of the shift, produces
infinite and finite signal models, and that reveals the degrees of
freedom that are available in this construction (see
Figure~\ref{shiftsm}). This procedure provides two important insights:
\begin{enumerate}
\item How to derive signal models based on shifts other than the standard
time shift; and 
\item The role of boundary conditions and signal
extensions in finite signal models.
\end{enumerate}

\begin{figure}\centering
\begin{picture}(110,20)
\put(-5,0){\makebox(0,14){definition of shift(s)}}
\put(38,7){\vector(1,0){34}}
\thicklines
\put(72,0){\framebox(66,14){\bf signal model}}
\thinlines
\end{picture}
\caption{A signal model can be derived from the definition of the
shift operation in two steps that we call {\em linear extension} and
{\em realization}. Using this procedure, we can derive signal models
that match our intuition by properly defining the shift.
Many shifts besides the standard time shift are possible.\label{shiftsm}}
\end{figure}

For example, regarding 1), when starting with an abstract definition 
of the standard 1-D time shift ($q$ is the shift
operator, $\diamond$ the shift operation, $t_n$ are discrete time
points)
\begin{equation}\label{tshift}
\text{\bf time:}\quad q\diamond t_n = t_{n+1},
\end{equation}
we obtain the well-known time models with the associated infinite and
finite $z$-transform. The very same procedure, when applied to a {\em
different definition of the shift}, namely to what we call the 1-D
{\em space shift},
\begin{equation}\label{sshift}
\text{\bf space:}\quad q\diamond t_n = \tfrac 12(t_{n-1} + t_{n+1}),
\end{equation}
leads to the infinite and finite $C$-transform (that we will define)
and, in the finite case, to the DCTs and DSTs (see
Table~\ref{fourmodels}). In other words, different shifts lead to
different signal models with different associated Fourier transforms;
in particular, the DCTs or DSTs are Fourier transforms in this
sense.

\begin{figure*}\centering
\subfigure[Two 2-D time shifts (separable)]{
\begin{picture}(90,50)
\put(20,10){$\bullet$}
\put(40,10){\makebox(0,0)[b]{$\xy\ar (10,0)\endxy$}}
\put(56,10){$\bullet$}
\put(20,46){$\bullet$}
\put(22.5,13){\makebox(0,0)[b]{$\xy\ar@{.>} (0,10)\endxy$}}
\end{picture}\qquad}
\quad
\subfigure[Two 2-D space shifts (separable)]{
\begin{picture}(100,60)
\put(5,30){$\bullet$}
\put(23,30){\makebox(0,0)[b]{$\xy\ar (-8,0)\endxy$}}
\put(36,30){$\bullet$}
\put(53,30){\makebox(0,0)[b]{$\xy\ar (8,0)\endxy$}}
\put(66,30){$\bullet$}
\put(36,62){$\bullet$}
\put(38,35){\makebox(0,0)[rb]{$\xy\ar@{.>} (0,8)\endxy$}}
\put(36,-2){$\bullet$}
\put(38,27){\makebox(0,0)[rt]{$\xy\ar@{.>} (0,-8)\endxy$}}
\end{picture}\qquad}
\quad
\subfigure[Two 2-D hexagonal space shifts (non-separable)]{
\begin{picture}(90,30)
\put(0,30){$\bullet$}
\put(18,59){$\bullet$}
\put(18,1){$\bullet$}
\put(20,30){\makebox(0,0)[b]{$\xy\ar@{.>} (-10,0)\endxy$}}
\put(22,33){\makebox(0,0)[lb]{$\xy\ar (-5,8)\endxy$}}
\put(22,30){\makebox(0,0)[lt]{$\xy\ar (-5,-8)\endxy$}}
\put(36,30){$\bullet$}
\put(56,30){\makebox(0,0)[b]{$\xy\ar (10,0)\endxy$}}
\put(54,59){$\bullet$}
\put(54,1){$\bullet$}
\put(72,30){$\bullet$}
\put(40,34){\makebox(0,0)[lb]{$\xy\ar@{.>} (5,8)\endxy$}}
\put(40,29){\makebox(0,0)[lt]{$\xy\ar@{.>} (5,-8)\endxy$}}
\end{picture}\qquad}
\quad
\subfigure[Two 2-D quincunx space shifts (non-separable)]{
\begin{picture}(110,30)
\put(0,30){$\bullet$}
\put(46,30){\makebox(0,0)[rb]{$\xy\ar@{.>} (-14,0)\endxy$}}
\put(50,30){$\bullet$}
\put(58,30){\makebox(0,0)[lb]{$\xy\ar@{.>} (14,0)\endxy$}}
\put(100,30){$\bullet$}
\put(25,55){$\bullet$}
\put(75,55){$\bullet$}
\put(50,33){\makebox(0,0)[rb]{$\xy\ar (-7,7)\endxy$}}
\put(55,33){\makebox(0,0)[lb]{$\xy\ar (7,7)\endxy$}}
\put(25,5){$\bullet$}
\put(75,5){$\bullet$}
\put(50,30){\makebox(0,0)[rt]{$\xy\ar (-7,-7)\endxy$}}
\put(55,30){\makebox(0,0)[lt]{$\xy\ar (7,-7)\endxy$}}
\end{picture}}
\caption{In 2-D signal models are derived from 2 shifts.  Examples
are shown here; the operation of the two shifts in each case is
represented by solid and dotted arrows, respectively (scaling factors
are omitted). The first two choices of shifts lead to separable
models, the others do not.\label{2dshifts}}
\end{figure*}
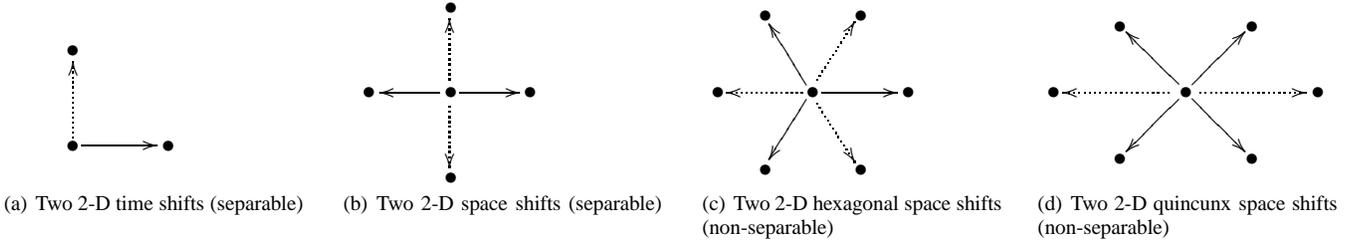

Other shifts are possible (see Figure~\ref{1dshifts}), and our
methodology produces the corresponding signal model and thus the
appropriate notion of filtering (or convolution), spectrum, and
Fourier transform for each of them. The method is the same for higher
dimensional signal models. For example, in 2-D, two shifts have to be
considered; possible choices are shown in Figure~\ref{2dshifts}.  The
first two lead to the known separable 2-D time and 2-D space models,
whereas the remaining two choices produce novel 2-D signal models (and
thus associated notions of $z$-transform, filtering, and Fourier
transform) for the finite spatial hexagonal and quincunx lattice
respectively, \cite{Pueschel:04a,Pueschel:05a}. It turns out that the
signal models arise again from polynomial algebras (and thus
they are shift invariant), but in two variables in this case.

\begin{figure}\centering
\subfigure[1-D time shift]{
\qquad
\begin{picture}(80,25)
\put(0,10){$\bullet$}
\put(0,0){$t_{n}$}
\put(20,10){\makebox(0,0)[b]{$\xy\ar (10,0)\endxy$}}
\put(36,10){$\bullet$}
\put(36,0){$t_{n+1}$}
\end{picture}}
\qquad
\subfigure[1-D space shift]{
\begin{picture}(90,25)
\put(0,10){$\bullet$}
\put(0,0){$t_{n-1}$}
\put(20,10){\makebox(0,0)[b]{$\xy\ar (-10,0)_{\frac{1}{2}}\endxy$}}
\put(36,10){$\bullet$}
\put(36,0){$t_{n}$}
\put(56,10){\makebox(0,0)[b]{$\xy\ar (10,0)^{\frac{1}{2}}\endxy$}}
\put(72,10){$\bullet$}
\put(72,0){$t_{n+1}$}
\end{picture}}
\subfigure[1-D generic next neighbor shift]{
\qquad
\begin{picture}(120,50)
\put(0,10){$\bullet$}
\put(0,0){$t_{n-1}$}
\put(20,10){\makebox(0,0)[b]{$\xy\ar (-10,0)_{a_n}\endxy$}}
\put(36,10){$\bullet$}
\put(38.5,30){\makebox(0,0){$\xy\ar @(ul,ur)^{b_n}\endxy$}}
\put(36,0){$t_{n}$}
\put(56,10){\makebox(0,0)[b]{$\xy\ar (10,0)^{c_n}\endxy$}}
\put(72,10){$\bullet$}
\put(72,0){$t_{n+1}$}
\end{picture}}
\caption{Examples of 1-D shifts considered by the algebraic
theory.\label{1dshifts}}
\end{figure}
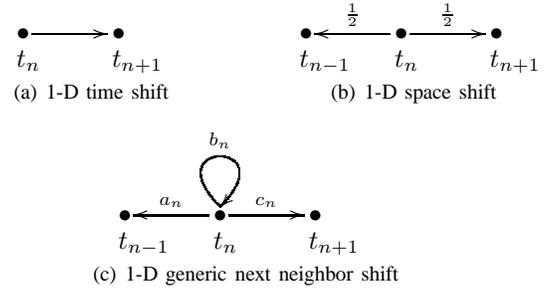

Regarding 2), the role of boundary conditions and signal extension,
our signal model derivation explains why they are unavoidable (under
certain assumptions) and what the choices are. For example, why is the
periodic extension the usual choice for finite time, and why is the
symmetric or antisymmetric extension an appropriate choice for the
DCTs and DSTs? This insight is very relevant when deriving the novel
2-D models mentioned above. Also, it is interesting to note that for
example in finite time other signal extensions besides periodic are
possible, which then produce different signal models and thus a
different associated ``DFT.''

\mypar{Fast algorithms} One important application of the algebraic
theory is in the discovery, derivation, and classification of fast
transform algorithms. There are many different transforms used in
signal processing (e.g., DFT, DCTs/DSTs, discrete Hartley transforms,
and variants thereof) and because of their importance there are
hundreds of publications on their fast algorithms. Most of these
algorithms are derived by ingenious manipulation of the transform
coefficients. These derivations, however, are usually tedious and
provide no insight into the structure nor the existence of these
algorithms. Further, it is not clear whether important algorithms may
not have been found. The exception is the DFT, for which the theory of
algorithms is well-understood due to early works like
\cite{Nicholson:71,Winograd:78,Auslander:84,Beth:84} and others. As a
result, very accessible standard books on DFT algorithms are now
available for application developers
\cite{Nussbaumer:82,VanLoan:92,Tolimieri:97,Tolimieri:97a}. We will
show that by extending ideas from the work on DFTs, the theory of
algorithms becomes a natural part of the algebraic theory of signal
processing.

The basic idea is to derive algorithms from the signal model
underlying a transform rather than from the transform itself.  We
briefly sketch how it works in a simple case. We consider a signal
model with $\alg = \md = \C[x]/p(x)$. To derive algorithms for the
associated Fourier transform $\four$, we first state what $\four$
actually does in this case. Namely, $\four$ decomposes the signal
module into its irreducible components, called its spectrum. This is
akin to decomposing vector spaces into invariant subspaces with
respect to some linear mapping. In the case of a polynomial algebra
this decomposition is an instantiation of the {\em Chinese remainder
theorem (CRT)} and looks as follows:
\begin{multline}\label{fourexample}
\four:\ \C[x]/p(x)\rightarrow\isdraft{}{\\}
  \C[x]/(x-\alpha_0)\dirsum\dots\dirsum\C[x]/(x-\alpha_{N-1}).
\end{multline}
Here $\deg(p) = N$ and the $\alpha_n$ are the zeros of $p$, assumed to
be distinct. The important point is that each of the summands on the
right side has dimension 1, i.e., $\C[x]/p(x)$ is fully decomposed.
Intuitively, algorithms are now derived by performing this
decomposition in steps. This is possible for example, if $p(x) =
q(r(x))$ decomposes (note that this is different from
factorization). In this case, we can perform \eqref{fourexample} in
two steps using again the CRT, namely
\begin{eqnarray*}
\C[x]/p(x) & \rightarrow & 
  \C[x]/(r(x)-\beta_0)\dirsum\dots\dirsum\C[x]/(r(x)-\beta_{K-1})\\
& \rightarrow & 
  \C[x]/(x-\alpha_0)\dirsum\dots\dirsum\C[x]/(x-\alpha_{N-1})
\end{eqnarray*}
Here, the $\beta_k$ are the zeros of $q(x)$ and $\deg(q) = K$.  A
general theorem that we already showed in \cite{Pueschel:03a}
provides an algorithm for $\four$ in this case. 

For the DFT, $p(x) = x^N - 1 = (x^M)^K - 1$ indeed decomposes if $N =
KM$ and the method yields the famous Cooley-Tukey FFT. For the DCT,
type 3, $p(x) = T_N(x) = T_K(T_M(x))$ and we obtain algorithms of a
structure very similar to the Cooley-Tukey FFT \cite{Pueschel:03}.
Interestingly, most of them are novel. Using the same method we can
derive many known and novel algorithms for all 16 DCTs/DSTs
\cite{Pueschel:05} and for other transforms in forthcoming papers.

The idea of decomposition generalizes to higher dimensions. For
example, reference \cite{Pueschel:04b} derives a Cooley-Tukey type
algorithm for the new discrete triangle transform for finite spatial
hexagonal lattices introduced in \cite{Pueschel:04a}. This transform
is a Fourier transform for a finite 2-D signal model based on the
definition of shifts in Figure~\ref{2dshifts}(c).

Using the algebraic approach we will also generalize the well-known
prime-factor FFT and Rader FFT, by identifying the algebraic
principles they are based on. We already showed first ideas along
these lines in~\cite{Pueschel:03a}.

\mypar{Algebraic theory: Forward and inverse problems} In this paper,
we apply the algebraic theory to address, in a sense, a forward 
problem and an inverse problem:
\begin{inparaenum}[(1)]\item 
\emph{Forward problem}: Derive a signal model $\left(\alg,\md,
\Phi\right)$ for a given application from basic principles, and find
the appropriate SP concepts like filtering, convolution, spectrum,
signal extension, Fourier transform and its fast algorithms, and
others;
\item \emph{Inverse problem}: Given a linear transform (say the DCT,
type~3), find the corresponding signal model, for which the transform
is a Fourier transform. From the signal model, we can then derive all
SP concepts mentioned in the forward problem above, including fast
algorithms for that transform.
\end{inparaenum}

The algebraic theory enables the solution of both problems. In the
paper we focus on \emph{linear}, \emph{shift invariant} signal
processing since this is a simple context where the algebraic approach
can provide immediate meaningful results. Also, rather than assuming
more general constructions, e.g., discrete spaces, metric spaces, or
Polish spaces, we restrict ourselves to scalar signals that are
rational, real, or complex valued---this more restrictive approach
still applies to many relevant linear transforms and signal models as
we consider here.

\mypar{Accessibility of the algebraic theory} Algebra is not among the
mathematical disciplines commonly taught or used in SP. However, the
major parts of the algebraic theory of SP can be developed from
working knowledge with series and polynomials and from basic linear
algebra techniques, each of which is common knowledge in SP.  We will
introduce several algebraic concepts; however, they describe existing
concepts in SP. For example, referring to the space of filters and
signals as an algebra and a module, respectively, does not impose new
structure; rather, it makes explicit the structure commonly adopted.

\mypar{Summary: Scope of the algebraic theory} In summary, the scope
of the algebraic theory of SP as we present it and plan to further
develop it can be visualized as an expansion of
Table~\ref{fourmodels}.  

First, we develop the general theory (first column) and then we apply
or instantiate this theory to expand the table with additional columns
by deriving relevant signal models. We start with 1-D SP and fill in
the models for most of the existing spectral transforms\footnote{The
impatient reader may want to check Table~\ref{overviewsm} for the
result.}, which include practically all known trigonometric
transforms. Future papers will then further expand the table through
separable and non-separable SP in higher dimensions.

Second, we expand Table~\ref{fourmodels} along the rows, for the
generic theory and for all signal models introduced.  We start again
with the most basic concepts such as spectrum, frequency response,
Fourier transform, diagonalization properties, and convolution
theorems.  Forthcoming papers will then develop the algebraic theory
of fast transform algorithms, subsampling, uncertainty relations,
filterbanks, multiresolution analysis, frame theory, and other
important concepts in~SP.

\subsection{Background: Algebra in Signal Processing}

In this section, we review prior and related work using algebraic
techniques in signal processing, and discuss the particular thread of
research that led to the work in this paper.  Existing work is mostly
focused on the derivation of algorithms for the DFT (or a few other
transforms) or on the construction of very specialized signal
processing schemes, such as Fourier analysis on groups.  This
restriction to specialized applications has put algebra and
representation theory essentially in the ``blind spot'' of signal
processing theory. The situation is different for the related
field of system theory, which we discuss first.

\mypar{Algebraic system theory} This paper describes an algebraic
view to explain the mathematical structure underlying signal
processing. Kalman in his seminal work on linear system theory (see
chapter~10 in his co-authored book~\cite{Kalman:69} from 1968) went
beyond vector spaces. His concern was to show that ``the entire theory
of the regulator problem $\ldots$ depends on the algebraic properties
of the [system] matrices $F$, $G$, and $H$ satisfying these two
conditions.''  The conditions Kalman refers to are the controllability
and observability conditions
\begin{eqnarray}
\label{controlability}
\mbox{rank}\left[G,FG,\cdots,F^{n-1}G\right]&=&n\\
\label{observability}
\mbox{rank}\left[H^T,F^TH^T,\cdots,{(F^T)}^{n-1}H^T\right]&=&n,
\end{eqnarray}
where $n$ is the dimension of the state space $\left(F: n\times n, G:
n\times m, H: l\times n\right)$. He interpreted~\eqref{controlability}
``algebraically by viewing it as a condition for generating a module
over polynomials in the matrix~$F$.''

Kalman used the algebraic approach to study the realization problem in
linear systems.  This is an inverse problem: How to go from a suitable
rational function that is the \emph{external} system description to
the state variable model, the matrices $F$, $G$, and $H$ above, that
is the \emph{internal} system description. Important concepts, like
invariance and invariant subspaces, more appropriately dealt with in
the framework of algebras and modules, play an important role in
realization theory and are common staple of linear systems and control
theory, starting with the early work of Basile and
Marro~\cite{Basile:69} and Wonham and Morse~\cite{Wonham:70} (see also
the critical paper by Willems and Mitter~\cite{Willems:71}). The
unpublished work in the PhD thesis of Johnston \cite{Johnston:73} and
Fuhrman \cite{Fuhrman:76} and his subsequent work, for example
\cite[ch.~10]{Fuhrman:96}, as well as others, have further developed
the algebraic theory of linear systems with emphasis on the system
realization problem.  Extensive work by Fuhrman focuses again on
linear time invariant systems because their algebraic properties can
provide significant insight. Many more references and authors have
explored these ideas in linear systems and related areas; it is not
our intention nor can we even provide here a fair coverage of this
literature, so we will not discuss this topic further and refer the
readers to the relevant literature.
 
In parallel with Kalman's and Fuhrman's module perspective on linear
systems, this paper explores the theory of algebras and modules, i.e.,
the representation theory of algebras, as a basic framework for
digital linear signal processing. In realization theory and in the
algebraic linear system theory the shift operator, its matrix
representations, and its irreducible components play a central
role. Likewise, in the direct and inverse SP problems that we are
interested in (see above), the shift operator and the decomposition of
modules in irreducibles play a very important role in identifying the
spectrum, the linear transform, and its fast algorithms associated
with a linear signal model.

\mypar{Algebraic methods for DFT algorithms} The advent of digital
signal processing is often attributed to the rediscovery of the fast
Fourier transform (FFT) by Cooley and Tukey in 1965
\cite{Cooley:65,Heideman:85}. In the following years, the recognition
of the importance of fast algorithms for the DFT led to the
first---and to date arguably most important---application of algebraic
methods in mainstream signal processing. Namely, it was already known
in the 19th century that the DFT can be described in the framework of
the representation theory of groups and, more specifically, can be
related to the cyclic group.  This connection was used to derive and
explain existing FFT algorithms including the Cooley-Tukey FFT
\cite{Apple:70,Cairns:71,Nicholson:71,Beth:84,Auslander:84}, but is
also the foundation of Winograd's seminal work on the multiplicative
complexity of bilinear forms in general and the DFT in
particular. This work provided an entirely new class of DFT algorithms
that, surprisingly and among other things, showed that the DFT can be
computed with only a linear number of (non-rational) multiplications
\cite{Winograd:78,Winograd:79,Winograd:80,Auslander:84a}.

\mypar{Fourier analysis and fast Fourier transforms on groups} The
connection between the DFT and the cyclic group made it natural to
explore the applicability of other ``group Fourier transforms'' in
signal processing. The general topic of Fourier analysis on groups
dates back to the early days (19th century) of the representation
theory of groups with major contributions by Gauss, Frobenius,
Burnside, Brauer, and Schur. For the infinite (additive) cyclic group
of integers $\Z$, the area of Fourier analysis is equivalent to the
theory of Fourier series, which is standard in functional analysis
\cite{Edwards:67,Edwards:67a} and in discrete-time signal
processing. Generalizations to other infinite, commutative groups have
also been extensively studied (e.g., \cite{Rudin:62}). In signal
processing, general commutative finite groups were considered already
in \cite{Apple:70} (including fast algorithms). The first proposition
of non-commutative groups is due to Karpovsky
\cite{Karpovsky:77}. This development raised the question of fast
algorithms for these transforms, starting a new area with the
pioneering work of Beth \cite{Beth:84,Beth:87}. The field was further
extended by Clausen \cite{Clausen:88,Clausen:89,Clausen:93}, and by
the large body of work by Rockmore et al., which shaped the field as
it stands today; examples include
\cite{Maslen:95,Rockmore:90,Diaconis:90,Maslen:00}.

The infinite cyclic group $\Z$ and the finite cyclic group lead to
discrete-time signal processing and finite time signal processing with
periodic boundary conditions, respectively. Every finite commutative
group is a direct product of cyclic groups, and can thus be viewed as
a multi-dimensional torus, which leads to separable multi-dimensional
finite signal processing. Beyond that, Fourier analysis on
non-commutative finite groups has found next to no applications in
signal processing.  There are a few notable exceptions. The work by
Diaconis \cite{Diaconis:89,Diaconis:88} identifies the symmetric group
as the proper structure to analyze ranked statistical data.  Driscoll
and Healy develop Fourier analysis for signals given on the 2-sphere
(the surface of a three-dimensional ball) \cite{Driscoll:94}.  More
recently, Foote et al.~propose groups that are wreath products for
signal processing \cite{Foote:00,Mirchandini:00}. These groups offer a
structure that naturally provides a multi-resolution scheme for finite
signals. Intriguingly, these wreath product group transforms
generalize the well-known Haar transform that is different from the
one in standard wavelet theory. We want to mention that groups do play
a role in standard wavelet analysis but in a sense different from the
work above \cite{Resnikoff:98}. Finally, and somewhat unrelated, we
want to mention the work by Shokrollahi et al.~\cite{Shokrollahi:01},
which provides a striking application of the representation theory of
groups in multiple-antenna signal processing.

\mypar{Background of this paper} The particular thread of research
that led to the present paper can also be traced back to the work of
Beth on fast Fourier transforms for groups~\cite{Beth:84} and to the
quest of a general theory of fast transform algorithms.  While group
theory provides a set of transforms and (in many cases) their fast
algorithms, many of the transforms used in signal processing, such as
the DCTs and DSTs, were not captured in this framework. In the search
for the algebraic properties of these transforms, Minkwitz, in his
PhD.~work, relaxed the idea of signals on groups to signals on sets on
which groups act via permutations (similar to \cite{Foote:00}
mentioned above) and found that indeed some of the DCTs could be
described as generalized group Fourier transforms this
way. Furthermore, he showed that, in these cases, fast algorithms for
these transforms can also be constructed by pure algebraic means
\cite{Minkwitz:93,Minkwitz:95}. Minkwitz' work was further extended by
Egner and P{\"u}schel in their PhD.~work including an automatic method
to analyze a given transform for group properties, and, in the
affirmative case, to automatically construct a fast algorithm
\cite{Egner:01,Egner:04,Pueschel:02}. Application to various signal
transform showed that several, but not all transforms, could be
characterized this way \cite{Egner:01}. Further, among the many
existing DCT/DST algorithms, only few could be derived and explained
this way. The conclusion was: if the DCTs/DSTs had a defining
algebraic property, it had to be outside the group framework. This
paper addresses precisely this issue and show that the DCTs and DSTs
can all be characterized in the framework of {\em polynomial} algebras
instead of {\em group} algebras.  Valuable hints in the search for
this structure were provided by \cite{Steidl:91,Moura:98,Strang:99}.
Using the polynomial algebras underlying the DCTs and DST, we showed
how to derive, explain, and classify most of the existing fast DCT/DST
algorithms \cite{Pueschel:03a} and we also derived \emph{new} fast
algorithms, not available in the literature or found with previous
methods \cite{Pueschel:03,Pueschel:05}.

\subsection{Organization}

This paper is divided into three main parts:
\begin{itemize}
\item Algebra and signal processing;
\item Discrete infinite and finite signal models and trigonometric
transforms; and
\item Algebraic signal models, graphs, Markov chains,
and Gauss-Markov random fields or processes.
\end{itemize}

\mypar{Algebra and signal processing}
The first part consists of Sections~\ref{asm}--\ref{wherearewe}. In Section~\ref{asm}, we introduce
 background on algebras and modules and establish their
connection to signal processing.  We define the concept of signal
model and explain the algebraic interpretation of the shift and
shift-invariant signal models.  In the finite case, these models
correspond to polynomial algebras, for which the signal processing is
developed in Section~\ref{polyalgs}. Finally, Section~\ref{wherearewe}
provides a short summary of the first part of the paper and connects
to the second part.

\mypar{Discrete infinite and finite signal models and trigonometric
transforms}
The second part of the paper consists of
Sections~\ref{ztrafo}--\ref{higherdim}. In each section (except the
last two) we derive an infinite or finite signal model from a
definition of the shift, following the same high-level steps. Thus,
these sections are organized very similarly, with subsections
corresponding to the derivation of the signal model, the derivation of
spectrum and Fourier transform, the model's visualization,
diagonalization properties of the Fourier transform, and, optionally,
convolution theorems, orthogonal Fourier transforms, and other
important properties of the model. Section~\ref{ow} gives an overview
of the finite signal models presented so far. Section~\ref{higherdim} concludes
this part with the algebraic theory of higher-dimensional signal
models.

\mypar{Algebraic signal models, graphs, Markov chains,
and Gauss-Markov random fields or processes}
The third part of the paper, consisting of Sections~\ref{mc} and
\ref{randomfield}, investigates the general connection between signal
models based on polynomial algebras, graphs, Markov chains, and
Gauss-Markov random fields. In particular, we show under which
conditions a random field is equivalent to a signal model, and thus
the concepts of Fourier transform and Karhunen-Lo\`eve transform
coincide.

Finally, we offer conclusions in Section~\ref{conclusions}.

In the appendix we provide some additional mathematical background that 
is used in this paper.


\section{Algebras, Modules, and Signal Models}\label{asm}

This section introduces the mathematical framework of the algebraic
theory of signal processing. As said in the introduction, by signal
processing, or SP, we mean linear signal processing.  We start with
relating algebras and modules to signal processing. Then, we introduce
modules and algebras more rigorously and establish the connection
between basic concepts in the representation theory of algebra (i.e.,
the theory of algebras and their modules) and SP.  Next, we formally
define the concept of a signal model, which is at the heart of the
algebraic theory, as a triple of an algebra, a module, and a bijective
map.  Instantiation of the signal model leads to different ways of
doing SP (as discussed in the context of
Table~\ref{fourmodels}). After that, we identify in the algebraic
theory the role of the shift(s) as the generator(s) of the filter
algebra and explain that shift-invariant signal models are precisely
those with a commutative filter algebra. Finally, we introduce the
notion of visualization of a signal model as a graph and introduce
module manipulation as a useful tool in working with signal models.

This section is mathematical by nature. We recommend that the reader
 consider it as a reference for the more concrete examples developed
 in the sequel.

\subsection{Motivation}

In SP (linear signal processing), the set of signals is considered to
be a {\em vector space}, like $\C^{\N}$, the set of one-sided
complex valued sequences. With vector spaces, signals can be added and
can be multiplied by a scalar $\alpha$ (from the base field), to yield
a new signal. Formally,
\begin{eqnarray*}
\text{signal}+\text{signal} & = & \text{signal},\\
\alpha\cdot\text{signal} & = & \text{signal}.
\end{eqnarray*}
The structure of a vector space gives access to the notions of
dimension, basis, linear mapping, and subspace. Because of our focus
on linear SP, and unless stated otherwise, we restrict the discussion
to vector spaces that are product spaces of the reals numbers $\R$ or
the complex numbers $\C$, i.e., $\R^I$ or $\C^I$, where the indexing
set $I$ is either countable or finite and the underlying field is
either~$\R$ or~$\C$.

From a mathematical point of view, the structure of vector spaces is
simple. Namely, any two vector spaces defined over the same field and
of the same dimension are isomorphic, i.e., structurally identical. As
we explain next, the signal models used in SP are actually algebraic
objects that have more structure than vector spaces.  Indeed, in SP,
signals interact with linear systems\footnote{We only consider
single-input single-output linear~(SISO) systems in this
paper. Extensions to multiple-input multiple-output~(MIMO) systems are
under research.}, commonly called {\em filters}.

This is represented in block diagram form by
\begin{equation}\label{filterblock}
\begin{picture}(140,10)
\put(10,0){\makebox(0,10){signal}}
\put(30,5){\vector(1,0){20}}
\put(50,-2){\framebox(30,14){filter}}
\put(80,5){\vector(1,0){20}}
\put(120,0){\makebox(0,10){signal}}
\end{picture}
\end{equation}

The operation of filters on signals imposes additional structure on
the signal space, namely, that of a module. This additional structure
casts linear signal processing in the framework of the representation
theory of algebras.  To recognize the additional structure, we first
denote the filter operation as multiplication~$\cdot$ and represent
\eqref{filterblock} as
$$
\text{filter }\cdot\text{ signal }=\text{ signal}.
$$
The multiplication is meant here in an abstract sense, i.e., it can
take different forms depending on the representation of filters and
signals, e.g., convolution (in the time domain) or standard
multiplication (in the $z$-transform domain) or any other adequate
form, as long as certain properties are satisfied, e.g., the
distributivity law:
$$
\begin{array}{rl}
& \text{filter }\cdot\ (\text{ signal }+\text{ signal })\\
= & \text{filter }\cdot\text{ signal }+\text{ filter }\cdot\text{ signal}.
\end{array}
$$ 
Furthermore, filters themselves can be combined to form new filters:
namely added, multiplied, and multiplied by a scalar $\alpha$ from the
base field, i.e.,
\begin{eqnarray*}
\text{filter}+\text{filter} & = & \text{filter}
\quad\text{(parallel connection),}\\
\text{filter}\cdot\text{filter} & = & \text{filter}
\quad\text{(series connection),}\\
\alpha\cdot\text{filter} & = & \text{filter}
\quad\text{(amplification).}
\end{eqnarray*}
Multiplication of two filters and the multiplication of
a filter and a signal, though written using the same symbol $\cdot$,
are conceptually different. 

Parallel connection and amplification, or, more generally, linear
combinations of filters makes the filter space (as the signal space) a
vector space. But, multiplication of filters or multiplication of a
signal by filters shows that there is more structure in the linear SP
that goes beyond vector spaces.  

Mathematically, the above structure is described by regarding the
filter space as an {\em algebra} $\alg$ that operates on the signal
vector space $\md$, thus making the signal space $\md$ an {\em
$\alg$-module}:

\begin{center}
\framebox[1.05\width]{
\begin{tabular}{r@{\quad\bf=\quad}l}
\bf set of filters/linear systems & \bf an algebra \boldmath$\alg$\\
\bf set of signals & \bf an \boldmath$\alg$-module $\md$
\end{tabular}
}
\end{center}

The signal module $\md$ as an $\alg$-module allows for
``multiplication,'' i.e., filtering, of an element of the module (the
signal) by an element of the algebra (the filter).  Given an algebra
$\alg$ and an associated $\alg$-module, the well-developed
mathematical theory of $\alg$-modules (or representation theory of
algebras) provides besides filtering access to a larger set of
concepts than linear algebra. Examples include the notions of
spectrum, irreducible representation (i.e., frequency response as we
explain later), and Fourier transform.

We remark that the structures of the signal space~$\md$ and the filter
space~$\alg$ are actually different. For example, signals can not be
multiplied, while filters can, and filters operate on signals, but not
vice-versa.

This paper addresses questions like: \begin{inparaenum}[(1)]
 \item How to connect existing signal processing concepts and theory 
with algebraic concepts and theory?
\item Which algebras and modules naturally occur in signal processing 
and why?
\item What benefits can we derive from this connection?
\end{inparaenum}

We already considered the first item by revealing the algebraic nature
of filters and of signals. Next section extends the connection between
algebra and signal processing to include concepts like spectrum,
frequency response, and Fourier transform. In Section~\ref{modsigmod}
we then introduce the definition of a signal model, which formalizes
the connection between algebras, modules, and linear SP.

To address the second item we will characterize at a very high level
those algebras that provide shift-invariant filters. In the finite
case, i.e., for finite-length signals, this will lead to polynomial
algebras. Later, we derive the algebras and modules associated with
infinite and finite discrete-time and infinite and finite
discrete-space SP. As mentioned in Section~\ref{intro}, the
distinction between time and space is not due to 1-D versus 2-D but
due to directed versus undirected in a sense that will be defined
rigorously later.  In the finite case these constructions lead
naturally to the discrete Fourier transform (DFT) and the discrete
cosine and sine transforms (DCTs/DSTs), respectively.  Further, we
will reveal the algebraic structure behind practically all known
trigonometric transforms and also extend this class by introducing new
transforms.

Regarding the third item, we mention that the algebraic theory
provides, as already mentioned, the common underpinning for many
different infinite and finite linear signal processing schemes,
showing, for example, that the spectral transforms arise as
instantiations of the same common theory. Further, the connection
between algebras/modules and signal processing goes in both
directions, namely, in the direct direction (see discussion in
Section~\ref{intro}) specifying an algebra $\alg$ and an $\alg$-module
$\md$ provides the ingredients to develop extensions to the existing
signal processing schemes. As examples, we briefly discussed in
Section~\ref{intro} non-separable 2-D SP on a finite quincunx or
hexagonal lattice.  Finally, also briefly mentioned in
Section~\ref{intro}, is the subject future papers
(e.g., \cite{Pueschel:05}), which extend the algebraic theory of signal
processing to the derivation and discovery of fast algorithms. The
algebraic theory makes the derivation of algorithms concise and
transparent, gives insight into the algorithms' structure, enables the
classification of the many existing algorithms, and enables the
discovery of \emph{new} algorithms for existing transforms and for new
linear transforms.

\subsection{Algebras, Modules, and Signal Processing}\label{algmodsm}

In this section, we introduce the concepts from the theory of algebras
and modules that are needed to formulate the algebraic theory of
SP. Formal definitions are in Appendix~\ref{algdefs}. For a more
thorough introduction to module theory, we refer to, e.g.,
\cite{Jacobson:74,James:93,Curtis:62}.  We will provide a short
dictionary between algebraic and signal processing concepts.
Section~\ref{modsigmod} formally defines the concept of a signal
model.

\mypar{Algebras (filter spaces)} We denote by $\C$ the set of complex
numbers. A $\C$-algebra $\alg$ is a $\C$-vector space that is also a
ring, i.e., the multiplication of elements in the $\C$-vector space is
defined (see Definition~\ref{algebradef} in Appendix~\ref{algdefs} for
the formal definition).  Examples of algebras include $\C$, the set
$\C^{n\times n}$ of complex $n\times n$ matrices, and the set of
polynomials $\C[x]$ in the indeterminate~$x$ and with coefficients in
$\C$.  We can choose a base field different from $\C$, for example the
real numbers $\R$ or the rational numbers $\Q$, and will do so
occasionally but then explicitly say so. Note that it has to be a
field, otherwise the vector space structure is lost. Since $\alg$ is a
vector space, concepts that only require this structure, such as basis
and dimension, are well-defined.

As we mentioned in the previous section, algebras serve as spaces of
filters in signal processing, with the filters (or linear systems)
being the elements of the algebra. Thus, in this paper, elements of
the algebra are to be considered filters.  To ease this
identification, we represent the elements of the algebra by~$h$,
a common symbol for filter in signal processing.

We say that elements $h_1,\dots,h_k\in\alg$ {\em generate} $\alg$, if
every element in $\alg$ can be written as a multivariate polynomial or
series in $h_1,\dots,h_k$, or, equivalently, by repeatedly forming
sums, products, and scalar multiples from these elements. Most
algebras considered in this paper are generated by one element; its
special role will be discussed in Section~\ref{shiftcomm}.

\mypar{Modules (signal spaces)}
If $\alg$ is an algebra, then a (left) $\alg$-module $\md$ is a vector
space, over the same base field (we assume $\C$) as $\alg$, that 
admits an operation of $\alg$ from the left\footnote{It can also be
defined with the algebra operating from the right, which leads to a
dual theory.}. We write this operation  as multiplication:
\begin{equation}\label{algop}
(h,s)\rightarrow h\cdot s\in\md,\quad
\text{for }h\in\alg,\ s\in\md.
\end{equation}
This ensures that $\md$ is {\em closed} or {\em invariant}
under the operation of $\alg$.  In addition, this operation satisfies
several properties including, for $h,h'\in\alg$, $s, s'\in\md$,
$\alpha\in\C$,
\begin{equation}\label{moduleprop}
\begin{array}{rcl}
h\cdot(s+s') & = & h\cdot s + h\cdot s',\\
h\cdot(\alpha s) & = & \alpha(h\cdot s),\\
h'\cdot(h\cdot s) &=& (h'\cdot h)\cdot s.
\end{array}
\end{equation}
The formal definition of an $\alg$-module is given in
Definition~\ref{moduledef} in Appendix~\ref{algdefs}. 

The multiplication $h\cdot s$ of elements of $h\in\alg$ with elements
of $s\in\md$ captures in the algebraic theory of signal processing the
concept of filtering: the elements $h\in\alg$ are the filters and the
elements $s\in\md$ are the signals. We emphasize that the definition
of a module always implies an associated algebra; viewed by itself, a
module is only a vector space.

In the algebraic theory of signal processing, modules are the signal
spaces and elements of modules are the signals. To help with this
identification, we denote signals with the symbol~$s$ whenever possible.
In this paper, we focus on {\em discrete} linear signal
processing\footnote{We deliberately use the term ``discrete'' instead
of ``discrete-time,'' since one of our goals is to identify signal
models for discrete space.} and thus on modules in which the elements
have the form of series or linear combinations $s =
\sum_{i\in I} s_ib_i$, where the index domain $I$ is discrete (e.g., 
$I$ could be finite or $I = \N, \Z$). The coordinate vector of a
signal $s$ is written as $\coord{s} = (s_i| i\in I)$. The base vectors
$b_i$ (which, as elements of $\md$, are signals) are in signal
processing called \emph{impulses} (also unit pulses or delta
pulses). Given a filter $h\in\alg$, its impulse response, i.e., the
response of the filter~$h$ to an input which is the impulse
$b_i\in\md$, is given by $h\cdot b_i$.  The definition of $\md$
assures that $h\cdot b_i$ is well-defined and again a signal, i.e., an
element of $\md$.

\mypar{Regular module (filter space = signal space)} An important
example of a module is the {\em regular} $\alg$-module. This is the
case when the module and the algebra are equal as sets: $\md = \alg$,
with the multiplication operation in \eqref{algop} given by the
ordinary multiplication in $\alg$. Even though the {\em sets} $\md =
\alg$ may be equal, their algebraic structures are different; for
example, elements in $\md$ cannot be multiplied. In this paper, we
will distinguish between elements in $\md$ and elements in $\alg$.

\mypar{Representations (filters as matrices)}
As a consequence of the properties in \eqref{moduleprop}, every filter
$h\in\alg$ defines a linear mapping on $\md$:
\begin{equation}\label{aop}
s\mapsto h\cdot s.
\end{equation}
If $\md$ has finite dimension~$n$ and we choose a basis $b =
(b_0,\dots,b_{n-1})$\footnote{We write bases always as lists in
parentheses, not as sets in curly braces, since the chosen order of
the base vectors is important.} in $\md$, this linear mapping is
represented by a complex $n\times n$ matrix $M_h$, which is a
\emph{matrix representation} of the filter~$h$. As usual with 
linear mappings, $M_h$ is obtained by applying $h$ to each base vector
$b_i$; the coordinate vector of the result $hb_i$ is the
$i$th column of $M_h$.

By constructing $M_h$ for every filter $h\in\alg$, we obtain a
mapping~$\phi$ from the algebra~$\alg$ (the set of filters) to the
algebra of $n\times n$ matrices $\C^{n\times n}$:
\begin{equation}\label{Aop}
\phi:\ \alg\rightarrow\C^{n\times n},\ 
h\mapsto\phi(h) = M_h.
\end{equation}
The mapping $\phi$ is a {\em homomorphism} of algebras, i.e., a mapping
that preserves the algebra structure (see Definition~\ref{alghomdef}
in Appendix~\ref{algdefs}). In particular,
$$
\phi(h+h') = \phi(h) + \phi(h')\quad\text{and}\quad
\phi(hh') = \phi(h)\phi(h').
$$
$\phi$ is called the {\em (matrix) representation of $\alg$ afforded
by the $\alg$-module $\md$ with basis $b$}. The
representation is fixed by the choice of the module~$\md$
\emph{and} the basis~$b$ of~$\md$. Different choices of~$\md$ and~$b$
lead to different matrix representations of~$\alg$.  However, $\phi$
is independent of the basis chosen in $\alg$.

The set of matrices $\phi(\alg)$ is an algebra that is structurally
identical to $\alg$. Correspondingly, if $s = \sum_{i=0}^{n-1}
s_ib_i\in\md$, i.e., $\coord{s} = (s_0,\dots,s_{n-1})^T$ is the
coordinate vector for $s$, then the abstract notion of filtering
(multiplication of $s\in\md$ by $h\in\alg$) becomes in coordinates
the matrix-vector multiplication:
\begin{equation}\label{am}
h\cdot s\Leftrightarrow\phi(h)\cdot\coord{s}.
\end{equation}
This coordinatization of filtering also shows the fundamental
difference between signals and filters; namely, in coordinates,
signals become vectors, and filters (as linear operators on signals)
become matrices.

If $\md$ is not of finite dimension, but still discrete, i.e.,
consisting of infinite series, say, of the form $s =
\sum_{i\in\Z}s_ib_i$, we still obtain a matrix representation $\phi$,
but the matrices are now infinite. If $\md$ is continuous, there is no
matrix representation, rather, an operator representation.

\mypar{Irreducible submodule (spectral component)} 
If $\md$ is an $\alg$-module, then
a subvector space $\md'\leq\md$ is an $\alg$-submodule of $\md$ if
$\md'$ is itself an $\alg$-module. Equivalently, $\md'$ is closed or
invariant under the operation of $\alg$. Most subvector spaces fail to
be $\alg$-submodules, because, intuitively, the smaller the vector
space $\md'$ is, the harder it is to remain invariant under $\alg$.

A submodule $\md'\leq\md$ is {\em irreducible} if it contains no
proper submodules, i.e., no submodules besides the trivial
submodules~$\md=\{0\}$ and~$\md$ itself.

In particular, every one-dimensional submodule $\md'$ is irreducible
and is a simultaneous eigenspace of all filters $h\in\alg$, i.e., $hs
= \lambda_hs$ for all $s\in\md$ with a suitable $\lambda_h\in\C$.

In signal processing, an irreducible module corresponds to a
spectral component. This will become clear in the next paragraph.

\mypar{Module decomposition, spectrum, Fourier transform} In signal 
processing, Fourier analysis involves the decomposition of signals
into spectral components.  The algebraic theory gives a 
general definition. Namely, Fourier analysis decomposes an
$\alg$-module into a direct sum\footnote{By ``direct sum'' of vector
spaces, we usually mean the more general ``outer'' direct sum rather
than the ``inner'' direct sum. See~Definition~\ref{directsumV} in
Appendix~\ref{algdefs} for an explanation.} of irreducible
$\alg$-submodules $\md_\omega$, where $\omega\in W$ (some index
domain).  We call the corresponding mapping $\Delta$ the {\em Fourier
transform} for the $\alg$-module $\md$:
\begin{equation}\label{moduledec}
\begin{array}{rrcl}
\Delta: & \md & \rightarrow & \bigdirsum_{\omega\in W}\md_\omega,\\
& s & \mapsto & (s_\omega)_{\omega\in W}.
\end{array}
\end{equation}
The existence of such a decomposition is not guaranteed and
depends on $\alg$ and $\md$.

In \eqref{moduledec}, each submodule $\md_\omega$ is called a {\em
spectral component} of the signal space $\md$, and each projection
$s_\omega\in\md_\omega$ is a spectral component of the signal $s$. The
collection of all $\md_\omega$ and all $s_\omega$, $\omega\in W$, is
called the {\em spectrum} of $\md$ and $s$, respectively. The spectrum
of $s$ is a list with index set $W$, and as such can be viewed
equivalently as a function on $W$:
$$
(s_\omega)_{\omega\in W} = \omega\mapsto s_\omega.
$$

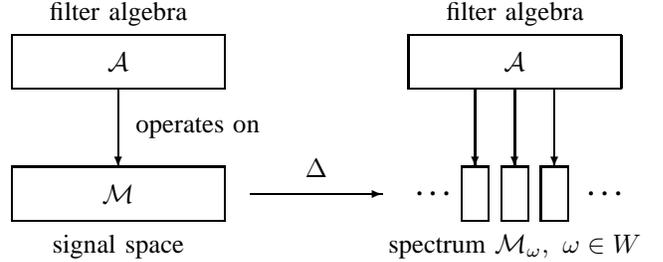
\begin{figure}\centering
\begin{picture}(250,108)
\put(0,0){\makebox(80,20){signal space}}
\put(0,88){\makebox(80,20){filter algebra}}
\put(0,20){\framebox(80,20){$\md$}}
\put(0,70){\framebox(80,20){$\alg$}}
\put(40,70){\vector(0,-1){30}}
\put(40,40){\makebox(60,30){operates on}}
\put(90,30){\vector(1,0){50}}
\put(90,30){\makebox(50,20){$\Delta$}}
\put(150,70){\framebox(80,20){$\alg$}}
\put(150,20){\makebox(20,20){\boldmath$\cdots$}}
\put(170,20){\framebox(10,20){}}
\put(175,70){\vector(0,-1){30}}
\put(185,20){\framebox(10,20){}}
\put(190,70){\vector(0,-1){30}}
\put(200,20){\framebox(10,20){}}
\put(205,70){\vector(0,-1){30}}
\put(215,20){\makebox(20,20){\boldmath$\cdots$}}
\put(150,0){\makebox(80,20){spectrum $\md_\omega,\ \omega\in W$}}
\put(150,88){\makebox(80,20){filter algebra}}
\end{picture}
\caption{A visualization of the concept Fourier transform, which
decomposes the $\alg$-module $\md$ into a direct sum of irreducible
(minimal) $\alg$-invariant subspaces, i.e., $\alg$-submodules. The
latter are called the spectrum of $\md$.\label{ftvis}}
\end{figure}

\begin{figure}\centering
\begin{picture}(250,110)
\put(0,0){\makebox(80,10){$\coord{s}$}}
\put(0,14){\line(1,0){80}}
\put(0,12){\line(0,1){4}}
\put(80,12){\line(0,1){4}}
\put(0,30){\framebox(80,80){$\phi(h)$}}
\put(100,60){\vector(1,0){30}}
\put(100,60){\makebox(30,15){$\four(\cdot )\four^{-1}$}}
\put(100,14){\vector(1,0){30}}
\put(100,14){\makebox(30,15){$\four$}}
\put(150,0){\makebox(80,10){$(\coord{s}_\omega)_{\omega\in W}$}}
\put(150,14){\line(1,0){80}}
\put(150,12){\line(0,1){4}}
\put(160,12){\line(0,1){4}}
\put(170,12){\line(0,1){4}}
\put(180,12){\line(0,1){4}}
\put(200,12){\line(0,1){4}}
\put(230,12){\line(0,1){4}}
\put(150,30){\dashbox{5}(80,80){}}
\put(150,110){\line(1,0){10}}
\put(150,110){\line(0,-1){10}}
\put(150,100){\line(1,0){20}}
\put(160,110){\line(0,-1){20}}
\put(160,90){\line(1,0){20}}
\put(170,100){\line(0,-1){20}}
\put(170,80){\line(1,0){30}}
\put(180,90){\line(0,-1){30}}
\put(180,60){\line(1,0){50}}
\put(200,80){\line(0,-1){50}}
\put(200,30){\line(1,0){30}}
\put(230,60){\line(0,-1){30}}
%
\put(180,80){\makebox(50,30){$\bigdirsum_\omega\phi_\omega(h)$}}
\end{picture}
\caption{Visualization of the convolution theorem \eqref{repdec} in
the finite case. On the left, filtering is in coordinates equivalent
to $\phi(h)\coord{s}$. In the Fourier domain on the right, the filter
operates on invariant subspaces; thus the matrix is now block-diagonal
with the blocks $\phi_\omega(h)$ being the frequency response of $h$. 
\label{repdecvis}}
\end{figure}
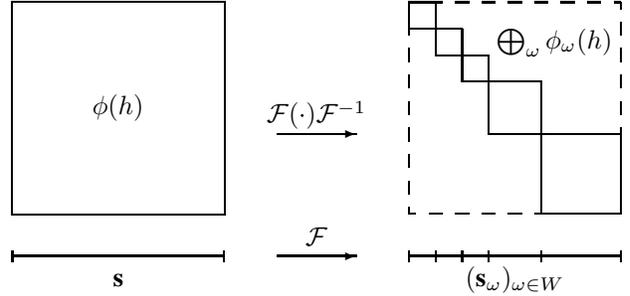

See Figure~\ref{ftvis} for a visualization of the Fourier transform.
Its definition is intuitive from a signal processing point of view;
the Fourier transform decomposes the signal space into the smallest
components that are invariant, no matter how the signal is filtered.
Another requirement usually imposed is that $\Delta$ be invertible,
i.e., that every signal can be reconstructed from its spectrum.

The Fourier transform is an $\alg$-module homomorphism (see
Definition~\ref{hommoddef} in Appendix~\ref{algdefs}), which 
means that $\Delta(h\cdot s) = h\cdot \Delta(s)$ for $h\in\alg,\ 
s\in\md$. In words, this means
that filtering in the signal space $\md$ is equivalent to parallel
filtering in the spectrum (as visualized in Figure~\ref{ftvis}), or
\begin{equation}\label{hsomega}
(h\cdot s)_\omega = h\cdot s_\omega,\quad\text{for all }
\omega\in W.
\end{equation}
If $\Delta$ is invertible, then $\Delta(h\cdot s) = h\cdot \Delta(s)$
also yields a general convolution theorem:
\begin{equation}\label{convtheorem}
h\cdot s = \Delta^{-1}(h\cdot\Delta(s)).
\end{equation}

The definition of $\Delta$ in \eqref{moduledec} (and thus also the
convolution theorem \eqref{convtheorem}) is {\em coordinate-free},
i.e., formulated independently of the bases chosen in $\md$ and the
$\md_\omega$, $\omega\in W$. We coordinatize $\Delta$ in two
steps. First, we choose bases on the right side, i.e., in the
$\md_\omega$'s.  Then $\Delta$ takes the form
\begin{equation}\label{moduledec1}
\begin{array}{rrcl}
\Delta: & \md & \rightarrow & 
  \bigdirsum_{\omega\in W}\C^{\dim(\md_\omega)},\\
& s & \mapsto & (\coord{s}_\omega)_{\omega\in W}.
\end{array}
\end{equation}
(By abuse of notation we use the same letter $\Delta$.)  
Again, the spectrum, now in its coordinate form, 
can be viewed as a function on $W$:
$$
(\coord{s}_\omega)_{\omega\in W} = \omega\mapsto\coord{s}_\omega.
$$

Choosing also a basis in $\md$, we obtain the coordinate form of the
Fourier transform, denoted by $\four$:\footnote{Note that we do not
write $\four:\ \C^{\dim(\md)}\rightarrow\bigdirsum_{\omega\in
W}\C^{\dim(\md_\omega)}$, since for $\dim(\md)=\infty$, the coordinate
space for $\md$ does not need to have the form $\C^{\dim(\md)}$; e.g., it 
could be $\ell^2(\Z)$.}
\begin{equation}\label{ftcoord}
\four:\ \coord{s}\mapsto(\coord{s}_\omega)_{\omega\in W}.
\end{equation}
We refer to both $\Delta$ and $\four$ as Fourier transform, and we
also refer to $(\coord{s}_\omega)_{\omega\in W}$ as the spectrum of
$s$.  In signal processing, the Fourier transform is usually thought
of either as $\Delta$ in \eqref{moduledec1}, or as $\four$ in
\eqref{ftcoord}.

In particular, if $\md$ is of finite dimension~$n$ and if the Fourier
transform exists, then $W$ is also finite and usually chosen as $W =
(0,\dots,k-1)$, $k\leq n$. In this case, $\four$ is an $n\times n$
matrix. If $k = n$, then all irreducible modules are of dimension 1.

\mypar{Irreducible representations (frequency response)} In the
decomposition \eqref{moduledec}, each irreducible $\md_\omega$ affords
an {\em irreducible representation} $\phi_\omega$ of $\alg$ with
respect to a chosen basis $b_\omega$. Namely, if
$\coord{s}_\omega\in\C^{\dim(\md_\omega)}$ is the coordinate vector of
the spectral component $s_\omega\in\md_\omega$, then for every filter
$h\in\alg$, by \eqref{am},
\begin{equation}\label{irrop}
(h\cdot s)_\omega = h\cdot s_\omega
  \Leftrightarrow \phi_\omega(h)\cdot \coord{s}_\omega,
\end{equation}
where $\phi_\omega(h)$ is a
$\dim(\md_\omega)\times\dim(\md_\omega)$ matrix.  For a fixed
filter~$h$, the collection
\begin{equation}\label{fouralg}
(\phi_\omega(h))_{\omega\in W} = \omega\mapsto\phi_\omega(h)
\end{equation}
is in signal processing called the {\em frequency response} of $h$
and can be viewed as a matrix-valued function on $W$. If the
irreducible submodule $\md_\omega$ is one-dimensional, then, by
invariance, it is an eigenspace for every $h\in\alg$, and
$\phi_\omega(h)$ is the corresponding eigenvalue.

We could call the mapping
$$
h\mapsto(\phi_\omega(h))_{\omega\in W}
$$ 
that maps every filter to its frequency response the Fourier
transform of $\alg$ (w.r.t.~the module $\md$), but we refrain from
doing so in this paper and reserve the term Fourier transform to
the decomposition of the module or of signals into their spectrum.

If $\md$ is of finite dimension $n$ and with $k$ spectral components
$\md_i$ of dimension $d_i$, $0\leq i<k$, then $\sum d_i = n$.  If we
choose bases in $\md$ and in the $\md_i$'s, then $\four$ takes the
form in \eqref{ftcoord} and is an $n\times n$ matrix. 
Filtering in $\md$ is in 
coordinates given by the matrix $\phi(h)$. Filtering 
in the decomposed module $\bigdirsum_{0\leq i<k}\md_i$ 
is in coordinates given by
$$
\phi_1(h)\dirsum\dots\dirsum\phi_k(h) =
\begin{bmatrix}
\phi_1(h)\\
&\ddots\\
&&\phi_k(h)
\end{bmatrix},
$$
where $\phi_i(h)$ is a $d_i\times d_i$ matrix and
\begin{equation}\label{dirsumdef}
A\dirsum B = \diag(A,B) =
\begin{bmatrix}A\\&B\end{bmatrix}
\end{equation}
denotes the direct sum of matrices.
Since $\four$ maps the underlying vector spaces, we get 
\begin{equation}\label{repdec}
\four\cdot\phi(h)\cdot\four^{-1} = 
  \phi_0(h)\dirsum\dots\dirsum\phi_{k-1}(h),
\end{equation}
which is visualized in Figure~\ref{repdecvis}.
In other words, the matrices $\four\phi(h)\four^{-1}$ are block
diagonal, with the sizes of the blocks given by the dimensions $d_i$
of the irreducible modules $\md_i$. In particular, if all $\md_i$ are
one-dimensional, $d_i = 1$, then $k = n$ and $\four\phi(h)\four^{-1}$
is diagonal, i.e., \eqref{repdec} gives the diagonalization property
of $\four$ and is a coordinatized version and a special case of the
convolution theorem~\eqref{convtheorem}.

\begin{table*}\centering
\caption{Correspondence between discrete signal processing concepts and
  algebraic concepts.}\label{algsp}
\isdraft{\renewcommand{\baselinestretch}{1}\small}{}
\ra{1.2}
\begin{tabular}{@{}lll@{}}\toprule
signal processing concept & algebraic concept (coordinate free)
& in coordinates \\ \midrule
filter & $h\in\alg$ \text{ (algebra) }
  & $\phi(h)\in\C^{I\times I}$\\
signal & $s = \sum s_ib_i\in\md$ \text{ ($\alg$-module)}
  & $\coord{s} = (s_i)_{i\in I}\in\C^I$\\
filtering & $h\cdot s$ & $\phi(h)\cdot \coord{s}$ \\
impulse & base vector $b_i\in\md$ & 
  $\coord{b}_i = (\dots,0,1,0,\dots)^T\in\C^I$\\
impulse response of $h\in\alg$ & 
  $h\cdot b_i\in\md$ & $\phi(h)\cdot\coord{b}_i\in\C^I$ \\
Fourier transform &
  $\Delta:\ \md\rightarrow\bigdirsum_{\omega\in W}\md_\omega$ &
  $\four:\ \C^I\rightarrow\bigdirsum_{\omega\in W}\C^{d_\omega}$ 
  $\Leftrightarrow \phi\rightarrow\bigdirsum_{\omega\in W}\phi_\omega$ \\
spectrum of signal & 
  $\Delta(s) = (s_\omega)_{\omega\in W} = \omega\mapsto s_\omega$ & 
  $\four(\coord{s}) = (\coord{s}_\omega)_{\omega\in W} = 
  \omega\mapsto\coord{s}_\omega$ \\
frequency response of $h\in\alg$ & n.a. & 
  $(\phi_\omega(h))_{\omega\in W} = \omega\mapsto\phi_\omega(h)$\\
\bottomrule
\end{tabular}
\isdraft{\renewcommand{\baselinestretch}{1.5}\normalsize}{}
\end{table*}

\mypar{Summary} We summarize the correspondence between algebraic
concepts and signal processing concepts in Table~\ref{algsp}. The
signal processing concepts are given in the first column and their
algebraic counterparts in the second column.  If we choose bases in
the occurring modules, we obtain the corresponding coordinate version
given in the third column.  In coordinates, the algebraic objects,
operations, and mappings become vectors and matrices and thus allow
for actual computation. This is the form used in signal
processing. However, the coordinate version hides the underlying
module structure, which cannot be easily recovered if it is not known
beforehand.

\mypar{Example: infinite discrete time} Often, in linear signal
processing, the index time is assumed to be continuous and taking
values on the real line~$\R$ or nonnegative reals~$\R^+$, or to be
discrete and taking values in~$\N$ (non-negative integers) or~$\Z$. We
will refer to these cases as the infinite continuous or infinite
discrete time case.  We develop here an example for the algebraic
framework for infinite discrete time and consider filters and signals
to be represented in the $z$-domain. The $z$-transform is denoted by
$\Phi$. For example, the set of all two-sided infinite series, called
Laurent series, is represented by $\Phi(\C^\Z) =
\{\sum_{n\in\Z}a_nz^{-n}|a_n\in\C\}$.

We find a suitable algebra of filters. A first attempt is to choose
the vector space $\Phi(\C^\Z)$, but this set is not an algebra since
multiplication, i.e., convolution is in general not possible. 
Namely, if $a,a'\in\Phi(\C[\Z])$, then 
\begin{equation}\label{cauchy}
aa' = \sum_{n\in\Z}\bigl(\sum_{k+\ell = n}a_ka'_\ell\bigr)z^{-n}.
\end{equation}
The inner sum has infinitely many terms and does not converge in
general. Thus we need to consider a smaller space.

To do this, we use Figure~\ref{map}, which shows in a block diagram
the sets that are commonly considered in infinite discrete time signal
processing and their algebraic structure.  We use the following
mnemonics. We indicate in each box the set it represents by a symbol
such as $\Phi(\C^\Z)$. Table~\ref{mapsymbol} explains these symbols by
giving a generic element.  Solid boxes in Figure~\ref{map} are vector
spaces, algebras are marked bold, and dashed boxes are multiplicative
groups that are not vector spaces. In each box we indicate a short
mathematical description~(M) of the set it represents or a
characterization of the set if viewed as a set of filters~(F) or if
viewed as a set of signals~(S) (see also the legend in
Figure~\ref{map}). The middle column is for two-sided series; the
right column is for one-sided series; and the left column is for
series expansions of rational functions\footnote{Note that for a
rational function, various expansions are possible in general. Also
note that we use the symbol $\C(z^{-1})$ for rational functions and
there expansions likewise.}.  The arrows between different boxes
depict various inclusion relationships (the tip of the arrow points
towards the smaller set). Finally, the dark-gray area indicates for
which modules the Fourier transform exists.

\begin{table}\centering
\caption{The generic element for the sets used in 
Figure~\ref{map}. From top to bottom: Laurent series, power series,
rational function, Laurent polynomial, polynomial.}\label{mapsymbol} 
$
\ra{1.3}
\begin{array}{@{}ll@{}}\toprule
\text{symbol} & \text{generic element} \\ \midrule
\Phi(V) & \sum_{n\in\Z} a_nz^{-n},\ (a_n)_{n\in\Z}\in V \\
\C[[z^{-1}]] & \sum_{n\geq 0} a_nz^{-n},\ a_n\in\C\\
\C(z^{-1}) & p(z^{-1})/q(z^{-1}),\text{ $p,q$ polynomials}\\
\C[z^{-1},z] & \sum_{-\ell\leq n\leq k}a_nz^{-n},\ a_n\in\C\\
\C[z^{-1}] & \sum_{0\leq n\leq k}a_nz^{-n},\ a_n\in\C\\ 
\bottomrule
\end{array}
$
\end{table}

\newcommand{\ratcol}[0]{40}
\newcommand{\noncausalcol}[0]{200}
\newcommand{\causalcol}[0]{360}
\newcommand{\thicker}[0]{\linethickness{1.2mm}}
%
\isdraft{%
\newcommand{\mybox}[5]{%
\put(#1,#2){\framebox(80,40){}}
\thicklines
\put(#1,#3){\framebox(80,15){#4}}
\put(#1,#2){\makebox(80,25){%
  \renewcommand{\arraystretch}{0.4}
  \begin{scriptsize}
  \begin{tabular}{@{\,\,}l@{ }l}
  #5
  \end{tabular}\end{scriptsize}}}
}
\newcommand{\myboxa}[5]{%
\put(#1,#2){\dashbox{4}(80,40){}}
\thicklines
\put(#1,#3){\makebox(80,15){#4}}
\put(#1,#3){\line(1,0){80}}
\put(#1,#2){\makebox(80,25){%
  \renewcommand{\arraystretch}{0.4}
  \begin{scriptsize}
  \begin{tabular}{@{\,\,}l@{ }l}
  #5
  \end{tabular}\end{scriptsize}}}
}
}{%
\newcommand{\mybox}[5]{%
\put(#1,#2){\framebox(80,40){}}
\thicklines
\put(#1,#3){\framebox(80,15){#4}}
\put(#1,#2){\makebox(80,25){%
  \begin{footnotesize}
  \renewcommand{\arraystretch}{0.6}
  \begin{tabular}{@{\,\,}l@{ }l}
  #5
  \end{tabular}\end{footnotesize}}}
}
\newcommand{\myboxa}[5]{%
\put(#1,#2){\dashbox{4}(80,40){}}
\thicklines
\put(#1,#3){\makebox(80,15){#4}}
\put(#1,#3){\line(1,0){80}}
\put(#1,#2){\makebox(80,25){%
  \begin{footnotesize}
  \renewcommand{\arraystretch}{0.6}
  \begin{tabular}{@{\,\,}l@{ }l}
  #5
  \end{tabular}\end{footnotesize}}}
}
}
\newcommand{\vecdown}[3]{%
  \put(#1,#2){\vector(0,-1){#3}}
}
\newcommand{\vecleft}[3]{%
  \put(#1,#2){\vector(-2,-1){#3}}
}
\newcommand{\vecright}[3]{%
  \put(#1,#2){\vector(2,-1){#3}}
}
\isdraft{\thispagestyle{empty}}{}
\begin{figure*}\centering
\isdraft{\vspace*{-10mm}}{}
\isdraft{\renewcommand{\baselinestretch}{1}\footnotesize}{}
%
\isdraft{\includegraphics[width=1.03\textwidth]{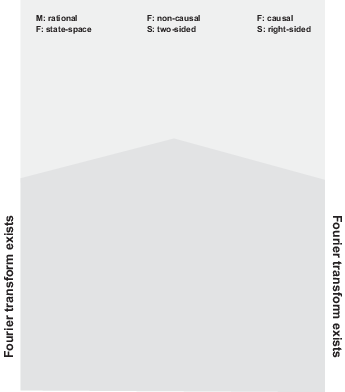}}
  {\includegraphics[width=0.945\textwidth]{figures/back}}

\isdraft{\vspace*{10pt}}{}
\begin{picture}(480,30)\thicklines
%
\mybox{\ratcol}{450}{475}{$\C(z^{-1})$}{%
  \rule{0mm}{3mm}M:&series expansions\\
  &of rational functions\\F:&state space models}
\thicker
\mybox{\ratcol}{210}{235}{$\C(z^{-1})\!\cap\!\Phi(\ell^1(\Z))$}{%
  \rule{0mm}{3mm}F:&rational and\\
  & BIBO stable}
\myboxa{\ratcol}{130}{155}{}{%
  \rule{0mm}{3mm}F:&minimum phase}
%
%
\vecdown{80}{450}{198}
\vecdown{80}{210}{40}
\vecright{121}{230}{77}
%
\mybox{\noncausalcol}{490}{515}{$\Phi(\C^\Z)$}{%
  \rule{0mm}{3mm}M:&Laurent series\\F:&LTI systems\\&IIR filters}
\mybox{\noncausalcol}{410}{435}{$\Phi(\ell^\infty(\Z))$}{%
  \rule{0mm}{3mm}S:&bounded}
\mybox{\noncausalcol}{330}{355}{$\Phi(\ell^2(\Z))$}{%
  \rule{0mm}{3mm}M:&square summable\\S:&finite energy}
\thicker
\mybox{\noncausalcol}{250}{275}{$\Phi(\ell^1(\Z))$}{%
  \rule{0mm}{3mm}M:&absolute summable\\F:&BIBO stable}
\thicker
\mybox{\noncausalcol}{170}{195}{$\C[z^{-1},z]$}{%
  \rule{0mm}{3.5mm}M:&Laurent polynomials\\F:&FIR filters\\S:&finite support}
\thicklines
\myboxa{\noncausalcol}{90}{115}{$\{z^{-k}| k\in\Z\}$}{%
  \rule{0mm}{3.5mm}M:&monomials\\F:&delays\\S:&(unit) impulses}
%
%
\vecdown{240}{490}{40}
\vecdown{240}{410}{40}
\vecdown{240}{330}{38}
\vecdown{240}{250}{38}
\vecdown{240}{170}{40}
\vecleft{199}{510}{78}
\vecleft{199}{270}{77}
\vecright{281}{510}{77}
\vecright{281}{430}{78}
\vecright{281}{350}{78}
\vecright{281}{270}{77}
\vecright{281}{190}{77}

\thicker
\mybox{\causalcol}{450}{475}{$\Phi(\C^\N)=\C[[z^{-1}]]$}{%
  \rule{0mm}{3mm}M:&formal power series\\F:&causal\\S:&right-sided}
\mybox{\causalcol}{370}{395}{$\Phi(\ell^\infty(\N))$}{%
  \rule{0mm}{3.5mm}S:&right-sided, bounded}
\mybox{\causalcol}{290}{315}{$\Phi(\ell^2(\N))$}{%
  \rule{0mm}{3.5mm}S:&right-sided,\\&finite energy}
\thicker
\mybox{\causalcol}{210}{235}{$\Phi(\ell^1(\N))$}{%
  \rule{0mm}{3.5mm}F:&causal, stable}
\thicker
\mybox{\causalcol}{130}{155}{$\C[z^{-1}]$}{%
  \rule{0mm}{3.5mm}M:&polynomials\\F:&causal FIR filters}
\thicker
\mybox{\causalcol}{50}{75}{$\C\cdot z^0$}{%
  \rule{0mm}{3.5mm}F:&memoryless\\S:&(time zero) impulses}
%
%
\vecdown{400}{450}{40}
\vecdown{400}{370}{40}
\vecdown{400}{290}{38}
\vecdown{400}{210}{38}
\vecdown{400}{130}{38}
\end{picture}
\noindent
\isdraft{\vspace*{-48pt}}{\vspace*{-25pt}}

\isdraft{\hspace{17pt}}{\hspace{2.5pt}}
\isdraft{\includegraphics[width=0.915\textwidth]{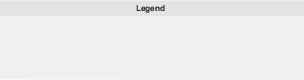}}
  {\includegraphics[width=0.84\textwidth]{figures/back-legend}}
\isdraft{\vspace*{-23pt}}{\vspace*{-20pt}}

\begin{picture}(480,0)
\thicklines
\thicker
\put(40,60){\framebox(30,15){}}
\thicklines
\put(40,30){\framebox(30,15){}}
\put(40,00){\dashbox{3}(30,15){}}
\put(90,60){\makebox(30,15)[l]{algebra}}
\put(90,30){\makebox(30,15)[l]{vector space only (no algebra)}}
\put(90,0){\makebox(30,15)[l]{multiplicative group (no vector space)}}
\put(240,60){\framebox(30,15){$A$}}
\put(270,67){\vector(1,0){30}}
\put(300,60){\framebox(30,15){$B$}}
\put(340,60){\makebox(0,15)[r]{\large\bf :}}
\put(340,30){\makebox(0,15)[r]{\large\bf M\ :}}
\put(340,15){\makebox(0,15)[r]{\large\bf F\ :}}
\put(340,0){\makebox(0,15)[r]{\large\bf S\ :}}
\put(345,60){\makebox(0,15)[l]{$B\subset A$}}
\put(345,30){\makebox(0,15)[l]{mathematical description}}
\put(345,15){\makebox(0,15)[l]{filter/system description}}
\put(345,0){\makebox(0,15)[l]{signal description}}
\end{picture}
\vspace*{5pt}
\isdraft{\renewcommand{\baselinestretch}{1.5}\small}{}
\caption{A block diagram identifying the algebraic structure of sets commonly
used in discrete-time signal processing. Note that the module property
is not captured in this diagram since it depends on the choice
of algebra.\label{map}}
\end{figure*}

We now return to the search for an algebra. We have ruled out already
above the set $\Phi(\C^\Z)$ as a possibility for $\alg$, because
multiplication of two-sided infinite series is not possible in
general. Among the sets in Figure~\ref{map}, the next largest
candidate for~$\alg$ is the set of right-sided series $\alg =
\Phi(\C^\N)$. In mathematics, this is the set of {\em formal power
series}, written as $\C[[z^{-1}]]$, and it is an algebra (in
\eqref{cauchy} the inner sum has only finitely many terms and is thus
always well-defined). In looking for an $\alg$-module $\md$ for this
algebra, it is easy to verify again that the set $\Phi(\C^\Z)$ is not
an $\alg$-module (again, for $a\in\alg$, $a'\in\md$, the inner sum in
\eqref{cauchy} will have infinitely many terms; thus, the product does
not exist in general).  The next candidate for the corresponding
$\alg$-module is the regular module $\md = \alg$. However, there are
problems with choosing for algebra of filters
$\alg=\C[[z^{-1}]]$. First, this set contains only causal or one-sided
filters. Second, the Fourier transform for the regular module $\md =
\alg$ does not exist (note that it lies outside the dark-gray area in
Figure~\ref{map}).

The next largest choice for~$\alg$ is the space of bounded-input,
bounded-output~(BIBO) stable systems $\alg = \Phi(\ell^1(\Z))$. This
is the set usually chosen in signal processing. As module for $\alg$,
we can consider again the regular module $\md = \alg$; in alternative,
we can actually choose a larger space of signals. A well-known
theorem\footnote{Theorem~\ref{lpmod}, provided with proof in
Appendix~\ref{modproplp}.} states that $\Phi(\ell^p(\Z))$ is a
$\Phi(\ell^1(\Z))$-module for $p\geq 1$. Thus, we could attempt to
select as a module the largest of such sets $\md =
\Phi(\ell^\infty(\Z))$. The problem is that, again, as it is
well-known, the Fourier transform does not exist. The choice commonly
made in signal processing is $\md = \Phi(\ell^2(\Z))$, the space of
finite-energy signals\footnote{We could actually choose larger modules
$\md = \Phi(\ell^p(\Z))$, $p<\infty$,
if the proper definition of convergence for the Fourier transform is
chosen~\cite{Edwards:67,Edwards:67a}. However, we will work here with
finite energy signals.}. With these choices, we emphasize that for
the discrete time example under consideration we chose {\em the sets}
$\alg$ and $\md$ to be {\em different}, i.e., the sets of filters and
of signals have not only different algebraic structures (one is an
algebra and the other is a module) but, as sets, they are actually
different, a fact that is rarely explicitly stated in the signal
processing literature. Finally, we note that because of the need for
efficient implementations of filters, usually only the smaller algebra
$\C(z^{-1})\!\cap\!\Phi(\ell^1(\Z))$ of BIBO stable series that are
also expansions of rational functions is considered.

In summary, with these choices $\alg=\Phi(\ell^1(\Z))$ and
$\md=\Phi(\ell^2(\Z))$, filters $h$ and signals $s$ take the form $h =
H(z) = \sum_{n\in\Z}h_nz^{-n}$ and $s = S(z) =
\sum_{n\in\Z}s_nz^{-n}$, respectively. The basis elements $z^{-n}$
are called delays if they are filters and impulses if they are
signals.

The representation $\phi$ of $\alg$ afforded by $\md$ with
basis\footnote{Note that we do not use the term basis in the strictest
mathematical sense, which requires the linear combination to be
finite. However, the notion of basis can be generalized to the way it
is used here, if the space is a Banach space \cite{Kashin:89}, which
it is for $\ell^2$ or $\ell^1$ coefficient sequences.} $b =
(\dots,z^{-1},z^0,z^1,\dots)$ maps filters $h$ to a doubly infinite
matrix with Toeplitz structure.

The next task is to identify the irreducible modules $\md_\omega$,
i.e., the spectrum of $\md$. It is well-known that each
\begin{equation}\label{eigenfunction}
E_\omega(z) = \sum_{n\in\Z}e^{j\omega n}z^{-n},\quad
  \omega\in W = (-\pi,\pi],
\end{equation}
is a simultaneous eigenvector for all filters $h=H(z)\in\alg$, namely
\begin{equation}\label{Homega}
H(z)E_\omega(z) = H(e^{j\omega})E_\omega(z),\quad\omega\in W.
\end{equation}
This implies that the one-dimensional space $\md_\omega$ spanned by
$E_\omega$ is an $\alg$-module and irreducible (since of dimension 1).
Further, \eqref{Homega} shows that
$$
\phi_\omega:\ H(z)\mapsto H(e^{j\omega})\in\C
$$
is the irreducible representation afforded by $\md_\omega$ if
the list of length 1 $(E_\omega)$ is chosen as basis. Note that
$\phi_\omega(h)=H(e^{j\omega})$ is a scalar because~$\md_\omega$ is
one-dimensional.

The corresponding Fourier transform is called the discrete-time
Fourier transform (DTFT) and, since the $E_\omega(z)$ are orthogonal,
it takes the form
\begin{equation*}
\begin{array}{rrcl}
\Delta: & \md & \rightarrow & \bigdirsum_{\omega\in W}\md_\omega,\\
& s = S(z) & \mapsto & (S(e^{j\omega})E_\omega(z))_{\omega\in W}.
\end{array}
\end{equation*}
This matches \eqref{moduledec}, but there is one problem. The spectral
components $E_\omega$ are not in $\md$, but only in
$\Phi(\ell^\infty(\Z))$.  So, the $\md_\omega$ are still irreducible
$\alg$-modules, but not {\em sub}modules of $\md$. This is one of the
problems that can arise in the infinite case; in this paper we are
mostly concerned with finite-dimensional modules where this problem
does not occur. In the present example however, $\Delta$ still exists,
and its coordinatized form
\eqref{moduledec1} is the one actually called DTFT in signal
processing:
\begin{equation*}
\begin{array}{rrcl}
\Delta: & \md & \rightarrow & 
  \bigl(\bigdirsum_{\omega\in W}\C\bigr) = \C^W,\\
& s = S(z) & \mapsto & (S(e^{j\omega}))_{\omega\in W} = 
  \omega\mapsto S(e^{j\omega}),
\end{array}
\end{equation*}
where $W = (-\pi,\pi]$.  $\Delta(s)$ is usually viewed as function on
the circle.\footnote{In fact, it is an $L^2$-function on the circle
\cite{Rudin:62}, but this fact is not of importance in our
discussion.} The operation of $\alg$ on $\md$, i.e., convolution,
$H(z)S(z)$ becomes a set of pointwise multiplications
$H(e^{j\omega})S(e^{j\omega})$ in the Fourier domain.  Further,
$\Delta$ can be inverted, i.e., the signal can be reconstructed from
its spectrum.

Finally, the frequency response for a filter $h=H(z)\in\alg$ is given
by the collection of all irreducible representations evaluated at $h$:
$$
(\phi_\omega(H(z)))_{\omega\in W} = (H(e^{j\omega}))_{\omega\in W} =
\omega\mapsto H(e^{j\omega}).
$$
Note that the frequency response of the filter $h$ is obtained in the
same way as the spectrum of the signal $s$, namely by evaluating at
$e^{j\omega}$, i.e., by ``applying the Fourier transform'' to the
filter $h\in\alg$ (we use double quotes since we defined the Fourier
transform only for $\md$). This is due to the special structure of the
algebra and module and may be misleading; in general, the spectrum (in
coordinate form) consists of {\em vectors} of length $d_\omega$ (the
dimension of $\md_\omega$), and the frequency response consists of
$d_\omega\times d_\omega$ {\em matrices} (the representations afforded
by $\md_\omega$). They coincide in dimensionality only for
$d_\omega=1$.

Figure~\ref{map} suggests that many other combinations of filter algebra
and signal module are possible, and this is indeed the
case.  For example, we can keep the signal module $\md =
\Phi(\ell^2(\Z))$ and restrict the filter algebra to a smaller
algebra, e.g., to causal FIR filters $\alg = \C[z^{-1}]$.  Choosing
now this $\alg$, we can reduce the signal module, for example, to the
signals with finite support, $\md = \C[z,z^{-1}]$, or, as another
example, to the right-sided signals $\md = \alg$. In the algebraic
framework, these would be different signal models; however, the
associated spectrum and Fourier transform in these cases is
essentially equivalent to the more general model considered above. We
will later consider models, which have a substantially different
notion of spectrum and thus also Fourier transform (as example see
also Table~\ref{fourmodels}).

At this point we hope to have conveyed to the reader that important
concepts from discrete-time signal processing are equivalent to the
more general concepts from the theory of algebras and modules. This
correspondence enables us to port linear signal processing to other
algebras and modules. A more immediate question, however, is whether
other modules and algebras are actually used in standard signal
processing without being explicitly stated. This is indeed the case as
 hinted at in Section~\ref{intro}, and the main
motivation for developing this algebraic theory. Before we consider
these models, we introduce the central concept in the algebraic theory
of signal processing: the formal, algebraic definition of a signal
model.

\subsection{Algebraic Definition of Signal Model}\label{modsigmod}

In the previous section we asserted that the assumptions underlying SP
naturally make the filter space an algebra $\alg$ and the signal space
an associated $\alg$-module $\md$. Conversely, if any $\alg$-module
$\md$ is given, filtering is automatically defined, and the
well-established module theory can be applied to rigorously derive the
spectrum, the Fourier transform, and the other concepts in signal
processing.

However, signal processing does not commonly consider modules. In
particular, signals are not viewed as elements of a module, but, in
the discrete case considered here, as sequences of numbers from the
base field over some index range. If the index range is fixed, e.g.,
$I = \{0,\dots,n-1\}$, then the corresponding set of signals, e.g.,
$\C^n$, naturally is a vector space. The question is: How do we
formally associate a module to this vector space?  The answer is given
by the following definition of a (linear) signal model, which is
the central concept in the algebraic theory of signal processing.

We consider discrete complex signals $\coord{s}$, i.e., sequences
$\coord{s}\in\C^I$ of complex numbers over some index range
$I\subseteq\Z$. The set of signals is a vector space $V\leq\C^I$. For
finite~$I = \{0,\dots,n-1\}$, typically, $V =
\C^{\{0,\dots,n-1\}}=\C^n$.  If $I = \N,\Z$, we usually consider $V =
\ell^1(I)$, $V=\ell^2(I)$, or $V = \C^I$.

\begin{definition}[Linear Signal Model]\label{sigmoddef}
Let $V\leq\C^I$ be a vector space of complex signals over a discrete
index domain~$I$.  A {\em discrete linear signal model}, or just
signal model, for $V$ is a triple $(\alg,\md,\Phi)$, where $\alg$ is
an algebra of filters, $\md$ is an $\alg$-module of signals with
$\dim(\md) = \dim(V)$, and
\begin{equation}\label{signalmodel}
\Phi:\ V\rightarrow\md
\end{equation}
is a bijective linear mapping. If $\alg,\md$ are clear from the
context, we simply refer to $\Phi$ as the signal model.

Further, we transfer properties from $\md$ to the signal model.
For example, we say the signal model is regular or finite, if
$\md$ is regular or finite (-dimensional), respectively.
\end{definition}

Note that the definition of the signal model has linearity built in (due
to the operation of $\alg$ on $\md$) in accordance with the algebraic
theory being a theory of {\em linear} signal processing.

\mypar{Remarks on signal model}
Intuitively, a signal model endows the vector space $V\leq\C^I$ with
the structure of the module $\md$ as graphically displayed in
Figure~\ref{modelfig}. Via the signal model we can then identify
$\coord{s}\in V$ with the element in the module $s =
\Phi(\coord{s})$. As a consequence, filtering is now well-defined
and we get immediate access to all module-theoretic concepts
introduced in Section~\ref{algmodsm}, including spectrum, Fourier
transform, and several others  not yet introduced.

For example, if $\md$ is of dimension $n$ with
basis $b = (b_0,\dots,b_{n-1})$\footnote{In this paper $b$ will always
denote a basis and $b_i$ always basis elements, which should not be
confused with scalars such as $s_i,h_i$.} and $\coord{s}\in\C^n$, then
\begin{equation}\label{finsigmodform}
\Phi(\coord{s}) = s = \sum_{i=0}^{n-1}s_ib_i
\end{equation}
defines a signal model for $V = \C^n$. Conversely, if $\Phi$ is {\em
any} signal model for $V$ with canonical basis $e_i$ ($i$th element in
$e_i$ is 1; all other elements are 0), then the list of all
$b_i=\Phi(e_i)$ is a basis of $\md$ (since $\Phi$ is bijective) and
thus $\Phi$ has the form in \eqref{finsigmodform}. In other words, the
definition of signal model implicitly chooses a basis in $\md$ and
$\Phi$ is dependent on this basis. In fact, we will later see examples
of signal models (associated to the DCTs) that differ {\em only} in this
choice of basis or $\Phi$, i.e., have the same algebra and module.

Definition~\ref{sigmoddef} makes it possible to apply different signal
models to the same vector of numbers. For example, we will later learn
that by applying a DFT or a DCT to a vector of length $n$ one is
implicitly adopting different signal models for the same finite-length vector.

From a strictly mathematical point of view, and in the algebraic
definition of signal model, the bijection~$\Phi$, in other words, the
usual $z$-transform in signal processing, serves simply to track the
basis chosen for the signal module~$\md$. This basis determines the
operation of the algebra on the vector space~$V$. 

We remark that Definition~\ref{sigmoddef} of the signal model and the
algebraic theory extends to the case of continuous signals. However,
in this ,we will not pursue this extension and limit ourselves to
discrete signals.

As an example, we show next that the $z$-transform, is the linear
mapping~$\Phi$ of a signal model in the sense of
Definition~\ref{sigmoddef}. For this reason, we will refer to the
linear mapping in other signal models as transforms, such as the
$C$-transform or the $P$-transform that we will introduce.

\mypar{Example: \boldmath$z$-transform}
We present the signal model for the $z$-transform. We choose as algebra 
$$
\alg =
\{\sum_{n\in\Z}h_nz^{-n}| (\dots,h_{-1},h_0,h_1,\dots)\in\ell^1(\Z)\}
$$ 
the set of all Laurent series with $\ell^1$ coefficient sequences, and
as module
$$
\md =
\{\sum_{n\in\Z}s_nz^{-n}| (\dots,s_{-1},s_0,s_1,\dots)\in\ell^2(\Z)\}
$$ 
the set of all Laurent series with $\ell^2$ coefficient sequences.
$\md$ is indeed an $\alg$-module as we discussed in
Section~\ref{algmodsm}.  We complete the definition of the signal
model by identifying the bijective linear mapping~$\Phi$ in
Definition~\ref{sigmoddef}. The ordinary $z$-transform will do
\begin{equation}\label{ztrafodef}
\begin{array}{rrcl}
\Phi: & \ell^2(\Z) & \rightarrow & \md = \Phi(\ell^2(\Z)),\\
& \coord{s} & \mapsto & s = S(z) = \sum_{n\in\Z}s_nz^{-n}.
\end{array}
\end{equation}
In summary, $(\alg,\md,\Phi)$ is a signal model for the vector space
$V = \ell^2(\Z)$. This signal model is, of course well-known and the
one commonly adopted in mainstream discrete-time signal processing.

After the $z$-transform is chosen, it becomes clear how to do
filtering. Namely, if $h=H(z)\in\alg$ and $s=S(z)\in\md$, then the
result $r = R(z)$ of filtering $s$ with $h$ is simply the product of
Laurent series
\begin{equation}\label{r-filtering}
R(z) = H(z)S(z).
\end{equation}
If we  work with the respective coefficient sequences, then the
$n$th coefficient $r_n$ of $r$ follows from~(\ref{r-filtering})
\begin{equation}\label{filterdef}
r_n = \sum_{i\in\Z}h_is_{n-i}.
\end{equation}
The signal model, the operation of~$\alg$ on~$\md$, and the choice of
transform makes clear the definition of~(\ref{filterdef})
or~(\ref{r-filtering}).  Without making explicit the algebraic
structure , the origin of~(\ref{filterdef}) as filtering is obscured.
The problem is that in~\eqref{filterdef} filtering is defined in terms
of {\em coordinates} with respect to a \emph{basis}, but the basis,
which explains the structure of~\eqref{filterdef}, is not provided.

Signal processing books emphasize the usefulness of the $z$-transform,
since common signal processing operations are conveniently expressed
in the $z$-domain. In algebraic terms this means that it is more
convenient to work with the explicit algebra and module rather than
with the vector spaces of coefficient sequences.  For this reason, we
believe it is necessary to identify the signal models
$(\alg,\md,\Phi)$ for all the spectral\footnote{We use the word
``spectral'' here for transforms such as DFT, DCT, and others, to
distinguish from other transforms (such as the $z$-transform), which
do not compute a spectrum of some sort.} linear transforms $\four$,
thus identifying $\four$ as the Fourier transform (in the algebraic
sense) for $\md$. This is one of the goals achieved by the algebraic
theory.

\subsection{Shifts, Shift-Invariance, and Commutative Algebras}
\label{shiftcomm}

So far, the only examples of algebras and modules used in signal
processing that we provided are those shown in Figure~\ref{map}. These
are associated with infinite discrete-time signal processing.  An
important question is which other algebras and modules actually occur
in discrete signal processing and why. It is possible to give a
preliminary answer to this question by introducing and requiring the
concept of {\em shift-invariance}.  We start by understanding what
``shift'' and ``shift-invariance'' means in our algebraic theory by
focusing first on the case where only one shift is available, i.e., on
1-D signals\footnote{In the sequel, we use 1-D and $m$-D to refer to
one-dimensional and $m$-dimensional signals with respect to the number
of indexing parameters of the signal. For example, a standard time
signal is 1-D, while signal like an image is a 2-D. We reserve the
word ``dimension'' to refer to the dimension of the signal space when
viewed as vector space.}. Then we extend the discussion to multiple
shifts.

\mypar{Shift} Defining transforms and processes on groups is common in
many areas. For example, in ergodic theory or in dynamical systems, a
probability space is associated with a mapping, which can be a shift,
that can take many different abstract forms depending on the
underlying space.  To be more specific, and following \cite{Gray:88},
the usual model in ergodic theory is a probability space $(\Omega ,
\mathcal{F}, {\cal P})$ and a measurable transformation ${\cal T}$
(often called the shift). The set $\Omega$ is assumed to consist of
infinite or finite duration sequences or waveforms and often assumed
to be a product space of the real line (or more generally a Polish
space). Measurable maps~$f$ are then defined on it. Of particular
interests are maps taking $\Omega$ into the real line or some subset
thereof. With such a mapping $f$, then $y(t)=f({\cal T}^t\omega)$ for
$t$ in some group gives a sequence or waveform for every $\omega \in
\Omega$. The mapping $f$ produces a shift-invariant mapping from
sequence (or waveform) to sequence (or waveform), which leads to a
general theory for general alphabets based on when ${\cal T}$ is
measure preserving (and hence the processes stationary). This general
setup works for time shifts and space shifts and most signals likely
to be of interest in applications.

This paper considers specific instantiations of this general theory
and looks for very particular forms of the shift as they have been
used in linear signal processing, or that may explain existing linear
transforms or may lead to new linear transforms. To achieve this, we
show that, in the algebraic theory, the shift has a particularly
simple interpretation.  The {\em shift} operator is a special filter,
and thus is an element\footnote{We write $x$ instead of $z^{-1}$ to
emphasize the abstract nature of the discussion. Later, this will
enable us to introduce without additional effort other shifts as
well.} $x\in\alg$. Further, it is common to require that {\em every}
filter $h\in\alg$ be expressed as a polynomial or series in the shift
operator $x$. Mathematically, this means that the shift operator
generates\footnote{This is not entirely correct, as, in a strict
sense, one element $x$ can only generate polynomials, not infinite
series. However, by completing the space with respect to some norm the
notion of generating can be expanded. We gloss over this detail to
focus on the algebraic nature of the discussion.} the
algebra~$\alg$. Since a similar statement holds also for multiple
shifts (discussed below):

\begin{center}
\framebox[1.1\width]{\rule[-1.3mm]{0mm}{5mm}
\bf shift(s)\quad
\boldmath$=$\quad chosen generator(s) of \boldmath$\alg$
}
\end{center}

\mypar{Shift-invariant algebras} A key concept in signal processing is
{\em shift-invariance}. In the algebraic theory this property takes a
very simple form. Namely, if $x$ is the shift operator and $h$ a filter, then
$h$ is shift-invariant, if, for all signals $s$, $h(xs) = x(hs)$,
which is equivalent to $hx = xh$. Requiring shift-invariance for all
filters $h$ thus means
\begin{equation}\label{shiftinv}
x\cdot h = h\cdot x,\quad
\text{for all }h\in\alg.
\end{equation}
Since  $x$ generates $\alg$, $\alg$ is necessarily
commutative, and~\eqref{shiftinv} is of
course guaranteed. Conversely, if $\alg$ is a commutative algebra and $x$
generates $\alg$, then all filters $h\in\alg$ are
shift-invariant.\footnote{The requirement of ``$x$ generating $\alg$''
is indeed necessary as there are linear shift-invariant systems that
cannot be expressed as convolutions, i.e., as series in $x$; see
\cite{Sandberg:98}.} This observation is simple but crucial, and it also 
holds for multiple shifts (discussed below):
%
\begin{center}
\framebox[1.06\width]{\rule[-1.3mm]{0mm}{5mm}
\bf shift-invariant signal model\quad
\boldmath$\Leftrightarrow$\ \ $\alg$ is commutative
}
\end{center}
In particular, shift-invariance is a property of the algebra, and not
of the chosen module (signal space) in a signal model. However,
different choices of modules will, in general, produce different
signal models as we will see later.

{\bf Which algebras are shift-invariant?} We can now ask which
algebras lead to shift-invariant signal models, or equivalently, which
algebras $\alg$ are commutative and generated by one element $x$? In
fact, if $\alg$ is generated by one element it is necessarily
commutative; in other words, signal models with just one shift are
always shift-invariant. This is different in the case of multiple
shifts discussed below.

In the case of one shift, we have to identify those algebras that are
generated by one element $x$. In the infinite-dimensional case, we get
algebras of series in $x$ or polynomials of arbitrary degree in $x$.
In the finite-dimensional case, these algebras are precisely the {\em
polynomial algebras}
$$
\alg = \C[x]/p(x),\quad\text{$p$ a polynomial of degree $n$}.
$$ 
$\C[x]/p(x)$ is the set of all polynomials of degree less than $n$ with
addition and multiplication modulo $p(x)$. As a vector space, $\alg$ has
dimension $n$.

Thus, using only shift-invariance as a requirement, we have identified
one of the key players in the algebraic theory of signal processing,
namely polynomial algebras. They provide the signal
models  for many  transforms,
such as the DFT, DCT, and others, and  for several new
transforms.  This observation motivates our
Section~\ref{polyalgs}, which develops the general theory of signal
processing using polynomial algebras by specializing the general
algebraic theory in Section~\ref{algmodsm}.

In the remaining discussion on shift-invariance, we consider the
situation where several shifts are available and the relationship
between polynomial algebras and group algebras. The reader may want to
skip this part at  first reading and proceed with
Section~\ref{modmanipulation}.

\mypar{Multiple shifts} In general, if $m$-D signals are considered,
$m$ shift operators $x_1,\dots,x_m$ are available. These may operate
along different dimensions of the signal as in the usual separable
case, but can also take different forms as shown in
Figure~\ref{2dshifts} for non-separable models that we derived using
the present algebraic theory.

The above discussion on one shift is readily extended to multiple
shifts but there are some differences. Again, the $x_1,\dots,x_m$
generate $\alg$ and shift-invariance becomes
$$
x_i\cdot h = h\cdot x_i\quad\text{for all}\,\,\,h\in\alg,1\leq i\leq k,
$$ 
which is  equivalent to $\alg$ being commutative. 
We can reduce this condition to 
\begin{equation}\label{commpairs}
x_i\cdot x_j = x_j\cdot x_i.
\end{equation}
In words, a signal model with $m$ shifts is shift-invariant if and only if
the shifts commute in pairs. 

Commutative algebras generated by $m$ elements include multivariate
series (e.g., Laurent series in more than one variables).

For an exact classification, we restrict ourselves to algebras
generated by $x_1,\dots,x_m$ in the strict sense, i.e., those
containing only multivariate polynomials, no series. In signal
processing terms, this is equivalent to $\alg$ containing only 
FIR filters. In particular, every signal model for a finite set of 
samples, i.e., with $\dim(\md)<\infty$ falls into that class.

Commutative algebras generated by $\multvar{x} = (x_1,\dots,x_m)$ are
precisely all multivariate {\em polynomial algebras} (the notation is
explained in Appendix~\ref{algdefs} together with the Chinese
remainder theorem)
\begin{equation}\label{poly-algebras}
\alg = 
  \C[\multvar{x}]/
    \langle p_1(\multvar{x}),\dots,p_k(\multvar{x})\rangle,
\end{equation}
where $p_i(\multvar{x})$, $1\leq i \leq k$, are polynomials in $m$
variables. In words, $\alg$ is the algebra of all polynomials in $m$
variables with addition and multiplication defined modulo the $k$
polynomials $p_i$.  Equivalently, $\alg$ is the algebra of {\em all}
polynomials in $m$ variables, with the restriction that the equations
$p_1(\multvar{x}) = \dots = p_k(\multvar{x}) = 0$ have been
introduced. Mathematically, $\langle
p_1(\multvar{x}),\dots,p_k(\multvar{x})\rangle$ is the {\em ideal} of
$\C[\overline{x}]$ generated by the $p_i$, and the polynomial algebra
is  called a {\em quotient algebra}. 
Note that if $k = 1$, i.e., $p = p_1$, then we write
simply $\C[x]/p(x)$ instead of $\C[x]/\langle p(x)\rangle$ as we 
did already above.

As a remark, we observe that the polynomial algebra $\alg$ can be of
infinite or finite dimension. For example, for $m = 1, k = 0$, we get
$\C[x]$, which is of infinite dimension but with countable basis. The
primary example in this paper, discussed above, is the case $m=1,
k=1$, i.e., $\alg = \C[x]/p(x)$, for some finite degree polynomial
$p(x)$. This algebra is of finite dimension.

Intuitively, if $m$ is given, we need at least $k=m$ polynomials $p_i$
in \eqref{poly-algebras} to make the dimension finite. However,
conversely, choosing $k=m$ polynomials does not guarantee the
polynomial algebra~$\alg$ to be finite-dimensional, unless $k = m =
1$. Also, it is known that for $m>1$ a polynomial algebra can have
arbitrary large $k$, no matter how the polynomials $p_i$ are chosen
\cite{Cox:97}.

The 2-D signal models referred to in Figure~\ref{2dshifts}, namely for
spatial signals residing on a finite hexagonal or quincunx lattice,
are indeed shift-invariant (and regular). The associated polynomial
algebras have $k = m = 2$.

Next, we briefly discuss Fourier analysis on groups to put it
into the context of the algebraic theory.

\mypar{Fourier analysis on groups} Let $G$ be a finite group. Take the
elements of the group to be a basis for the following vector space
$$
\C[G] = \myleft\{\sum_{g\in G}a_gg\mid a_g\in\C\myright\}.
$$ 
Clearly, $\C[G]$ is a vector space, spanned by the group
elements. It is also clear that we can define in a standard way
multiplication of elements in $\C[G]$ by using the distributive law
and the multiplication of group elements. Thus, $\C[G]$ is an algebra.
Another point of view is to regard $\C[G]$ as the set of complex
functions $g\mapsto a_g$ on the group $G$. The regular module $\md =
\alg = \C[G]$ provides a signal model in the sense of
Definition~\ref{sigmoddef}. Namely, if $G$ has $n$ elements, we can 
set $\alg = \md = \C[G]$, and 
\begin{equation}\label{groupsigmod}
\begin{array}{rrcl}
\Phi: & \C^n & \rightarrow & \C[G]\\
& \coord{s} & \mapsto & \sum_{g\in G} s_g g.
\end{array}
\end{equation}
In particular, both signals and filters are elements of the group
algebra in this case. The study of the signal models $(\alg,\md,\Phi)$
in \eqref{groupsigmod} is the area of Fourier analysis on finite
groups (briefly discussed in the introduction), which thus becomes an
instantiation of the algebraic theory of signal processing.

According to the notion of shift introduced above, we also have shift
operators in a group $G$, namely the elements of the chosen generating
set for $G$. Note that, unless the group is cyclic, $G$ and thus
$\C[G]$ requires at least two generators, i.e., shifts. If $G$ is
not commutative, then the generators will not commute in pairs, i.e.,
violate \eqref{commpairs}. Thus the signal model \eqref{groupsigmod}
is {\em shift-variant}.

An immediate question is how polynomial algebras and group algebras
differ. Since polynomial algebras are always commutative, it is clear
that for a non-commutative group the associated group algebra cannot
be a polynomial algebra. On the other hand, it is known that {\em
every} group algebra for a commutative group is a polynomial algebra
with a very specific structure. Namely, a commutative group $G$
is always the direct product of cyclic groups $G =
Z_{n_1}\times\dots\times Z_{n_m}$, where $Z_{n_i}$ is of size $n_i$
and is generated by $x_i$; thus we get
\begin{equation}\label{polygroup}
\C[G] = \C[x_1,\dots,x_m]/\langle \{x_i^{n_i}-1 | 1\leq i\leq m\}\rangle.
\C[G] = \C[x_1,\dots,x_m]/\langle \{x_i^{n_i}-1 | 1\leq i\leq m\}\rangle.
\end{equation}
In the case of one variable (one-dimensional signals), $G$ is
necessarily cyclic, $G = Z_n$, and we have
\begin{equation}\label{polygroup1}
\C[Z_n] = \C[x]/(x^n-1).
\end{equation}
This algebra is known to be associated to the DFT (of size $n$) as we
will discuss later. Comparing \eqref{polygroup} and
\eqref{polygroup1} to general polynomial algebras in
\eqref{poly-algebras}, it becomes clear that only 
very few polynomial algebras are also group algebras.

Historically, the observation that the DFT is associated to $\C[Z_n]$
spawned a significant effort in developing Fourier analysis for other,
non-commutative groups and in deriving their fast algorithms.  The
observation that non-commutative groups produce shift-variant signal
models may explain why these groups have to date not found many
applications in signal processing.

In this paper, we investigate polynomial algebras, i.e.,
shift-invariant systems, and we will show that algebras other than
$\C[x]/(x^n-1)$, and thus not related to groups, are indeed relevant in
signal processing.  Before we identify these algebras, we will provide
a general discussion of signal processing on polynomial algebras
$\C[x]/p(x)$ in Section~\ref{polyalgs} that specializes the algebraic
concepts from Section~\ref{algmodsm} to this specific case.

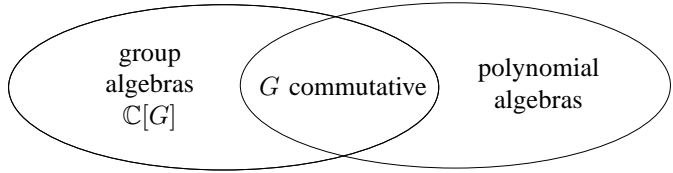
\begin{figure}[ht]\centering
\isdraft{\renewcommand{\baselinestretch}{1}\small}{}
\begin{picture}(250,50)
\put(125,25){\makebox(0,0){\includegraphics[scale=0.55]{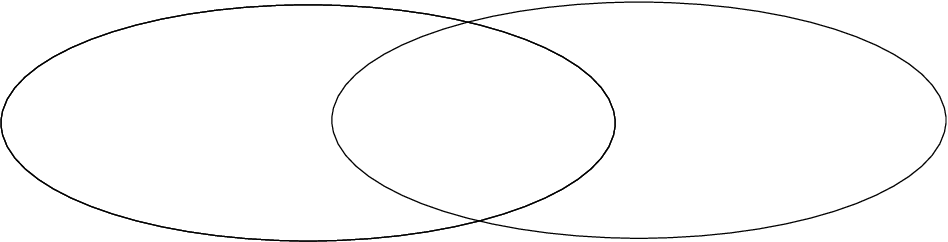}}}
\put(53,25){\makebox(0,0)
  {\begin{tabular}{c}group\\algebras\\$\C[G]$\end{tabular}}}
\put(200,25){\makebox(0,0)
  {\begin{tabular}{c}polynomial\\algebras\end{tabular}}}
\put(126,25){\makebox(0,0){%
$G$ commutative}}
\end{picture}
\caption{The intersection of the set of group algebras $\C[G]$ and the
set of polynomial algebras is precisely the group algebras for
commutative groups $G$. In particular, for one shift (one variable,
one group generator) there is only the common element $\C[Z_n] =
\C[x]/(x^n-1)$, which is associated to the
$\DFT_n$.\label{grouppolyalg}}
\end{figure}

\subsection{Visualization of a Signal Model}

A given signal model can be visualized by a graph (see
Appendix~\ref{algdefs}, Definition~\ref{graph}), which provides an
intuitive understanding of the model. We will use these visualizations
later.

\begin{definition}[Visualization of Signal Model]\label{visdef}
Assume that a signal model $(\alg,\md,\Phi)$ is given and 
$$
\Phi:\ \coord{s}\mapsto\sum_{n\in I}s_nb_n,
$$ 
where the $b_n$ form a basis $b$ of $\md$. Denote the chosen shift
operators, i.e., generators, of the corresponding algebra $\alg$ by
$x_1,\dots,x_n$.  Further, assume that $\phi$ is the representation of
$\alg$ afforded by $\md$ with basis $b$.  Then each $\phi(x_i)$ is an
infinite or finite matrix (which we call {\em shift matrix}) and can
be viewed as the adjacency matrix of a weighted graph ${\cal
G}_i$. Each of these graphs has the same vertices corresponding to
$b$. Thus we can join these graphs by adding the adjacency matrices of
the ${\cal G}_i$ to obtain a graph ${\cal G}$.

We call the graph ${\cal G}$ a {\em visualization} of the signal
model $(\alg,\md,\Phi)$.
\end{definition}

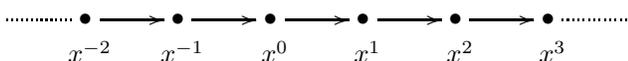
\begin{figure}[ht]\centering
\isdraft{\renewcommand{\baselinestretch}{1}\normalsize}{}
{
\centerline{
\xymatrix{
\ar@{.}[r] & \bullet \ar[r] & \bullet \ar[r] & 
\bullet \ar[r] & \bullet \ar[r] &
\bullet \ar[r] & \bullet \ar@{.}[r] &
}}
\centerline{
\isdraft{\setlength{\unitlength}{1.15pt}}{}
\begin{picture}(175,5)
\put(0,0){\makebox(0,0){$x^{-2}$}}
\put(35,0){\makebox(0,0){$x^{-1}$}}
\put(70,0){\makebox(0,0){$x^0$}}
\put(105,0){\makebox(0,0){$x^1$}}
\put(140,0){\makebox(0,0){$x^2$}}
\put(175,0){\makebox(0,0){$x^3$}}
\end{picture}}
}
\caption{Visualization of the signal model $z$-transform as a graph.
\label{timegraph}}
\end{figure}

Intuitively, the graph provides the topology imposed by the signal
model. This will become clear by looking at the various examples shown
in this paper. As a first example, we use the only signal model we
have identified so far, the $z$-transform defined in
\eqref{ztrafodef}. The chosen basis in $\md$ consists of the monomials
$b = (x^n)_{n\in\Z}$. The one available shift operator $x$ operates on $b$ as
$x\cdot x^n = x^{n+1}$. In coordinates, this means that $\phi(x)$ is a
doubly infinite matrix with ones on the lower diagonal and zeros else.
The graph that has $\phi(x)$ as adjacency matrix is shown in
Figure~\ref{timegraph}. The vertices are the base elements $b_n$, the
edges show the shift.

\subsection{Module Manipulation and Commutative Diagrams}
\label{modmanipulation}

A convenient tool when working with modules are $\alg$-module
homomorphisms (Definition~\ref{hommoddef} in Appendix~\ref{algdefs}). In
particular, these mappings are linear mappings, and, in many cases
considered in this paper simply base changes within one module. Since
we always work with explicitly chosen bases, in the finite-dimensional
case these mappings are represented by matrices. Formally, if
$\md_1,\md_2$ are $\alg$-modules with chosen bases, and $B$ is the
matrix representing the $\alg$-module homomorphism w.r.t.~these bases,
then we write
$$
\md_1\stackrel{B}{\longrightarrow}\md_2
$$
Above, we already encountered an example for this arrow notation: the
first row in \eqref{moduledec} maps $\md$ to its decomposition;
w.r.t.~chosen bases, the mapping corresponds to the matrix form of the
Fourier transform $\four$ (again, we consider only the
finite-dimensional case here). If two or more mappings between modules
are connected, then the associated matrices obey the following rule
(the $B_i$ are the base change matrices):
$$
\md_1\stackrel{B_1}{\longrightarrow}\md_2
  \stackrel{B_2}{\longrightarrow}\md_3
$$
implies
$$
\md_1\stackrel{B_2B_1}{\longrightarrow}\md_3.
$$ 
This identity shows that a given mapping (or matrix) from $\md_1$ to
$\md_3$ can be factorized via a module $\md_2$. With the proper choice
of inserted module, this will be a crucial tool for the derivation of
fast algorithms \cite{Pueschel:05}. In this paper, we use it for the
commutative diagrams introduced below.

Another important identity is
$$
\md_1\stackrel{B}{\longrightarrow}\md_2
\quad\text{implies}\quad
\md_2\stackrel{B^{-1}}{\longrightarrow}\md_1
$$
provided that $B$ is invertible.

The second important algebraic tool is the use of {\em commutative
  diagrams}, which are built from the above module mappings. A typical
  example looks like
\begin{equation}\label{commdef}
\algogen{\md_1}{B_1}{\md_2}{B_2}{B_3}{\md_3}{B_4}{\md_4}
\end{equation}
The term ``commutative'' signifies that different paths connecting
two given modules yield the same associated matrix. For example, in
the above diagram, we can connect $\md_1$ to $\md_4$ in two different
ways: via $\md_2$ or via $\md_3$. The diagram implies the identity
$$
B_4B_2 = B_3B_1.
$$

\section{Signal Processing on Polynomial Algebras}\label{polyalgs}

In Section~\ref{shiftcomm} we have learned that shift-invariance leads
naturally to polynomial algebras $\alg$ in the signal model
$(\alg,\md,\Phi)$.  In particular, in the case of finite 1-D signal
models, these algebras are necessarily of the form $\C[x]/p(x)$. With
this motivation, we investigate what it means to do signal processing
using these algebras. We do this by specializing the general theory
from Section~\ref{algmodsm} to $\alg =\C[x]/p(x)$ focusing on the
regular case $\md = \alg$.

We start with the definition of polynomial algebras in one variable,
then we investigate their spectral decomposition, their associated
Fourier transforms with diagonalization properties, and formulate a
general convolution theorem. The results provide the general framework
for signal processing using the DFTs, DCTs, and DSTs.

\subsection{Polynomial Algebras in One Variable} 

Let $p(x)$ be a polynomial of degree $\deg(p) = n$.  Then, $\alg =
\C[x]/p(x) = \{h(x)\mid \deg(h) < n\}$, the set of residue classes
modulo $p$, is an algebra with respect to the addition of polynomials,
and the polynomial multiplication modulo $p$. We call $\alg$ a {\em
polynomial algebra (in one variable)}. Polynomial algebras are always
{\em cyclic}, i.e., generated by one suitable element, usually chosen
to be $x$, which is then called the shift operator (see
Section~\ref{shiftcomm}). This means that all elements in $\alg$ are
obtained by repeatedly forming powers, sums, and scalar multiples from
$x$. In other words, the elements are polynomials in $x$ as implicit
from the definition above.

\mypar{Example}
As a simple instructive example, we consider $p(x) = (x-1)(x+1) = x^2
- 1$. Multiplying the two elements $x, x+1\in\alg$, we get
\begin{equation}\label{modexample}
x(x+1) = x^2 + x \equiv x + 1\text{ mod }(x^2 - 1)
\end{equation}
by replacing $x^2$ with $1$. We read \eqref{modexample}
as ``$x^2+x$ is congruent (or equal) to $x+1$ modulo
$x^2-1$.'' Thus, we do not use ``mod'' as an operator, as in
$$
(x^2+x)\text{ mod }(x^2-1) = x+1,
$$
but to denote equality of two polynomials modulo a third polynomial.

\subsection{Signal Model} 

We choose as vector space $V=\C^n$, as algebra a polynomial algebra
$\alg = \C[x]/p(x)$, $\deg(p) = n$, and as module the
regular\footnote{A more general discussion would consider arbitrary
modules for $\alg$; however, it turns out that most of the signal
models actually used in signal processing have regular modules. The
non-regular cases occurring in this paper will be studied as they
arise.} module $\md = \alg$. Further, we choose a basis $b =
(p_0,\dots,p_{n-1})$ of $\md$. As we show next, this provides a
finite, $n$-dimensional signal model $(\alg,\md,\Phi)$ in the sense of
Definition~\ref{sigmoddef}.  Namely, if $\coord{s} =
(s_0,\dots,s_{n-1})^T\in\C^n$, we can define the bijective linear
map~$\Phi$ as
\begin{equation}\label{polysigmod}
\Phi:\ \C^n\rightarrow\md,\ \coord{s}\mapsto \sum_{0\leq\ell<n}s_\ell p_\ell.
\end{equation}
Note that $\Phi$ depends on the chosen basis $b$. The basis elements
$p_i$ are the unit impulses in $\md$, i.e., those with coordinate
vector $s_\ell=0$ for $\ell\neq i$, and $s_i = 1$. The impulse
response of a filter $h\in\alg$ for the impulse $p_i$ is $hp_i\in\md$.

The mapping $\Phi$ endows the vector space $\C^n$ with the structure
of the $\alg$-module $\md$ with basis $b$.  Thus, we can now identify
$\C^n$ and $\md$, which becomes our signal space.  $\Phi$ in
\eqref{polysigmod} is the equivalent of the $z$-transform for this
model.

\mypar{Example} Continuing our previous example, we choose the basis 
$b=\{1,x\}$ in $\md = \alg = \C[x]/(x^2-1)$. A signal model for the
vector space $\C^2$ is now provided by
$$
\Phi:\ \C^2\rightarrow\md,\ (s_0,s_1)\mapsto s_0 + s_1x.
$$

\subsection{Filtering} 

As seen before, the signal model defines filtering on the
signal space $\md$ through the operation of $\alg$ on $\md$. The
algebra $\alg$ is the space of filters and the $\alg$-module~$\md$ the
space of signals. We mention again that even though the sets $\md$ and $\alg$ are equal,
their algebraic structure (i.e., the role or structure assigned to the
sets) is not. For example, $\alg$ operates on $\md$, not vice-versa,
and filters (elements of $\alg$) can be cascaded, i.e., multiplied,
which signals (elements of $\md$) cannot.

We can represent filtering in either a direct way or in
coordinates. Let $s =
\Phi(\coord{s}) = \sum_{0\leq\ell<n}s_\ell p_\ell\in\md$ be a signal and
$h\in\alg$ be a filter. Then, filtering $s$ with $h$ is simply the
product
\begin{equation}\label{filt1}
h\cdot s\in\md,
\end{equation}
i.e., the product of the polynomials $h$ and $s$ modulo $p$.  Note
that since the result is in $\md$, it is again a signal, as desired.
Filtering, or multiplication by $h$, is a linear mapping, and so it
has a matrix representation w.r.t.~the basis $b$.  This matrix is
given by $\phi(h)$, where $\phi$ is the representation of $\alg$
afforded by $\md$ with basis $b$ (see \eqref{Aop}). Thus, 
\eqref{filt1} becomes in coordinate form
\begin{equation}\label{filt2}
\phi(h)\cdot\coord{s}\in\C^n.
\end{equation}
In particular, we call the matrix $\phi(x)$ corresponding to the shift
$x$ the {\em shift matrix}.

By itself, \eqref{filt2}  does not  reveal the underlying
structure provided by the $\alg$-module $\md$. This
structure is explicit in the coordinate-free representation of
filtering given in \eqref{filt1}.

\mypar{Example}
In our example, let $h = h_0 + h_1x\in\alg=\C[x]/(x^2-1)$ be an
arbitrary filter\footnote{Note that the following derivation {\em does
not} depend on the chosen basis in $\alg$; in fact, we view $h$ here
as a polynomial, not as a filter expressed in the basis $\{1,x\}$
(admittedly a subtle difference).}. To compute its matrix
representation $\phi(h)$ w.r.t.~the basis $b = (1,x)$ of $\md$, we
apply the filter to the base vectors (unit impulses) to obtain their
impulse responses; the coefficient vectors of these responses are the
columns of $\phi(h)$. We have $h\cdot 1 = h_0+h_1x\in\md$ and $h\cdot
x = h_0x + h_1x^2\equiv h_1 + h_0x\text{ mod }(x^2-1)$. Thus
$$
\phi(h) = 
\begin{bmatrix}
h_0 & h_1\\h_1 & h_0
\end{bmatrix}
$$
and
$$
h\cdot s\Leftrightarrow\phi(h)\cdot\coord{s}.
$$
In particular, the shift matrix is given by
$$
\phi(x) = 
\begin{bmatrix}
0 & 1\\1 & 0
\end{bmatrix}.
$$

\subsection{Visualization}

The visualization of the signal model $(\alg,\md,\Phi)$ with $\Phi$ in \eqref{polysigmod} is
the graph with $n$ vertices that has the shift matrix $\phi(x)$ as
adjacency matrix (see Definition~\ref{visdef}). In the general case
considered in this section, this matrix has no apparent structure.

The signal models that are actually used in signal processing,
however, do have structure. In particular, $\phi(x)$ is always very
sparse for these models. Deriving and explaining this structure is one
of the tasks that we solve in this paper.

\subsection{Spectrum and Fourier Transform} 

Unless stated otherwise, we assume that $p(x)$ is a {\em separable}
polynomial, i.e., 
$$
p(x) = \prod_{k=0}^{n-1} (x - \alpha_k),\quad
\alpha_k\neq\alpha_\ell,\text{ for }k\neq\ell.
$$
We set $\alpha = (\alpha_0,\dots,\alpha_{n-1})$. In words,
separability means that $p$ has no zeros of multiplicity larger than
1. We will see below that this property ensures that the spectrum 
of $\md$ consists exclusively of one-dimensional spectral components.

The Fourier transform, or spectral decomposition, of the regular
module $\md = \C[x]/p(x)$ is given by the Chinese remainder theorem
(CRT; stated in Theorem~\ref{crt} in Appendix~\ref{algdefs}) and, as
filtering above, can be expressed in a direct, coordinate-free way,
or, alternatively, using coordinates w.r.t.~a given basis.

In a coordinate-free form, the Fourier transform is given by the
mapping
\begin{equation}\label{polyalgdec}
\begin{array}{r@{\ }c@{\ }l}
\Delta:\ \C[x]/p(x) & \rightarrow &
\C[x]/(x-\alpha_0)\dirsum\dots\dirsum\C[x]/(x-\alpha_{n-1}), \\
s = s(x) & \mapsto & (s(\alpha_0),\dots,s(\alpha_{n-1})).
\end{array}
\end{equation}
Each $\md_k = \C[x]/(x-\alpha_k)$ is of dimension~1. So the elements
(vectors) of $\C[x]/(x-\alpha_{k})$ are polynomials of degree 0 or
scalars $c\in\C$. Further, $\md_k$ is an $\alg$-module, since for
$h=h(x)\in\alg$ and $c\in\md_k$,
$$
h(x)\cdot c\equiv h(\alpha_k)\cdot c\text{ mod }(x-\alpha_k),
$$
i.e., the result is again in $\md_k$. Since $\md_k$ is of dimension 1, it 
is irreducible.

The scalars $s(\alpha_k)$ in \eqref{polyalgdec} are the spectral
components of~$s$.  The mapping in \eqref{polyalgdec} simultaneously
projects a signal (i.e., polynomial) $s\in\C[x]/p(x)$ into the 
modules $\C[x]/(x-\alpha_k)$. This projection is precisely the
evaluation
$$
s(x)\equiv s(\alpha_k)\text{ mod }(x-\alpha_k).
$$
The set of one-dimensional irreducible submodules $\md_k =
\C[x]/(x-\alpha_k)$ is the {\em spectrum} of the signal space $\md$.
Each submodule $\md_k$ is a simultaneous eigenspace to all filters (or
linear systems) in $\alg$. The spectrum of a signal $s\in\md$ is the
vector $\Delta(s) = (s(\alpha_0),\dots,s(\alpha_{n-1}))$.

\mypar{Example}
In our running example, the Fourier transform is given by
\begin{equation*}
\begin{array}{r@{\ }c@{\ }l}
\Delta:\ \C[x]/(x^2-1) & \rightarrow &
\C[x]/(x-1)\dirsum\C[x]/(x+1), \\
s = s(x) & \mapsto & (s(1),s(-1)).
\end{array}
\end{equation*}
The Fourier transform of the signal $s(x)=1+3x$ is $(4,-2)$.
 
\subsection{Frequency Response}

Filtering in the regular module $\md = \C[x]/p(x)$ becomes parallel
filtering in the frequency domain, i.e., on the irreducible
$\alg$-modules $\md_k$. Namely, let $h\in\alg$ be any filter and let
$s(\alpha_k)\in\md_k$ be a spectral component of a signal $s$. Then
the filtering by~$h$ of the signal given by this spectral component
$s(\alpha_k)$ is
$$
h(x)\cdot s(\alpha_k)\equiv h(\alpha_k)s(\alpha_k)\text{ mod }(x-\alpha_k).
$$
This shows that $\md_k$ affords the irreducible representation
$\phi_k$ that maps
\begin{equation}\label{polyirred}
\phi_k:\ h=h(x)\mapsto h(\alpha_k).
\end{equation}
The collection of the $\phi_k(h)$, namely
$(h(\alpha_0),\dots,h(\alpha_{n-1}))$ is the frequency response of the
filter~$h$. This means that the $k$th spectral component $s(\alpha_k)$
of a {\em signal} $s = s(x)$ is obtained in the same way as the
frequency response $h(\alpha_k)$ at $\alpha_k$, namely by evaluating
polynomials.  This is due to the simple structure of polynomial
algebras; in general, this is not the case. In general, the spectral
component is a $d$-dimensional vector ($d$ the dimension of the
spectral component $\md_k$), while the frequency response of a
filter~$h(x)$ at this component is a $d\times d$ matrix.  They have
the same dimensionality only for $d=1$.

The irreducible representations $\phi_k$ of $\alg =
\C[x]/p(x)$ are all different and one-dimensional, since $p$ is separable.
The following lemma states that these are indeed all one-dimensional
representations.

\begin{lemma}\label{allirreds}
Let $\alg = \C[x]/p(x)$, where $p$ is separable with zeros
$\alpha_0,\dots,\alpha_{n-1}$ and let $\phi:\ \alg\mapsto\C$ any
representation, i.e., homomorphism of algebras. Then $\phi$ is one of
the $\phi_k$ in
\eqref{polyirred}.
\end{lemma}
\begin{proof}
Let $\phi:\ \alg\mapsto\C$ be a representation. Then, because $\phi$
is a homomorphism of algebras (Definition~\ref{alghomdef} in
Appendix~\ref{algdefs}), $p(\phi(x)) =
\phi(p(x)) = \phi(0)=0$. Thus $\phi(x) = \alpha_k$ for a suitable $k$ and
$\phi(h(x)) = h(\phi(x)) = h(\alpha_k)$ as desired. 
\end{proof}

\mypar{Example}
The frequency response of $h\in\C[x]/(x^2-1)$ is $(h(1), h(-1))$.
Filtering $h\cdot s$ in $\md$ is equivalent to the point-wise product
$(h(1)s(1), h(-1)s(-1))$ in the frequency domain.

\subsection{Fourier Transform as a Matrix}

The Fourier transform $\Delta$ is a linear mapping, which can thus be
expressed by a matrix $\four$ after bases are chosen.  We will call
this matrix also a Fourier transform for $\md$. To compute this
matrix, we choose the basis $b = (p_0,\dots,p_{n-1})$, provided by
$\Phi$ in \eqref{polysigmod} for $\md$, and the basis $b_k = (1)$ (the
list containing the polynomial $x^0 = 1$) for each summand
$\C[x]/(x-\alpha_k)$. To compute $\four$, we apply $\Delta$ to the
base vectors $p_\ell$; the coordinate vectors of the result constitute
the columns of $\four$. We compute $\Delta(p_\ell)$ as the projection
$$
p_\ell(x)\equiv p_\ell(\alpha_k)\text{ mod }(x-\alpha_k).
$$
Consequently, the $\ell$th-column of $\four$ is the result
of applying $\Delta$ to the base vector~$p_\ell$, namely
$(p(\alpha_0),\dots,p(\alpha_{n-1}))^T$.
Taken together, $\four$ has the form
\begin{equation}\label{polytrafo}
\four = \poly_{b,\alpha} = [p_\ell(\alpha_k)]_{0\leq k,\ell < n},
\end{equation}
We call $\poly_{b,\alpha}$ a {\em polynomial transform}. It is
uniquely determined by the signal model $\alg=\md=\C[x]p(x)$
and $\Phi$ in \eqref{polysigmod}, which fixes $b$.

This definition coincides with the notion of a polynomial transform in
\cite{Driscoll:97,Potts:98} and is related but different from the use
in \cite{Nussbaumer:79}. In \cite{Kailath:97}, polynomial transforms
are called polynomial Vandermonde matrices.

Note that $\poly_{b,\alpha}$ can have entries equal to zero, but, as
an isomorphism (as stated by the CRT), it is necessarily invertible.

Let $s = s(x) = \sum s_\ell p_\ell(x)\in\md$ be a signal.
Then, in coordinates, $\Delta$ in
\eqref{polyalgdec} becomes the matrix-vector product
\begin{equation}\label{eval}
\Delta(s)\Leftrightarrow
\poly_{b,\alpha}\cdot\coord{s} =
(s(\alpha_0),\dots,s(\alpha_{n-1}))^T\in\C^n.
\end{equation}
In other words, \eqref{eval} computes the spectrum of $\coord{s}$
w.r.t.~to the signal model $(\C[x]/p(x),\C[x]/p(x),\Phi)$ with $\Phi$
defined in \eqref{polysigmod}.

Like we saw with filtering before, the coordinate form \eqref{eval} of
the Fourier transform alone does not readily reveal the underlying
signal model, i.e., the algebra and module.

The Fourier transform (in matrix form) for $\md = \C[x]/p(x)$ with
basis $b$ is not uniquely determined. The degree of freedom is in the
choice of bases in the irreducible submodules $\C[x]/(x-\alpha_k)$ of
$\md$ in \eqref{polyalgdec}. If we choose generic bases $b_k = (a_k)$,
$a_k\neq 0$, in $\C[x]/(x-\alpha_k)$, $0\leq k < n$, then the
corresponding Fourier transform is given by the {\em scaled}
polynomial transform
$$
\diag(1/a_0,\dots,1/a_{n-1})\cdot\poly_{b,\alpha}.
$$
We restate this in the following theorem.

\begin{theorem}[Fourier Transforms]\label{fourtrafo}
The matrix $\four$ is a Fourier transform for the regular module $\md
= \alg = \C[x]/p(x)$ ($p$ separable), if and only if $\four$ is a scaled
polynomial transform of the form
\begin{equation}\label{scaledpoly}
\four = \diag(1/a_0,\dots,1/a_{n-1})\cdot\poly_{b,\alpha},
\end{equation}
where $a_k\neq 0$ and $b$ is a basis for $\md$. As a consequence,
any such $\four$ is invertible.
\end{theorem}

Theorem~\ref{fourtrafo} is represented by the following commutative
diagram (introduced in \eqref{commdef}):
\begin{equation*}
\xymatrix{
\C[x]/p(x) \ar[dd]^{\dps\four} \ar[rr]^{\dps\one_n} &&
  \C[x]/p(x) \ar[dd]^{\dps\poly_{b,\alpha}} \\ \\
\bigdirsum\C[x]/(x-\alpha_k) \ar[rr]^{\dps\diag(a_k)} && 
  \bigdirsum\C[x]/(x-\alpha_k)
}
\end{equation*}

In the top row, we have two times the module $\C[x]/p(x)$ with the
same basis $b$, thus connected by the base change matrix given by the
identity $\one_n$.  We decompose the modules in two different ways: on
the right by $\poly_{b,\alpha}$, which implies the bases~$(1)$ in the
irreducible modules; and, on the left, by the scaled polynomial
transform $\four$ given in Theorem~\ref{fourtrafo}, which implies the
bases $(a_k)$ in the irreducible modules. The bottom row shows the
base change between the decomposed modules, which is
diagonal. Connecting the top right with the bottom left corner in two
different ways yields
\eqref{scaledpoly}. In short, the diagram shows that an arbitrary
Fourier transform differs from the unique polynomial transform only by
a diagonal matrix.

\mypar{Example}
We continue our previous example $p(x) = (x-1)(x+1)$ and choose the
basis $b = (1,x)$ in $\md$. The generic Fourier transform for $\md$ is
given by Theorem~\ref{fourtrafo} as
\begin{equation}\label{exfour}
\begin{array}{rcl}
\four & = & \diag(1/a_0, 1/a_1)\cdot
\begin{bmatrix}1&\phantom{-}1\\1&-1\end{bmatrix}\\
& = & \diag(1/a_0, 1/a_1)\cdot\DFT_2,
\end{array}
\end{equation}
where
$\DFT_2=[\begin{smallmatrix}1&\phantom{-}1\\1&-1\end{smallmatrix}]$.
In particular, $\DFT_2$ is the polynomial transform for $\md =
\C[x]/(x^2-1)$ with basis $b$.

Furthermore if $\coord{s}$ is the coordinate vector of a signal
$s\in\md$ w.r.t.~$b$, then
$$
\DFT_2\cdot\coord{s} = (s(1),s(-1))^T.
$$

\subsection{Diagonalization Properties and Convolution Theorems} 

The diagonalization property of any Fourier transform $\four$ of the
regular module $\md = \C[x]/p(x)$ is obtained as a special case of
\eqref{repdec} (see also Figure~\ref{repdecvis}).

\begin{theorem}[Diagonalization Properties]\label{diagprop}
Let $\four$ be a Fourier transform for the regular $\alg$-module
$\C[x]/p(x)$ with basis $b$ and corresponding representation $\phi$ of
$\alg$. Then 
\begin{equation}\label{polyeval}
\four\cdot A\cdot\four^{-1} = \diag(a_0,\dots,a_{n-1}),
\end{equation}
if and only if $A = \phi(h)$ for a filter $h\in\alg$. In this case $a_k =
h(\alpha_k)$, $0\leq k<n$, is the frequency response of $h$. 

In particular, $\four$ diagonalizes the shift matrix $\phi(x)$. The
shift operator $x$ has the frequency response
$(\alpha_0,\dots,\alpha_{n-1})$.
\end{theorem}
\begin{proof} Let $A=\phi(h)$. 
Then $\four\phi(h)\four^{-1}$ is diagonal, since it is the coordinate
representation of the filter $h$ in the frequency domain, which is the
diagonal matrix with the frequency response on the diagonal (all blocks
on the right in Figure~\ref{repdecvis} are $1\times 1$).

Conversely, the set of diagonal matrices
$\diag(a_0,\dots,a_{n-1})$ is an $n$-dimensional vector space. Since
$\four$ is invertible, the set of all matrices $A$ diagonalized
by $\four$ is also $n$-dimensional. Since $\alg$ is of dimension $n$,
and $\phi$ is injective, the set of all matrices $\phi(h)$ is a vector
space of dimension $n$ and thus the set of all matrices diagonalized
by $\four$.
\end{proof}

We also note that, using Theorem~\ref{diagprop}, we get immediately
the characteristic polynomial, trace, and determinant for every matrix
$\phi(h)$, since it is similar to the diagonal matrix
$\diag(h(\alpha_0),\dots,h(\alpha_{n-1}))$. In particular, the
characteristic polynomial of $\phi(x)$ is $p(x)$.

\mypar{Example} In our example, we obtain the diagonalization property 
of the $\DFT_2$, namely, for $h = h_0+h_1x\in\alg$,
$$
\DFT_2\cdot
\begin{bmatrix}
h_0 & h_1\\h_1 & h_0
\end{bmatrix}\cdot
\DFT_2^{-1} =
\diag(h(1), h(-1)).
$$
The characteristic polynomial of the shift matrix $\phi(x) = \phi(0+1\cdot
x) = [\begin{smallmatrix}0&1\\1&0\end{smallmatrix}]$ is $p(x) =
x^2-1$.

Theorem~\ref{diagprop} provides the essential step to obtaining a
convolution theorem in $\md$, i.e., a way to perform filtering in the
spectral domain. Namely, starting from the coordinate form
\eqref{filt2} of filtering, and using \eqref{polyeval}, we get
for all $\coord{a}\in\C^n$
\begin{equation}\label{conv1}
\begin{array}{rcl}
\phi(h)\cdot\coord{s} & = & 
  \four^{-1}\four\phi(h)\four^{-1}\four\cdot\coord{s} \\
& = & 
  \four^{-1}\diag(h(\alpha_0),\dots,h(\alpha_{n-1}))\four\cdot\coord{s},
\end{array}
\end{equation}
which again shows that multiplication by $\phi(h)$ is the same as
multiplying by $\diag(h(\alpha_0),\dots,h(\alpha_{n-1}))$, the
frequency response, in the spectral domain.
This is illustrated in the following commutative diagram:
\begin{equation*}
\xymatrix{
\C[x]/p(x) \ar[dd]^{\dps\four} \ar[rr]^{\dps\phi(h)} &&
  \C[x]/p(x) \ar[dd]^{\dps\four} \\ \\
\bigdirsum\C[x]/(x-\alpha_k) \ar[rr]^{\dps\diag(h(\alpha_k))} && 
  \bigdirsum\C[x]/(x-\alpha_k)
}
\end{equation*}
Filtering in the signal space (top row) is
equivalent to pointwise multiplication in the spectral domain (bottom
row). Connecting the top left with the top right corner in two
different ways yields \eqref{conv1}.

Next, we derive the general form of the convolution theorem.
Since, as said above, the frequency response is obtained in the same
way as the spectrum (evaluation of polynomials at the $\alpha_k$), we
can also compute it using a Fourier transform, even though we did not
introduce a Fourier transform for elements in the algebra.

Namely, we choose a basis $b'$ in $\alg$ and let $\coord{h}$ be the
coordinate vector of $h$ w.r.t.~$b'$. Then we choose a second
arbitrary Fourier transform $\four' = D\cdot\poly_{b',\alpha}$, where
$D$ is diagonal. Using
\eqref{eval},
$$
\poly_{b',\alpha}\cdot\coord{h} = D^{-1}\cdot
  (h(\alpha_0),\dots,h(\alpha_{n-1}))^T.
$$
Combining this equation with \eqref{conv1} yields the following theorem.

\begin{theorem}[Convolution Theorem]\label{convolution}
Using previous notation, let $\md$ have basis $b$ and $\alg = \md$
have basis $b'$. Let $s\in\md$ and $h\in\alg$. 
Further, let $\four$ 
and $\four' = D\cdot\poly_{b',\alpha}$ be any Fourier transforms of $\md$
w.r.t.~the bases $b$ and $b'$, respectively. Then
$$
\phi(h)\cdot\coord{s} =
  (D\four)^{-1}\cdot((\four'\cdot\coord{h})\odot(\four\cdot\coord{s})),
$$
where $\odot$ denotes the pointwise product of vectors. 
\end{theorem}

\mypar{Example} We illustrate Theorem~\ref{convolution} using our example
$\C[x]/(x^2-1)$ with basis $(1,x)$, choosing $\four' = \four = \DFT_2$.
We get the well-known convolution theorem for the DFT (of size 2):
$$
\begin{bmatrix}
h_0 & h_1\\h_1 & h_0
\end{bmatrix}\cdot\coord{s} =
\DFT_2^{-1}\cdot(\DFT_2\coord{h}\odot\DFT_2\coord{s}).
$$

\subsection{Example: Vandermonde matrix}\label{vander}

As an example, we consider a generic separable polynomial $p(x)$ with
zeros $\alpha_k$, $0\leq k < n$, and choose the basis $b =
(1,x,\dots,x^{n-1})$ in $\md = \C[x]/p(x)$ and the bases $b_k = (1)$
in $\C[x]/(x-\alpha_k)$. The corresponding polynomial transform
is the Vandermonde matrix
$$
\four = \poly_{b,\alpha} = [\alpha_k^\ell]_{0\leq k,\ell < n}.
$$
We evaluate the associated representation $\phi$ at the shift operator
$x\in\alg$.  Let $p(x) = \sum\beta_i x^i$, $\beta_n = 1$. Then $x\cdot
x^j \equiv x^{j+1}\text{ mod }p(x)$ for $0\leq j < n-1$, and $x\cdot
x^{n-1} = x^n \equiv \sum-\beta_i x^i\text{ mod }p(x)$. Thus,
the shift matrix is
\begin{equation}\label{companion}
\phi(x) =
\left[\begin{array}{rrrrc}
0 & & & & -\beta_0\\
 1 & 0 & & & -\beta_1 \\
  & \ddots & \ddots & & \vdots\\
  & & 1 & 0 & -\beta_{n-2}\\
  & &   & 1 & -\beta_{n-1}
\end{array}\right],
\end{equation}
which is the transpose of the companion matrix of $p$.
Using~\eqref{polyeval},
$$
\four\cdot\phi(x)\cdot\four^{-1} = \diag(\alpha_0,\dots,\alpha_{n-1}).
$$
For a convolution theorem, we choose $\four' =
\four$, and $D = \one_n$, to get
$$
\phi(h)\cdot\coord{s} =
  \four^{-1}\cdot((\four\cdot\coord{h})\odot(\four\cdot\coord{s})).
$$

\section{Where are we now?}\label{wherearewe}

Up to this point we have accomplished the following.
\begin{itemize}
\item We gave evidence that the basic assumptions underlying linear signal 
processing make the set of filters and signals not only vector spaces,
but an algebra and a module, respectively. This places SP into the
context of the representation theory of algebras, which includes but
goes beyond linear algebra. In particular, filtering produces the
module structure and the Fourier transform is usually not thought of
as a concept from linear algebra.

\item We elaborated on the correspondence, or better, equivalence, of
signal processing and the representation theory of algebras, by
providing a small dictionary that translates between signal processing
concepts and algebraic concepts.

\item We formally defined the notion of a signal model for a vector
space $V$ of signals as a triple $(\alg,\md,\Phi)$, where $\alg$ is
the chosen filter algebra, $\md$ is an $\alg$-module of the same
dimension as $V$, and $\Phi$ is a bijective linear mapping from $V$
onto $\md$. Once the signal model is defined, all main ingredients to
signal processing are immediately available. Further, different models
can be applied to the same signal space $V$ or signal $\coord{s}\in
V$, reflecting the fact that different transforms (such as the DFT and
the DCT) can be used for analysis. We will see later that both the DFT
and DCT are Fourier transforms for suitable signal models.

\item We identified the shift operator as generator of the filter algebra
$\alg$ and asserted that a signal model is shift-invariant if and only
if $\alg$ is commutative.  By further requiring that $\alg$ consist
exclusively of FIR filters---which is necessarily the case for finite
signal models ($\dim\md<\infty$)---commutative algebras are equivalent
to polynomial algebras.

\item With this as a motivation, we developed signal processing on
polynomial algebras, focusing on one variable $\alg = \C[x]/p(x)$
equivalent to one shift operator $x$ or 1-D signals, and to the
regular case $\md = \alg$.  We showed which specific form filtering,
spectrum, Fourier transform, and other signal processing concepts take
in this case.
\end{itemize}

In summary, our approach so far has been ``top-down:'' we considered
existing concepts in signal processing and established their algebraic
interpretation. We used repeatedly the example of infinite
discrete-time signal processing. In fact, in discrete signal
processing, concepts like filter, shift, and Fourier transform are
usually thought of in the context of infinite discrete time. The
question that may arise now is if the algebraic theory goes beyond
this example of infinite discrete-time signals. The answer is yes, as
we have hinted at already in Section~\ref{intro} and will show in
detail in the remainder of the paper.

In addressing this question, and in contrast to what was done before,
we now take a ``bottom-up'' perspective: we make assumptions only as
necessary and {\em derive} signal models from basic principles. This
is important and beneficial from several points of view and addresses
several fundamental questions such as:
\begin{itemize}
\item How to construct a signal model for \emph{finite} signals from a
corresponding signal model for infinite signals?
\item How to derive a signal model for discrete and finite
\emph{space} analogous to the corresponding signal models for time?
\item Why and when do \emph{boundary conditions} and \emph{signal
extensions} arise, e.g., why is the periodic signal extension
important and are others possible?
\item What type of objects are the \emph{discrete trigonometric
transforms}, in particular, the discrete cosine transforms~(DCTs)
and the discrete sine transforms~(DSTs)?
\end{itemize}
Finally, we believe the bottom-up derivation of signal models to be of
educational value as it explains (and visualizes) the models and
provides a recipe on how to possibly construct new models for
different applications.

The remaining part of this paper uses the algebraic concepts
introduced before. The reader may wish to revisit frequently the
previous sections, connecting the following concrete examples to the
previous, more theoretical discussion.

\section{Modeling Time: The z-Transform}\label{ztrafo}

Let $\coord{s} = (s_n\mid n\in\Z)$ be a discrete signal. The
$z$-transform of $\coord{s}$, in standard notation, is given by the
mapping
\begin{equation}\label{ztrafodef1}
\Phi:\ \coord{s}\mapsto s = S(z) = \sum_{n\in\Z} s_nz^{-n}.
\end{equation}
In Section~\ref{modsigmod} we showed that the $z$-transform is the
linear mapping of a signal model in the sense of
Definition~\ref{sigmoddef} for the vector space $\ell^2(\Z)$ of finite
energy sequences. The corresponding algebra is usually chosen as the
set of all Laurent series with coefficient sequences in $\ell^1(\Z)$.

Naturally, the following fundamental question arises: Why is $\Phi$
the appropriate linear mapping for the signal model for discrete time,
or, equivalently, why is discrete time filtering defined the way it
is? We give an answer to this question by {\em deriving} the
$z$-transform from basic principles. In other words, we identify the
assumptions that have to be made to obtain the $z$-transform. This
knowledge will then enable us, for example, to derive the
space-analogue of the $z$-transform, the $C$-transform in
Section~\ref{Ctrafo}. In fact, since infinite discrete time is
standard in classical signal processing theory, the whole purpose of
this section is to identify the recipe for constructing a signal
model.

The section is structured as follows. First, we build the signal model
in the sense of Definition~\ref{sigmoddef} in three steps: 1)
definition of the shift; 2) linear extension; and 3) realization.
Second, after the signal model is determined, we provide the
associated spectrum and Fourier transform. And third, we give a
visualization of the signal model in the sense of
Definition~\ref{visdef}. The derivation and analysis of the discrete
space models and discrete generic next neighbor model in
Sections~\ref{Ctrafo}, \ref{altreal}, and \ref{genshiftmodel} will
follow the same steps.

For notational convenience, we set in the following\footnote{Note that
the choice of $z^{-1}$ instead of $z$ in the definition
\eqref{ztrafodef1} is a {\em convention}, not a mathematical
necessity; choosing $z$ leads to equivalent properties and an
equivalent theory for the $z$-transform. In fact, the choice of
$z^{-1}$ in signal processing is in contrast to the previous
mathematical work on Laurent series. The reason may be the fact that
the shift operator $z^{-1}$ causes a delay of the signal. However, we
will see that, equivalently, this shift operator {\em advances} what
we call the time marks.} $x = z^{-1}$.

\subsection{Building the Signal Model}\label{buildztrafo}


\mypar{Definition of the shift}
Following Kalman~\cite{Kalman:69}, when considering time, we need two
ingredients: {\em time marks} $t_n$ and a {\em shift operator} $q$.

The time marks are symbolic independent variables $t_n, n\in\Z$; $t_n$
is associated to ``time $n$.'' Using these time marks, we can write
every signal $\coord{s} = (s_n)_{n\in\Z}\in\C^\Z$ as the symbolic
sum $s = \sum_{n\in\Z}s_nt_n$. The set of all these sums is again a
vector space isomorphic to $\C^\Z$. In other words, no new structure
was introduced in the signal space. At this point there is no
interaction between time marks at different times $n$, in particular
no notion of past and future, nor any notion of equidistance between
consecutive time marks. 

\begin{figure}[ht]\centering
\begin{picture}(100,35)
\multiput(-40,10)(4,0){8}{$\cdot$}
\put(0,10){$\bullet$}
\put(0,0){$t_{n-1}$}
\put(20,18){\makebox(0,0){\xy\ar @/^1ex/ (10,0)\endxy}}
\put(20,28){\makebox(0,0){$q\diamond$}}
\put(36,10){$\bullet$}
\put(36,0){$t_{n}$}
\put(56,18){\makebox(0,0){\xy\ar @/^1ex/ (10,0)\endxy}}
\put(56,28){\makebox(0,0){$q\diamond$}}
\put(72,10){$\bullet$}
\put(72,0){$t_{n+1}$}
\multiput(86,10)(4,0){8}{$\cdot$}
\end{picture}
\caption{The time shift $q\diamond t_n$.\label{timeshift}}
\end{figure}
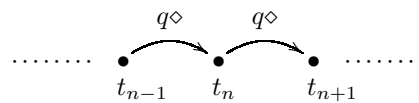

To address this problem, we introduce the shift operator $q$ and
the shift operation $\diamond$ by
\begin{equation}\label{opq}
\framebox[1.2\width]{\rule[-1.3mm]{0mm}{5mm}%
$\text{\bf time model:}\quad q\diamond t_n = t_{n+1}$}
\end{equation}
for $n\in\Z$. Figure~\ref{timeshift} shows a graphical representation
of the time shift.

The operator $q$ naturally expresses a direction from past to future
and the equidistance of the time marks $t_n$. 

Next, we extend the operator domain from a single shift operator $q$ to 
{\em $k$-fold shift operators} $q_k$, defined by
$$
q_k\diamond t_n = t_{n+k}.
$$
Clearly, $q_k = q^k$.

At this point of the construction, working only with $t_n$ and $q_k$,
there is no notion of linearity.

\mypar{Linear extension} Since we are interested in obtaining a {\em
linear} signal model, we carry out two extensions:
\begin{inparaenum}[1)]
\item we extend the 
operation~$\diamond$ of the shift operator~$q$ from the set of the
$t_n$ to the set of all formal sums $\sum s_nt_n$, by requiring
linearity, i.e., $q\diamond s=\sum s_n(q\diamond t_n)$; and
\item we extend the operating set from the set of $k$-fold shift operators
$q^k$ to the set of all formal sums $\sum h_kq^k$.
\end{inparaenum}
The first set will become the module of signals, while the second set
will become the algebra of filters.

In other words, starting with~\eqref{opq}, we first linearly extend the
operation~$\diamond$ to the vector space $\md = \{s = \sum
s_nt_n\}$, and then we extend the operator domain to obtain $\alg =
\{h = \sum h_kq^k\}$. Because the series in either set have an
infinite number of terms, we need to make sure that filtering is 
well defined: the multiplication of an infinite
series---the signal $s\in\md$---by another infinite series---the
filter $h\in\alg$---has to exist. We consider this in the next step,
called ``realization.''

\mypar{Realization}
To obtain the signal model, we first consider the ``realization'' of
the abstract model, which replaces the abstract objects~$t_n$ and~$q$
and the operation~$\diamond$ by objects we can compute with. To this
end, we choose a variable $x$ and set $q = x$, and $\diamond = \cdot$,
the ordinary multiplication of series. Then \eqref{opq} becomes
\begin{equation}\label{twoterm}
t_{n+1} = x\cdot t_n.
\end{equation}
This two-term recurrence, when started with $t_0=1$, has the unique solution
\begin{equation}\label{realtimeop}
t_n = x^n.
\end{equation}
In other words, the realization is essentially (up to a common
scaling factor for all $x^n$) unique.

As a result, we obtain $\md = \{s = \sum s_n x^n\}$ and $\alg = \{h =
\sum h_kx^k\}$. Since the series are infinite,
we have to ensure convergence as part of the realization; namely, that
filtering, the operation of $\alg$ on $\md$, is well-defined. This is
achieved, for example, by requiring $\coord{s}\in\ell^2(\Z)$ and
$\coord{h}\in\ell^1(\Z)$, as explained in Section~\ref{algmodsm}. Now
$\alg$ becomes the filter space and $\md$ becomes the signal space.

Table~\ref{corrtime} shows the correspondence between the abstract and
the realized concepts.

\begin{table}\centering
\caption{Realization of the abstract time model.\label{corrtime}}
\ra{1.2}
\begin{tabular}{@{}lll@{}}\toprule
concept & abstract & realized \\ \midrule
shift operator & $q$ & $x$ \\
shift operation & $\diamond$ & $\cdot$ \\
time mark & $t_n$ & $x^n$ \\
$k$-fold shift operator & $q_k=q^k$ & $x^k$ \\
shift & $q\diamond t_n = t_{n+1}$ & $x\cdot x^{n} = x^{n+1}$ \\
signal & $\sum s_nt_n$ 
  & $\sum s_nx^n$ \\
filter & $\sum h_k q^k$ 
  & $\sum h_kx^k$ \\ \bottomrule
\end{tabular}
\end{table}

At this point it seems that the only reason for doing a
``realization'' is to handle convergence issues. We will see later in
the derivation of the space model that the realization may not be
unique and thus may lead to different final signal models.

\mypar{Signal model: \boldmath$z$-transform} Now we can formally
define the signal model $(\alg,\md,\Phi)$ for discrete infinite time.
The filter algebra $\alg$ is the set of Laurent series $h$ with
coefficients $\coord{h}\in\ell^1(\Z)$; the signal module is the set of
Laurent series $s$ with $\coord{s}\in\ell^2(\Z)$.  The third component
of the signal model is the bijective linear mapping $\Phi$, which maps
signals from the space $V = \ell^2(\Z)$ into $\md$:
$$
\Phi:\ \coord{s}\mapsto \sum_{n\in\Z}s_nx^n
$$
is the $z$-transform.

Note that signals and filters are conceptually different (as pointed
out several times before) but {\em look the same (both are Laurent
series in $x$) because the realization maps both $q^n$ and $t_n$ to
$x^n$}.

\subsection{Spectrum and Fourier Transform: DTFT}\label{zfourtrafo}

We discussed the spectrum and the Fourier transform for the
$z$-transform already in Section~\ref{algmodsm}, so we are brief here.  

The $\alg$-module $\md=\Phi(\ell^2(\Z))$ decomposes into continuously
many one-dimensional irreducible modules $\md_\omega$, $\omega\in W =
(-\pi,\pi]$, which constitute the spectrum of $\md$.  Each spectral
component $\md_\omega$ is spanned by $E_\omega$ in
\eqref{eigenfunction}. As we stated before, the $E_\omega$ are not
contained in $\Phi(\ell^2(\Z))$, but only in $\Phi(\ell^\infty(\Z))$,
so the $\md_\omega$ are not submodules of $\md$. Choosing $E_\omega$
as basis in $\md_\omega$, the Fourier transform $\Delta$ (in the form
\eqref{moduledec1}) is called the discrete-time Fourier transform
(DTFT) and is given by
\begin{equation*}
\begin{array}{rrcl}
\Delta: & \md & \rightarrow & 
  \bigl(\bigdirsum_{\omega\in W}\C\bigr) = \C^W,\\
& s = S(x) & \mapsto & (S(e^{-j\omega}))_{\omega\in W} = 
  \omega\mapsto S(e^{-j\omega}).
\end{array}
\end{equation*}
Note that we have to write $S(e^{-j\omega})$ since our variable is $x$
instead of $z^{-1}$.  The spectrum $\Delta(s)$ of a signal can be
viewed as a (complex $L^2$-) function on the unit circle.

Every irreducible module $\md_\omega$ affords a one-dimensional
irreducible representation $\phi_\omega$. Namely, if $h=H(x)\in\alg$
is a filter, then
$$
H(x)E_\omega(x) = H(e^{-j\omega})E_\omega(x),
$$
which shows that
$$
\phi_\omega(H(x)) = H(e^{-j\omega}).
$$

The frequency response of $h=H(x)$ is the collection, for
$\omega\in(-\pi,\pi]$,
$$
(\phi_\omega(h))_{\omega} = (H(e^{-j\omega}))_\omega =
\omega\mapsto H(e^{-j\omega}).
$$
Filtering $h\cdot s = H(x)S(x)$ in the time-domain becomes the
parallel multiplication $H(e^{-j\omega})S(e^{-j\omega})$, $\omega\in
W$ by the frequency response in the frequency domain.

\subsection{Visualization}\label{timegraphrep}

Because of the simple shift operation associated with the
$z$-transform, the structure of $\md$ can be conveniently represented
as a graph, shown before in Figure~\ref{timegraph}, which expresses
the shift, i.e., the multiplication by $x$ in the chosen
realization.  Formally, the graph is obtained by using our
Definition~\ref{visdef} of the visualization of a signal model. We
have one shift operator $x$, and, if $\phi$ is the representation afforded by
$\md$ with the time marks $x^k$ as basis, then $\phi(x)$ is the
adjacency matrix of the graph in Figure~\ref{timegraph}.

The bullets in Figure~\ref{timegraph} represent the time marks to
which the signal values are associated via the signal model, i.e., via
the $z$-transform
\eqref{ztrafodef1}. If the signal space is considered as a vector space
these bullets become the basis, but are unconnected. The connecting
edges arise from the shift operation, which makes the signal space a
module.

Figure~\ref{timegraph} is helpful to obtain an intuitive understanding
of the module structure and is clearly familiar to most readers.
We will show similar graphs for the finite $z$-transform and their
space model analogues below. Comparing these graphs is the 
a simple way to intuitively understand the differences between the 
associated signal models.

\section{Finite z-Transform and DFT}\label{ztrafofinite}

In real applications, usually only a finite subsequence $\coord{s} =
(s_0,\dots,s_{n-1})$ is available, not the entire (sampled) sequence
$(s_k\mid k\in\Z)$. Thus, the question is how to develop a {\em finite
version} of the $z$-transform, i.e., with a finite number of terms.
We will see that the problem, when restricted to a finite sequence, is
in preserving the module property to have access to filters and a
notion of spectrum and Fourier transform. It turns out that to obtain
the module structure, we need to introduce boundary conditions
(\bc's). The \bc's~in turn determine a polynomial algebra and the
associated module. This gives a ``bottom-up'' justification for
polynomial algebras in addition to the motivation presented in
Section~\ref{shiftcomm}, which identified them as those algebras
providing finite shift-invariant signal models. The bijective linear
map~$\Phi$ in the signal model we obtain, i.e., the associated finite
$z$-transform, maps signals $\coord{s}\in\C^n$ into this module. The
chosen \bc's~determine the signal extension, and, by requiring a {\em
simple} signal extension (in a sense defined below), we naturally
obtain the DFT (and its variants) as Fourier transform for the
associated module.

Interestingly, the concept of a finite $z$-transform in the sense
defined below is not used in signal processing, even though it is the
precise finite analogue of the standard $z$-transform. Further, it
explains the DFT and the role of boundary conditions and signal
extension, and facilitates the derivation of various DFT algorithms,
just to name a few benefits.

For our investigation, we first need a formal notion of {\em signal
extension}, which we define now.

\begin{definition}[Signal Extension]\label{sedef}
Let $\coord{s} = (s_k\mid k\in I)$ be a signal given on an index set
$I\subset\Z$.  A (linear) signal extension of $\coord{s}$ is the
sequence of linear combinations (only finitely many summands are
nonzero)
$$
s_k = \sum_{i\in I}\beta_{k,i}s_i,\quad
\text{for }k\not\in I.
$$
The signal extension is called {\em monomial}, if, for each $k$,
the sum has only one summand.
\end{definition}
In other words, in a monomial signal extension, every signal value
outside the signal scope is assumed to be a multiple of a signal
value inside the signal scope. 

\subsection{Building the Signal Model}\label{buildfinztrafo}

\mypar{Shift, linear extension, realization} To construct a finite
time signal model, we follow the exact same steps as in
Section~\ref{buildztrafo}. We start with the time marks, of which we
have now only a finite set $t_0,\dots,t_{n-1}$.  As before, we
consider the shift operator~$q$, the operation~$\diamond$ and the
realization of the shift operator by the variable~$x$. However, as we
will see next, this leads to one important difference.

Let $\coord{s} = (s_0,\dots,s_{n-1})\in\C^n$ be a sampled signal. To
realize time, we could attempt to realize as before the two step
recursion~\eqref{twoterm} and to define the ``finite'' $z$-transform,
by mapping $\coord{s}$ to the polynomial of degree less than~$n$
$$
s = s(x) = \sum_{0\leq k< n}s_kx^k.
$$
Proceeding with the definition of the signal model, we attempt to
identify the module and the algebra. Clearly, the set $\C_n[x]$ of the
polynomials~$s(x)$ (of degree less than~$n$) is a vector space with
the natural basis $b = (x^0,\dots, x^{n-1})$.  The problem, however,
arises from the operation of the (realized) time shift
operator~$x$: the set of polynomials of degree less than~$n$ is {\em
not closed} under multiplication by~$x$. More precisely, the root of
the problem is
\begin{equation}\label{rightproblem}
x\cdot x^{n-1} = x^n\not\in\C_n[x],
\end{equation}
and, if non-causal filters are considered,
\begin{equation}\label{leftproblem}
x^{-1}\cdot x^0 = x^{-1}\not\in\C_n[x].
\end{equation}
Thus, the time shift as has been defined is not a valid operation on
$\C_n[x]$, which implies that we cannot define filtering in $\C_n[x]$,
or, algebraically, $\C_n[x]$ is not a module.  Without filtering,
there is also no notion of spectrum or Fourier transform. To resolve
this we need to take care of the problems raised
by~\eqref{rightproblem} and~\eqref{leftproblem}, which we do by
introducing boundary conditions.

\mypar{Boundary condition and signal extension}
To remedy the first problem \eqref{rightproblem}, we have to make sure
that $x^n$ can be expressed as a polynomial of degree $n-1$. This is
achieved by introducing an equation
\begin{equation}\label{polyintro}
x^n = r(x) = \sum_{0\leq k < n}\beta_kx^k,
\quad\text{or}\quad x^n - r(x) = 0.
\end{equation}
This equation is equivalent to the right boundary condition
$$
s_n = \sum_{0\leq k < n}\beta_ks_k.
$$
As a consequence of \eqref{polyintro}, we get the series of
equations
$$
x^k(x^n - r(x)) = x^k\cdot 0 = 0,\quad
k\geq 0.
$$
Thus, the boundary condition $x^n = r(x)$ determines the entire
right signal extension that is obtained by reducing $x^{k+n}$ modulo
$(x^n - r(x))$ to a polynomial $r_k(x)$ of degree less than $n$, i.e.,
\begin{equation}\label{rightsepoly}
x^{k+n} \equiv r_k(x)\text{ mod }(x^n - r(x)).
\end{equation}
Algebraically, the boundary condition replaces the vector space
$\C_n[x]$ by the vector space $\md = \C[x]/(x^n - r(x))$, which is of
the same dimension, but {\em closed} under multiplication by the time
shift operator $x$ and is thus a module. The corresponding algebra $\alg$,
generated by $x$, is identical to $\md$.  The remaining question to
consider is \eqref{leftproblem}. There are two cases.

Case 1: $x|r(x)$.  Then also $x|(x^n - r(x))$, and thus $x$ (the shift
operator) is not invertible\footnote{A polynomial $q(x)\in\C[x]/p(x)$
is invertible if and only if $\gcd(q(x),p(x))=1$, since in this case
there are polynomials $r(x),s(x)$ such that $1 = s(x)q(x) + r(x)p(x)$,
which implies that $s(x)\equiv q(x)^{-1}\text{ mod }p(x)$.}  in $\alg
= \C[x]/(x^n-r(x))$ and
\eqref{leftproblem} does not need to be considered: the signal has no
left \bc, since ``the past'' is not accessible without an invertible
$x$.  

Case 2: $x\not| r(x)$. Then, from~\eqref{polyintro}, we get
$$
x^{-1} = -\frac{1}{\beta_0}
(\beta_1 + \beta_2 x + \dots + \beta_{n-1}x^{n-2} - x^{n-1}),
$$ 
which is the left boundary condition. Similar to above, the left
signal extension can be determined by multiplying by $x^{-k}$ and
reducing modulo $x^n - r(x)$.  Thus, the signal extension in {\em
both} directions is determined by {\em one}
equation~\eqref{polyintro}, which provides the left and the right 
\bc:
$$
\text{\bc }\Longrightarrow\text{ right and left signal extension.}
$$

By assuming the generic boundary condition $x^n = r(x)$, we obtain a
valid signal model. However, the corresponding signal extension
\eqref{rightsepoly} has in general no simple structure.
To obtain a module that is reasonable for applications, we thus require
\begin{itemize}
\item the shift operator $x\in\alg$ to be invertible; and
\item the signal extension to be monomial (see Definition~\ref{sedef}).
\end{itemize}
A monomial signal extension is the condition that leads to the signal
model for the DFT in the \emph{finite time} case and for the 16 DCTs
and DSTs in the \emph{finite space} case. In
Section~\ref{skewdttssec}, we will slightly relax this condition and
obtain a new class of transforms.

We can now explicitly determine the polynomials $x^n-r(x)$ that satisfy the
above two conditions.

\begin{lemma}\label{monse}
The boundary condition $x^n = r(x)$ makes $\alg = \C[x]/(x^n - r(x))$
an algebra in which $x$ is invertible and determines a monomial signal
extension in $\md = \alg$, if and only if the polynomial $r(x)$ is a
nonzero constant, i.e., $r(x) = a \neq 0$.  The signal extension in
this case is given by $x^k = a^{k_2}x^{k_1}$, where $k\in\Z$ is
expressed as $k = k_1+k_2n$, with $0\leq k_1 < n$.
\end{lemma}
\begin{proof}
Let $r(x) = a$, $a\neq 0$, and let $k\in\Z$. We write $k = k_1 +
k_2n$, with $0\leq k_1 < n$, and thus $x^k \equiv a^{k_2}x^{k_1}
\text{ mod }x^n - a$, which is a monomial signal
extension. Conversely, let $x^n = r(x)$ determine a monomial signal
extension. This implies $x^n = ax^\ell$, for some $0\leq
\ell<n$. Since $x$ is by assumption invertible modulo $x^n-ax^\ell$,
it follows $\ell = 0$ and $a\neq 0$ as desired.
\end{proof}

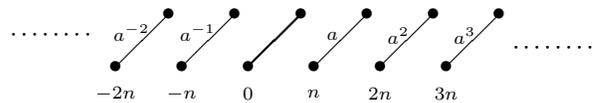
\begin{figure}\centering
\begin{picture}(150,40)
\multiput(-40,20)(4,0){8}{$\cdot$}
\put(0,10){\usebox{\upline}}
\put(6,23){\makebox(0,0){$\sst a^{-2}$}}
\put(0,0){\makebox(0,0){$\sst -2n$}}
\put(25,10){\usebox{\upline}}
\put(31,23){\makebox(0,0){$\sst a^{-1}$}}
\put(25,0){\makebox(0,0){$\sst -n$}}
\put(50,10){\usebox{\upbline}}
\put(50,0){\makebox(0,0){$\sst 0$}}
\put(75,10){\usebox{\upline}}
\put(82,22){\makebox(0,0){$\sst a$}}
\put(75,0){\makebox(0,0){$\sst n$}}
\put(100,10){\usebox{\upline}}
\put(107,22){\makebox(0,0){$\sst a^2$}}
\put(100,0){\makebox(0,0){$\sst 2n$}}
\put(125,10){\usebox{\upline}}
\put(132,22){\makebox(0,0){$\sst a^3$}}
\put(125,0){\makebox(0,0){$\sst 3n$}}
\multiput(150,15)(4,0){8}{$\cdot$}
\end{picture}
\caption{The signal extension of $\C[x]/(x^n - a)$.\label{sigextxna}}
\end{figure}

It is instructive to graphically display the signal extension
associated to $\md = \C[x]/(x^n-a)$. There are several ways to do
this. We choose to display it in a virtual coordinate system. The
x-axis carries the time marks $x^i$, $i\in\Z$, and the y-axis carries
the basis $x^0,\dots,x^{n-1}$ of $\md$. For every $i\in\Z$, we express
$x^i$ in the basis and enter the coefficients in the
graph. Figure~\ref{sigextxna} shows the result (with the y-axis
coordinates omitted). Within the signal scope (shown by the bold
line) $0\leq i < n$, we have the identity (since $x^i = 1\cdot x^i$);
for $n\leq i <2n$ the line is scaled by $a$ (since $x^i\equiv
ax^{i-n}\text{ mod }(x^n - a)$), and so on. The bullets mark starting
points (at multiples of $n$) and end points of the lines to enhance
the presentation.

Using Lemma~\ref{monse} we observe that the signal extension
corresponding to the \bc $x^n = a$ is eventually periodic, if and only
if $a$ is an arbitrary $m$th root of unity. The period in this case
has length $mn$.

\mypar{Signal model: finite \boldmath$z$-transform} In summary, the
 signal space so obtained is the regular module $\md = \C[x]/(x^n - a)$
with algebra of filters $\alg = \md$. The final component of the
associated signal model for $V =
\C^n$ is the {\em finite $z$-transform}
$$
\Phi:\ \coord{s}\mapsto \sum_{0\leq k < n}s_kx^k\in\md
$$ 
that endows the signal $\coord{s}\in\C^n$ with the structure of the
module~$\md$.  The associated algebra $\alg = \md$ of filters has the
same basis $b = (1,x,\dots,x^{n-1})$ as $\md$; in $\md$ the basis
polynomials (or impulses) $x^k$ represent time marks; in $\alg$ they
represent ($k$-fold) time shift operators.

Note that we restricted this definition of a finite $z$-transform to
the cases of monomial signal extensions only, since these are the only
ones we consider in this paper. Of course, the definition could be
generalized to any $p(x) = x^n - r(x)$.

\subsection{Spectrum and Fourier transform: DFT and variants}
\label{dftvars}

Using the general results from Section~\ref{polyalgs}, we now derive
the spectrum and the Fourier transform for the signal model finite
$z$-transform. In other words, we consider the regular module 
$\md = \alg = \C[x]/(x^n - a)$ with the same basis 
$b = (x^0,x^1,\dots x^{n-1})$ chosen in $\md$ and $\alg$.
We use the notation $j = \sqrt{-1}$ in this section.

Since $a\neq 0$, $p(x) = x^n - a$ is separable. Let $a =
|a|e^{\nu j}$ in polar coordinates.
The zeros $\alpha_k$ of $x^n - a$ are given by
\begin{equation}\label{zerosofxna}
\alpha_k = |a|^{\ts\frac{1}{n}}e^{\ts\frac{(\nu-2k\pi)j}{n}} =
|a|^{\ts\frac{1}{n}}e^{\ts\frac{\nu j}{n}}\omega_n^k,
\end{equation}
where $\omega_n = e^{-2\pi j/n}$. 
Thus, spectrum and Fourier transform of $\C[x]/(x^n-a)$ are given by
$$
\begin{array}{rrcl}
\Delta: & \C[x]/(x^n-a) & \rightarrow & 
  \bigdirsum_{0\leq k <n}\C[x]/(x-\alpha_k)\\
& s = s(x) & \mapsto & (s(\alpha_0),\dots,s(\alpha_{n-1})).
\end{array}
$$

In matrix form, the polynomial transform corresponding to $\Delta$ is
given by
\begin{equation}\label{polygendft}
\poly_{b,\alpha} = \DFT_n\cdot
  \diag_{0\leq\ell<n}\bigl(|a|^{\ts\frac \ell n}
    e^{\ts\frac{\nu\ell j}{n}}\bigr),
\end{equation}
where
$$
\DFT_n = [\omega_n^{k\ell}]_{0\leq k,\ell<n}
$$
is the (standard) discrete Fourier transform.
Using Theorem~\ref{fourtrafo}, {\em all} Fourier transforms
are given by
\begin{equation}\label{gendft}
\four = D\poly_{b,\alpha},
\end{equation}
where $D$ is any invertible diagonal matrix $D$, which is determined
by the basis chosen in the spectrum.

This derivation shows the origin of the DFT in signal processing
terms, namely, the DFT arises by constructing a finite time model
under the assumption of a monomial signal extension.  The standard DFT
(without scaling factors) arises from the special case $a = 1$ and
$D=\one_n$, which implies that the $\DFT_n$ is a polynomial transform.

\mypar{Generalized DFTs} The class of DFTs \eqref{gendft} also
includes as special cases what have sometimes been called
``generalized Fourier transforms'' (see
\cite{Bongiovanni:76,Martucci:94,Britanak:99}).  We prefer
``generalized DFTs,'' since, as we have explained in
Section~\ref{algmodsm}, the concept of Fourier transform is far more
general, encompassing decompositions of arbitrary modules.

The generalized DFTs are matrices of the form
$$
\four_{c,d} = [\omega_n^{(k+c)(\ell+d)}]_{0\leq k,\ell<n},
$$ 
where $c,d\in\R$. We briefly investigate the 4 special cases given
by $c,d\in\{0,1/2\}$, which in \cite{Britanak:99} are called
DFTs of types 1--4, written as $\DFTt{1},\dots,\DFTt{4}$. Namely,
\begin{gather}
\DFTt{1}_n = \four_{0,0} = \DFT_n,\nonumber\\
\DFTt{2}_n = \four_{0,1/2} = 
  \diag_{0\leq k<n}(\omega_n^{k/2})\DFT_n,\label{dft2}\\
\DFTt{3}_n = \four_{1/2,0},\label{dft3}\\
\DFTt{4}_n = \four_{1/2,1/2} = 
  \diag_{0\leq k<n}(\omega_n^{(k+\fh)/2})\DFTt{3}_n.\label{dft4}
\end{gather}
We identify the signal models for which these transforms are Fourier
transforms, by comparing their definitions to \eqref{gendft} and
\eqref{polygendft}.  The $\DFTt{1}_n = \DFT_n$ is, as said above, a
polynomial transform for $\C[x]/(x^n-1)$. The $\DFTt{2}_n$ in
\eqref{dft2} is also a Fourier transform, but not the polynomial
transform, for $\C[x]/(x^n-1)$. The $\DFTt{3}_n$ in \eqref{dft3} is
the polynomial transform for $\C[x]/(x^n+1)$, since
$\omega_n^{k+1/2}$, $0\leq k<n$, are precisely the zeros of $x^n+1$.
Finally, the $\DFTt{4}_n$ in \eqref{dft4} is also a Fourier
transform, but not the polynomial transform, for $\C[x]/(x^n+1)$.
This means, these DFTs cover the two  important cases of boundary
conditions $x^n = a = \pm 1$.

The $\DFT_n = \DFTt{1}_n$ is the matrix containing the evaluation of
the polynomials $x^\ell$ at the roots of unity
$\omega_n^k$. Similarly, the matrix $\DFTt{2}_n$ consists of the
evaluations of the ``fractional'' polynomials $x^{\ell+1/2}$ at the
$\omega_n^k$. Thus, in a sense, and leaving our framework of
polynomial algebras, $\DFTt{2}_n$ could be viewed as a Fourier
transform for the $\alg$-module $\md$ spanned by the basis
$\{x^{1/2},x^{3/2},\dots,x^{n-1/2}\}$ and with periodic boundary
condition $x^{n+1/2} = x^{1/2}$, and $\alg =\C[x]/(x^n-1)$ as
operating algebra.  In other words, $\DFTt{2}_n$ can be viewed as a
``shifted'' version of $\DFT_n$. This shifting, as we showed above,
does not change the underlying signal model. 

A similar relationship exists between $\DFTt{3}_n$ and $\DFTt{4}_n$.

\mypar{Other boundary conditions and effect on spectrum}
At this point it is instructive to investigate what problems arise if
we slightly relax the conditions in Lemma~\ref{monse} by dropping
the requirement that the shift operator $x$ is invertible in the
constructed algebra. In particular, this includes, as we show, the zero
extension and the constant extension.

The proof of Lemma~\ref{monse} shows that a monomial signal extension
requires a \bc of the form $x^n = ax^\ell$, i.e., $p(x) =
x^n-ax^\ell$. Conversely, each such \bc determines a monomial signal
extension. A simple choice is $a=0$ yielding the regular module
$\C[x]/x^n$, which realizes a right zero extension ($x^n = 0$ implies
$x^{n+k}=0$ for $k \geq 0$). There is no concept of a left extension,
since the shift operator $x$ is not invertible in $\alg$. The problem of this
model is the spectrum: $\C[x]/x^n$ cannot be decomposed by the Chinese
Remainder Theorem~(CRT), or, in other words, the model is not
appropriate for spectral analysis.  If $\phi$ is the representation
afforded by $\md$, then this can also be seen from the shift matrix
$\phi(x)$, which is the (lower) Jordan block (a special case of
\eqref{companion})
$$
\phi(x) = 
\begin{bmatrix}
0 \\
1 & 0 \\
& \ddots & \ddots \\
&& 1 & 0
\end{bmatrix}.
$$
As a Jordan block, this matrix cannot be block diagonalized
any further.

Another simple choice is the symmetric \bc $x^n = x^{n-1}$, i.e.,
$p(x) = x^n - x^{n-1}$. This choice implies a constant right signal
extension, since $x^n = x^{n-1}$ implies $x^{n+k} = x^{n-1}$ for all
$k\geq 0$. In this case, the CRT yields
$$
\C[x]/(x^n-x^{n-1})\cong\C[x]/(x-1)\dirsum\C[x]/x^{n-1},
$$
and the rightmost module, of dimension $n-1$, is again indecomposable,
making spectral analysis trivial. We will see later, that a symmetric
signal extension in a discrete finite {\em space} model leads to a
monomial signal extension {\em and} to a separable polynomial $p$, and
thus provides a useful model for signal processing.

As a final example, we consider the generic right \bc $x^n = r(x)$,
such that $p(x) = x^n-r(x)$ is separable.  This leads to a generic
regular module $\C[x]/p(x)$ with basis $(1,x,\dots,x^{n-1})$. As we
have seen in Section~\ref{vander}, the (polynomial) Fourier transform
in this case is a Vandermonde matrix and the shift matrix $\phi(x)$ is
the companion matrix
\eqref{companion}.  In other words, Vandermonde matrices are precisely
the (polynomial) Fourier transforms for (separable) finite time models.

The above discussion shows that the choice of boundary condition
affects both, the signal extension and the notion of spectrum. Models
that have ``good'' properties with respect to both are useful in
signal processing.

Signal extensions that are not linear (for example second order
polynomial extensions) are sometimes considered in signal processing.
These do not produce signal models (in particular, filtering becomes
non-linear) and are thus not covered by our theory.

\subsection{Visualization}

\begin{figure}[t]
\quad
\begin{minipage}{0.9\textwidth}
\vspace*{10mm}
{
\xymatrix{
\bullet \ar@{->}[r] & \bullet \ar@{->}[r] & 
\bullet \ar@{.}[rr] && \bullet \ar@{->}[r] &
\bullet \ar@{->}[r] & \bullet \ar@{->}@(ul,ur)[llllll]
}
\begin{picture}(210,4)
\put(7,0){\makebox(0,0){$x^0$}}
\put(42,0){\makebox(0,0){$x^1$}}
\put(77,0){\makebox(0,0){$x^2$}}
\put(142,0){\makebox(0,0){$x^{n-3}$}}
\put(177,0){\makebox(0,0){$x^{n-2}$}}
\put(212,0){\makebox(0,0){$x^{n-1}$}}
\put(107,38){\makebox(0,0){$a$}}
\end{picture}
}
\end{minipage}
\vspace*{2mm}
\caption{Visualization of the finite discrete time model in the
case of a monomial signal extension $\alg=\md=\C[x]/(x^n - a)$.
\label{fintimegraph}}
\end{figure}
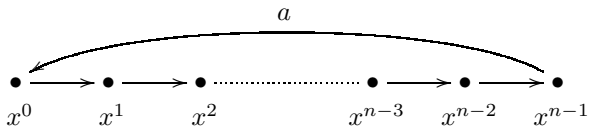

The structure of the signal model given by the finite $z$-transform is
visualized by the graph in Figure~\ref{fintimegraph}. Following
Definition~\ref{visdef}, the adjacency matrix of this graph is the
shift matrix $\phi(x)$ ($\phi$ is the representation afforded by
$\md$):
\begin{equation}\label{ashift}
\phi(x) = 
\begin{bmatrix}
0 &&& a\\
1 & 0 \\
& \ddots & \ddots \\
&& 1 & 0
\end{bmatrix},
\end{equation}
which is again a special case of \eqref{companion}.  The boundary
condition $x^n = a$ is represented by the weighted edge from $x^{n-1}$ to
$x^0$.

\subsection{Diagonalization Properties}

To determine the diagonalization properties of $\four$ in
\eqref{gendft}, we need to calculate the coordinate form $\phi(h)$ of
filters $h\in\alg$ (see Theorem~\ref{diagprop}).

We consider first $h = x^\ell$. To compute $\phi(h)$,
we determine how $h$ operates on the basis $b$ of $\md$. We get,
for $0\leq k < n$,
$$
x^\ell\cdot x^k = 
\left\{
\begin{array}{ll}
x^{\ell+k}, & \ell+k < n\\
ax^{\ell+k-n}, & \ell+k \geq n
\end{array}\right..
$$ 
Thus, $\phi(h)$ has the entry $1$ at positions $(i,j)$ with
$i-j=\ell$, the entry $a$ at positions $(i,j)$ with $i-j=\ell-n$, and
zero entries else ($0\leq i,j < n$). The special case $\ell=1$ yields 
the shift matrix \eqref{ashift} for $a=1$, also called cyclic shift.
In the general case $h = \sum
h_\ell x^\ell$, 
\begin{equation}\label{ksh}
\phi(h) = \phi\bigl(\sum h_\ell x^\ell\bigr) = \sum h_\ell\phi(x^\ell)
\end{equation}
is the generic
matrix diagonalized by $\four$ in \eqref{gendft} and given by

\begin{equation}\label{gencirc}
\phi(h) =
\left[
\begin{array}{ccccc}
h_0 & ah_{n-1} & ah_{n-2} & \hdots & ah_{1} \\
h_1 & h_0 & ah_{n-1} & \hdots & ah_{2} \\
h_2 & h_1 & h_0  & \ddots & \vdots \\
 \vdots & \ddots & \ddots & \ddots & ah_{n-1} \\
h_{n-1} & \hdots & h_{2} & h_{1} & h_0
\end{array}
\right].
\end{equation}
Further, by Theorem~\ref{diagprop},
$$
\four\cdot\phi(h)\cdot\four^{-1} =
\diag(h(\alpha_0),\dots,h(\alpha_{n-1})),
$$
where the $\alpha_k$ are the zeros of $x^n-a$ given by
\eqref{zerosofxna}. The most prominent special case is 
obtained for $h = x$ and $a = 1$, and shows that the DFT diagonalizes
the cyclic shift.

All entries in $\phi(h)$ above the main diagonal are due to the signal
extension. Matrices with this structure are sometimes called
$a$-circulant (e.g., \cite{Pan:01}). For $a=1$, $\phi(h)$ is an
ordinary circulant matrix, which is diagonalized by $\four =
\DFT_n$. The structure of the matrix $\phi(h)$ reflects the 
choice of basis $b$ {\em in the algebra} $\alg$, namely, $\phi(h)$ is
a sum of weighted $k$-fold time shifts as seen from~\eqref{ksh}.

\subsection{Convolution Theorem}

Filtering in the signal space $\md$ is the multiplication of two
polynomials $s\in\md$ (the signal) and $h\in\alg$ (the filter) modulo
$x^n - a$. Using Theorem~\ref{convolution}, we obtain the following
known convolution theorem, where $\four$ is any matrix in
\eqref{gendft}:
\begin{eqnarray*}
h\cdot s\text{ mod }(x^n-a) & \Leftrightarrow & \phi(h)\cdot\coord{s}\\
 & \Leftrightarrow & (D\four)^{-1}((\four\coord{h})\odot(\four\coord{s})),
\end{eqnarray*}
where, as usual, $\coord{h}, \coord{s}$ are the coefficient vectors of
$h, s$ with respect to the common basis $b$ of $\alg$ and $\md$,
respectively.

\subsection{Unitary Transform}

The matrix $\DFT_n$ diagonalizes the matrix $\phi(x)$, which is
unitary and has pairwise distinct eigenvalues. Thus, a diagonal matrix
$D$ exists such that $D\DFT_n$ is unitary. It is well-known (and
easily computable) that $D = 1/\sqrt{n}\one_n$. Since the scaling is
from the left, and using Theorem~\ref{fourtrafo}, it follows that the
unitary DFT is also a Fourier transform for
$\C[x]/(x^n-1)$. Similarly, for all generalized DFTs of types
$t\in\{1,2,3,4\}$ (see Section~\ref{dftvars}),
\begin{equation}\label{orthgendft}
\frac{1}{\sqrt{n}}\DFTt{t}_n
\end{equation}
is a unitary matrix, and is a Fourier transform
for the same signal model as its non-unitary counterpart.

\subsection{Real Signal Model: RDFTs and DHTs}\label{realdfts}

One question that arises is how to include the real versions of the
DFT, in particular, the real discrete Fourier transform (RDFT) and the
discrete Hartley transform (DHT), in the algebraic theory of signal
processing. In other words, for which signal models are the RDFT and
the DHT Fourier transforms? It turns out that different signal models
can be associated to these transforms, which give rise to different
interpretations. In this paper we discuss two interpretations.  The
first interpretation, in this section, identifies these transforms
according to their name as {\em real} DFTs, i.e., as Fourier transforms
for the real analogue of the finite time model. The second
interpretation identifies them as Fourier transforms for a particular
choice of finite space model and is discussed in
Section~\ref{altrealfin}.

The algebraic interpretation of the DHT as real DFT in this section is
equivalent to recognizing the DHT as a special case of an ADFT
(algebraic discrete Fourier transform), a general concept introduced
in \cite{Beth:84,Beth:83} and rediscovered (using a different name) in
\cite{Hong:94} to map the DFT and its algorithms into a basefield
smaller than $\C$, i.e., a basefield in which the $n$th roots of unity
are not available. Using this method, DHT algorithms are readily
obtained from their DFT counterparts \cite{Beth:89,Hong:94}.

\mypar{Real signal model}
To derive the real DFTs, we start with the finite time model (see
Section~\ref{buildfinztrafo}), but restrict it to {\em real} signals
$\coord{s}\in\R^n$:
\begin{equation}\label{realfintime}
\Phi:\ \coord{s}\mapsto\sum_{0\leq k<n}s_kx^k\in\md = \alg = 
\R[x]/(x^n-1).
\end{equation}
This model has the same visualization as its complex counterpart and
thus imposes the same structure. This implies that filtering is defined
as in the complex case but restricted to real filters.

\mypar{Spectrum and Fourier transform}
The difference arises when computing the spectrum. Since only real
numbers are available, and the roots of $x^n-1$ are complex, $\md$
cannot be decomposed into one-dimensional irreducible modules. In
algebraic terms, $\R$ is not a {\em splitting field} for the
$\alg$-module $\md$. Over $\R$, the irreducible factors of $x^n-1$ are
polynomials of degree 1 or 2. Namely, if $\omega_n^k,
\omega_n^{n-k}$ are conjugated complex roots of $x^n-1$, i.e., 
$k\neq 0,n/2$, then
$$
(x-\omega_n^k)(x-\omega_n^{n-k}) = x^2 - \cos(2k\pi/n)x + 1
$$
is irreducible over $\R$. In other words, if
$$
\C[x]/(x-\omega_n^0),\C[x]/(x-\omega^1_n),\dots,\C[x]/(x-\omega_n^{n-1})
$$ 
is the spectrum of the complex finite time model, then the spectrum
of the real finite time model consists of one or two irreducible
modules of dimension 1: 
$$
\R[x]/(x-1)\text{ and }\R[x]/(x+1) = \R[x]/(x-\omega_n^{n/2}),
\text{ ($n$ even)},
$$
and the remaining spectral components are of dimension 2:
$$
\R[x]/(x^2 - 2\cos(2k\pi/n) + 1),\quad 1\leq k < n/2.
$$
Each of these
spectral components affords a two-dimensional representation $\phi_k$;
if $h$ is a filter, then $\phi_k(h)$ is the frequency response at
frequency $k$. The real spectrum is obtained from the complex spectrum
by {\em fusing} every complex spectral component with its conjugate
counterpart. In the two-dimensional spectral components any real basis
can be chosen. Thus, any {\em real} matrix of the form
\begin{equation}\label{realdft}
\four = X\cdot\DFT_n
\end{equation}
with an invertible matrix $X$ of the x-shaped form
\begin{equation}\label{xshape}
X = 
\left[
\begin{array}{ccccc}
\ast & 0 & \cdots & \cdots & 0\\
0 & \ast & & & \ast \\
\vdots & & \ddots & \oddots \\
\vdots & & \oddots & \ddots \\
0 & \ast & & & \ast
\end{array}
\right],
\end{equation}
is a Fourier transform for the real finite time model. The matrix $X$
consists of one $1\times 1$ block at entry with row and column
indices~0 and one $1\times 1$ block at entry with row and column
indices $n/2$ (if $n$ is even). The remaining $2\times 2$ blocks in
\eqref{xshape} occur at index sets $(k,n-k)\times (k,n-k)$, 
for $1\leq k<n/2$; $k, n-k$ are the indices of conjugate pairs of
complex spectral components of $\C[x]/(x^n-1)$.  Examples of such
transforms include the real discrete Fourier transform (RDFT) and the
discrete Hartley transform (DHT) \cite{Bracewell:83}, defined
respectively by
$$
\RDFT_n = [r_{k\ell}]_{0\leq k,\ell < n},
$$
with
$$
r_{k\ell} =
\begin{cases}
\phantom{-}\cos\ts\frac{2\pi k\ell}{n}, &
0\leq k\leq n/2,\\
-\sin\ts\frac{2\pi k\ell}{n}, & n/2 < k < n.
\end{cases}
$$
and
$$
\begin{array}{rcl}
\DHT_n & = & [\cos\ts\frac{2k\ell\pi}{n} + 
  \sin\frac{2k\ell\pi}{n}]_{0\leq k,\ell,n}\\
& = & [\cas\ts\frac{2k\ell\pi}{n}]_{0\leq k,\ell,n}
\end{array}
$$
Compare this to the DFT, defined as
$$
\DFT_n = [\cos\ts\frac{2k\ell\pi}{n} -
  j\sin\frac{2k\ell\pi}{n}]_{0\leq k,\ell,n}.
$$

The RDFT replaces two conjugate base vectors with indices $k,n-k$ in
the DFT with real and imaginary part as ($j = \sqrt{-1}$)
\begin{equation}\label{rdftmethod}
\begin{pmatrix}a\\b\end{pmatrix} =
\fh \begin{bmatrix}1&1\\-j&j\end{bmatrix}
\begin{pmatrix}a+bj\\a-bj\end{pmatrix},
\end{equation}
which determines the matrix $X = X^{\RDFT}$. Similarly, the DHT
replaces two conjugate base vectors by the difference and sum of their
real and imaginary part as
\begin{equation}\label{dhtmethod}
\begin{pmatrix}a-b\\a+b\end{pmatrix} =
\fh \begin{bmatrix}1+j&1-j\\1-j&1+j\end{bmatrix}
\begin{pmatrix}a+bj\\a-bj\end{pmatrix},
\end{equation}
which again determines $X = X^{\DHT}$.

We can now  represent the set of all Fourier transforms
for the real finite time model more conveniently by the set of all matrices
\begin{equation}\label{xdegfreedom}
\four = X\cdot\RDFT_n,
\end{equation}
where $X$ is any real, invertible matrix of the shape in
\eqref{xshape}. For example, 
$$
\DHT_n = X^{\DHT}(X^{\RDFT})^{-1}\RDFT_n.
$$
The reader should compare \eqref{xdegfreedom} to the set of all
Fourier transforms in Theorem~\ref{fourtrafo}. There, the spectrum
consisted only of one-dimensional spectral components; thus, the
degree of freedom in choosing a basis in the spectrum yielded a
diagonal matrix. Here, two-dimensional modules occur; thus, the degree
of freedom in choosing bases is larger and leads to the x-shaped
matrix~$X$.

Both transforms, the RDFT and the DHT, are special among the class of
all possible real DFTs. The RDFT appears to have the lowest arithmetic
complexity\footnote{We do not have a proof. The assertion is
based on the best known algorithms.} and the DHT is uniquely
determined (up to a diagonal matrix $D$ with diagonal entries 1 or -1)
by being equal to its inverse.

In Section~\ref{dftvars} we discussed 4 types of DFTs and identified
them as Fourier transforms (scaled or unscaled polynomial transforms)
for $\C[x]/(x^n-1)$ and $\C[x]/(x^n+1)$. Using the above methods
\eqref{rdftmethod} and \eqref{dhtmethod}, each of these 4 types has 
a corresponding ``RDFT'' and ``DHT,'' which we denote accordingly with
$\RDFTt{1},\dots,\RDFTt{4}$, and with $\DHTt{1},\dots,\DHTt{4}$,
respectively. By construction, the RDFT and the DHT of type $t$,
$t\in\{1,2,3,4\}$, are Fourier transforms for the real counterpart of
the complex signal model for the DFT of type $t$. For completeness, we
provide the definitions for types 2--4:
$$
\RDFTt{t} = [r_{k,\ell}]_{0\leq k,\ell<n},
$$
with
$$
\renewcommand{\arraystretch}{3}
\begin{array}{rl}
$t=2:$ &
r_{k\ell} =
\begin{cases}
\phantom{-}\cos\ts\frac{2\pi k(\ell+1/2)}{n}, &
0\leq k\leq (n-1)/2,\\
-\sin\ts\frac{2\pi k(\ell+1/2)}{n}, & 
 (n-1)/2 < k < n.
\end{cases}\\
$t=3:$ &
r_{k\ell} =
\begin{cases}
\phantom{-}\cos\ts\frac{2\pi (k+1/2)\ell}{n}, &
0\leq k\leq (n-1)/2,\\
-\sin\ts\frac{2\pi (k+1/2)\ell}{n}, & 
 (n-1)/2 < k < n.
\end{cases}\\
$t=4:$ &
r_{k\ell} =
\begin{cases}
\phantom{-}\cos\ts\frac{2\pi (k+1/2)(\ell+1/2)}{n}, &
0\leq k\leq n/2-1,\\
-\sin\ts\frac{2\pi (k+1/2)(\ell+1/2)}{n}, & 
 n/2-1 < k < n.
\end{cases}\\
\end{array}
$$
And,
$$
\renewcommand{\arraystretch}{2}
\begin{array}{rcl}
\DHTt{2}_n & = & [\cas\ts\frac{2k(\ell+1/2)\pi}{n}]_{0\leq k,\ell,n},\\
\DHTt{3}_n & = & [\cas\ts\frac{2(k+1/2)\ell\pi}{n}]_{0\leq k,\ell,n},\\
\DHTt{4}_n & = & [\cas\ts\frac{2(k+1/2)(\ell+1/2)\pi}{n}]_{0\leq k,\ell,n}.
\end{array}
$$

Further, for $t\in\{1,2,3,4\}$, we have relations of the form
\begin{equation}\label{relrecom}
\begin{array}{rcl}
\RDFTt{t} & = & X^{(t)}\DFTt{t}_n,\\
\DHTt{t} & = & X^{(t)'}\DFTt{t}_n,
\end{array}
\end{equation}
where $X^{(t)}$, and $X^{(t)'}$ are of the form \eqref{xshape} for
$t=1,2$, and of the form
\begin{equation}\label{xshape1}
X = 
\left[
\begin{array}{ccccc}
\ast & & & \ast \\
 & \ddots & \oddots \\
 & \oddots & \ddots \\
\ast & & & \ast
\end{array}
\right]
\end{equation}
for $t=3,4$.

The four transforms $\DHTt{t}$ were introduced (in their orthogonal
form) in \cite{Wang:81,Wang:81a,Wang:81b,Wang:85}, where they were
called discrete W transforms (DWTs) of type 1--4.  Our above suggestion
to renaming these transforms to DHTs of type 1--4 is motivated by 1)
the name DHT (for type 1) is much more commonly used than DWT, and the
types 2--4 are just variants; and 2) even though the DHT and the DWT
were introduced at the about the same time (\cite{Bracewell:83} and
\cite{Wang:85}), the continuous counterpart was
introduced by Hartley already in 1942 \cite{Hartley:42}.

\mypar{Diagonalization properties} The above discussion gives
immediately the ``diagonalization'' properties of the RDFT and DHT. We
use double quotes, since these properties are not actually a
diagonalization. If $h\in\alg = \R[x]/(x^n-1)$ is any filter, then
$\phi(h)$ is a real circulant matrix, i.e, of the form \eqref{gencirc}
with $a = 1$. Then
\begin{equation}\label{rdftdiag}
\RDFT_n\phi(h)\RDFT_n^{-1} = X,
\end{equation}
where $X$ is real and of the form \eqref{xshape}. The same holds, if
we replace RDFT by any other real DFT including the DHT.  Of course,
the RDFT and DHT have also true diagonalization properties (as every
invertible matrix), but they do not arise from their interpretation in
this section, but from their different interpretation in
Section~\ref{altrealfin}.

Further, \eqref{rdftdiag} generalizes to the RDFTs and DHTs of types
2--4. For type 3 and 4, $X$ in \eqref{rdftdiag} has the form
\eqref{xshape1}.

Similarly, convolution theorems can be derived. Also, the above
discussion can be easily generalized to every real signal model for
the case of a generic monomial signal extension $\R[x]/(x^n-a)$,
$a\in\R$.

\mypar{Orthogonal transform}
The orthogonal version of the RDFTs and DHTs of types 1--4 follows 
directly from \eqref{orthgendft} and \eqref{relrecom}, namely, for 
$t\in\{1,2,3,4\}$,
$$
\sqrt{\frac 2n}D\cdot\RDFTt{t}_n\quad\text{and}
\quad\sqrt{\frac 1n}\cdot\DHTt{t}_n
$$ 
are orthogonal, where 
$$
D = 
\begin{cases}
\diag(\sqrt{2},1,\dots,1,\sqrt{2},1,\dots,1), & t=1,2,\ n\text{ even},\\
\diag(\sqrt{2},1,\dots,1), & t=1,2,\ n\text{ odd},\\
\one_n, & t=3,4,\ n\text{ even}\\
\diag(1,\dots,1,\sqrt{2},1,\dots,1), & t=3,4,\ n\text{ odd}.
\end{cases}
$$

\subsection{Rational Signal Model}\label{rationaldfts}

At this point it is interesting to extend the above discussion of real
DFTs by further reducing the base field from $\R$ to the field of
rational numbers $\Q$.

\mypar{Rational signal model} The linear mapping for the signal model is now
\begin{equation}\label{eq:rationalsigmodel}
\Phi:\ \coord{s}\mapsto\sum_{0\leq k<n}s_kx^k\in\md = \alg = 
\Q[x]/(x^n-1).
\end{equation}
As in the real case before, the model imposes the same structure
(visualization, notion of filtering) as its complex counterpart with
the restriction that signals and filters have coefficients in $\Q$.

\mypar{Spectrum and Fourier transform}
Reducing the basefield in the real case above had the effect of fusing
spectral components. This effect is even more pronounced in the
rational case as we show next.

We restrict ourselves to a 2-power size $n = 2^k$. Then, $x^n-1$
decomposes over $\Q$ into irreducible factors as
$$
x^{2^k}-1 = (x-1)(x+1)(x^2+1)\cdots(x^{2^{k-1}}+1).
$$
These factors determine the spectrum of the rational signal model
$\Phi$; namely, the spectral components are
$$
\C[x]/(x-1)\quad\text{and}\quad
\C[x]/(x^{2^i}+1),\ 0\leq i<k.
$$
Clearly, there is now a large degree of freedom in choosing bases in
the spectral components, i.e., in defining a Fourier transform
$\four$.  In the following, we assume the standard monomial basis in
each spectral component, and derive $\four$ recursively.  We will call
this transform $\QDFT_{2^k}$.  For $k = 1$, $x^2-1$ decomposes over
$\Q$ as over $\C$,
$$
\C[x]/(x^2-1)\rightarrow \C[x]/(x-1)\dirsum\C[x]/(x+1).
$$
Thus, a Fourier transform is $\QDFT_2 = \DFT_2$.

For an arbitrary size $2^{k+1}$, we use the CRT to get
the partial decomposition
\begin{equation}\label{rhtdec}
\C[x]/(x^{2^{k+1}}-1)\rightarrow
\C[x]/(x^{2^{k}}-1)\dirsum\C[x]/(x^{2^{k}}+1).
\end{equation}
To compute the base change matrix for \eqref{rhtdec}, we determine the
coordinate vector of each basis elements $x^\ell$ of the left-hand
side in \eqref{rhtdec} w.r.t.~the basis on the right-hand side. These
coordinate vectors are the columns of the base change matrix.  Namely,
for $0\leq \ell< 2^k$,
$$
\begin{array}{rcl}
x^\ell\equiv x^\ell\text{ mod }x^{2^k}-1,\\
x^\ell\equiv x^\ell\text{ mod }x^{2^k}+1,
\end{array}
$$
and
$$
\begin{array}{rcl}
x^{2^k+\ell}\equiv x^\ell\text{ mod }x^{2^k}-1,\\
x^{2^k+\ell}\equiv -x^\ell\text{ mod }x^{2^k}+1.
\end{array}
$$
The resulting matrix is
$$
\begin{bmatrix}
\one_{2^{k}} & \phantom{-}\one_{2^{k}}\\
\one_{2^{k}} & - \one_{2^{k}}
\end{bmatrix} =
\DFT_2\tensor\one_{2^{k}}.
$$
The left side in \eqref{rhtdec} is decomposed over $\Q$ by
$\QDFT_{2^{k+1}}$, the right side is decomposed recursively by
$\QDFT_{2^{k}}\dirsum\one_{2^{k}}$ (since $\C[x]/(x^{2^{k}}+1)$ is
irreducible). Thus, the recursion for $\QDFT$ is
\begin{equation}\label{recqdft}
\QDFT_{2^{k+1}} = (\QDFT_{2^{k}}\dirsum\one_{2^{k}})
  (\DFT_2\tensor\one_{2^{k}}).
\end{equation}
In summary, $\QDFT_{2^k}$ is a Fourier transform for the rational
finite time model with $\Phi$ in \eqref{eq:rationalsigmodel}, with the
monomial basis chosen in each spectral component. Note that $\QDFT_n$
is, up to a permutation of the columns, equal to the rationalized Haar
transform (RHT). The RHT is recursively defined as \eqref{recqdft},
but with a permutation multiplied from the right. This implies that
the RHT could be seen as a Fourier transform for a rational signal
model with $\Phi'$ as linear mapping, which arises from $\Phi$ by
permuting the monomial basis. However, we do not pursue further this
interpretation here since this signal model and transform $\Phi'$ are
better characterized in the context of wavelets and filterbanks.

Since, the degree of freedom in choosing a Fourier transform for the
rational signal model with $\Phi$ in \eqref{eq:rationalsigmodel} is in
the choice of bases in the spectrum, {\em all} its Fourier transforms
are given by rational matrices of the form
$$
B\cdot\QDFT_{2^k},
$$
where $B$ is rational and a direct sum of invertible matrices of
increasing block sizes $1,1,2,4,\dots,2^{k-1}$.

\mypar{Diagonalization properties}
If $h\in\Q[x]/(x^n-1)$ is a rational filter, i.e., $\phi(h)$ is a
rational circulant matrix, then
$$
\QDFT_n\phi(h)\QDFT_n^{-1} = B,
$$
where $B$ has the same block structure as above.

\section{Modeling Space: The C-Transform}\label{Ctrafo}

Signal models realized by the infinite and finite $z$-transforms are
{\em time} models; the signal samples are along an oriented time axis,
i.e., with intrinsic direction (from past to future).  Algebraically,
this direction is described by the operator $q$ and its action
$\diamond$ on the time-marks $t_n$ (see \eqref{opq} and
Figure~\ref{timeshift}).

In many applications, however, the signal is not sampled along time,
but along {\em space}. Important examples are sequences of pixels in
images. Space has no intrinsic direction.  In the absence of further
conditions, space is inherently {\em symmetric}. Thus, a model based
on the $z$-transform is not appropriate.

We will now develop a {\em space model} for signals following the
exact same steps as in Section~\ref{ztrafo}. However, the starting
point will be different: a shift operation defined to model space.  It
is worth emphasizing this seemingly small but crucial change, since
the shift in classical signal processing has one and only one meaning
\eqref{opq}.

We will see that the realization of this model in terms of ordinary
multiplication leads naturally to the Chebyshev polynomials and the
$C$-transform that we will introduce. In the following section, we
will then show that the 16 discrete trigonometric transforms~(DTT)
arise from this model in the same way as the DFT arises from the
$z$-transform.

\subsection{Building the Signal Model}\label{buildCtrafo}

\begin{figure}[ht]\centering
\begin{picture}(100,35)
\multiput(-40,10)(4,0){8}{$\cdot$}
\put(0,10){$\bullet$}
\put(0,0){$t_{n-1}$}
\put(20,18){\makebox(0,0){$\xy\ar @/^-1ex/ (-10,0)|{\frac{1}{2}}\endxy$}}
\put(36,10){$\bullet$}
\put(36,28){\makebox(0,0){$q\diamond$}}
\put(36,0){$t_{n}$}
\put(56,18){\makebox(0,0){$\xy\ar @/^1ex/ (10,0)|{\frac{1}{2}}\endxy$}}
\put(72,10){$\bullet$}
\put(72,0){$t_{n+1}$}
\multiput(86,10)(4,0){8}{$\cdot$}
\end{picture}
\caption{The space shift $q\diamond t_n$.\label{spaceshift}}
\end{figure}
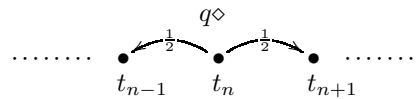

\mypar{Definition of the shift}
Analogously to Section~\ref{buildztrafo}, we consider discrete signals
$\coord{s}\in\C^\Z$, i.e., we consider the vector space $V=\C^\Z$. We define
now {\em space} marks $t_n$ and an appropriate {\em space} shift
operator $q$ and its operation~$\diamond$ on the space marks.  As
mentioned above, $q$ should operate symmetrically. We adopt the  definition is
\begin{equation}\label{spaceop}
\framebox[1.1\width]{\rule[-1.3mm]{0mm}{5mm}%
$\text{\bf space model:}\quad q\diamond t_n = (t_{n+1} + t_{n-1})/2$}
\end{equation}
for $n\in\Z$. Figure~\ref{spaceshift} shows a graphical representation
of the space shift and should be compared to Figure~\ref{timeshift}.

As in Section~\ref{buildztrafo}, we proceed by extending the operator
domain from $q$ to $k$-fold shift operators $q_k$. A natural
definition of the $k$-fold space shift is
\begin{equation}\label{kspaceshift}
q_k\diamond t_n = (t_{n+k} + t_{n-k})/2,
\end{equation}
since $t_{n+k}$ and $t_{n-k}$ are those space marks at distance $k$
from $t_k$.

Here we have the first interesting difference with respect to the time model
derivation, since clearly $q_k\neq q^k$. Furthermore,
\eqref{kspaceshift} implies $q_k = q_{-k}$; hence, it is sufficient to
consider only shift operators $q_k$ with $k\geq 0$. This agrees with our
intuition: $q$ operates symmetrically on $t_n$, hence there are no
negative $k$-fold space shifts. Thus, the natural representation of a
filter is $\sum_{k\geq 0}h_kq_k$. The following lemma shows that the
$q_k$ are given by the Chebyshev polynomials of the first kind $T_k$
(see Appendix~\ref{chebs}) in the variable $q$.

\begin{lemma}\label{kspaceshiftT}
The $k$-fold space shift operator is given by $q_k = T_k(q)$.
\end{lemma}
\begin{proof}
Induction on $k$. By definition $q_0 = 1$, and $q_1 = q$.  Also by
definition, $q_{k+1}\diamond t_n = (t_{n+k+1} + t_{n-k-1})/2 =
(t_{n+k+1}+t_{n+k-1}+t_{n-k+1}+t_{n-k-1})/2 - (t_{n+k-1}+t_{n-k+1})/2
= 2q\diamond(t_{n+k}+t_{n-k})/2 - (t_{n+k-1}+t_{n-k+1})/2 =
(2qq_k-q_{k-1})\diamond t_n$, for $n\in\Z$. From the induction hypothesis,
$q_k = T_k(q),\ q_{k-1} = T_{k-1}(q)$, and thus, using the recurrence
of the Chebyshev polynomials introduced below in~\eqref{gencheb},
$q_{k+1} = T_{k+1}(q)$, as desired.
\end{proof}

The Chebyshev polynomials, which just surprisingly emerged as the
$k$-fold space shift operator~$q_k$, will play a central role in the
definition of the space model. For this reason, we provide the
necessary background on Chebyshev polynomials in Appendix~\ref{chebs},
which we encourage the reader to briefly review at this point.

\mypar{Linear extension} To construct a linear signal model we
 extend by linearity the operation of~$q$ to the entire set $\md =
\{s=\sum_{n\in\Z}s_nt_n\}$, and extend linearly the operator domain to $\alg =
\{h = \sum_{k\geq 0}h_kT_k(q)\}$. Note that we used the result of Lemma~\ref{kspaceshiftT} in~$\alg$.  

\mypar{Realization} Analogous to Section~\ref{buildztrafo}, we
determine a ``realization'' of the model introduced in the previous
section.  We set in~\eqref{spaceop} $q = x$, $\diamond = \cdot$, and
determine polynomials $C_n$ that replace the space marks $t_n$ in
\eqref{spaceop}, i.e., that satisfy
\begin{equation}\label{realspaceop}
x\cdot C_n = (C_{n+1} + C_{n-1})/2.
\end{equation}
Since \eqref{realspaceop} is equivalent to \eqref{gencheb} (in
Appendix~\ref{chebs}), the solution is given by a sequence of
Chebyshev polynomials.

We immediately notice differences with respect to the corresponding
derivation in Section~\ref{buildztrafo}. These differences are intrinsic to
the space model:
\begin{itemize}
\item Equation \eqref{realspaceop} is a three-term recurrence for the
space marks, whereas \eqref{twoterm} is a two-term recurrence for the
time marks.
\item Only the $C_n$, $n\geq 0$, are linearly independent; the $C_n$,
$n < 0$, are polynomials in $x$ and can thus be expressed as linear
combinations of $\{C_n\mid n\geq 0\}$. In other words, the realization
of the space model introduces a starting point in space, given by $C_0
= 1$. Fixing $C_1$ determines the left boundary condition and the left
signal extension.
\item As a consequence, even after normalizing $C_0 = 1$, the sequence
$C_n$ of Chebyshev polynomials is not uniquely determined. The degree
of freedom is given by the choice of $C_1$ as a polynomial of
degree~1.
\item Again, we note that in the time model, a $k$-fold shift operator is given
by $x^k$:
$$
x^k\cdot x^n = x^{n+k},
$$ 
in contrast to the space model, where, by Lemma~\ref{kspaceshiftT},
the $k$-fold shift operator is given by $T_k(x)$, independent of $C$
(compare with Lemma~\ref{chebprop} iv)):
\begin{equation}\label{Top}
T_k\cdot C_n = (C_{n+k} + C_{n-k})/2.
\end{equation}
\end{itemize}

As a result of this discussion, we obtain the spaces $\alg =
\{h=\sum_{k\geq 0}h_kT_k\}$ and $\md = \{s=\sum_{n\geq 0}s_nC_n\}$,
i.e., the signal model that we obtain later will be only for
right-sided sequences.

Table~\ref{corrspace} shows the correspondence between abstract and
realized concepts.

\begin{table*}\centering
\caption{Realization of the abstract space model.\label{corrspace}}
\ra{1.2}
\begin{tabular}{@{}lll@{}}\toprule
concept & abstract & realized \\ \midrule
shift operator & $q$ & $T_1(x) = x$ \\
shift operation & $\diamond$ & $\cdot$ \\
space mark & $t_n$ & $C_n$ \\
$k$-fold shift operator & $q_k=T_k(q)$ & $T_k(x)$ \\
space shift & $q\diamond t_n = \fh(t_{n+1} + t_{n-1})$ &
$x\cdot C_{n} = \fh(C_{n+1} + C_{n-1})$ \\
signal & $\sum s_nt_n$ & $\sum s_n C_n(x)$ \\
filter & $\sum h_k T_k(q)$ & $\sum h_k T_k(x)$ \\ \bottomrule
\end{tabular}
\end{table*}

To ensure convergence, we would like to require as before
$\coord{h}\in\ell^1(\N)$ and $\coord{s}\in\ell^2(\N)$. However, to
prove convergence we have first to choose proper boundary conditions,
i.e., we have to choose the proper Chebyshev polynomials $C$. We
analyze the boundary conditions in the next paragraph. This discussion
has no counterpart in the time model derivation in
Section~\ref{ztrafo}.

\mypar{Left boundary condition and left signal extension}
The degree of freedom for choosing a Chebyshev sequence $C$,
normalized by $C_0 = 1$, is given by the choice of $C_1$, or,
equivalently, by the choice of $C_{-1}$, since the entire
sequence is then obtained by applying the Chebyshev recursion
\eqref{gencheb} in both directions (see Lemma~\ref{chebprop}, i)).
Fixing either $C_1$ or $C_{-1}$ is equivalent to choosing a {\em left
boundary condition (b.c.)} for the signal $\coord{s} = (s_0, s_1,
\dots)$. For example, setting $C_1 = x$ implies $C_{-1} = x$, and thus
$C_{-1} = C_1$, which imposes for the signal $s$ the left \bc $s_{-1}
= s_1$. Using Table~\ref{4cheb}, the corresponding sequence is $C =
T$.

To determine the left \bc~in the general case, we set $C_0 = 1$ and
$C_1 = ax+b$, $a\neq 0$ (to satisfy $\deg(C_1) = 1$). Then, by
applying \eqref{gencheb} backwards, we get
\begin{equation}\label{genleftbc}
C_{-1} = 2x - (ax+b) = \frac{2-a}{a}C_1 - \frac{2b}{a}C_0.
\end{equation}
Since $C_{-1}$ is of degree not larger than 1, every polynomial
$C_{-n}$, $n > 0$, obtained by the recursion \eqref{gencheb}, is of
degree not larger than $n$, and thus a linear combination of the
polynomials $C_0,\dots,C_n$,
\begin{equation}\label{leftse}
C_{-n} = \sum_{0\leq i\leq n}\beta_i\cdot C_i,
\quad n>0,
\end{equation}
which is the {\em left} signal extension associated with the sequence
$C$. On the other hand, by comparing the degrees of freedom, it is
obvious that not every signal extension can be obtained by choosing a
suitable \bc Thus,
$$
(C\Longleftrightarrow\text{ left b.c.})\ 
\Longrightarrow\text{ left signal extension.}
$$
For a generic left b.c., the left signal extension~\eqref{leftse} has
no simple structure; in particular, it is not monomial.  Similar to
Section~\ref{buildfinztrafo}, we determine now those left \bc that
yield a {\em monomial} left signal extension
\eqref{leftse}. The answer is provided in the following lemma.

\begin{lemma}[Monomial left signal extension]\label{leftbc}
Let $C = (C_n\mid n\in\Z)$ be a sequence of Chebyshev polynomials with
$C_0 = 1$ and $\deg(C_1) = 1$. Then the left signal extension
associated with $C$ is monomial, i.e., every $C_k$, $k<0$, is a multiple
of a $C_n$, $n\geq 0$, if and only if $C\in\{T,U,V,W\}$. The
corresponding left \bc's are given by $C_{-1} = C_1, C_{-1} = 0, C_{-1}
= C_0$, and $C_{-1} = -C_0$, respectively.
\end{lemma}
\begin{proof}
If $C\in\{T,U,V,W\}$, then the assertion holds as shown in the
``symmetry'' column of Table~\ref{4cheb}. It remains to show the
converse. We start with the generic left \bc in \eqref{genleftbc}.
Because the signal extension associated with $C$ is monomial, one
of the two summands in \eqref{genleftbc} has to vanish.

Case~1: $C_{-1}$ is a multiple of $C_0$, i.e., constant. It follows $a
= 2$, $C_1 = 2x+b$, $C_{-1} = -b$, $C_{-2} = -2bx-1$. Now, either
$C_{-2}$ is constant, i.e., $b = 0$, which implies $C = U$, or
$C_{-2}$ is a multiple of $C_1$, which implies $b = \pm 1$, or
$C\in\{V,W\}$. 

Case~2: $C_{-1}$ is a multiple of $C_1$. It follows $b
= 0$, $C_1 = ax$, $C_2 = 2ax^2-1$, $C_{-1} = (2-a)x$, $a\neq 2$, and
$C_{-2} = 2(2-a)x^2 - 1$. Since $C_{-2}$ has to be a multiple of $C_2$,
we get $a = 1$ and thus $C = T$. This completes the proof.
\end{proof}

The four boundary conditions derived in Lemma~\ref{leftbc} are the
discrete versions of the so-called Dirichlet \bc (``zero value'') and
von-Neumann \c (``zero slope''), e.g., \cite{Moura:92}. In
each case, the symmetry point is either a ``whole'' sample point, or a
``half'' sample point, i.e., is located between two sample points.

After we identified the suitable \bc's, we can show that filtering is
well-defined (i.e, converges). We assume $\coord{s}\in\ell^2(\N)$ and
$\coord{h}\in\ell^1(\Z)$ and consider the example $C = T$. Using the
power form of $T$ in \eqref{powerform} in Appendix~\ref{chebs},
\begin{eqnarray*}
h\cdot s 
& = & \sum_{n\geq 0}h_nT_n \cdot \sum_{n\geq 0}s_nT_n\nonumber \\
& = & \bigl(h_0 + \sum_{n\geq 1}\frac{h_n}{2}(u^n + u^{-n})\bigr)\isdraft{}{\\}
\isdraft{}{&&\phantom{==}}
  \bigl(s_0 + \sum_{n\geq 1}\frac{s_n}{2}(u^n + u^{-n})\bigr),\label{Cz}
\end{eqnarray*}
which is the ordinary convolution and exists, since the coordinate 
sequences are in $\ell^1(\Z)$ and $\ell^2(\Z)$, respectively.
The resulting Laurent series is again symmetric of the form
$$
t = \bigl(t_0 + \sum_{n\geq 1}\frac{t_n}{2}(u^n + u^{-n})\bigr),
$$
and thus the result is $t = \sum_{n\geq 0}t_nT_n$ and has a coefficient
sequence $\coord{t}\in\ell^2(\N)$.

Similar computations confirm convergence for $C = U,V,.W$.  Note that
the existence of the power form and thus the monomial signal extension
is crucial for this proof.

\mypar{Signal model: \boldmath$C$-transform} Let $(C_n\mid n\in\N)$,
$C\in\{T,U,V,W\}$, be a sequence of Chebyshev polynomials.  We have
constructed a signal model for $V=\ell^2(\N)$, which we call the
$C$-transform, given by
$$
\Phi:\ \coord{s}\mapsto \sum_{n\geq 0} s_nC_n.
$$ 
The module is given by $\md=\Phi(\ell^2(\N))$ and, independent of
the polynomials~$C$, the algebra consists of all series
$\alg=\{\sum_{k\geq 0}h_kT_k\}$ with coefficients
$\coord{h}\in\ell^1(\N)$.

We use~$C$ as a generic notation, but will replace it by
either $T,U,V$, or~$W$, when appropriate, and, accordingly, refer to
the $T$-, $U$-, $V$-, or $W$-transform.

\subsection{Spectrum and Fourier Transform: DSFT}

Again, we consider the case $C=T$. First we identify the spectrum,
i.e., the irreducible modules. Straightforward computation shows that
each series
\begin{equation}\label{eomega}
E_\omega(x) = s_0/2 + \sum_{n\geq 1}\cos n\omega T_n(x)
\end{equation}
is an eigenvector of the shift operator $x$ and thus for all filters in $\alg$.
Namely, $xE_\omega(x) = \cos\omega E_\omega(x)$, or, more general,
\begin{equation}\label{eigenf}
H(x)E_\omega(x) = H(\cos\omega)E_\omega(x).
\end{equation}
As in the time case, the $E_\omega$ are not in $\Phi(\ell^2(\Z))$, but only
in $\Phi(\ell^\infty(\Z))$; thus, the $\alg$-modules $\md_\omega$ are
not submodules of $\md$.

Another way of obtaining the $E_\omega$ is to use again the power form
\eqref{powerform} of $T$. Namely,
every $s=S(x)\in\md$ (and every $h=H(x)\in\alg$) can be written as
\begin{gather*}
S(x) = \sum_{n\in\Z}s_nT_n(x) = 
  s_0 + \sum_{n\geq 1}\frac{s_n}{2}(u^n + u^{-n}),
  \isdraft{\quad}{\\} x = \frac{u^{-1}+u}{2},
\end{gather*}
which has the form of a $z$-transform and shows that $E_\omega$ in
\eqref{eomega} is the sum of two conjugate spectral components
\eqref{eigenfunction} (in the variable $u$) in the time case, and, as
such, is invariant under $u$ and thus under $x=(u^{-1}+u)/2$.

Since $u\mapsto e^{-j\omega}$ implies $x = (u^{-1}+u)/2\mapsto
\cos\omega$, we get the following Fourier transform associated to the
$T$-transform.
$$
\begin{array}{rrcl}
\Delta: & \md & \rightarrow & 
  \bigl(\bigdirsum_{\omega\in [0,\pi]}\C\bigr) = \C^{[0,\pi]}\\
& s = S(x) & \mapsto & S(\cos\omega)_{\omega\in [0,\pi]} = 
  \omega\mapsto S(\cos\omega).
\end{array}
$$ 
The above derivation also shows the existence of this Fourier
transform via the existence of the DTFT. The spectrum of $s$ can be
viewed alternatively as an even function on the circle, since
$S(\cos\omega) = S(\cos(-\omega))$, or as a function on the half
circle.

Further, every spectral component $\md_\omega$ affords a
one-dimensional irreducible representation $\phi_\omega$ of
$\alg$. Namely, from \eqref{eigenf},
$$
\phi_\omega(H(x)) = H(\cos\omega).
$$
The collection, for $\omega\in[0,\pi]$,
$$
(\phi_\omega(H(x)))_\omega = (H(\cos\omega))_\omega = 
  \omega\mapsto H(\cos\omega)
$$
is the frequency response of the filter $h = H(x)$.

Similar derivations provide the Fourier transforms for the cases $C =
U,V,W$, however the existence for $\coord{s}\in\ell^2(\N)$ is not as
easily guaranteed. We refer to books on general orthogonal series, for
example \cite{Alexitis:61,Kashin:89}.

Independent of $C$, we call this Fourier transform, in analogy to the
DTFT, the discrete-space Fourier transform (DSFT).

\subsection{Visualization}\label{spacegraphrep}

\begin{figure}[ht]\centering
\isdraft{\renewcommand{\baselinestretch}{1}\normalsize}{}
{
\centerline{
\xymatrix{
\ar@{.}[r] & \bullet \ar@{<->}[r] & \bullet \ar@{<->}[r] & 
\bullet \ar@{<->}[r] & \bullet \ar@{<->}[r] &
\bullet \ar@{<->}[r] & \bullet \ar@{.}[r] &
}}
\centerline{
\isdraft{\setlength{\unitlength}{1.15pt}}{}
\begin{picture}(175,5)
\put(0,0){\makebox(0,0){$t_{-2}$}}
\put(35,0){\makebox(0,0){$t_{-1}$}}
\put(70,0){\makebox(0,0){$t_0$}}
\put(105,0){\makebox(0,0){$t_1$}}
\put(140,0){\makebox(0,0){$t_2$}}
\put(175,0){\makebox(0,0){$t_3$}}
\end{picture}}
}
\caption{The abstract space model (graphically).\label{abstractspacegraph}}
\end{figure}
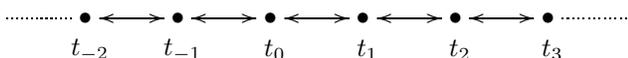

We visualize the space model by a graph using
Definition~\ref{visdef}. However, in contrast to before, there is a
difference between the graph suggested by the abstract model and its
realization. Namely, from Figure~\ref{spaceshift}, the graph for the
space model should look like Figure~\ref{abstractspacegraph}.  As in
Figure~\ref{timegraph}, the edges represent the space
shift (we drop the common weight factor $1/2$ of all edges).

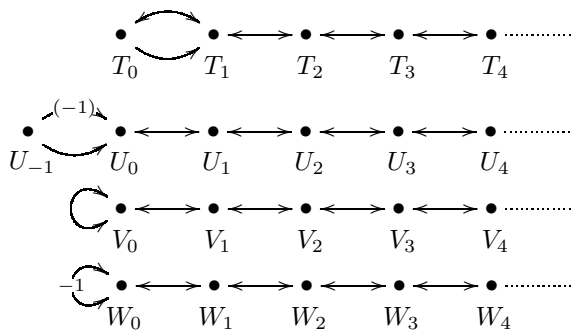
\begin{figure}[ht]\centering
\isdraft{\hspace*{37mm}}{\hspace*{15mm}}
\begin{minipage}{0.8\textwidth}
{
\xymatrix{
\bullet \ar@/_2ex/[r] & \ar@{<->}@/_2ex/[l] \bullet \ar@{<->}[r] &
\bullet \ar@{<->}[r] & \bullet \ar@{<->}[r] & \bullet \ar@{.}[r] &
}
\isdraft{\setlength{\unitlength}{1.15pt}}{}
\begin{picture}(150,5)
\put(7,0){\makebox(0,0){$T_0$}}
\put(42,0){\makebox(0,0){$T_1$}}
\put(77,0){\makebox(0,0){$T_2$}}
\put(112,0){\makebox(0,0){$T_3$}}
\put(147,0){\makebox(0,0){$T_4$}}
\end{picture}
}
\end{minipage}

\bigskip
\isdraft{\hspace*{15mm}}{\hspace*{2.6mm}}
\begin{minipage}{0.8\textwidth}
{
\xymatrix{
\bullet \ar@/^2ex/[r]|{(-1)}
\ar@/_2ex/[r] &
\bullet \ar@{<->}[r] &  \bullet \ar@{<->}[r] &
\bullet \ar@{<->}[r] & \bullet \ar@{<->}[r] & \bullet \ar@{.}[r] &
}
\isdraft{\setlength{\unitlength}{1.15pt}}{}
\begin{picture}(150,5)
\put(7,0){\makebox(0,0){$U_{-1}$}}
\put(42,0){\makebox(0,0){$U_0$}}
\put(77,0){\makebox(0,0){$U_1$}}
\put(112,0){\makebox(0,0){$U_2$}}
\put(147,0){\makebox(0,0){$U_3$}}
\put(182,0){\makebox(0,0){$U_4$}}
\end{picture}
}
\end{minipage}

\bigskip
\isdraft{\hspace*{38mm}}{\hspace*{15mm}}
\begin{minipage}{0.8\textwidth}
{
\xymatrix{
\bullet \ar@{<->}@(ul,dl) \ar@{<->}[r] & \bullet \ar@{<->}[r] & 
\bullet \ar@{<->}[r] & \bullet \ar@{<->}[r] & \bullet \ar@{.}[r] &
}
\isdraft{\setlength{\unitlength}{1.15pt}}{}
\begin{picture}(150,5)
\put(7,0){\makebox(0,0){$V_0$}}
\put(42,0){\makebox(0,0){$V_1$}}
\put(77,0){\makebox(0,0){$V_2$}}
\put(112,0){\makebox(0,0){$V_3$}}
\put(147,0){\makebox(0,0){$V_4$}}
\end{picture}
}
\end{minipage}

\bigskip
\isdraft{\hspace*{32mm}}{\hspace*{8.5mm}}
\begin{minipage}{0.8\textwidth}
{
\xymatrix{
\bullet \ar@{<->}@(ul,dl)|{-1} \ar@{<->}[r] & \bullet \ar@{<->}[r] & 
\bullet \ar@{<->}[r] & \bullet \ar@{<->}[r] & \bullet \ar@{.}[r] &
}
\isdraft{\setlength{\unitlength}{1.12pt}}{}
\begin{picture}(170,5)
\put(25,0){\makebox(0,0){$W_0$}}
\put(60,0){\makebox(0,0){$W_1$}}
\put(95,0){\makebox(0,0){$W_2$}}
\put(130,0){\makebox(0,0){$W_3$}}
\put(165,0){\makebox(0,0){$W_4$}}
\end{picture}
}
\end{minipage}

\medskip
\caption{Visualization of the space models given by the $C$-transforms 
for $C\in\{T,U,V,W\}$.\label{spacegraph}}
\end{figure}

However, due to the boundary conditions that we needed to introduce in
the realization, the graph of the realized model looks different and
is in particular one-sided, i.e., has a left border.  Corresponding to
the four types of \bc's that lead to a monomial signal extension
(Lemma~\ref{leftbc}), we get the four graphs in
Figure~\ref{spacegraph}.

For example, in the topmost graph, the extra arrow from $T_0$ to $T_1$
arises from $T_{-1} = T_1$ and the arrow that would go from $T_0$ to
$T_{-1}$. In the second graph, we could have omitted the two arrows from
$U_{-1}$ to $U_0$, since their weights add up to zero, which is also reflected 
by  $U_{-1} = 0$.
Including $U_{-1}$ gives a better understanding of the signal
extension, since, following the arrows, each graph not only represents
the boundary condition, but also the entire left signal extension.

We will show in Section~\ref{altreal} that it is possible to choose a
different realization that has Figure~\ref{abstractspacegraph} as
visualization; however, the realization in Section~\ref{altreal} will
raise a different difficulty that we will discuss there.

\section{Finite C-Transform and DTTs}\label{finCtrafo}

In Section~\ref{ztrafofinite}, we derived the finite
$z$-transform, the finite signal model for time, by choosing boundary
conditions and constructing a polynomial algebra of the form $\alg =
\C[x]/p(x)$ with basis $b = (1,x,\dots,x^{n-1})$. By requiring a
monomial signal extension, we obtained $p(x) = x^n - a$ and thus the
DFT (and its variants) as the associated Fourier transform for the
regular module $\md = \alg$.

Analogously, we derive now the \emph{finite} versions of the
$C$-transform. By identifying those boundary conditions that lead to a
monomial signal extension, we obtain 16 specific polynomial algebras
with bases, whose Fourier transforms are precisely the 16 types of
discrete trigonometric transforms (DTTs) comprising 8 discrete cosine
transforms (DCTs) and 8 discrete sine transforms (DSTs).  Using the
general results in Section~\ref{polyalgs}, we then explain DTT domain
filtering and derive the DTT's diagonalization properties and
convolution theorems. We also explain why the DTTs are almost
orthogonal and many other of their properties.  In particular, we show
that the associated signal models reveal close relationships
between certain DTTs, which allow us to divide them into four groups
of four each. DTTs within the same group can be translated into each
other at the expense of $O(n)$ operations.  An even stronger
relationship exists between ``dual'' DTTs, a notion that we will
define.

The derivation follows the same structure as
Section~\ref{ztrafofinite}.  However, as in the infinite cases (see
Sections~\ref{ztrafo} and \ref{Ctrafo}), there will be important
inherent differences between the finite time model and the finite
space model.

\subsection{Building the Signal Model}\label{buildfinCtrafo}

\mypar{Shift, linear extension, realization} 
We consider a finite number of space marks $t_0,\dots,t_{n-1}$ and
adopt the space shift operator~$q$ in Figure~\ref{spaceshift} and its
realization by setting $q=x$, and $t_k = C_k$ (a generic sequence of
Chebyshev polynomials), as derived in Section~\ref{buildCtrafo}. These
definitions will need to be complemented by appropriate boundary
conditions, as we discuss next.

Let $\coord{s} = (s_0,\dots,s_{n-1})\in\C^n$ be a finite sampled
signal and $C$ a sequence of Chebyshev polynomials. As in
Section~\ref{buildfinztrafo}, a straightforward realization seems to
lead to the set of all polynomials
$$
\sum_{0\leq k<n}s_kC_k.
$$ 
The set of these sums is the vector space $\C_n[x]$ (with
basis polynomials $C_k$); however, this space is not closed under
multiplication by the shift operator $x$, and thus it is not a module,
which means filtering is not well-defined. In particular, the problem is with
\begin{equation}\label{rightproblemC}
x\cdot C_{n-1} = (C_{n-2} + C_{n})/2\not\in\C_n[x],
\end{equation}
since $C_n\not\in\C_n[x]$. Note that, in contrast to
\eqref{leftproblem}, multiplying the first space mark 
by $T_{-1} = T_1 = x$ yields
$$
x\cdot C_0 = (C_{-1} + C_1)/2\in\C_n[x],
$$
since, by \eqref{genleftbc}, the choice of $C$ already implies a left
\bc So the remaining task is to determine the proper 
right boundary conditions.

\mypar{Boundary condition and signal extension} To solve the problem
in \eqref{rightproblemC}, we introduce an equation
\begin{equation}\label{polyintroC}
C_n = r = \sum_{0\leq k < n}\beta_kC_k,
\quad\text{or}\quad C_n - r = 0,
\end{equation}
which is equivalent to the right \bc
$$
s_n = \sum_{0\leq k < n}\beta_ks_k.
$$ 
As a consequence of \eqref{polyintroC}, using the $k$-fold space shift
operator $T_k$ (see Lemma~\ref{kspaceshiftT}), we get the series of equations
$$
T_k\cdot (C_n - r) = T_k\cdot 0 = 0,\quad
k\geq 0,
$$
which determine the entire right signal extension. It is obtained
by applying Lemma~\ref{chebprop}, iv) and reducing $C_{n+k}$ modulo
$(C_n-r)$. 

Algebraically, the right boundary condition replaces the vector space
$\C_n[x]$ (with basis $b = (C_0,\dots,C_{n-1})$) by the polynomial
algebra $\md = \C[x]/(C_n-r)$ (also with basis $b$), viewed as a
regular module, i.e., the algebra is $\alg=\md$. The natural basis of
$\alg$ is given by $(T_0,\dots,T_{n-1})$, regardless of the choice of
$C$.

For a general choice of left \bc (given by the choice of $C$) and
right \bc (given by the choice of $r$), the corresponding signal
extension has a complicated structure.  As in
Section~\ref{buildfinztrafo}, we identify those \bc that lead to a
simple, i.e., {\em monomial} signal extension. Lemma~\ref{leftbc}
gives already the left \bc for a monomial left signal extension and
shows that they are obtained by choosing $C\in\{T,U,V,W\}$. It remains
to identify the admissible right \bc We do this in Lemma~\ref{rightbc}
and show that, again, there are 4 choices, which give rise to a total
number of 16 possibilities---corresponding to the 16 types of DTTs as
we will see below.

\begin{lemma}[Monomial right signal extension]\label{rightbc}
To assure a monomial left signal
extension, let $C\in\{T,U,V,W\}$. The only four right \bc~that yield a
monomial signal extension for $\md = \C[x]/p(x)$ are $C_n = C_{n-2}$,
$C_n = 0$, and $C_n = \pm C_{n-1}$, which implies
$p\in\{C_n-C_{n-2},C_n,C_n\pm C_{n-1}\}$.
\end{lemma}
\begin{proof}
Necessarily, the \bc has the form $C_n = aC_k$, $0\neq k<n$. By
multiplying by $x$ on both sides, we obtain $C_{n+1} = a(C_{k+1} +
C_{k-1}) - C_{n-1}$. We determine under which conditions the three
summands on the right reduce to at most one summand. 

Case~1: $k \neq n-1$. Then either $a = 0$, or $k = n-2$ and $a = 1$.

Case~2: $k =
n-1$. Then $aC_{k+1} = aC_n = a^2C_{n-1}$ and thus $a = \pm 1$. 

It remains to show that these four
\bc~yield a monomial signal extension, which is done by induction. We
omit the details.
\end{proof}
It is interesting to note that the right \bc's in Lemma~\ref{rightbc}
are the reflections of the left \bc's in Lemma~\ref{leftbc}.

We investigate the structure of the signal extension. The four left
signal extensions, corresponding to setting $C = T,U,V,W$, can be
displayed (from top to bottom, respectively) as follows,
using `$|$' to denote the boundaries:
\begin{equation}\label{leftse1}
\isdraft{\renewcommand{\baselinestretch}{1}\normalsize}{}
\begin{array}{r@{\mid}l@{\ \dots\ }r@{\mid}l}
\dots & \markit{C_n}\ C_{n-1} & C_{1} \\
\dots\markit{-C_n}\ {-C_{n-1}} & {-C_{n-2}} & -C_0\ 0 \\
\dots\markit{C_n} & C_{n-1} & C_0 & \lifta{C_0\ \dots\ C_{n-1}}\\
\dots\markit{-C_n} & -C_{n-1} & -C_0 \\
\multicolumn{3}{r@{\,\,}}{\underbrace{%
\rule{55mm}{0mm}}_{\text{left signal extension}}} & 
\multicolumn{1}{@{\,}l}{\underbrace{%
\rule{20mm}{0mm}}_{\text{signal scope}}} 
\end{array}
\end{equation}

We observe that the left signal extension is completely determined
by the left \bc, up to the occurrence
of $C_n$ (underlined), which is determined by the right \bc
Similarly, we can display the right signal extensions as
\begin{equation}\label{rightse1}
\isdraft{\renewcommand{\baselinestretch}{1}\normalsize}{}
\begin{array}{c@{\mid}l@{\ \dots\ }r@{\mid}l}
& C_{n-2} & C_0\ \markit{C_{-1}} & \dots \\
& 0\ {-C_{n-1}} & -C_1 & -C_0\ \markit{{-C_{-1}}}\dots \\
\lifta{C_0\ \dots\ C_{n-1}} & C_{n-1} & C_0 & \markit{C_{-1}}\dots \\
& -C_{n-1} & -C_0 & \markit{-C_{-1}}\dots \\
\multicolumn{1}{r@{\,\,}}{\underbrace{%
\rule{20mm}{0mm}}_{\text{signal scope}}} &
\multicolumn{3}{@{\,}l}{\underbrace{%
\rule{55mm}{0mm}}_{\text{right signal extension}}}
\end{array}
\end{equation}
Again, the right signal extension is completely determined by the
right \bc, up to the occurrence of $C_{-1}$ (underlined), which is
determined by the left \bc In the literature \cite{Martucci:94}, the
four signal extensions in \eqref{leftse1} and \eqref{rightse1} are
sometimes called, from top to bottom: whole point symmetry (WS), whole
point antisymmetry (WA), half point symmetry (HS), and half point
antisymmetry (HA).

Taken together, the left and right \bc~determine both the left and
right signal extension,
$$
\text{left and right \bc }
\Longrightarrow\text{ left and right signal extension.}
$$

By combining the left and right \bc, we get, in all 16 cases, an
eventually periodic signal extension. The period lengths are displayed
in Table~\ref{periodlength}.

\begin{table}\centering
\caption{Period lengths for the 16 monomial signal extensions.
\label{periodlength}}
\ra{1.2}
$\begin{array}{@{}lllll@{}}\toprule
C & C_n = C_{n-2} & C_n = 0 & C_n = C_{n-1} & C_n = -C_{n-1} \\ \midrule
T  & 2n-2 & 4n & 2n-1 & 4n-2 \\
U & 4n & 2n+2 & 4n+2 & 2n+1 \\
V & 2n-1 & 4n+2 & 2n & 4n \\
W & 4n-2 & 2n+1 & 4n & 2n \\ \bottomrule
\end{array}$
\end{table}

We will show a visualization of the signal extensions, similar to
Figure~\ref{sigextxna}, after the 16 DTTs have been introduced.

\mypar{Signal model: finite \boldmath$C$-transform}  Let 
$C_0,\dots,C_{n-1}$, $C\in\{T,U,V,W\}$, be a sequence of Chebyshev
polynomials. Further, let $p(x)$ be one of the four choices in
Lemma~\ref{rightbc}.  The {\em finite $C$-transform} is the signal
model $(\alg,\md,\Phi)$ for $V =\C^n$ with $\md = \alg = \C[x]/p(x)$
and the bijective linear map~$\Phi$ defined by
$$
\Phi:\ \coord{s}\mapsto\sum_{0\leq k < n}s_kC_k\in\md.
$$ 
This definition requires that the module explicitly be stated,
since the 16 candidate modules have only 4 choices of bases.
As in the infinite case, if $C$ is fixed, we will also refer to
the finite $T$-,$U$-,$V$-, and $W$-transform.

It is important to note that in the algebra $\alg = \md$ of filters,
the natural basis is given by the $k$-fold space shift operators
$(T_0,\dots,T_{n-1})$, and thus independent of the choice of $C$,
i.e., of the basis (or impulses) $C_k$ in the signal space $\md$.
This is different from the finite $z$-transform associated with the
DFT (see the end of Section~\ref{buildfinztrafo}), where both $\md$
and $\alg$ had the same basis consisting of polynomials $x^k$
(representing time marks and $k$-fold time shift operators, respectively).

The above definition of the finite $C$-transform includes 16 variants,
corresponding to the 16 DTTs introduced next.

\subsection{Spectrum and Fourier Transform: DCTs and DSTs}
\label{dtttransform}

In this section, we derive the Fourier transforms for the 16 types of
finite $C$-transforms defined above and show that they are given by
the 16 types of DCTs and DSTs. As an aside, in doing that, we settle
the question why there are 16 DTTs to begin with, as the original
derivation of the full set of all 16 \cite{Wang:85} does not provide
an explanation.

To compute the spectrum and a Fourier transform using
Theorem~\ref{fourtrafo}, we have to determine the zeros of the 16
polynomials $p$ given by $C_n, C_n\pm C_{n-1}, C_n-C_{n-2}$, for
$C\in\{T,U,V,W\}$. In all cases, the zeros can be expressed in closed
form using the identities given in Table~\ref{chebids} and
Table~\ref{4cheb} in Appendix~\ref{chebs}. We note that
Table~\ref{chebids} is a consequence of well-known trigonometric
identities.

\begin{table}\centering
\caption{Identities among the four series of Chebyshev polynomials;
$C_n$ has to be replaced by $T_n$, $U_n$, $V_n$, $W_n$ to obtain rows
$1,2,3,4$, respectively.\label{chebids}}
$
\ra{1.2}
\begin{array}{@{}lllll@{}}\toprule
 & C_n - C_{n-2} & C_n & C_n - C_{n-1} & C_n + C_{n-1}\\ \midrule
T_n & 2(x^2-1)U_{n-2} & T_n & (x-1)W_{n-1} & (x+1)V_{n-1}\\
U_n & 2T_n & U_n & V_n & W_n\\
V_n & 2(x-1)W_{n-1} & V_n & 2(x-1)U_{n-1} & 2T_n\\
W_n & 2(x+1)V_{n-1} & W_n & 2T_n & 2(x+1)U_{n-1}\\ \bottomrule
\end{array}
$
\end{table}

Instead of computing the spectrum and a Fourier transform in detail
for all 16 cases, we consider only one representative example and then
state the result for all 16 DTTs. But first, we introduce the DTTs as they
are defined in the literature.

\mypar{DTT definitions} There are 16 types of discrete trigonometric
transforms (DTTs): 8 types of discrete cosine transforms (DCTs) and 8
types of discrete sine transforms (DSTs). The most important to date
is the DCT of type~2, which was first introduced in \cite{Ahmed:74}
and is used in the JPEG image compression standard. A complete
introduction to all 16 types is in~\cite{Wang:85}. Table~\ref{dttdefs}
gives the definitions of the {\em unscaled} version of the 16 DTTs. We
note that the DTTs of type 1, 4, 5, 8 are symmetric, and that the DTTs
of type 2 and 3, 6 and 7, respectively, are transposes of each other.
We use arabic instead of roman numbers to denote the type, following
\cite{Strang:99}, since it is more convenient when dealing with all 
8 types of DCTs and DSTs. For example, we write $\DCTt{2}_n$ instead of
$\DCT^{\text{(II)}}_n$.

\begin{table}\centering
\caption{8 types of DCTs and DSTs (unscaled) of size $n$. The entry at
row $k$ and column $\ell$ is given for $0\leq k,\ell <
n$.\label{dttdefs}}
$
\ra{1.2}
\begin{array}{@{}cll@{}}\toprule
\text{type} & \multicolumn{1}{c}{\text{DCT}} &
\multicolumn{1}{c}{\text{DST}} \\ \midrule
\text{1} & \cos k\ell\frac{\pi}{n-1} &
\sin (k+1)(\ell+1)\frac{\pi}{n+1}\\
\text{2} & \cos k(\ell+\fh)\frac{\pi}{n} &
\sin (k+1)(\ell+\fh)\frac{\pi}{n}\\
\text{3} & \cos (k+\fh)\ell\frac{\pi}{n} &
\sin (k+\fh)(\ell+1)\frac{\pi}{n}\\
\text{4} & \cos (k+\fh)(\ell+\fh)\frac{\pi}{n} &
\sin (k+\fh)(\ell+\fh)\frac{\pi}{n}\\
\text{5} & \cos k\ell\frac{\pi}{n-\fh} &
\sin (k+1)(\ell+1)\frac{\pi}{n+\fh}\\
\text{6} & \cos k(\ell+\fh)\frac{\pi}{n-\fh} &
\sin (k+1)(\ell+\fh)\frac{\pi}{n+\fh}\\
\text{7} & \cos (k+\fh)\ell\frac{\pi}{n-\fh} &
\sin (k+\fh)(\ell+1)\frac{\pi}{n+\fh}\\
\text{8} & \cos (k+\fh)(\ell+\fh)\frac{\pi}{n+\fh} &
\sin (k+\fh)(\ell+\fh)\frac{\pi}{n-\fh}\\ \bottomrule
\end{array}$
\end{table}

\mypar{Example: Signal model for DCT, type 2} We work out this 
example in detail.  We choose the left \bc $s_{-1} = s_0$, i.e.,
$C_{-1} = C_0$, which is afforded by the base polynomials $C = V$ (see
Lemma~\ref{leftbc}). As right \bc we choose $s_n = s_{n-1}$, i.e.,
$C_n = C_{n-1}$, which implies 
$$
p = C_n - C_{n-1} = V_n - V_{n-1} = 2(x-1)U_{n-1}
$$ 
using row 3, column 3 in Table~\ref{chebids}.  Thus, we obtain the
regular module $\md =
\C[x]/(x-1)U_{n-1}(x)$ (the 2 can be dropped, as scalar factors do not
matter in $p$). The zeros of $p(x) = (x-1)U_{n-1}(x)$ are given by
$\alpha_k = \cos k\pi/n$, $0\leq k<n$ (from Table~\ref{4cheb}). Thus
the Fourier transform for $\md$ is given by
\begin{equation}\label{dct2exdec}
\begin{array}{rrcl}
\Delta: & \C[x]/(C_n-C_{n-1}) & \rightarrow & 
  \bigdirsum_{0\leq k <n}\C[x]/(x-\alpha_k)\\
& s = s(x) & \mapsto & (s(\alpha_0),\dots,s(\alpha_{n-1})).
\end{array}
\end{equation}
In matrix form, a Fourier transform for $\md$ is given by the 
polynomial transform with entries
$$
V_\ell(\alpha_k) = \frac{1}{\cos k\pi/(2n)}\cdot\cos k(\ell+1/2)\pi/n.
$$
To obtain a DCT, we need proper scaling. Namely, using
Table~\ref{dttdefs}, we have
\begin{equation}\label{dct2ex}
\DCTt{2}_n = \diag_{0\leq k < n}(\cos k\pi/(2n))\cdot[V_\ell(\alpha_k)],
\end{equation}
and thus, by Theorem~\ref{fourtrafo}, $\DCTt{2}_n$ is a Fourier
transform for the regular module $\md = \C[x]/(x-1)U_{n-1}(x)$.

The scaling diagonal in \eqref{dct2ex} shows the basis chosen on the
right hand side of \eqref{dct2exdec}, namely $1/(\cos k\pi/(2n))$ in
the one-dimensional module (spectral component) $\md_k = \C[x]/(x-\cos
k\pi/n)$, for $0\leq k < n$.

In other words, applying $\DCTt{2}_n$ to a signal $\coord{s}\in\C^n$
gives the spectrum of $\coord{s}$ with respect to the finite $V$-transform
$$
\Phi:\ \C^n\rightarrow\md,\quad 
\coord{s}\mapsto\sum_{0\leq k < n}s_kV_k\in\md,
$$
where $\md = \alg = \C[x]/(x-1)U_{n-1}$.

\mypar{All DTTs}
Similar computations for all 16 cases establishes the 16 DTTs as Fourier
transforms for the 16 finite $C$-transforms.

\begin{theorem}[DTTs and polynomial algebras]\label{main}
The 16 DTTs are the Fourier transforms for the 16 finite
$C$-transforms. The correspondence is given in Table~\ref{chebmods} as
follows. Let $(\alg,\md,\Phi)$ be a finite $C$-transform with $\md =
\C[x]/p$ with basis $b = (C_0,\dots,C_{n-1})$. The choice of $C$
(rows of Table~\ref{chebmods}) determines the left \bc and a scaling
function $f$. The choice of left \bc (four rightmost columns of
Table~\ref{chebmods}) then determines the polynomial $p$, given at the
intersection of row and column. The corresponding DTT is given above
$p$. If $\alpha = (\alpha_0,\dots,\alpha_{n-1})$ are the zeros of $p$, 
then
\begin{equation}\label{maindtt}
\DTT_n = \diag_{0\leq k<n}(f(\alpha_k))\cdot\poly_{b,\alpha},
\end{equation}
i.e., $\DTT_n$ is a scaled polynomial transform and thus a Fourier
transform for the associated signal model (see
Theorem~\ref{fourtrafo}).  Equation~\eqref{maindtt} implies that the
chosen basis in the spectral component $\bigdirsum_{0\leq k <
n}\C[x]/(x-\alpha_k)$ is $1/f(\alpha_k)$, $0\leq k<n$.
\end{theorem}

\newcommand{\mbold}[1]{\text{\boldmath$#1$}}
\begin{table*}\centering
\caption{Overview of the 16 DTTs and the associated modules
$\C[x]/p(x)$ with a basis of Chebyshev polynomials that admit a
monomial signal extension.  The left \bc (rows) determines a scaling
function $f$ ($\cos\theta = x$) and the Chebyshev polynomials
$C\in\{T,U,V,W\}$.  The right \bc (columns) then determines the DTT
and $p(x)$ (given below the DTT).\label{chebmods}}
\ra{1.2}
$
\begin{array}{@{}lllll@{\hspace*{10mm}}cc@{}}\toprule
& s_n - s_{n-2} & s_n & s_n - s_{n-1} & s_n + s_{n-1} & f & C\\ \midrule
s_{-1} = s_1 & 
  \mbold{\DCTt{\bf 1}} & \mbold{\DCTt{\bf 3}} & \mbold{\DCTt{\bf 5}} & 
  \mbold{\DCTt{\bf 7}} & 1 & T\\
& 
  2(x^2-1)U_{n-2} & T_n & (x-1)W_{n-1} & (x+1)V_{n-1} &
  \\ \addlinespace[1mm]
s_{-1} = 0
  & \mbold{\DSTt{\bf 3}} & \mbold{\DSTt{\bf 1}} & \mbold{\DSTt{\bf 7}} & 
  \mbold{\DSTt{\bf 5}} & \sin\theta & U \\
&
  2T_n & U_n & V_n & W_n & \\ \addlinespace[1mm]
s_{-1} = s_0 & 
  \mbold{\DCTt{\bf 6}} & \mbold{\DCTt{\bf 8}} & \mbold{\DCTt{\bf 2}} & 
  \mbold{\DCTt{\bf 4}} & \cos \fh\theta & V \\
& 
  2(x-1)W_{n-1} & V_n & 2(x-1)U_{n-1} & 2T_n & \\ \addlinespace[1mm]
s_{-1} = -s_0 & 
  \mbold{\DSTt{\bf 8}} & \mbold{\DSTt{\bf 6}} & \mbold{\DSTt{\bf 4}} & 
  \mbold{\DSTt{\bf 2}} & \sin\fh\theta & W \\
& 
  2(x+1)V_{n-1} & W_n & 2T_n & 2(x+1)U_{n-1} & \\ \bottomrule
\end{array}
$
\end{table*}

The DCT, type 3, was recognized as polynomial transform in
\cite{Steidl:91}. The DCTs and DSTs of types 1--4 where recognized
as (scaled) polynomial transform in \cite{Kailath:96}. In neither 
case any connection to signal processing was established.

\mypar{Polynomial DTTs} Theorem~\ref{main} shows that each DTT is a
Fourier transform for a suitable module $\C[x]/p$, but, in general,
not the polynomial transform. Thus, we associate to each DTT its
polynomial transform $\poly_{b,\alpha}$ obtained by omitting
the scaling factors in \eqref{maindtt}.
\begin{definition}[Polynomial DTTs]\label{pdtt}
Let $\DTT_n$ be given. We call the unique polynomial transform
$\poly_{b,\alpha}$ associated with $\DTT_n$ by \eqref{maindtt}
the ``polynomial DTT'' and denote it by $\pDTT_n$. Thus, \eqref{maindtt}
can be rewritten as
$$
\DTT_n = \diag_{0\leq k < n}(f(\alpha_k))\cdot\pDTT_n.
$$
We have $\DTT = \pDTT$ if and only if $\DTT$ appears in the first
row of Table~\ref{chebmods}, i.e., if $\DTT$ is one of $\DCTt{1}, \DCTt{3},
  \DCTt{5}, \DCTt{7}$.
\end{definition}
The polynomial DTTs will play an important role in the derivation of
fast DTT algorithms \cite{Pueschel:05}. In several cases it will be
natural to derive a fast algorithm for $\pDTT$ and apply a final
scaling to obtain a fast algorithm for the corresponding $\DTT$. As a
consequence, we will show that the polynomial DTTs are a suitable
choice of {\em scaled} DTTs in applications, where the DTT is followed
by scaling and can thus be replaced by any transform $D\cdot\DTT$ to
reduce the number of multiplications.

\mypar{Signal extension}
We display graphically the signal extension for the 16 DTTs in
Table~\ref{sigextdtts}. This is similar to Figure~\ref{sigextxna} for
the DFT variants. The dotted lines (including the adjacent hollow
bullets) signify a scaling by $-1$. A ``0'' signifies that the signal
model assumes a signal value equal to zero. In each case we display
four times the signal scope, which may be a single period or comprise
two periods, depending on the DTT. For some of the bullets, the labels
at the (virtual) x-axis are given.

\begin{table}\centering
\caption{Signal extension for the 16 DTTs.\label{sigextdtts}}
\begin{tabular}{@{}ll@{}}\toprule
\multicolumn{1}{c}{DCT, type 1--8} & 
\multicolumn{1}{c}{DST, type 1--8} \\ \midrule
\isdraft{\hspace*{10mm}}{\hspace*{1mm}}
\isdraft{\begin{minipage}[t]{0.3\linewidth}}{%
\begin{minipage}[t]{0.45\linewidth}}
\begin{picture}(100,40)
\put(0,10){\usebox{\upbline}}
\put(20,10){\usebox{\downline}}
\put(0,0){\makebox(0,0){$\sst 0$}}
\put(40,10){\usebox{\upline}}
\put(60,10){\usebox{\downline}}
\put(40,0){\makebox(0,0){$\sst 2n-2$}}
\put(80,0){\makebox(0,0){$\sst 4n-4$}}
\end{picture}

\begin{picture}(100,40)
\put(0,10){\usebox{\upbline}}
\put(25,10){\usebox{\downline}}
\put(0,0){\makebox(0,0){$\sst 0$}}
\put(50,10){\usebox{\upline}}
\put(75,10){\usebox{\downline}}
\put(47,0){\makebox(0,0){$\sst 2n-1,2n$}}
\put(95,0){\makebox(0,0){$\sst 4n-1$}}
\end{picture}

\begin{picture}(100,40)
\put(0,10){\usebox{\upbline}}
\put(25,30){\makebox(0,0){$\sst 0$}}
\put(30,10){\usebox{\downdline}}
\put(0,0){\makebox(0,0){$\sst 0$}}
\put(50,10){\usebox{\updline}}
\put(75,30){\makebox(0,0){$\sst 0$}}
\put(80,10){\usebox{\downline}}
\put(50,0){\makebox(0,0){$\sst 2n$}}
\put(100,0){\makebox(0,0){$\sst 4n$}}
\end{picture}

\begin{picture}(100,40)
\put(0,10){\usebox{\upbline}}
\put(25,10){\usebox{\downdline}}
\put(0,0){\makebox(0,0){$\sst 0$}}
\put(50,10){\usebox{\updline}}
\put(75,10){\usebox{\downline}}
\put(47,0){\makebox(0,0){$\sst 2n-1,2n$}}
\put(95,0){\makebox(0,0){$\sst 4n-1$}}
\end{picture}

\begin{picture}(100,40)
\put(0,10){\usebox{\upbline}}
\put(25,10){\usebox{\downline}}
\put(0,0){\makebox(0,0){$\sst 0$}}
\put(45,10){\usebox{\upline}}
\put(70,10){\usebox{\downline}}
\put(42,0){\makebox(0,0){$\sst 2n-1$}}
\put(90,0){\makebox(0,0){$\sst 4n-2$}}
\end{picture}

\begin{picture}(100,40)
\put(0,10){\usebox{\upbline}}
\put(20,10){\usebox{\downline}}
\put(0,0){\makebox(0,0){$\sst 0$}}
\put(45,10){\usebox{\upline}}
\put(65,10){\usebox{\downline}}
\put(42,0){\makebox(0,0){$\sst 2n-2,2n-1$}}
\put(85,0){\makebox(0,0){$\sst 4n-3$}}
\end{picture}

\begin{picture}(100,40)
\put(0,10){\usebox{\upbline}}
\put(25,10){\usebox{\downdline}}
\put(0,0){\makebox(0,0){$\sst 0$}}
\put(45,10){\usebox{\updline}}
\put(70,10){\usebox{\downline}}
\put(45,0){\makebox(0,0){$\sst 2n-1$}}
\put(90,0){\makebox(0,0){$\sst 4n-2$}}
\end{picture}

\begin{picture}(100,40)
\put(0,10){\usebox{\upbline}}
\put(25,30){\makebox(0,0){$\sst 0$}}
\put(30,10){\usebox{\downdline}}
\put(0,0){\makebox(0,0){$\sst 0$}}
\put(55,10){\usebox{\updline}}
\put(80,30){\makebox(0,0){$\sst 0$}}
\put(85,10){\usebox{\downline}}
\put(52,0){\makebox(0,0){$\sst 2n,2n+1$}}
\put(105,0){\makebox(0,0){$\sst 4n+1$}}
\end{picture}
\end{minipage}
\hspace*{-3mm}
&
%
\isdraft{\hspace*{9mm}}{}
\isdraft{\begin{minipage}[t]{0.32\linewidth}}{%
\begin{minipage}[t]{0.48\linewidth}}
\begin{picture}(105,40)
\put(0,10){\usebox{\upbline}}
\put(25,30){\makebox(0,0){$\sst 0$}}
\put(30,10){\usebox{\downdline}}
\put(0,0){\makebox(0,0){$\sst 0$}}
\put(55,10){\makebox(0,0){$\sst 0$}}
\put(60,10){\usebox{\upline}}
\put(85,30){\makebox(0,0){$\sst 0$}}
\put(90,10){\usebox{\downdline}}
\put(55,0){\makebox(0,0){$\sst 2n+1$}}
\put(115,0){\makebox(0,0){$\sst 4n+2$}}
\put(115,10){\makebox(0,0){$\sst 0$}}
\end{picture}

\begin{picture}(105,40)
\put(0,10){\usebox{\upbline}}
\put(25,10){\usebox{\downdline}}
\put(0,0){\makebox(0,0){$\sst 0$}}
\put(50,10){\usebox{\upline}}
\put(75,10){\usebox{\downdline}}
\put(47,0){\makebox(0,0){$\sst 2n-1,2n$}}
\put(95,0){\makebox(0,0){$\sst 4n-1$}}
\end{picture}

\begin{picture}(105,40)
\put(0,10){\usebox{\upbline}}
\put(20,10){\usebox{\downline}}
\put(0,0){\makebox(0,0){$\sst 0$}}
\put(45,10){\makebox(0,0){$\sst 0$}}
\put(50,10){\usebox{\updline}}
\put(70,10){\usebox{\downdline}}
\put(45,0){\makebox(0,0){$\sst 2n-1$}}
\put(95,0){\makebox(0,0){$\sst 4n-2$}}
\put(95,10){\makebox(0,0){$\sst 0$}}
\end{picture}

\begin{picture}(105,40)
\put(0,10){\usebox{\upbline}}
\put(25,10){\usebox{\downline}}
\put(0,0){\makebox(0,0){$\sst 0$}}
\put(50,10){\usebox{\updline}}
\put(75,10){\usebox{\downdline}}
\put(47,0){\makebox(0,0){$\sst 2n-1,2n$}}
\put(95,0){\makebox(0,0){$\sst 4n-1$}}
\end{picture}

\begin{picture}(105,40)
\put(0,10){\usebox{\upbline}}
\put(25,10){\usebox{\downdline}}
\put(0,0){\makebox(0,0){$\sst 0$}}
\put(50,10){\makebox(0,0){$\sst 0$}}
\put(55,10){\usebox{\upline}}
\put(80,10){\usebox{\downdline}}
\put(50,0){\makebox(0,0){$\sst 2n$}}
\put(105,0){\makebox(0,0){$\sst 4n+1$}}
\put(105,10){\makebox(0,0){$\sst 0$}}
\end{picture}

\begin{picture}(100,40)
\put(0,10){\usebox{\upbline}}
\put(25,30){\makebox(0,0){$\sst 0$}}
\put(30,10){\usebox{\downdline}}
\put(0,0){\makebox(0,0){$\sst 0$}}
\put(55,10){\usebox{\upline}}
\put(80,30){\makebox(0,0){$\sst 0$}}
\put(85,10){\usebox{\downdline}}
\put(52,0){\makebox(0,0){$\sst 2n,2n+1$}}
\put(105,0){\makebox(0,0){$\sst 4n+1$}}
\end{picture}

\begin{picture}(105,40)
\put(0,10){\usebox{\upbline}}
\put(25,10){\usebox{\downline}}
\put(0,0){\makebox(0,0){$\sst 0$}}
\put(50,10){\makebox(0,0){$\sst 0$}}
\put(55,10){\usebox{\updline}}
\put(80,10){\usebox{\downdline}}
\put(50,0){\makebox(0,0){$\sst 2n$}}
\put(105,0){\makebox(0,0){$\sst 4n+1$}}
\put(105,10){\makebox(0,0){$\sst 0$}}
\end{picture}

\begin{picture}(100,40)
\put(0,10){\usebox{\upbline}}
\put(20,10){\usebox{\downline}}
\put(0,0){\makebox(0,0){$\sst 0$}}
\put(45,10){\usebox{\updline}}
\put(65,10){\usebox{\downdline}}
\put(42,0){\makebox(0,0){$\sst 2n-2,2n-1$}}
\put(85,0){\makebox(0,0){$\sst 4n-3$}}
\end{picture}
\end{minipage} 
\\ 
\ & \\ \bottomrule
\end{tabular}
\end{table}

\mypar{Remarks and observations}
We make the following remarks.
\begin{itemize}
\item For each DTT, we have three relevant versions. First, the
polynomial version $\pDTT$, which is the unique polynomial transform
for its associated signal model (see Definition~\ref{pdtt} above). Second,
the unscaled or natural version, which has pure cosines (or sines) as
entries (see Table~\ref{dttdefs}). Third, the orthogonal version,
which arises from the other two by suitable scaling of rows and
columns, i.e., by slightly adjusting the signal model (explained below in
Section~\ref{orthodtts}).
\item The 16 DTTs can be divided into four groups of four each with
respect to the polynomial $p$ in the associated module $\C[x]/p$ (see
Table~\ref{chebmods}). For example, the $T$-group comprises all DTTs
of type 3 and 4, which have the same module $\md = \C[x]/T_n$.  The
modules within the other groups differ slightly, e.g., in the
$U$-group that comprises the DTTs on the main diagonal in
Table~\ref{chebmods}. The difference between the DTTs within the same
group is the choice of basis, which is one of $T, U, V, W$. As a
consequence, these transforms can be converted into each other using a
sparse base change (explained in Section~\ref{groupdtt}), and several
convolution theorems can be derived within a group (explained in
Section~\ref{filter}).
\item A closer relationship exists between the DTTs that are in the same
group but, in addition, on mirror positions w.r.t.~the diagonal.  In
other words, these DTTs have mirrored \bc We call such a pair {\em
dual} to each other and show a uniform relationship between dual DTTs
(see Section~\ref{dualsec}).
\end{itemize}

\subsection{Visualization}

Since the right \bc for the 16 DTTs are precisely the mirrored left
\bc, we obtain the visualizations (see Definition~\ref{visdef}) of their
associated signal models readily from Figure~\ref{spacegraph}. We show
four important cases in Figure~\ref{finspacegraph}, from top to bottom
the DCTs of type 1--4.

\begin{figure}[ht]
\isdraft{\hspace*{30mm}}{\hspace*{6mm}}
\begin{minipage}{0.8\textwidth}
{
\xymatrix{
\bullet \ar@/_2ex/[r] & \ar@{<->}@/_2ex/[l] \bullet \ar@{<->}[r] &
\bullet \ar@{.}[rr] && \bullet \ar@{<->}[r] &
\bullet \ar@/^2ex/@{<->}[r] & \ar@/^2ex/[l] \bullet
}
\isdraft{\setlength{\unitlength}{1.15pt}}{}
\begin{picture}(220,5)
\put(7,0){\makebox(0,0){$T_0$}}
\put(42,0){\makebox(0,0){$T_1$}}
\put(77,0){\makebox(0,0){$T_2$}}
\put(142,0){\makebox(0,0){$T_{n-3}$}}
\put(177,0){\makebox(0,0){$T_{n-2}$}}
\put(212,0){\makebox(0,0){$T_{n-1}$}}
\end{picture}
}
\end{minipage}

\bigskip
\isdraft{\hspace*{30mm}}{\hspace*{6mm}}
\begin{minipage}{0.8\textwidth}
{
\xymatrix{
\bullet \ar@{<->}@(ul,dl) \ar@{<->}[r] & \bullet \ar@{<->}[r] & 
\bullet \ar@{.}[rr] && \bullet \ar@{<->}[r] &
\bullet \ar@{<->}[r] & \bullet \ar@{<->}@(ur,dr)
}
\isdraft{\setlength{\unitlength}{1.15pt}}{}
\begin{picture}(220,5)
\put(7,0){\makebox(0,0){$V_0$}}
\put(42,0){\makebox(0,0){$V_1$}}
\put(77,0){\makebox(0,0){$V_2$}}
\put(142,0){\makebox(0,0){$V_{n-3}$}}
\put(177,0){\makebox(0,0){$V_{n-2}$}}
\put(212,0){\makebox(0,0){$V_{n-1}$}}
\end{picture}
}
\end{minipage}

\bigskip
\isdraft{\hspace*{30mm}}{\hspace*{6mm}}
\begin{minipage}{0.8\textwidth}
{
\xymatrix{
\bullet \ar@/_2ex/[r] & \ar@{<->}@/_2ex/[l] \bullet \ar@{<->}[r] &
\bullet \ar@{.}[rr] && \bullet \ar@{<->}[r] &
\bullet \ar@{<->}[r] & \bullet
}
\isdraft{\setlength{\unitlength}{1.15pt}}{}
\begin{picture}(220,5)
\put(7,0){\makebox(0,0){$T_0$}}
\put(42,0){\makebox(0,0){$T_1$}}
\put(77,0){\makebox(0,0){$T_2$}}
\put(142,0){\makebox(0,0){$T_{n-3}$}}
\put(177,0){\makebox(0,0){$T_{n-2}$}}
\put(212,0){\makebox(0,0){$T_{n-1}$}}
\end{picture}
}
\end{minipage}

\bigskip
\isdraft{\hspace*{30mm}}{\hspace*{6mm}}
\begin{minipage}{0.8\textwidth}
{
\xymatrix{
\bullet \ar@{<->}@(ul,dl) \ar@{<->}[r] & \bullet \ar@{<->}[r] & 
\bullet \ar@{.}[rr] && \bullet \ar@{<->}[r] &
\bullet \ar@{<->}[r] & \bullet \ar@{<->}@(ur,dr)|{-1}
}
\isdraft{\setlength{\unitlength}{1.15pt}}{}
\begin{picture}(220,5)
\put(7,0){\makebox(0,0){$V_0$}}
\put(42,0){\makebox(0,0){$V_1$}}
\put(77,0){\makebox(0,0){$V_2$}}
\put(142,0){\makebox(0,0){$V_{n-3}$}}
\put(177,0){\makebox(0,0){$V_{n-2}$}}
\put(212,0){\makebox(0,0){$V_{n-1}$}}
\end{picture}
}
\end{minipage}

\medskip
\caption{Visualizations of the space models given by the $C$-transforms 
  and associated with the DCTs of type 1--4 (from top to bottom) and size
  $n$.\label{finspacegraph}}
\end{figure}

Since the graph represents the operation of the space shift operator $x
= T_1$, the adjacency matrix of the graphs in
Figure~\ref{finspacegraph} is in each case given by the shift matrix
$\phi(x)$ ($\phi$ is the representation afforded by the respective
signal model).

\subsection{Diagonalization Properties}\label{DTTdiagprop}

Using the algebraic framework, the diagonalization properties of the
16 DTTs can be easily derived from Theorem~\ref{diagprop} and can be
stated in a unified way for all 16 DTTs. Let $\DTT_n$ be given and let
$\alg = \md = \C[x]/p(x)$ be the associated regular module with basis
$b$ and let $\alpha=(\alpha_0,\dots,\alpha_{n-1})$ be the vector of
zeros of $p$. Denote by $\phi$ the corresponding representation of
$\alg$. Then, for a filter $h\in\alg$,
\begin{equation}\label{diagpropDTT}
\DTT_n\cdot\phi(h)\cdot\DTT_n^{-1} = \diag_{0\leq k<n}(h(\alpha_k)).
\end{equation}
Conversely, the $\phi(h)$ are {\em all} the matrices diagonalized by
$\DTT$.  

We first investigate the special case of the space shift operator $h =
T_1 = x$. Since $\alg$, as any polynomial algebra, is generated by the
shift operator $x$, the diagonalization of the shift matrix $\phi(x)$
implies the diagonalization of all matrices $\phi(h)$,
$h\in\alg$. Using
\begin{equation}\label{xrec}
x\cdot C_\ell = (C_{\ell+1} + C_{\ell-1})/2,
\end{equation}
we obtain as the general structure of $\phi(x)$ for all 16 cases
\begin{equation}\label{T1diag}
\phi(x) = \fh\cdot\left[
\begin{array}{ccccccc}
\beta_1 & 1 \\
\beta_2 & 0 & 1 \\
0 & 1 & 0 & \cdot \\
 & & 1 & \cdot & 1\\
 & & & \cdot & 0 & \beta_3 \\
 & & & & 1 & \beta_4
\end{array}
\right]
\end{equation}
%
%
where the numbers $\beta_1,\beta_2$ and $\beta_3,\beta_4$ are
determined by the left and right \bc's, respectively, i.e., by
\eqref{xrec} for $\ell = 0, n-1$. For example, for $\DTT = \DCTt{2}$,
we have $C_{-1} = C_0$ (i.e., $C = V$) and thus $x\cdot C_0 =
(C_0+C_1)/2$ or $\beta_1 = \beta_2 = 1$. Further, $C_n = C_{n-1}$, and
thus $x\cdot C_{n-1} = (C_{n-2}+C_{n-1})/2$, or $\beta_3 = \beta_4 =
1$. The resulting shift matrix $\phi(x)$ is the adjacency matrix of the 
second graph in Figure~\ref{finspacegraph}.

Table~\ref{betas} lists the values of the $\beta_i$, $1\leq i\leq
4$, in all 16 cases. The property
$$
\DTT_n\cdot\phi(x)\cdot\DTT_n^{-1} =
  \diag(\alpha_0,\dots,\alpha_{n-1})
$$
is, in a strict mathematical sense, the analogue of the DFT
diagonalizing the cyclic shift.

\begin{table}\centering
\caption{The values $\beta_1,\beta_2,\beta_3,\beta_4$ from
\eqref{T1diag} for the 4 respective choices of left \bc
and right \bc\label{betas}}
\ra{1.2}
$
\begin{array}{@{}lrr@{}}\toprule
\text{left \bc} & \beta_1 & \beta_2 \\ \midrule
s_{-1} = s_1  & 0 & 2 \\
s_{-1} = 0    & 0 & 1 \\
s_{-1} = s_0  & 1 & 1 \\
s_{-1} = -s_0 & -1 & 1 \\ \bottomrule
\end{array}
\qquad
\ra{1.2}
\begin{array}{@{}lrr@{}}\toprule
\text{right \bc} & \beta_3 & \beta_4 \\ \midrule
s_n = s_{n-2}  & 2 & 0 \\
s_n = 0        & 1 & 0 \\
s_n = s_{n-1}  & 1 & 1 \\
s_n = -s_{n-1} & 1 & -1 \\ \bottomrule
\end{array}
$
\end{table}

Next, we investigate the special case of a $k$-fold space shift operator $h =
T_k$. We have, by Lemma~\ref{chebprop} iv),
\begin{equation}\label{Tprop}
T_k\cdot C_\ell = (C_{\ell+k} + C_{\ell-k})/2.
\end{equation}
Further, in the cases where $\ell+k$ or $\ell-k$ is outside the range
$0,\dots,n-1$, the monomial signal extension characteristic to the 16
DTTs assures that $T_kC_\ell$ is again a sum of only two basis
polynomials in $(C_0\dots C_{n-1})$. Thus, the matrix $\phi(T_k)$ has 
at most two entries per row and column, and further has, in all
16 cases, a ``rhombus-like'' shape. We illustrate this again for 
$\DTT = \DCTt{2}$. Using \eqref{Tprop} and row 3 of \eqref{leftse1} and
row 3 of \eqref{rightse1}, we get
\newcommand{\rbo}[1]{\raisebox{#1mm}[-#1mm]{$\cdot$}}
$$
\phi(T_k) = \fh\cdot\left[
\begin{array}{cccccc}
 && 1 \\
 & \oddots && \ddots \\
1 &&&& \ddots \\
 & \ddots &&&& 1\\
 && \ddots && \oddots \\
 &&& 1
\end{array}\right],
$$
where the dots signify 1's; the rest of the matrix entries are zero.
In this rhombus-shaped matrix, the upper left and lower right side of
the rhombus are due to the symmetric boundary conditions.

In the general case, $h = \sum a_k T_k$, we get by linearity $\phi(h)
= \sum a_k\phi(T_k)$. Thus $\phi(h)$ is a structured matrix obtained
by a generic linear combination of the matrices $\phi(T_k)$. Taken
together we obtain the diagonalization properties stated in
\cite{Sanchez:95} (which considers only the 8 DCTs) and, in addition,
give insight into the structure of the matrices and explicitly give
the obtained diagonal matrix in \eqref{diagpropDTT}. This makes it
possible, for example, to determine whether a given matrix $\phi(h)$
is positive definite. For example,
\cite{Strang:99} uses $h = 2 - 2x$ to illustrate the different types
of DCTs. Using \eqref{diagpropDTT}, the eigenvalues of $\phi(h)$ in
all 16 cases are of the form $2 - 2\alpha\geq 0$ since the zeros of
$p$ in Theorem~\ref{main} are all cosines. Thus $\phi(2-2x)$ is
positive semi-definite.  In general, we have the following result.

\begin{lemma}
Let $\md = \C[x]/p(x)$ with basis $b$ and $\phi$ the afforded
representation of $\alg = \md$. Let $h\in\alg$. Then $\phi(h)$ is
positive definite (semi-definite), if $h(\alpha_k) > 0$ ($\geq 0$) for
all zeros $\alpha_k$ of~$p$.
\end{lemma}

Finally, we note again that the structure of the matrices $\phi(h)$,
$h\in\alg$, reflects the basis $(T_0,\dots,T_{n-1})$ of $\alg$, chosen
independently of $\md$. The $T_k$ are the $k$-fold space shift operators.

\subsection{Convolution Theorems}\label{filter}

With the underlying signal models for the 16 DTTs identified, we
obtain a natural, unified description for DTT domain filtering. Let
$\DTT_n$ be given with associated regular module $\md = \alg =
\C[x]/p(x)$ with basis $b$ (Theorem~\ref{main}). Then filtering with
respect to the signal model given by the $\alg$-module $\md$ is, as
usual, the multiplication of a polynomial $s\in\md$ (the signal) by a
polynomial $h\in\alg$ (the filter) modulo $p$. In coordinate form, we
have, also as usual,
$$
h\cdot s\text{ mod }p\Leftrightarrow\phi(h)\cdot\coord{s},
$$ 
where $\phi$ is the representation of $\alg$ afforded by $\md$ with
basis $b$. We determined the structure of $\phi(h)$ in
Section~\ref{DTTdiagprop} w.r.t.~the (natural) $T$-basis in the 
algebra $\alg$ of filters.

A convolution theorem for DTT domain filtering is now obtained as a
special case of Theorem~\ref{convolution}. We illustrate with two
examples.  In the first example, we choose $p=T_n$ and $\md =
\C[x]/T_n$ with $V$-basis\footnote{The reader may have noticed that
the symbol~$V$ is used to represent either the vector space in the
signal model or one of the Chebyshev polynomials. The context should
make it clear which meaning is attached to~$V$.} and associated
Fourier transform $\four = \DCTt{4}_n$. As above, let $\coord{s}$ be
the coordinate vector of the signal $s$. Further, let $h\in\alg$ be a
filter. The Fourier transform $\four'$ in Theorem~\ref{convolution} is
w.r.t.~the basis in $\alg$, for which the natural choice, as we
learned, is always the $T$-basis. Thus $\four' = \DCTt{3}_n$, which
has no scaling diagonal. Let $\coord{h}$ be the coordinate vector of
$h$, then
\begin{equation}\label{convex}
\phi(h)\cdot\coord{s} = \DCTt{4}_n^{-1}\cdot(
  \DCTt{3}_n\cdot\coord{h}\odot\DCTt{4}_n\cdot\coord{s}).
\end{equation}
We can choose a basis in $\alg$ different from the natural $T$-basis,
to obtain variants, where $\DCTt{3}$ in \eqref{convex} is replaced 
by the DTTs in the $T$-group, and \eqref{convex} is modified to 
account for their corresponding scaling diagonals $D$ 
(Theorem~\ref{convolution}).

In the second example, we choose $p=(x-1)U_{n-1}$ and $\md =
\C[x]/(x-1)U_{n-1}$ with $V$-basis and associated Fourier transform
$\DCTt{2}_n$. Proceeding as above leads to the problem that for $\alg
= \md$ with $T$-basis there is no associated DTT.  Namely, the DTT in
the $U$-group with $T$-basis is the $\DCTt{1}$, which has the
associated module $\md = \C[x]/(x^2-1)U_{n-2}$, which differs from the
$\md$ above by a linear factor in $p$.

To obtain a Fourier transform $\four'$ for $\C[x]/(x-1)U_{n-1}$ with
$T$-basis, we thus split off the extra linear factor and start with the
module for $\DCTt{1}_{n+1}$ to obtain comparable sizes.  Namely, using
the CRT,
$$
\C[x]/(x^2-1)U_{n-1}\rightarrow
  \C[x]/(x-1)U_{n-1}\dirsum\C[x]/(x+1).
$$
We choose the $T$-basis in the smaller modules. The corresponding base 
change matrix $B_{n+1}$ has the form
$$
B_{n+1} =
\left[
\begin{array}{cccc}
1 &&& \ast \\
& \ddots && \vdots \\
&& 1 & \ast \\
\ast & \hdots & \ast & \ast
\end{array}
\right],
$$
where $\ast$ denotes entries whose exact form we do not need.
We get
$$
\DCTt{1}_{n+1} = (\four'\dirsum\one_1)B_{n+1}.
$$
The special form of $B_{n+1}$ allows us to use the $\DCTt{1}_{n+1}$ in
a convolution theorem as
\begin{eqnarray*}
\phi(h)\cdot\coord{s} & = & 
  \DCTt{2}_n^{-1}\cdot(\four'\cdot\coord{h}\odot\DCTt{2}_n\cdot\coord{s}) \\
& = & \DCTt{2}_n^{-1}\cdot(
  (\DCTt{1}_{n+1}\cdot\coord{h}')''\odot\DCTt{2}_n\cdot\coord{s}),
\end{eqnarray*}
where the $\coord{h}'$ arises from $\coord{h}$ by appending (padding) a zero
value, and $(\cdot)''$ signifies omitting the last value.

The DTTs can also be used to compute the ``ordinary'' linear
convolution, i.e., time domain filtering, if the signal or the filter
have a symmetry property compatible with a DTT.  This is the subject
of \cite{Martucci:94}. That paper also observes that the 16 DTTs
divide into four groups of four each of ``compatible'' DTTs, i.e.,
DTTs that can occur in one convolution theorem. These groups are the
same as we introduce in Section~\ref{groupdtt}, namely those with
(essentially) the same associated module, which explains this
observation.  Further, the necessity of zero padding and last value
omission in various cases in
\cite{Martucci:94} has the same origin as in our above $\DCTt{2}$ 
convolution theorem; it occurs when the modules for the two DTTs used
in the convolution (and thus being necessarily in the same group) do not 
have the exact same associated module.

\subsection{Orthogonal DTTs}\label{orthodtts}

It is well-known that the DTTs, as defined in Table~\ref{dttdefs}, are
``almost orthogonal,'' which means that after a suitable scaling of
rows and columns they become orthogonal.  Table~\ref{dttorth} gives
these orthogonal versions of the DTTs, which arise from the unscaled
version by scaling the first or last row or column by a factor
of $1/\sqrt{2}$, and by multiplying the entire matrix by a suitable
scalar factor. We call them {\em orthogonal DTTs}.

\begin{table}\centering
\caption{Definition of the orthogonal versions of the DCTs and DSTs;
$a_{k,l}$ is the entry at row $k$ and column $l$ of the respective
unscaled DTT as given in Table~\ref{dttdefs}.  The row/column scaling
factors are given by: $c_i = 1/\sqrt{2}$ for $i = 0$ and $= 1$ else;
$d_i = 1/\sqrt{2}$ for $i = n-1$ and $= 1$ else.\label{dttorth}}
\ra{1.3}
$
\begin{array}{@{}cll@{}}\toprule
\text{type} & \multicolumn{1}{c}{\text{DCT}} & 
\multicolumn{1}{c}{\text{DST}} \\ \midrule
\text{1} & \sqrt{\frac{2}{n-1}}\cdot c_kc_\ell d_kd_\ell\cdot a_{k,l} & 
\sqrt{\frac{2}{n+1}}\cdot a_{k,l}\\
\text{2} & \sqrt{\frac 2n}\cdot c_k\cdot a_{k,l} & 
\sqrt{\frac 2n}\cdot c_k\cdot a_{k,l}\\
\text{3} & \sqrt{\frac 2n}\cdot c_\ell\cdot a_{k,l} & 
\sqrt{\frac 2n}\cdot c_\ell\cdot a_{k,l}\\
\text{4} & \sqrt{\frac 2n}\cdot a_{k,l} & \sqrt{\frac 2n}\cdot a_{k,l}\\
\text{5} & \sqrt{\frac{2}{n-1/2}}\cdot c_kc_\ell\cdot a_{k,l} & 
\sqrt{\frac{2}{n+1/2}}\cdot a_{k,l}\\
\text{6} & \sqrt{\frac{2}{n-1/2}}\cdot c_kd_\ell\cdot a_{k,l} & 
\sqrt{\frac{2}{n+1/2}}\cdot a_{k,l}\\
\text{7} & \sqrt{\frac{2}{n-1/2}}\cdot d_kc_\ell\cdot a_{k,l} & 
\sqrt{\frac{2}{n+1/2}}\cdot a_{k,l}\\
\text{8} & \sqrt{\frac{2}{n+1/2}}\cdot a_{k,l} & 
\sqrt{\frac{2}{n-1/2}}\cdot d_kd_\ell\cdot a_{k,l}\\ \bottomrule
\end{array}
$
\end{table}

A natural question to ask is how these scaling factors are obtained.
In Theorem~\ref{main} we established that the 16 DTTs are Fourier
transforms for regular modules $\md = \C[x]/p$ with a basis of
Chebyshev polynomials. However, if $\md$ is given, the natural choice
for a Fourier transform is the polynomial transform $\pDTT$ introduced in Definition~\ref{pdtt}.
The scaling functions $f$ to obtain the DTTs from the $\pDTT$s were found by
observation, i.e, by comparing with their definition.

In this section we derive the orthogonal DTTs from their polynomial
counterparts, i.e., we compute diagonal matrices $D_1, D_2$, such that
\begin{equation*}
D_1\cdot\pDTT\cdot D_2
\end{equation*}
is orthogonal. In the case when  the matrix $D_2$ in this scaling is
necessary, it is immediately clear (from Theorem~\ref{fourtrafo}) that
the underlying signal model has to be modified to admit the orthogonal
DTT as Fourier transform. We derive these signal models and show that
they are symmetric in a sense that will be defined.

\mypar{Derivation of scaling factors} The root of the ``almost
orthogonality'' of the DTTs is the following {\em Christoffel-Darboux
formula} for orthogonal polynomials \cite{Szegoe:67}.

\begin{theorem}[Christoffel-Darboux formula]\label{cd}
Let $(P_k\mid k \geq 0)$ be a sequence of
orthogonal polynomials over $I\subset\R$ with respect to some weight
function $\omega(x)$, i.e., ($\delta$ denotes the Kronecker delta function)
$$
\int_{I}P_k(x)P_\ell(x)\omega(x)dx = \mu_k\delta_{k\ell},\quad
\mu_n > 0.
$$
Further, denote with $\beta_k$ the leading coefficient of $P_k$. Then
\isdraft{%
\begin{equation}\label{CD}
\sum_{0\leq k < n}\mu_k^{-1}P_k(x)P_k(y)
=
\begin{cases}
c_n\dps\frac{P_{n-1}(y)P_n(x) - P_n(y)P_{n-1}(x)}{x-y}, &x\neq y,\\
c_n(P_{n-1}(x)P_n'(x) - P_n(x)P_{n-1}'(x)), &x = y,
\end{cases}
\end{equation}
}{%
\begin{multline}\label{CD}
\sum_{0\leq k < n}\mu_k^{-1}P_k(x)P_k(y) \\
=
\begin{cases}
c_n\dps\frac{P_{n-1}(y)P_n(x) - P_n(y)P_{n-1}(x)}{x-y}, &x\neq y,\\
c_n(P_{n-1}(x)P_n'(x) - P_n(x)P_{n-1}'(x)), &x = y,
\end{cases}
\end{multline}
}
where the constant $c_n$ is given by $c_n =
\beta_{n-1}\beta_n^{-1}\mu_{n-1}^{-1}$, and $P'_k$ denotes the
derivative of $P_k$.
\end{theorem}

As a consequence of the Christoffel-Darboux formula, we get the
following construction method for orthogonal versions of polynomial
transforms. This theorem, as the Christoffel-Darboux formula, is more
general than needed here. We will need the full generality for the
generic next-neighbor model in Section~\ref{genshiftmodel}.  We also
note that the following theorem restates, in algebraic terms, the
construction method for unitary transforms from \cite{Yemini:79},
where it is called {\em Gauss-Jacobi procedure}.

\begin{theorem}\label{orthpolytrafo}
Let $(P_k\mid k \geq 0)$ as in Theorem~\ref{cd}, and let $\md =
\C[x]/P_n$ with basis $b = (P_0,\dots,P_{n-1})$. Since $P_n$
is an orthogonal polynomial, it is separable (see
\cite[p.~28]{Chihara:78}); we denote
its list of zeros by $\alpha = (\alpha_0,\dots,\alpha_{n-1})$. 
Further, we define the two diagonal matrices
\begin{equation}\label{diags}
\begin{array}{rcl}
D & = & 
  c_n^{-1}\diag_{0\leq k<n}((P_{n-1}(\alpha_k)P'_n(\alpha_k))^{-1}),\\
E & = & \diag_{0\leq k < n}(\mu_k^{-1}),
\end{array}
\end{equation}
where $c_n, \mu_k$ are defined as in Theorem~\ref{cd}. Then,
\begin{equation}\label{almostorth}
\sqrt{D}\cdot\poly_{b,\alpha}\cdot\sqrt{E}
\end{equation}
is orthogonal.
\end{theorem}
\begin{proof}
First, we note that $D$ is well-defined: $P_{n-1}(\alpha_k)\neq 0$,
since $P_{n-1}$ and $P_n$ have disjoint sets of zeros
(\cite[p.~28]{Chihara:78}); $P'_n(\alpha_k)\neq 0$, since $P_n$ is
separable. Now, we substitute the zeros $\alpha_i,\alpha_j$ for $x,y$ in
\eqref{CD} to get
$$
\poly_{b,\alpha}^{-1} = E\poly_{b,\alpha}^T D,
$$
which implies the desired result.
\end{proof}

This general property explains the form of the orthogonal versions of
the DTTs in all cases in which the associated module has the form
$\C[x]/C_n$ with basis $(C_0,\dots,C_{n-1})$, i.e., for all DTTs in
the second column in Table~\ref{chebmods}, i.e., for $\DCTt{3}$,
$\DSTt{1}$, $\DCTt{8}$, and $\DSTt{6}$.  For the other 12 DTTs, we
need to derive variants of \eqref{CD}. We have deferred this rather
technical derivation to Appendix~\ref{directorth}.

The above derivation of the orthogonal DTTs gives little intuition
into this, at first glance, surprising property. However, the property
is easy to understand and easier to derive by looking at the
diagonalization properties of the DTTs, as was pointed out in
\cite{Strang:99}. We explain this next, and give further insight by
constructing the signal models, for which the orthogonal DTTs are
Fourier transforms.

The orthogonal transform in \eqref{almostorth} is scaled from both
sides by a diagonal matrix. Scaling on the left side, as we have
shown, is equivalent to choosing a different basis in the spectrum,
i.e., in the decomposed module (Theorem~\ref{fourtrafo}). Scaling on the
right changes (by scaling) the basis in the module and thus the
signal model. We explain this using the $\DCTt{3}$ as an example, 
and, at the same time, motivate the origin of the diagonal matrices.

\mypar{Signal model for the orthogonal DCT, type 3}
The $\DCTt{3}_n$ is a Fourier transform for the signal model
$$
\Phi:\ \coord{s}\mapsto\sum_{0\leq k<n}s_kT_k\in\md = \C[x]/T_n,
\quad\alg = \md.
$$
If $\phi$ is the associated representation of $\alg$, then
$\DCTt{3}_n$ diagonalizes any matrix $\phi(h)$, $h\in\alg$. In particular,
\begin{equation}\label{T1diag3}
\phi(x) = \fh\cdot\left[
\begin{array}{ccccccc}
0 & 1 \\
2 & 0 & 1 \\
0 & 1 & 0 & \cdot \\
 & & 1 & \cdot & 1\\
 & & & \cdot & 0 & 1 \\
 & & & & 1 & 0
\end{array}
\right],
\end{equation}
which is a special case of \eqref{T1diag}.  

\begin{figure}[ht]
\isdraft{\renewcommand{\baselinestretch}{1}\normalsize}{}

\medskip
\isdraft{\hspace*{30mm}}{\hspace*{6mm}}
\begin{minipage}{0.8\textwidth}
{
\xymatrix{
\bullet \ar@{<->}[r] & \bullet \ar@{<->}[r] & 
\bullet \ar@{.}[rr] && \bullet \ar@{<->}[r] &
\bullet \ar@{<->}[r] & \bullet
}
\isdraft{\setlength{\unitlength}{1.15pt}}{}
\begin{picture}(220,5)
\put(22,20){\makebox(0,0){$\sqrt{2}$}}
\put(7,0){\makebox(0,0){$T_0$}}
\put(42,0){\makebox(0,0){$T_1$}}
\put(77,0){\makebox(0,0){$T_2$}}
\put(142,0){\makebox(0,0){$T_{n-3}$}}
\put(177,0){\makebox(0,0){$T_{n-2}$}}
\put(212,0){\makebox(0,0){$T_{n-1}$}}
\end{picture}
}
\end{minipage}

\medskip
\caption{Visualization of the symmetric signal model associated with
  the orthogonal version of the DCT, type 3.\label{visortho}}
\end{figure}
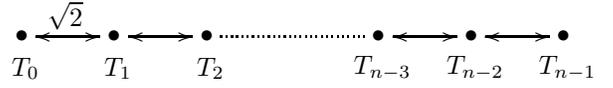

We observe that the matrix $\phi(x)$ in
\eqref{T1diag3} is ``almost'' symmetric.  In fact, symmetry can be 
readily established by conjugating $\phi(x)$ with the diagonal matrix
$$
E = \diag(\sqrt{2},1,\dots,1),
$$
i.e., $E\phi(x)E^{-1}$ is symmetric. 
This corresponds to the following
base change in the underlying module:
$$
b = (T_0,\dots,T_{n-1})\rightarrow
b' = (\ts\frac{1}{\sqrt{2}}T_0,T_1,\dots,T_{n-1}).
$$
If $\phi'$ is the representation afforded by $b'$, then
$$
\phi'(x) = E\phi(x)E^{-1} =
\fh\cdot\left[
\begin{array}{ccccccc}
0 & \sqrt{2} \\
\sqrt{2} & 0 & 1 \\
0 & 1 & 0 & \cdot \\
 & & 1 & \cdot & 1\\
 & & & \cdot & 0 & 1 \\
 & & & & 1 & 0
\end{array}
\right].
$$
The base change with $E$ also changed the underlying signal model, 
which is now given by
$$
\Phi':\ \coord{s}\mapsto s_0\frac{1}{\sqrt{2}}T_0 + \sum_{1\leq k<n}s_kT_k
\in\md = \alg = \C[x]/T_n.
$$ 
The change can also be seen in its visualization in
Figure~\ref{visortho} (as usual, the global factor $1/2$ was
omitted), which should be compared to the visualization of the
original model $\Phi$ in the third graph in
Figure~\ref{finspacegraph}. The new model $\Phi'$ has still the same
left \bc and signal extension, but,  in order to make it symmetric, the
space shift was locally redefined, which means, at these marks, the
model is now variant.
Namely, for 
$T_0' = 1/\sqrt{2}T_0,T_{-1}'=T_1',T_2'\in b'$ as
$$
\begin{array}{rcl}
xT'_0 & = & \fh(\frac{1}{\sqrt{2}}T'_{-1} + \frac{1}{\sqrt{2}}T'_1)
  = \fh\sqrt{2}T'_1,\\
xT'_1 & = & \fh(\sqrt{2}T'_0 + T'_2).
\end{array}
$$

To compute a Fourier transform for $\Phi'$, we use the following
commutative diagram. In the top row, the modules have the bases $b$
and $b'$, respectively.

\begin{equation*}
\algogen{\C[x]/T_n}{E}{\C[x]/T_n}%
  {\poly_{b,\alpha} = \DCTt{3}_n}{\poly_{b',\alpha}}
  {\bigdirsum\C[x]/(x-\alpha_k)}
  {\one_n}{\bigdirsum\C[x]/(x-\alpha_k)}
\end{equation*}

In particular, 
$$
\DCTt{3}_n = \poly_{b,\alpha} = \poly_{b',\alpha}E.
$$

At this point, we remind the reader that every symmetric matrix $A$
can be diagonalized by an orthogonal matrix $M$ (e.g.,
\cite{Gantmacher:59}): $MAM^{-1}$ is diagonal. 
Further, if $A$ has pairwise distinct eigenvalues, then {\em all}
matrices that diagonalize $A$ have the form $DM$, where $D$ is any
invertible diagonal matrix. In our particular case,
$\poly_{b',\alpha}$ diagonalizes a symmetric matrix with pairwise
distinct eigenvalues, thus, 
\begin{equation}\label{nowortho}
M = D\poly_{b',\alpha} = D\DCTt{3}_nE^{-1}
\end{equation}
is orthogonal for a suitable diagonal matrix $D$ that normalizes the
row vectors of $\poly_{b',\alpha}$ to have length 1.  The matrix $M$
is a Fourier transform, but not a polynomial transform, for the signal
model $\Phi'$.

\mypar{Signal models for all orthogonal DTTs} 
In the general case of a DTT, i.e., a $\phi(x)$ in \eqref{T1diag},
conjugation, $E\phi(x)E^{-1}$, with the matrix
$$
E = \diag(\sqrt{\beta_2},1,\dots,1,\sqrt{\beta_3})
$$
makes $\phi(x)$ symmetric and we get the following theorem.

\begin{theorem}[Signal models for orthogonal DTTs]\label{symsmdtt}
Let $\DTT_n$ be any of the 16 DTTs with signal model $(\alg,\md,\Phi)$
where $\md = \alg = \C[x]/p(x)$ with $p$ one of the polynomials in
Lemma~\ref{rightbc}, and
$$
\Phi:\ \coord{s}\mapsto\sum_{0\leq k<n}s_kC_k.
$$ 
Let $\phi(x)$ be the associated shift matrix given
by~\eqref{T1diag}. Further, let $\beta_2,\beta_3$ be the values
in~\eqref{T1diag} for $\phi(x)$.  If $\DTT'_n$ is the orthogonal
version of $\DTT_n$, then $\DTT'_n$ is a Fourier transform for the
signal model $(\alg,\md,\Phi')$ with
$$
\Phi':\ \coord{s}\mapsto s_0\frac{1}{\sqrt{\beta_2}}C_0 + 
  \sum_{1\leq k<n-1}s_kC_k + s_{n-1}\frac{1}{\sqrt{\beta_3}}C_{n-1}.
$$
\end{theorem}
We call the signal models $\Phi'$ {\em symmetric} in the following sense.

\begin{definition}[Symmetric signal model]\label{symsm}
Let $(\alg,\md,\Phi)$ be a finite shift-invariant signal model with
$\md = \C[x]/p(x)$, and let $\phi$ be the afforded representation of $\alg$.
We call this model {\em symmetric}, if $\phi(h)$ is symmetric for all 
filters $h\in\alg$.
\end{definition}

Note that if $\alg = \md$, then the model is symmetric if and only if
$\phi(x)$ is symmetric. (If $\alg$ is smaller than $\md$, then
$x\not\in\alg$; we will have an example of this situation later.)

The above discussion explains how to easily obtain, for a given DTT,
the diagonal matrix $E$ that scales from the right. The matrix $D$ is
then obtained from the norms of the DTT's row vectors using basic
trigonometric identities.

The adjustment of the signal model is necessary, because the matrices
$\phi(x)$ in \eqref{T1diag} are not all symmetric due to their
boundary conditions. Closer inspection shows that among the four types
of boundary conditions only one, namely $s_{-1} = s_1$ on the left,
and $s_n = s_{n-2}$ on the right, causes this problem. Thus, the only
DTTs that need an adjustment of the signal model are those in the
first row or column in Table~\ref{chebmods}. All other DTTs need to be
scaled only from the left as can be confirmed from
Table~\ref{dttorth}.

\subsection{Duality}\label{dualsec}

We observed before that the right \bc's for the DTTs are precisely the
mirrored versions of the left \bc's, a fact that meets our intuition
since the DTTs are based on  symmetric space models. However, the
construction of the module $\C[x]/p$ for a given DTT (see
Theorem~\ref{main}) deals differently with the left \bc~(which
determines the choice of the base sequence $C$) and the right
\bc~(which determines $p$); thus, we obtain different DTTs for a 
given pair of \bc's and for its mirrored counterpart. The associated
pair of DTTs occurs in positions in Table~\ref{chebmods} that are
mirrored at the main diagonal. We call such a pair {\em dual} and show
that dual DTTs have a very close relationship.

\begin{definition}[Duality]
Let $\DTT$ and $\DTT'$ be at mirrored positions in
Table~\ref{chebmods}, i.e., at positions $(i,j)$, $(j,i)$,
$1\leq i,j\leq 4$, respectively. We call $\DTT$ and $\DTT'$ {\em dual}
to each other. The DTTs on the main-diagonal are called self-dual.
Dual DTTs have the same associated regular module $\C[x]/p$.
\end{definition}

To derive the relationship between dual DTTs, we use $\DCTt{3}_n$ and
$\DSTt{3}_n$ as an example. The module $\md = \C[x]/p$ associated with
$\DCTt{3}_n$ carries the left \bc $s_{-1} = s_1$, determined by the
$T$-basis $b = (T_0,\dots,T_{n-1})$, and the right \bc $s_n = 0$,
determined by the equation $p = T_n = 0$. The zeros of $T_n$ are given
by $\alpha = (\cos(k+1/2)\pi/n\mid 0\leq k < n)$.
Now, we consider the same module $\md$, but choose a different basis,
namely $b' = (U_{n-1},\dots,U_0)$. Our goal is to construct the
following diagram; $\C[x]/p$ with basis $b'$ is the top left module.
\begin{equation*}
\algogen{\C[x]/p}{\oppone_n}{\C[x]/p}%
  {\DCTt{3}_n}{\DSTt{3}_n}{\bigdirsum\C[x]/(x-\alpha_k)}{D_n}%
  {\bigdirsum\C[x]/(x-\alpha_k)}
\end{equation*}
We know how to decompose $\md$ with $b'$, namely by first reversing
the order of the basis with a permutation $\oppone_n$ (the identity
matrix with the columns in reversed order) to obtain the top right
module, which in turn is decomposed by $\DSTt{3}_n$, where the
decomposed module has the basis $(1/f(\alpha_k)\mid 0\leq k < n)$,
where $f = \sin\theta$, $\cos\theta = x$ (see Theorem~\ref{main}).

On the other hand, $\md$ with $b'$ affords the same representation as
$\md$ with $b$. To see this, we list $b$ and $b'$ together with their
\bc's; the vertical lines indicate the boundaries of the signal scope:
$$
\begin{array}{c@{\ |\ }ccc@{\ |\ }c}
T_{-1} = T_1 & T_0 & \dots & T_{n-1} & T_n = 0\\
U_n = U_{n-2} & U_{n-1} & \dots & U_0 & U_{-1} = 0
\end{array}
$$
Note that $T_n = 0$, expressed in $U$-polynomials, becomes $U_n =
U_{n-2}$, which is the right \bc~of $\DSTt{3}$ and the mirror image of
the left \bc~$T_{-1} = T_1$ of $\DCTt{3}$. Observe, e.g., that in
$\md$ $xU_{n-1} = (U_n + U_{n-2})/2 = U_{n-2}$ and $xT_0 = (T_{-1} +
T_1)/2 = T_1$, i.e., $x$ has the same effect on $b$ as on $b'$.

Since $b'$ affords the same representation as $b$, $\md$ with $b'$ is
also decomposed by the $\DCTt{3}$ (left column in the diagram); it remains
to determine the basis in the decomposed module. Note that $\DCTt{3}$
is a polynomial transform for $\md$ and $b$, but not for $\md$ and
$b'$.

In fact, using
$$
U_{n-1}T_i = (U_{n-1-i} + U_{n-1+i})/2 \equiv U_{n-1-i}
  \text{ mod }T_n,
$$
which is due to the signal extension of $\DSTt{3}_n$, we get
$$
b' = U_{n-1}b.
$$
This shows that the basis in the decomposed module (bottom left) is
given by the numbers
$$
U_{n-1}(\alpha_k) = \sin(k+1/2)\pi/f(\alpha_k) = (-1)^k/f(\alpha_k),
$$
where $f$ is the scaling function of $\DSTt{3}$ (see above).
This completes our diagram with $D_n = \diag_{0\leq k < n}((-1)^k)$.
As an equation, we get
\begin{equation}\label{duality-a}
\diag_{k = 0}^{n-1}((-1)^k)\cdot\DCTt{3}_n = 
  \DSTt{3}_n\cdot\oppone_n.
\end{equation}
Analogous computations verify the same identity for all pairs of dual
DTTs. Interestingly, the
diagonal $D_n$ is in all cases the same.  Also, we note that
$\oppone_n^2 = D_n^2 = \one_n$.

\begin{theorem}[Duality Relationship]\label{duality}
Let $\DTT_n$ and $\DTT'_n$ be a pair of dual $\DTT$s. Then
$$
\diag_{0\leq k < n}( (-1)^k)\cdot\DTT_n = \DTT'_n\cdot\oppone_n,
$$
with corresponding diagram
\begin{equation*}
\algogen{\C[x]/p}{\oppone_n}{\C[x]/p}%
  {\DTT_n}{\DTT'_n}{\bigdirsum\C[x]/(x-\alpha_k)}{\diag( (-1)^k)}%
  {\bigdirsum\C[x]/(x-\alpha_k)}
\end{equation*}
\end{theorem}
As an important consequence of Theorem~\ref{duality}, dual DTTs have
the same arithmetic complexity. 

In the literature, the special case of the duality (as defined by us)
between $\DCTt{3}$ and $\DSTt{3}$ was the subject of \cite{Wang:82}.

\subsection{Groups of DTTs and Relationships}\label{groupdtt}

In Section~\ref{dualsec} we introduced the concept of dual DTTs, which
necessarily have the same associated module $\C[x]/p$. However, in
Table~\ref{chebmods}, we also have DTTs with the same module, which are not
dual. An example is given by the DCTs of type 3 and 4 with module
$\C[x]/T_n$. In fact, closer inspection of Table~\ref{chebmods} shows
that, barring linear factors, each of the four types of Chebyshev
polynomials occurs exactly four times as $p$. For example, $p = T_n$
occurs for all four DCTs and DSTs of type 3 and 4. Thus, we have a
natural classification of the 16 DTTs into four groups of four
each. We call these groups, depending on $p$, $T$-group, $U$-group,
$V$-group, and $W$-group. In particular, dual DTTs are in the same
group.

Further inspection shows that, in each group, all possible left and
right \bc's are present. Thus the DTTs in one group have (almost) the
same module, but with different bases. Thus, we can translate DTTs in
the same group into each other using a base change. Further, because
of Table~\ref{chebids}, the resulting base change matrices are sparse,
i.e., require $O(n)$ operations for multiplication.

Before we give two instructive examples, we remind the reader that
decomposing a regular module $\C[x]/p$ by a polynomial transform
implies that in  each one-dimensional irreducible summand
$\C[x]/(x-\alpha_k)$ the basis $(x^0) = (1)$ is chosen; other choices
of base elements $a_k\neq 1$ lead to a scaled polynomial transform
(see Theorem~\ref{fourtrafo}). The DTTs are scaled polynomial
transforms; their polynomial counterparts are denoted by $\pDTT$ (see
Definition~\ref{pdtt}). We have $\DTT = \pDTT$ if and only if DTT is a
DCT of odd type.

\mypar{Example: DCT, type 3 and 4}
First, we consider $\DCTt{3}_n$ and $\DCTt{4}_n$, which are both in
the $T$-group, i.e., the associated module is $\md = \C[x]/T_n$. The
difference is in the choice of bases.
$$
\begin{array}{lll}
\DCTt{3}_n: & \C[x]/T_n, & b = (T_0,\dots,T_{n-1}),\\
\DCTt{4}_n: & \C[x]/T_n, & b' = (V_0,\dots,V_{n-1}).
\end{array}
$$
Using
\begin{equation}\label{TV}
T_\ell = (V_\ell + V_{\ell-1})/2
\end{equation}
from Table~\ref{chebids} and $V_{-1} = V_0$, the corresponding base
change matrix $S'_n$ for $b\rightarrow b'$ is given by
\begin{equation}\label{summatrix}
S'_n = \fh\cdot\left[
\begin{array}{ccccc}
2 & 1 \\
0 & 1 & 1 \\
 & & \cdot & \cdot \\
 & & & 1 & 1\\
 & & & & 1
\end{array}
\right].
\end{equation}
We denote the zeros of $T_n$ by $\alpha_k = \cos(k+1/2)\pi/n$. 
As a consequence of the above, we get the commutative diagram
\begin{equation}\label{dct34diagram}
\algogen{\C[x]/T_n}{S'_n}{\C[x]/T_n}%
  {\DCTt{3}_n}{\pDCTt{4}_n}{\bigdirsum\C[x]/(x-\alpha_k)}
  {\one_n}{\bigdirsum\C[x]/(x-\alpha_k)}
\end{equation}
which implies the equation
$$
\DCTt{3}_n = \pDCTt{4}_n\cdot S'_n.
$$
Note that we have $\one_n$ in the bottom row of \eqref{dct34diagram}
since both $\DCTt{3}$ and $\pDCTt{4}$ are polynomial transforms and
thus use the same basis $(1,\dots,1)$ in the decomposed
module.

Introducing the scaling diagonal 
$$
D_n = \diag_{0\leq k<n}(\cos(2k+1)\pi/(4n))
$$
of the $\DCTt{4}$ (see Table~\ref{chebmods}), we get
$$
\algogen{\C[x]/T_n}{S'_n}{\C[x]/T_n}%
  {\DCTt{3}_n}{\DCTt{4}_n}{\bigdirsum\C[x]/(x-\alpha_k)}
  {D_n}{\bigdirsum\C[x]/(x-\alpha_k)}
$$
or, as an equation,
\begin{equation}\label{dct4and3}
D_n\cdot\DCTt{3}_n = \DCTt{4}_n\cdot S'_n.
\end{equation}

We use \eqref{dct4and3} to also show how formal manipulation derives new
relationships from known ones. Since it is known that the $\DCTt{4}$
is more expensive to compute than the $\DCTt{3}$, we multiply
\eqref{dct4and3} by $(S'_n)^{-1}$ to get
\begin{equation}\label{dct4and3a}
D_n\cdot\DCTt{3}_n\cdot(S'_n)^{-1} = \DCTt{4}_n.
\end{equation}
We can multiply by $(S'_n)^{-1}$ using $n-1$ recursive additions, but
this produces a critical path of length $n-1$. To solve the problem,
we invert both sides of \eqref{dct4and3a} using $\DCTt{3}_n^{-1} =
2/n\cdot\diag(2,1,\dots,1)\cdot\DCTt{2}_n$ and $\DCTt{4}_n^{-1} =
2/n\cdot\DCTt{4}_n$ (follows from their orthogonal versions in
Table~\ref{dttorth} and Table~\ref{dttdefs}).  It turns out that some
factors cancel each other, and we get
\begin{equation}\label{dct4and2}
S_n\cdot\DCTt{2}_n\cdot \frac 12 D_n^{-1} = \DCTt{4}_n
\end{equation}
with
\begin{equation}\label{Sn}
S_n =
\left[
\begin{array}{ccccc}
1 & 1 \\
0 & 1 & 1 \\
 & & \cdot & \cdot \\
 & & & 1 & 1\\
 & & & & 1
\end{array}
\right].
\end{equation}
Transposing \eqref{dct4and2} yields
\begin{equation}\label{dct4and3final}
\frac 12 D_n^{-1}\cdot\DCTt{3}_n\cdot S_n^T = \DCTt{4}_n,
\end{equation}
where we used that $\DCTt{2}_n^T = \DCTt{3}_n$ and that $\DCTt{4}_n$
is symmetric (Section~\ref{dtttransform}); \eqref{dct4and3final} is a clear
improvement over \eqref{dct4and3a} obtained by
inversion-transposition.

\mypar{Example: DCT, type 1 and 2}
As a second example, we derive a relationship between $\DCTt{1}_{n+1}$ and
$\DCTt{2}_n$, which, without the module framework, is less obvious to
derive. Note the difference in size ($n+1$ versus $n$) to obtain
comparable modules, which are
$$
\begin{array}{lll}
\DCTt{1}_{n+1}: & \C[x]/(x^2-1)U_{n-1}, & b = (T_0,\dots,T_n),\\
\DCTt{2}_n: & \C[x]/(x-1)U_{n-1}, & b' = (V_0,\dots,V_{n-1}).
\end{array}
$$
To translate a $\DCTt{1}_{n+1}$ into a $\DCTt{2}_n$, we have to
partially decompose the module $\C[x]/(x^2-1)U_{n-1}$ using the
Chinese remainder theorem,
$$
\C[x]/(x^2-1)U_{n-1} \rightarrow \C[x]/(x-1)U_{n-1} \dirsum \C[x]/(x+1).
$$ 
As bases in these three modules we choose, from left to right, $b$,
$b'$, and $(1)$. We compute the corresponding base change matrix by
using \eqref{TV} and $T_\ell\text{ mod }(x+1) = T_\ell(-1) =
(-1)^\ell$ from Lemma~\ref{cheb1}, iii) in Appendix~\ref{chebs}. The
result is
$$
B_{n+1} = \frac 12\cdot
\left[
\begin{array}{ccccccc}
2 & 1 & \\
0 & 1 & 1 \\
 & & \ddots & \ddots \\
 & & & 1 & 1 & 0\\
 & & & & 1 & 2 \\
2 & -2 & 2 & \cdot & \cdot 
\end{array}
\right].
$$ 
The last column uses the \bc~$V_n = V_{n-1}$.
Consequently, we get
$$
\DCTt{1}_{n+1} = (\pDCTt{2}_n\dirsum\one_1)\cdot B_{n+1}.
$$
Using the scaling diagonal $D_n = \diag_{0\leq
k<n}(\cos k\pi/(2n))$ of the $\DCTt{2}$, this can be translated into
\begin{equation}\label{dct321}
(D_n\dirsum\one_1)\cdot\DCTt{1}_{n+1} = 
(\DCTt{2}_n\dirsum\one_1)\cdot B_{n+1},
\end{equation}
or, as a commutative diagram,
$$
\algogen{\C[x]/(x^2-1)U_{n-1}}{B_{n+1}}%
  {{\begin{array}{r@{\ }l} & \C[x]/(x-1)U_{n-1}\\
      \dirsum & \C[x]/(x+1)\end{array}}}%
  {\DCTt{1}_{n+1}}{\DCTt{2}_n\dirsum\one_1}%
  {\bigdirsum\C[x]/(x-\cos\frac{k\pi}{n})}%
  {D_n\dirsum\one_1}{\bigdirsum\C[x]/(x-\cos\frac{k\pi}{n})}
$$
Transposition of \eqref{dct321} yields a relationship between
$\DCTt{1}_{n+1}$ and $\DCTt{3}_n$.

\mypar{Base change theorem}
Using base changes between DTTs of the same group, combined with 
transposition, and inversion, we get the following theorem.
Note that the sparsity of the base change matrix is guaranteed by
Table~\ref{chebids}.

\begin{theorem}[Base Change Theorem]\label{dtttranslate}
By a base change in the associated module, all DTTs of type 1--4, and
all DTTs of type 5--8 can be translated into each other using $O(n)$
arithmetic operations.
\end{theorem}

\subsection{Real and Rational Signal Model}

In Section~\ref{realdfts}, we showed the effect of reducing the base
field from $\C$ to $\R$ in the finite time model.  Doing the same in
the finite space models, in contrast, does not incur any change, since
all considered polynomials factorize completely over $\R$.  Thus, the
real finite space models will share all the properties from their
complex counterparts.

The situation is different, if we further restrict the base field to
$\Q$. The question becomes how the polynomials in Table~\ref{chebmods}
factor over $\Q$. This question is answered in \cite{Rayes:98} and can
be used to derive rational versions of the DTTs. An application could
be the derivation of algorithms, but we did not pursue this direction.

\section{Finite Skew C-Transform and Skew DTTs}
\label{skewdttssec}

In this section we introduce a new class of transforms that is closely
related to the DTTs. We call these transforms {\em skew DTTs}. More
specifically, the skew DTTs correspond to and generalize the DTTs in
the $T$-group, i.e., those with associated module $\C[x]/T_n$, which
are the DCTs and DSTs of type~3 and~4. 

We introduce the skew DTTs for the following reasons:
\begin{itemize}
\item They are interesting from a signal processing point of view. As
the DTTs, they provide a finite space model, their associated boundary
conditions are simple, and their signal extension is 2-monomial
(defined below) and also eventually periodic.
\item They are necessary building blocks in the Cooley-Tukey FFT type
DTT algorithms that we will derive and present in detail in our next
paper on the algebraic theory of signal processing,
\cite{Pueschel:05}. The special case of DCT type 3 (and 2) 
algorithms and the first skew DCT were introduced in
\cite{Pueschel:03}.
\end{itemize}

We follow our usual structure and derive first the signal model and then
the associated Fourier transform.

\subsection{Building the Signal Model}

\mypar{Shift, Linear Extension, Realization} The model we create is
based on the finite space shift and thus its derivation follows the
exact same steps as the derivation of the finite space model in
Section~\ref{buildfinCtrafo}. The model we create now generalizes
the signal model for the DTTs in the $T$-group, i.e., the DCTs and DSTs
of type 3 and 4 (see Table~\ref{chebmods}). The generalization is done
by modifying the right \bc's and thus the right signal extension as we
explain next.

\mypar{Boundary condition and signal extension} In the
previous finite space model, we chose the right \bc's to ensure a monomial
signal extension. Now, we just state the boundary conditions and 
derive the signal extension later.

The goal is to modify the right \bc of the DCTs and DSTs of type 3 and
4 such that the associated module is given by $\C[x]/(T_n - \cos
r\pi)$, $r\in\Q$, $0\leq r \leq 1$. In this case we can read off the
\bc's from Table~\ref{chebmods}:
\begin{equation}\label{skewbc}
\begin{array}{rl}
\DCTt{3}_n: & T_n = \cos r\pi, \\
\DSTt{3}_n: & U_n = U_{n-2} + 2\cos r\pi, \\
\DCTt{4}_n: & V_n = -V_{n-1} + 2\cos r\pi, \\
\DSTt{4}_n: & W_n = W_{n-1} - 2\cos r\pi. \\
\end{array}
\end{equation}
Thus, we obtain four equal modules $\C[x]/(T_n - \cos
r\pi)$ with $T$-basis, $U$-basis, $V$-basis, $W$-basis, 
respectively.

In the general case $r\neq 1/2$, these \bc lead to no monomial signal
extension, since this property defines the signal models for the 16
DTTs. However, it is intriguing that the signal extension is
``almost'' monomial in the following sense.

\begin{definition}
Using the notation from Definition~\ref{sedef}, we call a signal
extension
$$
s_k = \sum_{i\in I}\beta_{k,i}s_i,\quad\text{for }k\not\in I,
$$
{\em 2-monomial} if for each $k$ the sum has at most 2 summands.
\end{definition}

Now, we can explicitly state the signal extensions in these 
four cases.

\begin{lemma}
The module $\C[x]/(T_n-\cos r\pi)$ with $T$-, $U$-, $V$-, or $W$-basis
has a 2-monomial signal extension. More precisely, the signal
extension is displayed in Table~\ref{sigextskew} using the same method
as in Figure~\ref{sigextxna} and Table~\ref{sigextdtts}. The occurring
constants are $c_k = T_k(\cos r\pi) = \cos kr\pi$ and $u_k = U_k(\cos
r\pi)$. The constants $u_k$ scale the entire line they are adjacent to
with the exception of the signal extension for the $\DCTt{3}(r)$ (top
line), where the bottom values (bullets) are scaled by $c_k$.
\end{lemma}
\begin{proof}
The proof uses induction and the two-term recurrence of the Chebyshev
polynomials. We show the induction step for the case of a $V$-basis in a
boundary case. According to Table~\ref{sigextskew},
\begin{eqnarray*}
V_{kn-2} & = & u_{k-1}V_{n-2} - u_{k-2}V_1, \\
V_{kn-1} & = & u_{k-1}V_{n-1} - u_{k-2}V_0.
\end{eqnarray*}
We compute $V_{kn}$ using the Chebyshev recurrence and the \bc $V_n =
-V_{n-1} - 2\cos r\pi$,
\begin{eqnarray*}
V_{kn} & = & 2xV_{kn-1} - V_{kn-2} \\
& = & u_{k-1}(V_{n-2} + V_n) - u_{k-2}(V_0 + V_1) \\
& & -u_{k-1}V_{n-2} + u_{k-2}V_1 \\
& = & u_{k-1}(-V_{n-1} + 2\cos r\pi V_0)-u_{k-2}V_0\\
& = & -u_{k-1}V_{n-1} + u_k V_0
\end{eqnarray*}
where we used $2\cos r\pi\cdot u_{k-1} - u_{k-2} = u_k$, which is again the
recurrence for Chebyshev polynomials. The result coincides with
Table~\ref{sigextskew}, as desired.
\end{proof}

\begin{table}\centering
\caption{(Right) signal extensions for the four skew DTTs.\label{sigextskew}}
\ra{1.5}
\begin{tabular}{@{}ll@{}}\toprule
\raisebox{7mm}[-7mm]{$\DCTt{3}_n(r)$} &
\begin{picture}(180,45)
\put(0,10){\usebox{\bupbline}}
\put(35,40){\makebox(0,0){$\sst 0$}}
\put(40,10){\usebox{\bdownline}}
\put(70,10){\usebox{\bupline}}
\put(105,40){\makebox(0,0){$\sst 0$}}
\put(110,10){\usebox{\bdownline}}
\put(140,10){\usebox{\bupline}}
\put(175,40){\makebox(0,0){$\sst 0$}}
\put(35,10){\usebox{\bupline}}
\put(70,40){\makebox(0,0){$\sst 0$}}
\put(75,10){\usebox{\bdownline}}
\put(105,10){\usebox{\bupline}}
\put(140,40){\makebox(0,0){$\sst 0$}}
\put(145,10){\usebox{\bdownline}}
\put(0,0){\makebox(0,0){$\sst 0$}}
\put(35,0){\makebox(0,0){$\sst n$}}
\put(70,0){\makebox(0,0){$\sst 2n$}}
\put(105,0){\makebox(0,0){$\sst 3n$}}
\put(140,0){\makebox(0,0){$\sst 4n$}}
\put(175,0){\makebox(0,0){$\sst 5n$}}
\put(29,10){\makebox(0,0){$\sst c_1$}}
\put(64,10){\makebox(0,0){$\sst c_2$}}
\put(99,10){\makebox(0,0){$\sst c_3$}}
\put(134,10){\makebox(0,0){$\sst c_4$}}
\put(169,10){\makebox(0,0){$\sst c_5$}}
\put(38,20){\makebox(0,0){$\sst u_1$}}
\put(40,32){\makebox(0,0){$\sst -u_0$}}
\put(73,20){\makebox(0,0){$\sst u_2$}}
\put(75,32){\makebox(0,0){$\sst -u_1$}}
\put(108,20){\makebox(0,0){$\sst u_3$}}
\put(110,32){\makebox(0,0){$\sst -u_2$}}
\put(143,20){\makebox(0,0){$\sst u_4$}}
\put(145,32){\makebox(0,0){$\sst -u_3$}}
\end{picture}\\ \addlinespace[2mm]
\raisebox{7mm}[-7mm]{$\DSTt{3}_n(r)$} & 
\begin{picture}(180,40)
\put(0,10){\usebox{\bupbline}}
\put(30,10){\usebox{\bdownline}}
\put(65,10){\makebox(0,0){$\sst 0$}}
\put(70,10){\usebox{\bupline}}
\put(100,10){\usebox{\bdownline}}
\put(135,10){\makebox(0,0){$\sst 0$}}
\put(140,10){\usebox{\bupline}}
\put(30,10){\makebox(0,0){$\sst 0$}}
\put(35,10){\usebox{\bupline}}
\put(65,10){\usebox{\bdownline}}
\put(100,10){\makebox(0,0){$\sst 0$}}
\put(105,10){\usebox{\bupline}}
\put(135,10){\usebox{\bdownline}}
\put(170,10){\makebox(0,0){$\sst 0$}}
\put(38,20){\makebox(0,0){$\sst u_1$}}
\put(34,30){\makebox(0,0){$\sst u_0$}}
\put(73,20){\makebox(0,0){$\sst u_2$}}
\put(69,30){\makebox(0,0){$\sst u_1$}}
\put(108,20){\makebox(0,0){$\sst u_3$}}
\put(104,30){\makebox(0,0){$\sst u_2$}}
\put(143,20){\makebox(0,0){$\sst u_4$}}
\put(139,30){\makebox(0,0){$\sst u_3$}}
\put(0,0){\makebox(0,0){$\sst 0$}}
\put(30,0){\makebox(0,0){$\sst n-1$}}
\put(65,0){\makebox(0,0){$\sst 2n-1$}}
\put(100,0){\makebox(0,0){$\sst 3n-1$}}
\put(135,0){\makebox(0,0){$\sst 4n-1$}}
\put(170,0){\makebox(0,0){$\sst 5n-1$}}
\end{picture}\\ \addlinespace[2mm]
\raisebox{7mm}[-7mm]{$\DCTt{4}_n(r)$} &
\begin{picture}(180,40)
\put(0,10){\usebox{\bupbline}}
\put(35,10){\usebox{\bdownline}}
\put(70,10){\usebox{\bupline}}
\put(105,10){\usebox{\bdownline}}
\put(140,10){\usebox{\bupline}}
\put(35,10){\usebox{\bupline}}
\put(70,10){\usebox{\bdownline}}
\put(105,10){\usebox{\bupline}}
\put(140,10){\usebox{\bdownline}}
\put(0,0){\makebox(0,0){$\sst 0$}}
\put(35,0){\makebox(0,0){$\sst n$}}
\put(67,0){\makebox(0,0){$\sst 2n-1,2n$}}
\put(102,0){\makebox(0,0){$\sst 3n-1,3n$}}
\put(137,0){\makebox(0,0){$\sst 4n-1,4n$}}
\put(170,0){\makebox(0,0){$\sst 5n-1$}}
\put(38,20){\makebox(0,0){$\sst u_1$}}
\put(35,32){\makebox(0,0){$\sst -u_0$}}
\put(73,20){\makebox(0,0){$\sst u_2$}}
\put(70,32){\makebox(0,0){$\sst -u_1$}}
\put(108,20){\makebox(0,0){$\sst u_3$}}
\put(105,32){\makebox(0,0){$\sst -u_2$}}
\put(143,20){\makebox(0,0){$\sst u_4$}}
\put(140,32){\makebox(0,0){$\sst -u_3$}}
\end{picture}\\ \addlinespace[2mm]
\raisebox{7mm}[-7mm]{$\DSTt{4}_n(r)$} &
\begin{picture}(180,40)
\put(0,10){\usebox{\bupbline}}
\put(35,10){\usebox{\bdownline}}
\put(70,10){\usebox{\bupline}}
\put(105,10){\usebox{\bdownline}}
\put(140,10){\usebox{\bupline}}
\put(35,10){\usebox{\bupline}}
\put(70,10){\usebox{\bdownline}}
\put(105,10){\usebox{\bupline}}
\put(140,10){\usebox{\bdownline}}
\put(0,0){\makebox(0,0){$\sst 0$}}
\put(35,0){\makebox(0,0){$\sst n$}}
\put(67,0){\makebox(0,0){$\sst 2n-1,2n$}}
\put(102,0){\makebox(0,0){$\sst 3n-1,3n$}}
\put(137,0){\makebox(0,0){$\sst 4n-1,4n$}}
\put(170,0){\makebox(0,0){$\sst 5n-1$}}
\put(38,20){\makebox(0,0){$\sst u_1$}}
\put(37,32){\makebox(0,0){$\sst u_0$}}
\put(73,20){\makebox(0,0){$\sst u_2$}}
\put(72,32){\makebox(0,0){$\sst u_1$}}
\put(108,20){\makebox(0,0){$\sst u_3$}}
\put(107,32){\makebox(0,0){$\sst u_2$}}
\put(143,20){\makebox(0,0){$\sst u_4$}}
\put(142,32){\makebox(0,0){$\sst u_3$}}
\end{picture}\\
\bottomrule
\end{tabular}
\end{table}

\mypar{Signal model: finite skew C-transform} Consider the regular
module $\md=\C[x]/(T_n - \cos r\pi)$ with $r\in\Q$ and $0\leq r\leq
1$.  Let $C\in\{T,U,V,W\}$. The {\em finite skew $C$-transform} is the
mapping
$$
\Phi:\ \coord{s}\mapsto\sum_{0\leq k<n}s_kC_k\in\md,
$$ 
and is a signal model for $V = \C^n$.
As in the finite $C$-transform, the basis in the algebra is,
independent of $C$, the $T$-basis: $\alg = \{h = \sum_{0\leq k<n}h_kT_k\}$.
For $r = 1/2$ or $\cos r\pi = 0$, the skew $C$-transform reduces to its
ordinary counterpart.

\subsection{Spectrum and Fourier Transform}\label{skewdtttransform}

To compute the spectrum and a Fourier transform for the regular
module $\C[x]/(T_n - \cos r\pi)$ with the four different bases, we need
to determine the zeros of $T_n - \cos r\pi$ and fix a proper ordering.

\begin{lemma}\label{skewTzeros} Let $r\in\Q$, $0\leq r\leq 1$. 
We have the factorization
\begin{equation}\label{skewzeros}
T_{n} - \cos r\pi = 2^{n-1}\prod_{0\leq i < n}(x - \cos\ts\frac{r +
2i}{n}\pi),
\end{equation}
which determines the zeros of $T_{n} - \cos r\pi$. We order the zeros
as $\alpha = (\cos r_0\pi,\dots,\cos r_{n-1}\pi)$, such that $0\leq
r_i\leq 1$, and $r_i < r_j$ for $i < j$. The list $\alpha$ is given by
the concatenation
$$
\alpha = \bigcup_{0\leq i < n/2}
(\cos\ts\frac{r+2i}{n}\pi, \cos\frac{2-r+2i}{n}\pi)
$$
for $n$ even, and by
$$
\alpha = \myleft(\bigcup_{0\leq i < \frac{n-1}{2}}
(\cos\ts\frac{r+2i}{n}\pi, \cos\ts\frac{2-r+2i}{n}\pi)\myright)
\cup (\cos\ts\frac{r+n-1}{n}\pi)
$$
for $n$ odd. In the particular case of $r = 1/2$ or $\cos r\pi = 0$,
we thus have $\alpha = (\cos(i + 1/2)\pi/n\mid 0\leq i < n)$ as
in Table~\ref{4cheb}.
\end{lemma}
\begin{proof}
The zeros of $T_{n} - \cos r\pi$ are proved using the closed form of
$T_n$. The ordering of $\alpha$ is shown by inspection. We omit the
details.
\end{proof}
In words, the list $\alpha$ arises from the list $\gamma =
(\cos(r+2i)\pi/n\mid 0\leq i <n)$ in \eqref{skewzeros} by
interleaving the first half of $\gamma$ with the reversed (and reduced
modulo $\pi$) second half of $\gamma$.  

Lemma~\ref{skewTzeros} gives also the spectrum of $\md$, which we will not 
state explicitly. Instead we now formally define the skew DTTs.

\begin{definition}[Skew DTTs]\label{odddtts}
Let $p = T_n-\cos r\pi$, $0\leq r\leq 1$, and $\md = \C[x]/p$ with
basis $b = (C_0,\dots,C_{n-1})$, where $C$ is one of $T,U,V,W$.  Let
$\alpha = (\cos r_i\pi)_{0\leq i < n}$ denote the list of zeros of $p$
in the order specified in Lemma~\ref{skewTzeros}. We denote the
associated polynomial transforms $\poly_{b,\alpha}$ for $\md$ by
$\pDCTt{3}_n(r), \pDSTt{3}_n(r), \pDCTt{4}_n(r), \pDSTt{4}_n(r)$, for
$C = T, U, V, W$, respectively.  Further, we define for each of these four
 $\pDTT(r)$ the associated {\em scaled} polynomial transforms
$$
\DTT_n(r) = \diag_{0\leq i < n}(f(\cos r_i\pi))\cdot\pDTT_n(r),
$$ 
where $f$ is the scaling function associated with (ordinary) $\DTT$
(see Table~\ref{chebmods}). We call these transforms {\em skew
DTTs}. If $r=1/2$, then $\pDTT_n(1/2) =
\pDTT_n$ and $\DTT_n(1/2) = \DTT_n$ in all four cases. 
In the case of the $\DCTt{3}_n(r)
= \pDCTt{3}_n(r)$, we will omit the bar for the skew versions.

Equivalently we can define
\begin{eqnarray*}
\DCTt{3}_n(r) & = & [\cos r_k\ell\pi ]_{0\leq k,\ell < n}, \\
\DSTt{3}_n(r) & = & [\sin r_k(\ell+1)\pi ]_{0\leq k,\ell < n}, \\
\DCTt{4}_n(r) & = & [\cos r_k(\ell+1/2)\pi ]_{0\leq k,\ell < n}, \\
\DSTt{4}_n(r) & = & [\sin r_k(\ell+1/2)\pi ]_{0\leq k,\ell < n}.
\end{eqnarray*}
\end{definition}

As an example, we consider the $\DCTt{4}_3(1/3)$. Using
Lemma~\ref{skewTzeros}, the zeros of $T_3 - \cos(\pi/3) = T_3 - 1/2$
are given by $\alpha = (\cos(\pi/9), \cos(5\pi/9), \cos(7\pi/9))$.
We get
\begin{eqnarray*}
\DCTt{4}_3(1/3) & = & 
\renewcommand{\arraystretch}{1.2}
\begin{bmatrix}
\cos\ts\frac{1}{18}\pi & \cos\ts\frac{1}{6}\pi & \cos\ts\frac{5}{18}\pi \\
\cos\ts\frac{5}{18}\pi & \cos\ts\frac{5}{6}\pi & \cos\ts\frac{11}{18}\pi \\
\cos\ts\frac{7}{18}\pi & \cos\ts\frac{5}{6}\pi & \cos\ts\frac{1}{18}\pi
\end{bmatrix}.
\end{eqnarray*}

\subsection{Diagonalization Property} 

The representation $\phi$ afforded by $\C[x]/(T_n - \cos r\pi)$ with
the four $C$-bases, evaluated at the shift operator $x$, is obtained
from \eqref{T1diag} and \eqref{skewbc}.  Thus, the shift matrix
$\phi(x)$ arises from \eqref{T1diag} by adding in the upper right
corner $\beta(r) = \cos r\pi$ for $\DCTt{3}(r)$, and $\beta(r) = 2\cos
r\pi$ for the other skew transforms. We obtain
\begin{equation}\label{skewT1diag}
\phi(x) = \fh\cdot\left[
\begin{array}{ccccccc}
\beta_1 & 1 & & & & \beta(r) \\
\beta_2 & 0 & 1 \\
0 & 1 & 0 & \cdot \\
 & & 1 & \cdot & 1\\
 & & & \cdot & 0 & \beta_3 \\
 & & & & 1 & \beta_4
\end{array}
\right].
\end{equation}
The values for the $\beta_i$ coincide with the non-skew case given in
Table~\ref{betas}.  As a consequence, in the four cases,
$$
\DTT(r)\cdot\phi(x)\cdot\DTT(r)^{-1} = \diag(\alpha),
$$
with $\alpha$ denoting the zeros of $T_n - \cos r\pi$ provided by
Lemma~\ref{skewTzeros}.

We do not explicitly state a convolution theorem, which can be obtained
from Theorem~\ref{convolution}.

\subsection{Translation into Non-Skew DTTs}\label{transnonskew}

Each of the skew DTTs can be translated into their non-skew
counterpart using a sparse x-shaped matrix. 

\begin{lemma}\label{skewtranslate} Let $\DTT_n(r)$ be a skew DTT.
Then
\begin{eqnarray*}
\DTT_n(r) & = & \DTT_n\cdot X^{(\ast)}_n(r),\quad\text{and}\\
\pDTT_n(r) & = & \pDTT_n\cdot X^{(\ast)}_n(r).
\end{eqnarray*}
Here, $X^{(\ast)}_n(r)$ depends on the DTT and takes the following
forms, indicated by $\ast\in\{C3,S3,C4,S4\}$.
$$
X^{(C3)}_n(r) = 
\left[
\begin{array}{ccccc}
1 & 0 & \cdots & \cdots & 0\\
0 & c_1 & & & s_{n-1} \\
\vdots & & \ddots & \oddots \\
\vdots & & \oddots & \ddots \\
0 & s_1 & & & c_{n-1}
\end{array}
\right],
$$
$$
X^{(S3)}_n(r) = 
\left[
\begin{array}{ccccc}
c_1 & & & -s_{n-1} & 0 \\
& \ddots & \oddots && \vdots \\
& \oddots & \ddots && \vdots \\
-s_1 & & & c_{n-1} & 0 \\
0 & \cdots & \cdots & 0 & c_n
\end{array}
\right],
$$ 
with $c_\ell = \cos (1/2 - r)\ell\pi/n$ and $s_\ell = \sin(1/2 -
r)\ell\pi/n$.
$$
X_n^{(C4)}(r) = 
\left[
\begin{array}{ccccc}
c'_0 & & & s'_{n-1} \\
 & \ddots & \oddots \\
 & \oddots & \ddots \\
s'_0 & & & c'_{n-1}
\end{array}
\right],
$$ 
with $c'_\ell = \cos (1/2 - r)(2\ell+1)\pi/(2n)$ and $s'_\ell =
\sin(1/2 - r)(2\ell+1)\pi/(2n)$. For $\DSTt{4}(r)$, the sines $s'_\ell$
in $X_n^{(C4)}(r)$ are multiplied by $-1$.

In all four cases, if the lines intersect, the numbers are added at
the intersecting position.
\end{lemma}
\begin{proof}
Follows by direct computation, using the definitions of the matrices
and $\cos(x)\cos(y) = (\cos(x+y) + \cos(x-y))/2$.
\end{proof}

The $2\times 2$ blocks in the translation matrices $X_n(r)$ are not
rotations, which implies that the skew DTTs are not ``almost''
orthogonal in the sense of \eqref{almostorth}. However, using
Lemma~\ref{skewtranslate}, we can easily invert skew DTTs by inverting
$X_n(r)$.

\subsection{Translation between Skew DTTs}

All skew $\DTT(r)$ share the same associated module, but different
bases. Thus they can be translated into each other by a base change
similar to the ordinary DTTs in Section~\ref{groupdtt}. As in that
section, we consider the skew DCTs, type 3 and 4 as an example. The
base change matrix $S'_n$ we computed in \eqref{summatrix} did not
depend on the right \bc Thus, the diagram \eqref{dct34diagram}
generalizes, for arbitrary $r$, as
\begin{equation}\label{skewdct34diagram}
\algogen{\C[x]/(T_n-\cos r\pi)}{S'_n}{\C[x]/(T_n-\cos r\pi)}%
  {\DCTt{3}_n(r)}{\pDCTt{4}_n(r)}{\bigdirsum\C[x]/(x-\alpha_k)}
  {\one_n}{\bigdirsum\C[x]/(x-\alpha_k)}
\end{equation}
The first difference occurs when we extend \eqref{skewdct34diagram}
to the non-polynomial $\DCTt{4}_n(r)$, since the scaling diagonal
depends on $r$. Let $\alpha = (\alpha_0,\dots,\alpha_{n-1})$ denote
the zeros of $T_n-\cos r\pi$ and $f$ the scaling function of
$\DCTt{4}$ and let $D_n(r) = \diag_{0\leq k < n}(f(\alpha_k))$. Then 
$$
\algogen{\C[x]/(T_n-\cos r\pi)}{S'_n}{\C[x]/(T_n-\cos r\pi)}%
  {\DCTt{3}_n(r)}{\DCTt{4}_n(r)}{\bigdirsum\C[x]/(x-\alpha_k)}
  {D_n(r)}{\bigdirsum\C[x]/(x-\alpha_k)}
$$
or, as an equation,
\begin{equation}\label{skewdct3and4}
D_n(r)\cdot\DCTt{3}_n(r) = \DCTt{4}_n(r)\cdot S'_n,
\end{equation}
which generalizes \eqref{dct4and3}. 

In Section~\ref{groupdtt}, we continued by inverting (and transposing)
this equation, using the fact that the DTTs are almost orthogonal, to
derive the different relationship \eqref{dct4and3final}, which
requires a smaller number of operations. As mentioned in
Section~\ref{transnonskew}, the skew DCTs are not ``almost''
orthogonal in the sense of \eqref{almostorth}. However, it is still
desirable to invert \eqref{skewdct3and4}, since we will need it later
when we derive fast algorithms \cite{Pueschel:05}. For this purpose
we first define the proper ``inverse'' skew DTTs. The definition is
motivated by and a generalization of the equations
\begin{eqnarray*}
\DCTt{3}_n^{-1} & = & 2/n\cdot\diag(1/2,1,\dots,1)\cdot\DCTt{2}_n\\
\DSTt{3}_n^{-1} & = & 2/n\cdot\diag(1,1,\dots,1/2)\cdot\DSTt{2}_n\\
\end{eqnarray*}
and
$$
\DTT_n^{-1} = n/2\cdot\DTT^T_n = n/2\cdot\DTT_n 
$$
for $\DTT = \DCTt{4},\DSTt{4}$.

\begin{definition}[Inverse Skew DTTs]\label{invskewdctdef}
We define the {\em inverse skew DTTs} by
\begin{eqnarray*}
\iDCT{3}_n(r) & = & n/2\cdot\diag(2,1,\dots,1)\cdot\DCTt{3}_n(r)^{-1},\\
\iDST{3}_n(r) & = & n/2\cdot\diag(1,1,\dots,2)\cdot\DSTt{3}_n(r)^{-1},\\
\iDCT{4}_n(r) & = & n/2\cdot\DCTt{4}_n(r)^{-1},\\
\iDST{3}_n(r) & = & n/2\cdot\DSTt{4}_n(r)^{-1}.
\end{eqnarray*}
Thus, for $r = 1/2$, we have $\iDCT{3}_n(1/2) = \DCTt{2}_n$, $\iDST{3}_n(1/2)
= \DSTt{2}_n$, $\iDCT{4}_n(1/2) = \DCTt{4}_n$, $\iDCT{4}_n(1/2) =
\DCTt{4}_n$.
\end{definition}

Note that Definition~\ref{invskewdctdef} does not provide direct
knowledge about the matrix entries of the $\iDTT$s. These, however,
can be computed using Lemma~\ref{skewtranslate}. For example
\begin{equation}\label{iskewtranslate}
\begin{array}{rcl}
\iDCTt{3}_n(r) & = & \bigl(X^{(C3)}_n(r)\bigr)^{-1}\cdot\DCTt{2}_n,\\
\iDCTt{4}_n(r) & = & \bigl(X^{(C4)}_n(r)\bigr)^{-1}\cdot\DCTt{4}_n,
\end{array}
\end{equation}
and similarly for $\DSTt{3}$ and $\DSTt{4}$.  Note that
$\bigl(X_n^{(\ast)}(r)\bigr)^{-1}$ has in all four cases the same
x-shaped pattern as $X_n(r)$. Namely, the four inverses are derived
from
$$
\begin{bmatrix}
\cos a & \sin b\\
\sin a & \cos b 
\end{bmatrix}^{-1} =
\frac{1}{\cos(a+b)}
\begin{bmatrix}
\cos b & -\sin b\\
-\sin a & \cos a 
\end{bmatrix}.
$$
For example,
\begin{multline*}
\bigl(X^{(C3)}_n(r)\bigr)^{-1} = \isdraft{}{\\}
\frac{1}{\cos(1/2-r)\pi}
\left[
\begin{array}{ccccc}
c_n & 0 & \cdots & \cdots & 0\\
0 & c_{n-1} & & & -s_{n-1} \\
\vdots & & \ddots & \oddots \\
\vdots & & \oddots & \ddots \\
0 & -s_1 & & & c_{1}
\end{array}
\right].
\end{multline*}

Using Definition~\ref{invskewdctdef}, we can now invert
\eqref{dct4and3} to get a generalization of \eqref{dct4and2},
$$
S_n\cdot\iDCTt{3}_n(r)\cdot \frac 12 D_n(r)^{-1} = \iDCTt{4}(r)_n,
$$
where $S_n$ is given in \eqref{Sn}.

\section{Alternative Infinite Space Model}\label{altreal}

In the realization of the space model in \eqref{realspaceop}, we set
the space shift operator to $q = x$, which implied that the space marks were
realized by the Chebyshev polynomials, $t_n = C_n$. This, in turn,
produced a left boundary, and thus the need for left boundary
conditions.  Interestingly, and in contrast to the time model, it is
possible to choose a different realization, which we will develop in
this section. The visualization of this model will match
Figure~\ref{abstractspacegraph}, which seems desirable, but we will
encounter the different problem of collapsing frequency responses.

\subsection{Building the Signal Model}

\mypar{Shift, linear extension, realization}
We start with the same shift, the space shift, and the same linear
extension as in Section~\ref{buildCtrafo}. In that section, the
realization of the space model follows by setting the shift operator $q = x$,
which implied $t_n = C_n$, the Chebyshev polynomials. Now we choose a
different realization, namely by setting the space marks to $t_n =
x^n$ as in the time case. Necessarily, the shift now takes the form
\begin{equation}\label{altspace}
q = \frac{x^{-1} + x}{2}.
\end{equation}
The corresponding $k$-fold space shift can be obtained in two
different ways.  From
\eqref{kspaceshift} we can directly read off that $q_k =
(x^{-k}+x^k)/2$. Or, we use Lemma~\ref{kspaceshiftT} to get $q_k =
T_k(q) = (x^{-k}+x^k)/2$ as can be seen from the power form of $T_k$
in \eqref{powerform} in Appendix~\ref{chebs}. 
Table~\ref{normalalt} contrasts the two different realizations
of the abstract space model.

\begin{table}\centering
\caption{The different realizations in the space model and the
alternative space model.\label{normalalt}}
\ra{1.4}
\begin{tabular}{@{}cccc@{}}\toprule
 & space marks & shift operator & $k$-fold shift operator \\ \midrule
space model & $C_n$ & $x$ & $T_k(x)$\\ \addlinespace[1.0mm]
\ra{1.0}
\begin{tabular}{@{}c@{}}alternative\\space model\end{tabular} &
$x^n$ & $\dps\frac{x^{-1}+x}{2}$ & $\dps\frac{x^{-k}+x^k}{2}$ \\ 
\bottomrule
\end{tabular}
\end{table}

As a result, we get as signal space $\md = \{s =
\sum_{n\in\Z}s_nx^n\}$, and as filter space $\alg =
\{h = \sum_{k\geq 0}h_k(x^{-k}+x^k)/2\}$. Filtering is well-defined
for $\coord{s}\in\ell^2(\Z)$ and $\coord{h}\in\ell^1(\N)$.

\mypar{Signal model}
The alternative infinite space model $(\alg,\md,\Phi)$ is given by the
algebra of symmetric filters $\alg = \{h = \sum_{k\geq
0}h_k(x^{-k}+x^k)/2 \}$ with $\coord{h}\in\ell^1(\N)$ and the signal
module $\md$ consists of Laurent series $s$ with
$\coord{s}\in\ell^2(\Z)$.  The mapping $\Phi$ is, as in the time case,
the $z$-transform:
$$
\coord{s}\mapsto s = S(x) = \sum_{n\in\Z}s_nx^n\in\md.
$$
The important difference to the $z$-transform is the algebra, which is
now smaller: it consists only of the \emph{symmetric} filters.
As an aside, this motivates why a signal
model requires that we explicitly specify the algebra (see
Definition~\ref{sigmoddef}).

\subsection{Spectrum and Fourier Transform}

The impact of a smaller algebra when compared with the time case
becomes evident when we compute the spectrum, which we do next.  Since
the set of operating filters, the symmetric filters, is smaller than
in the time case, every eigenfunction $E_\omega(x)$ in the time case
is also an eigenfunction in the space model under consideration. This
implies that the Fourier transform for $\md$ can be chosen as in the
time case
$$
\Delta:\ s = S(x)\mapsto (S(e^{j\omega}))_{\omega\in [0,2\pi)}.
$$
Accordingly, the frequency response of the filters $h =
H((x^{-1}+x)/2)\in\alg$ at frequency $\omega$, or, in algebraic terms,
the irreducible representation $\phi_\omega$ afforded by the
one-dimensional module generated by $E_\omega(x)$ becomes
$$
\phi_\omega:\ H((x^{-1}+x)/2)\mapsto H(\cos j\omega).
$$
This shows that pairs of conjugate frequencies $\omega,
\overline{\omega}$ afford the same representation 
$$
\phi_\omega = \phi_{\overline{\omega}},
$$ 
i.e., produce the same frequency response. As we explain later,
this property of collapsing frequency responses carries over to the
finite-dimensional case, where it may serve as an explanation why
certain transforms work in practice better than others on ``space
signals,'' e.g., images.  Further, it shows that there is a larger
degree of freedom in choosing a Fourier transform, since in any of the
two-dimensional eigenspaces spanned by $E_\omega(x),
E_{\overline{\omega}}(x)$ we can choose {\em any} basis.

\subsection{Visualization}

The visualization of the associated signal model is given in
Figure~\ref{abstractspacegraph}, i.e., no boundary is intrinsic to
this model and thus it seems more natural at first glance. However,
this model has the different problem of collapsing frequency responses
as explained in the previous section.

\subsection{Remarks}

As a summary, we observe that different realizations of the same
abstract model may be possible, which motivates the concept of
realization.  Different realizations may have different properties and
shortcomings.  For example, in the case just studied, we trade the
need for a left boundary by the collapsing of conjugate
frequencies. We have no proof that these two realizations of the space
model we considered are the only ones possible. However, regardless of
the chosen realization, the Chebyshev polynomials  come into
play as a consequence of Lemma~\ref{kspaceshiftT}.

\section{Alternative Finite Space Model: RDFTs and DHTs}\label{altrealfin}

In Section~\ref{altreal}, we discussed an alternative realization of
the infinite space model. In contrast to the original realization of
the infinite space model presented in Section~\ref{Ctrafo}, the
alternative realization does not require a left boundary, but has the
counterpart of collapsing frequencies, i.e., the spectrum consists of
two-dimensional eigenspaces. As a consequence, the Fourier transform
was no longer uniquely determined since in each of these
two-dimensional spaces any basis can be chosen.

In this section, we discuss this alternative realization briefly in the
finite-dimensional case and derive the corresponding transforms. We
will see that, similar to the infinite case, there will be the problem
of collapsing frequencies and thus a larger degree of freedom in
choosing the Fourier transform. This degree of freedom interestingly
leads, for different reasons, practically to the same class of
transforms that arose as real Fourier transforms for the finite time
model in Section~\ref{realdfts}.

\subsection{Building the Signal Model}

\mypar{Shift, linear extension, realization} The alternative
realization of the signal model in Section~\ref{altreal} led, as in
the time case, to the $z$-transform.  Thus, the corresponding finite
signal modules also coincide with the finite time case, namely, they
are given by $\C[x]/p(x)$ with basis $b =
(x^0,x^1,\dots,x^{n-1})$. The question of the monomial signal
extension in this case has already been settled in Lemma~\ref{monse},
namely by requiring $p(x) = x^n - a$, where $a\in\C$, $a\neq 0$, is
any complex constant.  As an aside, this emphasizes that the signal
extension is a property of the module (including the chosen basis)
only, and independent of the chosen operating algebra.

The difference to its time counterpart lies in the operating algebra
of filters; the modules are in both cases equal, namely $\md =
\C[x]/(x^n-a)$. In the time model, this module was regular, $\alg =
\md$, since $\alg$ contained the time shift operator $q = x$, which generates
the entire algebra $\alg$. In the alternative space model, however,
the shift operator is given by $q = (x^{-1}+x)/2$; see \eqref{altspace}. The
algebra for the space model may now, as in the infinite case, be
smaller than $\C[x]/(x^n-a)$, namely consisting only of those filters
that are expressible as a polynomial in $q$. In algebraic terms, the
algebra for the alternative finite space model is the subalgebra of
$\C[x]/(x^n-a)$ generated by $q$:
$$
\alg = \langle(x^{-1}+x)/2\rangle\leq\C[x]/(x^n-a).
$$ 
Note that $x^{-1}$ is well-defined in $\C[x]/(x^n-a)$ (since $a\neq
0$) and given by $x^{-1} = x^{n-1}/a$. The question now is whether $\alg$
is equal to or smaller than $\C[x]/(x^n-a)$, in particular, for the most
interesting case $a=1$. As we show next, the answer depends on $a$,
and, in the case where $\alg$ is smaller, it is an algebra belonging to
a DCT.

\begin{lemma}[Algebras for the alternative finite space model]\label{altfin}
The subalgebra $\alg = \langle (x^{-1}+x)/2\rangle$ of $\C[x]/(x^n-a)$
is equal to $\C[x]/(x^n-a)$ if and only if $a\neq\pm 1$. If $a =\pm 1$,
then the structure of $\alg$ is given by:
$$
\alg\cong
\begin{cases}
\C[x]/(T_m-T_{m-2}),\ m = \frac n2+1, & n\text{ even, }a=1\\
\C[x]/T_m,\           m = \frac n2+1, & n\text{ even, }a=-1\\
\C[x]/(T_m-T_{m-1}),\ m=\frac{n+1}{2}, & n\text{ odd, }a=1\\
\C[x]/(T_m+T_{m-1}),\ m=\frac{n+1}{2}, & n\text{ odd, }a=-1
\end{cases},
$$
which are the algebras associated to the DCTs of types 1,3,5,7,
respectively. The number $m$ is their respective dimension.
\end{lemma}
\begin{proof}
A straightforward way to compute the algebra is to consider the
sequence of powers $1,q,q^2,\dots$, and to determine when they become
linearly dependent. We choose a different way by considering the
sequence of $k$-fold space shift operators (the natural basis in the space
model) $T_k(q) = (x^{-k}+x^k)/2$, $k\geq 0$, and focus on the case
where $n$ is even. Using $x^{-k} = x^{n-k}/a$, we get the following
$n$ elements in $\C[x]/(x^n-a)$:
\begin{gather*}
1,\frac{\frac 1a x^{n-1} + x}{2},\frac{\frac 1a x^{n-2} + x^2}{2},
\dots,\\
\frac{\frac 1a x^{n/2+1} + x^{n/2-1}}{2},
\frac{1+\frac 1a}{2}x^{n/2},
\frac{\frac 1a x^{n/2-1} + x^{n/2+1}}{2},\\
\dots,
\frac{\frac 1a x + x^{n-1}}{2}.
\end{gather*}
The next element would be $(1/a+a)/2$, which makes the set certainly
linearly dependent (also due to the length exceeding $n$).
The question is when this set is linearly independent. We observe that
the powers $x^k$ and $x^{n-k}$ always occur together. Thus the set is linearly
dependent if and only if there is an $\alpha\in\C$ such that for any $k$,
\begin{eqnarray*}
\frac 1a x^{n-k} + x^k & = & \alpha(x^k + \frac 1a x^{n-k})\\
\Leftrightarrow\quad\qquad\qquad a & = & \pm 1.
\end{eqnarray*}
Conversely, if $a = \pm 1$, then only the first $n/2+1$ polynomials
$T_0(q),T_1(q),\dots,T_{n/2}(q)$ are linearly independent. We focus on
the case $a = 1$ in which $T_{n/2+1}(q) = T_{n/2-1}(q)$. This shows
that
$$
\alg\cong\C[q]/(T_m(q) - T_{m-2}(q)),\quad m = n/2+1,
$$
as desired. From Table~\ref{chebmods}, we see that this algebra is 
associated to the $\DCTt{1}_m$.
The cases $a= -1$ and $n$ odd are derived analogously.
\end{proof}

\mypar{Signal model}
In summary, we obtain a signal model for $V = \C^n$ given by
$$
\Phi:\ \coord{s}\mapsto\sum_{0\leq k<n}s_kx^k\in\md,
$$ 
with module $\md = \C[x]/(x^n - a)$, and $\alg$ is given by
Lemma~\ref{altfin}. The natural basis in $\alg$ consists of the $k$-fold 
space shift operators $T_k((x^{-1}+x)/2) = (x^{-k}+x^k)/2$.

\subsection{Spectrum and Fourier transform}\label{altdtttransform}

Lemma~\ref{altfin} shows that for $a = \pm 1$, the algebra $\alg$ is
of smaller dimension than the module, namely $n/2 + 1$ (or
$(n+1)/2$). Because of Lemma~\ref{allirreds}, $\alg$ has only $n/2 +
1$ different irreducible representations. Thus, as in the infinite
case, we have the problem of collapsing frequencies: the $n$ spectral
components (eigenspaces) of $\md$ produce only $n/2+1$ (or $(n+1)/2$)
different frequency responses (irreducible representations). More
precisely, focusing on $a = 1$, $\md$ decomposes as in the time case
(regular case) as
\begin{equation*}
\C[x]/(x^n-1)\rightarrow 
  \C[x]/(x-\omega_n^0)\dirsum\dots\dirsum\C[x]/(x-\omega_n^{n-1}).
\end{equation*}
We set $\md_k = \C[x]/(x-\omega_n^k$). The irreducible modules
$\md_k$ are mutually different (as vector spaces), but their frequency
responses are not. Namely, because of
$$
(x^{-1}+x)/2\cdot x^0\equiv 
  \cos 2k\pi/n\mod(x-\omega_n^k)
$$ 
the spectral components $\md_k$ and $\md_{n-k}$ afford the same
irreducible representation $\phi_k = \phi_{n-k}$, i.e., produce the
same frequency response. In other words, conjugate frequencies in the
(regular) time model collapse to the same frequency if $\md$ is viewed
only as an $\alg$-module, $\alg < \C[x]/(x^n-1)$, and not as a regular
module.

As a consequence of the previous section, there is a larger degree of
freedom in choosing a Fourier transform for the $\alg$-module $\md$.
In the time model, the generic Fourier transform for the regular module
$\md$ was given by
$$
\four = D\cdot\DFT_n,
$$ 
where $D$ was the degree of freedom: any invertible diagonal
determined by the choice of bases in the irreducible modules
$\md_k$. In the present case, the degree of freedom is in choosing
bases in the two-dimensional spaces $\md_k\dirsum\md_{n-k}$, which
afford the same frequency response, and the remaining one-dimensional
spaces $\md_0$ and $\md_{n/2}$ (if $n$ is even). Regarding the Fourier
transform, this leads to a similar situation as in
Section~\ref{realdfts}, but for different reasons.  In
Section~\ref{realdfts}, we reduced the base field, which caused pairs
of conjugate spectral components to fuse to real spectral components
of dimension 2. In the present case, the base field is still $\C$, but
the smaller algebra causes pairs of conjugate frequency responses to
become equal, i.e., the spectral components still have all dimension
1, but afford the same representation.

As a result, the generic Fourier transform is given by any matrix of the form
\begin{equation}\label{altspacefour}
\four = X\cdot\DFT_n,
\end{equation}
where $X$ is any invertible matrix of the x-shaped form
\eqref{xshape}.  The difference between~\eqref{altspacefour} and
\eqref{realdft} is that every matrix in \eqref{realdft} is one in
\eqref{altspacefour} but not vice-versa, since the matrix
in~\eqref{realdft} has to be real valued. As a consequence, the RDFT
and the DHT, defined in Section~\ref{realdfts}, are Fourier transforms
for the finite alternative space model.

\subsection{Visualization}

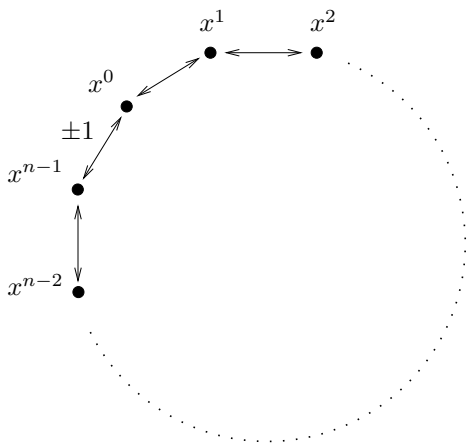
\begin{figure}[ht]\centering
\vspace*{3mm}
\begin{picture}(0,0)
\put(-10,60){\makebox(0,0){$x^{n-2}$}}
\put(-10,103){\makebox(0,0){$x^{n-1}$}}
\put(15,137){\makebox(0,0){$x^0$}}
\put(57,160){\makebox(0,0){$x^1$}}
\put(99,160){\makebox(0,0){$x^2$}}
\put(6,118){\makebox(0,0){$\pm 1$}}
\end{picture}
\includegraphics[scale=0.65]{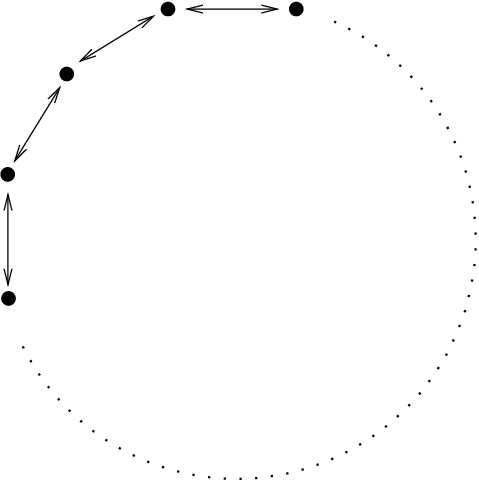}
\caption{Visualization of the alternative space model for $a=\pm 1$.
  \label{aspacetimegraph}}
\end{figure}

We visualize the alternative finite space model for $a = \pm 1$ using
Definition~\ref{visdef} by the graph in Figure~\ref{aspacetimegraph}
that has the shift matrix \eqref{aspacemat} below as adjacency
matrix. The graph is similar to Figure~\ref{fintimegraph} but is
undirected, i.e., all edges are now in both directions. If $a\neq\pm 1$, 
then the edge from $x^{n-1}$ to $x^0$ has weight $a$, while the reverse
edge has weight $1/a$.

\subsection{Diagonalization Properties}

Let $\phi$ be the representation afforded by the $\alg$-module $\md =
\C[x]/(x^n-1)$ and consider the shift operator $q =
(x^{-1}+x)/2$. It is
\begin{equation}\label{aspacemat}
\phi(q) = 
\fh\cdot\left[
\begin{array}{ccccccc}
0 & 1 & & & & a \\
1 & 0 & 1 \\
0 & 1 & 0 & \cdot \\
 & & 1 & \cdot & 1\\
 & & & \cdot & 0 & 1 \\
\frac 1a & & & & 1 & 0
\end{array}
\right],
\end{equation}
which should be compared to \eqref{T1diag}; the inner structure is the
same, determined by the space shift, but the boundary conditions are
different. Further, $\phi(q)$ and thus the signal model $\Phi$, is
symmetric (see Definition~\ref{symsm}) if and only if $a = \pm 1$.

From the above discussion, we know already that, for $a = \pm 1$,
$\phi(q)$ has duplicate eigenvalues in contrast to the matrix in
\eqref{T1diag}. In particular, for $a = 1$, we get the diagonalization
property
$$
\four\phi(q)\four^{-1} = 
  \diag(1,\cos\ts\frac{2\pi}{n},\cos\ts\frac{4\pi}{n},\dots,\dots,
    \cos\ts\frac{4\pi}{n},\cos\ts\frac{2\pi}{n}),
$$ 
where the value in the middle of the diagonal depends on whether
$n$ is even or odd. In this equation, $\four$ can be chosen, e.g., as
$\DFT_n,\RDFT_n,\DHT_n$.  The generic matrix diagonalized by this
transform is given by $\phi(h)$, $h\in\alg$. The structure of
$\phi(h)$ for $a=1$ is symmetric circulant.

As a summary, the RDFT and the DHT diagonalize symmetric circulant
matrices and ``almost'' diagonalize (i.e., reduce to x-shape)
circulant matrices, shown in \eqref{rdftdiag}.

\section{The Generic Nearest Neighbor Model}\label{genshiftmodel}

Signal models based on the time shift operator \eqref{opq} and, as we
have shown, the space shift operator \eqref{spaceop} are widely used
in signal processing, reflected by the common use of the DFT and the
DTTs. After recognizing that the concept of a shift is by no means
restricted to the time shift used in classical signal processing
theory, it is natural to start exploring other shift operators, their
associated infinite and finite signal models, and their use in signal
processing. In this section we briefly discuss a generalization of the
space shift: the generic nearest neighbor (GNN) shift\footnote{We
could call it the Markov shift for obvious reasons and call the
resulting model a Markov model. Since this term is already taken, we
refrain from doing so.  We will discuss the relationship between the
GNN model and Gauss-Markov random fields in
Section~\ref{randomfield}.}. Based on this shift, we first derive the
infinite and the finite signal model, following the same steps as used
before in the derivation of the time and space models.


\subsection{Building the Infinite GNN Signal Model}

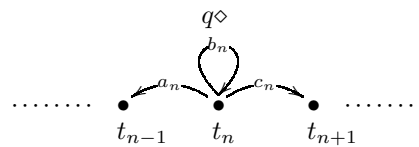
\begin{figure}[ht]\centering
\begin{picture}(100,45)
\multiput(-40,10)(4,0){8}{$\cdot$}
\put(0,10){$\bullet$}
\put(0,0){$t_{n-1}$}
\put(20,18){\makebox(0,0){$\xy\ar @/^-1ex/ (-10,0)|{a_n}\endxy$}}
\put(36,10){$\bullet$}
\put(38.5,28){\makebox(0,0){$\xy\ar @(ul,ur)|{b_n}\endxy$}}
\put(37,45){\makebox(0,0){$q\diamond$}}
\put(36,0){$t_{n}$}
\put(56,18){\makebox(0,0){$\xy\ar @/^1ex/ (10,0)|{c_n}\endxy$}}
\put(72,10){$\bullet$}
\put(72,0){$t_{n+1}$}
\multiput(86,10)(4,0){8}{$\cdot$}
\end{picture}
\caption{The generic nearest neighbor shift $q\diamond t_n$.\label{genshift}}
\end{figure}

\mypar{Definition of the GNN shift} As before, we denote the abstract
shift operator by $q$ and define the generic nearest neighbor shift by
\begin{equation}\label{gshift}
q\diamond t_n = a_nt_{n-1} + b_nt_n + c_nt_{n+1},
\end{equation}
which is depicted in Figure~\ref{genshift}. 

We require that $a_n,c_n\neq 0$. As usual, the $t_n$ denotes ``space
marks''\footnote{The word ``space'' is used in lack of a better
term; the model is only ``space'' in the sense that the shift
connects to both neighbors.}. Note that, in contrast to the time and
space shift, \eqref{gshift} is {\em variant} since the coefficients
depend on $n$.

In this generic case, it is not clear how to define a $k$-fold GNN
shift. A natural choice would be a polynomial $p_{k,n}$ of degree $k$
such that $p_{k,n}(q)t_n$ is a linear combination of only $t_{n-k}$
and $t_{n+k}$ (the marks at distance $k$ from $t_n$). However, this is
not possible in the general case as can be shown by direct computation
(compute $t_n, q\diamond t_n, q^2\diamond t_n, q^3\diamond t_n$; it is
not possible to linearly combine them so that the result has only
$t_{n-k}$ and $t_{n+k}$ as summands).

\mypar{Linear extension} As usual, we linearly extend the operation of
$q$ to the entire set $\md = \{s = \sum_{n\in\Z}s_nt_n\}$. Since we do
not have a notion of $k$-fold GNN shifts, it is not clear which basis
to choose in the associated algebra of filters.

\mypar{Realization: orthogonal polynomials} We realize the model by
setting $q = x$, which leads to the recurrence
\begin{equation}\label{orthgen}
P_{n+1} = \ts\frac{x-b_n}{c_n}P_n - \frac{a_n}{c_n} P_{n-1},
\end{equation}
which also motivates the requirement $c_n\neq 0$.
Normalizing $P_0 = 1$ and $\deg(P_1) = 1$ defines a sequence of
polynomials orthogonal with respect to some moment functional
(Favard's theorem), \cite[p. 21]{Chihara:78}. Conversely, every
sequence of orthogonal polynomials obeys a recurrence of the form
\eqref{orthgen}
\cite{Chihara:78}. The recursion \eqref{orthgen} can be run in the 
negative direction (since we required $a_n\neq 0$) to obtain the $P_n$
for $n < 0$. Every such $P_n$ is again a polynomial and can thus be
expressed as a linear combination in the $P_n$, $n\geq 0$.  As a
consequence, we consider only right-sided sequences, and $\md = \{s=
\sum_{n\geq 0}s_nP_n\}$. Consequently, $\alg$ also contains only 
right-sided sequences.

The sequence $P_n$, $n<0$ is the left signal extension, and expressing
$P_{-1}$ in $P_0,P_1$ is the left \bc

\mypar{Signal model}
As a result, we obtain an infinite discrete signal model
for a vector space $V\subset\C^\N$ given by the {\em $P$-transform}
$$
\Phi:\ \coord{s} = (s_0,s_1,\dots)\mapsto\sum_{k\geq 0}s_kP_k.
$$
Choosing $V = \ell^1(\N)$ ensures that $\Phi(\coord{s})$ defines a
function on the interval of orthogonality of the $P_n$. How much $V$
can be enlarged depends on $P$, see \cite{Kashin:89}. The algebra of
filters is a set of right sided series, but as said above, it is not
clear which basis polynomials to choose.

Finally, we conjecture that it is not possible to establish monomial
left \bc~unless $P_k\in\{T_k,U_k,V_k,W_k\}$.  Note that $P_k = x^k$
cannot be included here since we required $a_n\neq 0$ in the
recurrence \eqref{orthgen}.

We did not pursue the question of the exact form of the Fourier
transform, and we also omit the visualization of the $P$-transform.

\subsection{Building the Finite GNN Signal Model}

\mypar{Shift, linear extension, realization} As before a straightforward
realization leads to the set of all polynomials
$$
\sum_{0\leq k<n} s_kP_k,
$$
which is the vector space $\C_n[x]$, but not a module, since
$xP_{n-1}\not\in\C_n[x]$. To solve this problem we need a right \bc

\mypar{Boundary condition and signal extension}
We can choose any right \bc $P_n = \sum_{0\leq <k}\beta_kP_k$, but the relevant
question is which one is desirable. We cannot expect to establish a
monomial signal extension as in the special case of the Chebyshev
polynomials.

We argue that a natural choice is the \bc $P_n = 0$, since it achieves
three crucial properties.
\begin{itemize}
\item $P_n$ is separable, a general property of orthogonal polynomials
\cite[p.~44]{Szegoe:67}, which ensures one-dimensional spectral components.
\item The associated Fourier transform is ``almost'' orthogonal.
\item The Fourier transform has a fast algorithm.
\end{itemize}
The last two points are discussed in greater detail below.

In summary, we obtain the regular module $\md = \C[x]/P_n(x)$ with
$P$-basis.

\mypar{Signal model}
The corresponding signal model is given by
the {\em finite $P$-transform}
$$
\Phi:\ \coord{s}\mapsto\sum_{0\leq k<n}s_kP_k\in\C[x]/P_n(x),
$$ 
where $\md = \C[x]/P_n(x)$  is regular, i.e., $\alg = \md$.
We call this signal model also the {\em finite GNN model}.

\subsection{Spectrum and Fourier Transform}\label{dpt}

We obtain the (polynomial) Fourier transform for the regular module
$\C[x]/P_n(x)$ with basis $b = (P_0,\dots,P_{n-1})$ as a special case
of \eqref{polyalgdec} and Theorem~\ref{fourtrafo}.  We assume that
$\alpha = (\alpha_0,\dots,\alpha_{n-1})$ are the (mutually distinct as
mentioned above) zeros of $P_n$.  Thus, the Fourier transform is given
by
\begin{equation}
\begin{array}{rrcl}
\Delta: & \C[x]/P_n(x) & \rightarrow & 
  \bigdirsum_{0\leq k <n}\C[x]/(x-\alpha_k),\\
& s = s(x) & \mapsto & (s(\alpha_0),\dots,s(\alpha_{n-1})).
\end{array}
\end{equation}
In matrix form, we get the polynomial Fourier transform
\begin{equation}\label{almorthP}
\four = \poly_{b,\alpha} = [P_\ell(\alpha_l)]_{0\leq k,\ell<n}.
\end{equation}

\subsection{Diagonalization Properties} 

If $\phi$ is the representation of $\alg$ afforded by $\md$, then
$\four$ diagonalizes any matrix $\phi(h)$, $h\in\alg$. In particular,
the shift matrix $\phi(x)$ is tridiagonal:
\begin{equation}\label{gnnshiftrep}
\phi(x) =
\left[
\begin{array}{ccccccc}
b_0 & a_1 & & & & \\
c_0 & b_1 & a_2 \\
 & c_1 & b_2 & \cdot \\
 & & c_2 & \cdot & a_{n-2}\\
 & & & \cdot & b_{n-2} & a_{n-1} \\
 & & & & c_{n-2} & b_{n-1}
\end{array}
\right],
\end{equation}
where $b_0,c_0$ depend on the left \bc, or equivalently on $P_1$. 
Thus,
$$
\four\phi(x)\four^{-1} = \diag(\alpha_0,\dots,\alpha_{n-1}).
$$
A convolution theorem follows directly from Theorem~\ref{convolution}
and is not explicitly stated.

\subsection{Orthogonal Transform}\label{orthgnn}

The (polynomial) Fourier transform $\four$ for the regular module $\md
= \C[x]/P_n$ with basis $b = (P_0,\dots,P_{n-1})$ in \eqref{almorthP}
is almost orthogonal in the sense that there exist diagonal matrices
$D,E$ such that $D\four E$ is orthogonal. We showed this general
result before in Theorem~\ref{orthpolytrafo} and used it to derive the
matrices $D,E$ to obtain the orthogonal DTTs.  We further showed that
if a DTT is a Fourier transform for a signal model $\Phi$, then the
corresponding orthogonal DTT is a Fourier transform for a slightly
adjusted signal model $\Phi'$ (Theorem~\ref{symsmdtt}), which arises
from $\Phi$ by a scaling of the module basis, i.e., a base change
with a diagonal matrix $E$.

The derivation for the DTTs in Section~\ref{orthodtts} transfers
directly to the finite GNN model.  Let $\four$ be the Fourier
transform in \eqref{almorthP} for the finite GNN model
$(\alg,\md,\Phi)$ with shift matrix $\phi(x)$ in \eqref{gnnshiftrep}.
Further, let $D,E$ be diagonal matrices computed using the
Christoffel-Darboux formula (Theorem~\ref{cd}), to make
$$
\four' = D\four E^{-1}
$$ 
orthogonal ($E^{-1}$ is chosen to obtain the form in
equation~\eqref{nowortho}). Then, $E\phi(x)E^{-1}$ is
symmetric. Further, if $E = \diag(e_0,\dots,e_{n-1})$, then $\four'$
is a Fourier transform for the signal model $(\alg,\md,\Phi')$, with
$$
\Phi':\ \coord{s}\mapsto\sum_{0\leq k<n}s_k\frac{1}{e_k}P_k.
$$ 
This model is symmetric (see Definition~\ref{symsm}), since, if
$\phi'$ is its afforded representation, then the matrix $\phi'(x)$,
and thus all matrices $\phi'(h)$, $h\in\alg$, are symmetric.

As we mentioned before, this general construction of orthogonal
transforms from orthogonal polynomials has been proposed in signal
processing by \cite{Yemini:79}, where it was called Gauss-Jacobi
procedure and the resulting transforms Gauss-Jacobi transforms.  Our
theory identifies these transforms as Fourier transforms for 
finite GNN signal models and provides the associated filters, spectrum,
boundary conditions, signal extensions, and other concepts and
properties.

A few special cases, corresponding to special choices of orthogonal
polynomials, have been proposed in the signal processing
literature including for the Laguerre polynomials \cite{Mandyam:96},
for the Hermite polynomials \cite{Martens:90}, and for the so-called
discrete orthogonal polynomials
\cite{Aburdene:94,Haddad:88,Mukundan:01}.

With our algebraic theory, we get immediately all important concepts
associated to these transform including an understanding of the
underlying signal model, the notion of filtering, and several
properties of the transforms.

\subsection{Fast Algorithms}\label{fastgnn}

Although fast algorithms are not the subject of this paper, and will
be considered in~\cite{Pueschel:05}, we want to mention that every
polynomial transform constructed as above from orthogonal polynomials
has a fast algorithm that allows its computation using only
$O(n\log^2(n))$ operations (see \cite{Driscoll:97}, or the numerically
more stable version in \cite{Potts:98}). These algorithms are based on
the three-term recurrence characteristic for orthogonal
polynomials. Note that this cost is slightly worse than for the
special case of the DTTs, which are known to have a complexity of
$\Theta(n\log(n))$. This implies that DTT algorithms are due to
special properties of the Chebyshev polynomials, a fact that we will
confirm in a different paper, \cite{Pueschel:05}, in which we extend our
algebraic theory to the derivation and discovery of fast algorithms.

\section{Overview of Finite Signal Models}\label{ow}

In Table~\ref{overviewsm} we list all the finite signal models, and
their associated Fourier transforms, that we introduced in this paper.
The table is divided into complex time models, real time models,
complex/real space models (in contrast to time, for space the
restriction to a real base field does not change the spectrum or the
Fourier transform), and complex/real GNN models.  In each row, we list
in the first three columns the signal model as ($\Phi,\md,\alg$), in
the fourth column the associated unique polynomial Fourier transform,
and in the fifth column possibly other relevant Fourier transforms for
the model. Note that the notion of polynomial transform only exists
for regular modules of separable polynomial algebras; thus, the real
time models and the alternative space models have no polynomials
transforms (in the table indicated as n.a.).

\mypar{Orthogonal transforms} Each of the listed transforms has an
orthogonal counterpart, which is in each case obtained by proper
scaling of rows or columns.  In some cases, this scaling requires
an adjustment (namely a scaling of the module basis) of the signal
model. This is the case for certain DCTs/DSTs
(Section~\ref{orthodtts}) and for the GNN transforms
(Section~\ref{orthgnn}) based on general orthogonal polynomials.

\mypar{Discrete trigonometric transforms}
In this paper, we have used the term discrete trigonometric transforms
(DTTs) to denote the 16 DCTs and DSTs. In the literature, the DTTs are
often considered as the entire class of transforms whose entries are
expressible using cosines and sines, which includes also the DFT, DHT,
and RDFT. To our best knowledge, Table~\ref{overviewsm} contains all
1-D trigonometric transforms (in this sense) that have been introduced in
the literature, and extends this class by the RDFTs of types 2--4, the
polynomial DCTs and DSTs, and the four types of skew DCTs and DSTs.
Further, we suggest to rename the (rarely occurring) DWTs of type 1--4
to be called DHTs of type 1--4.

The Fourier transforms for the GNN model are not trigonometric
transforms and should not be considered as such.

\newcommand{\als}[0]{\addlinespace[4mm]}
\newcommand{\myh}[0]{\hspace{11mm}}
\ra{1.2}
\begin{table*}\centering
\caption{Overview of all finite signal models and associated Fourier
transforms discussed in this paper. All these Fourier transforms
(except those for the GNN model) are trigonometric
transforms.\label{overviewsm}}
\isdraft{\renewcommand{\baselinestretch}{1}\small}{}
{
\begin{tabular}{@{}l@{\myh}l@{\myh}l@{\myh}l@{\myh}l@{}}
%
\multicolumn{5}{@{}l@{}}{\bf Time (complex): complex finite z-transform 
  \hfill Section~\ref{dftvars}}\\ \toprule
$\Phi$ & $\md$ & $\alg$ & $\four = \poly_{b,\alpha}$ & other $\four$ 
  \\ \midrule
$\coord{s}\mapsto\sum s_kx^k$ & $\C[x]/(x^n-a)$ & regular & 
  $\DFT_n\cdot D$ & \da \\
 & $\C[x]/(x^n-1)$ & regular & $\DFT_n = \DFTt{1}_n$ & $\DFTt{2}_n$ \\
 & $\C[x]/(x^n+1)$ & regular & $\DFTt{3}_n$ & $\DFTt{4}_n$ \\
\midrule \als
%
\multicolumn{5}{@{}l@{}}{\bf Time (real): real finite z-transform 
  \hfill Section~\ref{realdfts}}\\ \toprule
$\Phi$ & $\md$ & $\alg$ & $\four = \poly_{b,\alpha}$ & other $\four$
  \\ \midrule
$\coord{s}\mapsto\sum s_kx^k$ & $\R[x]/(x^n-1)$ & regular & n.a.
 & $\RDFT_n = \RDFTt{1}_n$ \\
 & $\R[x]/(x^n-1)$ & regular  & n.a. & $\RDFTt{2}_n$ \\
 & $\R[x]/(x^n-1)$ & regular & n.a. & $\DHT_n = \DHTt{1}_n\ (\DWTt{1}_n)$ \\
 & $\R[x]/(x^n-1)$ & regular & n.a. & $\DHTt{2}_n\ (\DWTt{2}_n)$ \\
 & $\R[x]/(x^n+1)$ & regular & n.a. & $\RDFTt{3}_n$ \\
 & $\R[x]/(x^n+1)$ & regular & n.a. & $\RDFTt{4}_n$ \\
 & $\R[x]/(x^n+1)$ & regular & n.a. & $\DHTt{3}_n\ (\DWTt{3}_n)$ \\
 & $\R[x]/(x^n+1)$ & regular & n.a. & $\DHTt{4}_n\ (\DWTt{4}_n)$ \\ 
\midrule\als
%
\multicolumn{5}{@{}l@{}}{\bf Space (complex/real): 
  finite C-transform (C = T,U,V,W)\hfill 
  Sections~\ref{dtttransform}, \ref{skewdtttransform},
  \ref{altdtttransform}}\\ \toprule
$\Phi$ & $\md$ & $\alg$ & $\four = \poly_{b,\alpha}$ & other $\four$ 
  \\ \midrule
$\coord{s}\mapsto \sum s_kT_k$ & $\C[x]/(x^2-1)U_{n-2}$ & regular & 
  $\DCTt{1}_n = \pDCTt{1}_n$ & \da \\
 & $\C[x]/T_n$ & regular & 
  $\DCTt{3}_n = \pDCTt{3}_n$ & \da \\
 & $\C[x]/(x-1)W_{n-1}$ & regular & 
  $\DCTt{5}_n = \pDCTt{5}_n$ & \da \\
 & $\C[x]/(x+1)V_{n-1}$ & regular & 
  $\DCTt{7}_n = \pDCTt{7}_n$ & \da \\
 & $\C[x]/(T_n-\cos r\pi)$ & regular & 
  $\DCTt{3}_n(r) = \pDCTt{3}_n(r)$ & \da \\ \midrule
$\coord{s}\mapsto \sum s_kU_k$ & $\C[x]/T_n$ & regular & 
  $\pDSTt{3}_n$ & $\DSTt{3}_n$ \\
 & $\C[x]/U_n$ & regular & 
  $\pDSTt{1}_n$ & $\DSTt{1}_n$ \\
 & $\C[x]/V_n$ & regular & 
  $\pDCTt{7}_n$ & $\DCTt{7}_n$ \\
 & $\C[x]/W_n$ & regular & 
  $\pDSTt{5}_n$ & $\DSTt{5}_n$ \\
 & $\C[x]/(T_n-\cos r\pi)$ & regular & 
  $\pDSTt{3}(r)_n$ & $\DSTt{3}(r)_n$ \\ \midrule
$\coord{s}\mapsto \sum s_kV_k$ & $\C[x]/(x-1)W_{n-1}$ & regular & 
  $\pDCTt{6}_n$ & $\DCTt{6}_n$ \\
 & $\C[x]/V_n$ & regular & 
  $\pDCTt{8}_n$ & $\DCTt{8}_n$ \\
 & $\C[x]/(x-1)U_{n-1}$ & regular & 
  $\pDCTt{2}_n$ & $\DCTt{2}_n$ \\
 & $\C[x]/T_n$ & regular & 
  $\pDCTt{4}_n$ & $\DCTt{4}_n$ \\
 & $\C[x]/(T_n-\cos r\pi)$ & regular & 
  $\pDCTt{4}(r)_n$ & $\DCTt{4}(r)_n$ \\ \midrule
$\coord{s}\mapsto \sum s_kW_k$ & $\C[x]/(x+1)V_{n-1}$ & regular & 
  $\pDSTt{8}_n$ & $\DSTt{8}_n$ \\
 & $\C[x]/W_n$ & regular & 
  $\pDSTt{6}_n$ & $\DSTt{6}_n$ \\
 & $\C[x]/T_n$ & regular & 
  $\pDCTt{4}_n$ & $\DCTt{4}_n$ \\
 & $\C[x]/(x+1)U_{n-1}$ & regular & 
  $\pDSTt{2}_n$ & $\DSTt{2}_n$ \\
 & $\C[x]/(T_n-\cos r\pi)$ & regular & 
  $\pDSTt{4}(r)_n$ & $\DSTt{4}(r)_n$ \\ \midrule
$\coord{s}\mapsto \sum s_kx^k$ & $\C[x]/(x^n-1)$ & 
  $\langle (x^{-1}+x)/2\rangle$ & n.a. & $\RDFT_n=\RDFTt{1}_n$ \\
 & $\C[x]/(x^n-1)$ & $\langle (x^{-1}+x)/2\rangle$ & n.a. & $\RDFTt{2}_n$ \\
 & $\C[x]/(x^n-1)$ & $\langle (x^{-1}+x)/2\rangle$ & n.a. & $\DHT_n = \DHTt{1}_n$ \\
 & $\C[x]/(x^n-1)$ & $\langle (x^{-1}+x)/2\rangle$ & n.a. & $\DHTt{2}_n$ \\
 & $\C[x]/(x^n+1)$ & 
  $\langle (x^{-1}+x)/2\rangle$ & n.a. & $\RDFTt{3}_n$ \\
 & $\C[x]/(x^n+1)$ & 
  $\langle (x^{-1}+x)/2\rangle$ & n.a. & $\RDFTt{4}_n$ \\
 & $\C[x]/(x^n+1)$ & 
  $\langle (x^{-1}+x)/2\rangle$ & n.a. & $\DHTt{3}_n$ \\
 & $\C[x]/(x^n+1)$ & 
  $\langle (x^{-1}+x)/2\rangle$ & n.a. & $\DHTt{4}_n$ \\ \midrule\als
%
\multicolumn{5}{@{}l@{}}{\bf GNN (complex/real): 
  finite P-transform (P orthogonal polynomials)
  \hfill Section~\ref{dpt}}\\ \toprule
$\Phi$ & $\md$ & $\alg$ & $\four = \poly_{b,\alpha}$ & other $\four$
  \\ \midrule
$\coord{s}\mapsto \sum s_kP_k$ & $\C[x]/P_n$ & regular & 
no specific name  & \da \\ \midrule
\end{tabular}
}
\end{table*}

\section{Higher-Dimensional Signal Models}\label{higherdim}

In Section~\ref{shiftcomm}, we identified the equivalence (under some
weak assumptions) of signal models with shift-invariant systems (or
filters) and commutative algebras $\alg$.  More specifically, we
asserted that if $\alg$ consists exclusively of FIR filters, then
$\alg$ is necessarily a polynomial algebra, i.e., of the form
\begin{equation}\label{polyagain}
\alg = \C[\overline{x}]/
  \langle p_1(\overline{x}),\dots,p_k(\overline{x})\rangle
\end{equation}
where $\overline{x} = (x_1,\dots,x_m)$ and the $p_i$ are polynomials
in $n$ variables. Usually, the generators of this algebra are chosen
as $x_1,\dots,x_n$, which are the shift operators in this algebra.  The whole
discussion in this paper focused on the case of only one shift operator
$x=x_1$, which corresponds to signal models for sampled 1-D signals.
In this case, $\alg = \C[x]/p(x)$, which, as a regular module, provides
the underlying signal model for most 1-D transforms including the DFT
and the DTTs.


In signal processing, $m$-D signals are usually processed with $m$-D
versions of 1-D transforms. In matrix form, if $\four$ is some 1-D
transform, such as the DFT or a DCT, its $m$-D counterpart is simply the
$m$-fold tensor (or Kronecker) product of $\four$ with itself,
\begin{equation}\label{tens}
\four\tensor\dots\tensor \four.
\end{equation}
The resulting transform is called ``separable,'' since it 
operates independently along the different dimensions, according to
the formula
\begin{eqnarray*}
\four\tensor\dots\tensor \four & = &
  \phantom{\cdot}(\four\tensor\one_{n^{m-1}}) \\
&& \cdot(\one_n\tensor \four\tensor\one_{n^{m-2}}) \\
&& \dots \\
&& \cdot(\one_{n^{m-1}}\tensor \four).
\end{eqnarray*}
This also shows that the $m$-D version of $\four$ can be computed using
$mn^{m-1}$ many 1-D $\four$'s.
The question is: what is the underlying signal model, i.e., how to
represent $m$-D transforms in our algebraic theory? Clearly, the signal
model is shift-invariant, which implies that the algebra has to be of
the form \eqref{polyagain}. 

In this section, we explain that algebras for separable signal models
are a special case of \eqref{polyagain}, and that the algebraic
construction of a separable $m$-D signal model from its 1-D
counterpart is a natural one. We focus our discussion on the 2-D case;
the $m$-D case is completely analogous.

\mypar{Separable module and algebra}
Let $\md_x=\C[x]/p(x)$ be a regular module with basis $b_x =
(p_0(x),\dots,p_{n-1}(x))$.  The $x$ in the subscript is the name of
the variable or shift operator; we mention this since we will create copies with
different variable names.  Then, a corresponding regular module in two
variables can be constructed as
\begin{equation}\label{sep}
\md_{x,y} = \alg_{x,y} = \C[x,y]/\langle p(x), p(y)\rangle.
\end{equation}
Note that this is a special case of \eqref{polyagain} for $m = 2$,
and, more important, the two polynomials $p(x),p(y)$ depend on
different variables; this is equivalent to the separability. The
natural basis $b_{x,y}$ for $\md_{x,y}$ is the cross product of the basis $b_x$
with $b_y$, namely
\begin{multline}\label{bxy}
b_{x,y} = b_x\times b_y \isdraft{}{\\}
= 
\begin{array}{l}
(p_0(x)p_0(y), p_0(x)p_1(y),\dots,p_0(x)p_{n-1}(y), \\
p_1(x)p_0(y), p_1(x)p_1(y),\dots,p_1(x)p_{n-1}(y), \\
\dots\dots \\
p_{n-1}(x)p_0(y), p_{n-1}(x)p_1(y),\dots,p_{n-1}(x)p_{n-1}(y)),
\end{array}
\end{multline}
which shows that the dimensions multiply, i.e., $\md_{x,y}$ has
dimension $n\cdot n = n^2$. Omitting one of the polynomials in
\eqref{sep}, say $p(y)$, would make the module infinite-dimensional,
since it would now contain any polynomial in $y$.  Note that two
shift operators operate on the above basis: $x$ and $y$, the generators of
$\alg_{x,y}$.

\mypar{Signal model}
The regular module $\md_{x,y}$ provides a signal model for $\C^{n^2}$,
whose elements we assume to be arranged in a two-dimensional 
array $\coord{s} = (s_{k,\ell})$:
$$
\Phi:\ \coord{s}\mapsto \sum_{0\leq k<n \atop 0\leq\ell <n}
  s_{k,\ell}p_kp_\ell.
$$

\mypar{Decomposition into a tensor product}
Instead of working with \eqref{sep}, it is more convenient to exhibit 
the separable structure of $\md_{x,y}$, which is encapsulated algebraically
by decomposing the module into a {\em tensor product}:
$$
\md_{x,y} = \md_x\tensor\md_y,
$$ 
with associated algebra $\alg_{x,y} = \alg_x\tensor\alg_y$.  The
elements of a tensor product of vector spaces are generated by its
natural basis $b_{x,y}$ in \eqref{bxy}. Intuitively, the tensor
decomposition decouples the two variables and makes explicit the
separability. Since the tensor product is a ``natural'' construction
in algebra, most relevant properties are also ``naturally'' derived
from the 1-D counterpart.

\mypar{Spectrum and Fourier transform} 
Let $p(x) = \prod_{0\leq k<n}(x-\alpha_k)$. The
Fourier transform of the regular module $\md_x\times\md_y$ is given by
\isdraft{%
\begin{equation}\label{2ddec}
\Delta:\ \C[x]/p(x)\times\C[y]/p(y)\rightarrow
  \bigdirsum_{0\leq k_1<n \atop 0\leq k_2<n}
  \C[x]/(x-\alpha_{k_1})\times\C[y]/(y-\alpha_{k_2}),
\end{equation}
}{%
\begin{multline}\label{2ddec}
\Delta:\ \C[x]/p(x)\times\C[y]/p(y)\rightarrow \\
  \bigdirsum_{0\leq k_1<n \atop 0\leq k_2<n}
  \C[x]/(x-\alpha_{k_1})\times\C[y]/(y-\alpha_{k_2}),
\end{multline}
}
and for every signal the spectrum is computed as
$$
s=s(x,y)\mapsto\Delta(s) = 
  (s(\alpha_k,\alpha_\ell))_{0\leq k,\ell<n},
$$
using $s(x,y)\equiv s(\alpha_k,\alpha_\ell)\text{mod}
\langle x-\alpha_k, x-\alpha_\ell\rangle$.

Note that the irreducible modules in \eqref{2ddec} have all dimension
$1\cdot 1 = 1$.  If $\four$ is a Fourier transform for $\md_x$ with
basis $b_x$, then $\four\tensor\four$ is a Fourier transform for
$\md_x\times\md_y$ with basis $b_x\times b_y$.

\mypar{Representation} We evaluate the representation $\phi_{x,y}$
afforded by the regular module $\md_{x,y}$ with basis $b_{x,y}$ at the
shift operators $x$ and $y$. The operation of $x$ on $b_x$ yields the matrix
$\phi_x(x)$. Thus, from the special form of $b_{x,y}$ in \eqref{bxy},
we obtain
$$
\phi_{x,y}(x) = \phi_x(x)\tensor\one_n,
$$
and, analogously,
$$
\phi_{x,y}(y) = \one_n\tensor\phi_y(y).
$$
As usual, and obvious in this case, $\four\tensor\four$ diagonalizes
both matrices.

\begin{figure}\centering
\includegraphics[scale=0.6]{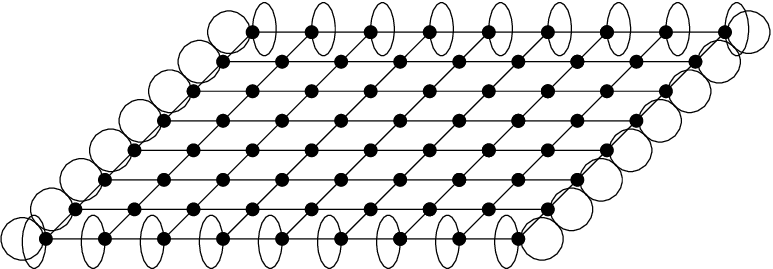}
\caption{Visualization of the 2-D space model corresponding to the 
  DCT, type~2.\label{2ddctgraph}}
\end{figure}

\mypar{Visualization} A visualization of the separable signal model is
obtained by constructing the graph with adjacency matrix
$\phi_{x,y}(x) + \phi_{x,y}(y) = \phi_{x,y}(x+y)$, which is exactly
the direct product of the graph for the 1-D model given by the
adjacency matrix $\phi_x(x)$ with itself.  For example, for the model
underlying the two-dimensional DCT of type 2 and size 8, which is used
in the JPEG image compression standard, we obtain this way
Figure~\ref{2ddctgraph} from the second graph in
Figure~\ref{finspacegraph}. All horizontal lines are incurred by the
$x$-shift, all vertical lines by the $y$-shift. The loops at the boundary 
visualize the symmetric \bc's of the DCT, type 2.

For the DFT, the direct product of the directed circle graph
(Figure~\ref{fintimegraph}) with itself is a directed torus.

\mypar{Summary on separability}
As a summary, it is worth to emphasize again that the 
separable construction of $m$-D signal models is quite natural, given the
equivalence of the following concepts:
\begin{center}
\begin{tabular}{rl}
& tensor product of algebras/regular modules \\
$\leftrightarrow$ & tensor product of Fourier transforms \\
$\leftrightarrow$ & direct product of visualizing graphs.
\end{tabular}
\end{center}

\mypar{Non-separable signal models} Separable signal models are
commonly used in signal processing for two main reasons:
\begin{itemize}
\item They produce models for square grids, which is by far 
the most common format for $m$-D signals such as images.
\item The associated Fourier transforms and its fast algorithms 
are immediately available from the 1-D counterparts.
\end{itemize}

An interesting question is whether there are any interesting
non-separable signal models. The above discussion shows that they
would likely model a non-square grid. Indeed, using the algebraic
theory, reference~\cite{Pueschel:04a} derives a non-separable 2-D transform, called discrete
triangle transform, that operates on a regular triangular grid. Further,   also using algebraic
methods, it can be shown that this transform possesses fast algorithms
\cite{Pueschel:04b}. A more detailed paper is in preparation.

\mypar{Remarks} In the case of one variable, or shift operator, we have argued
before that the algebra has always the form $\C[x]/p(x)$. This was due
to $\C[x]$ being a Euclidean ring, which allows us to always reduce
$\C[x]/\langle p_1(x),p_2(x)\rangle$ to $\C[x]/p(x)$, with $p =
\gcd(p_1,p_2)$. In the $m$-D case \eqref{polyagain} with more than one
variable $m>1$, this is in general no longer the case, i.e., the
number $k$ of polynomials ``factored out'' in \eqref{polyagain} can be
arbitrarily large. If $k < m$, then the dimension of $\alg$ is
infinite; if $k\geq m$ the dimension depends on the choice of he
$p_i$.  The investigation of these multivariate polynomial algebras is
the subject of the theory of varieties \cite{Cox:97}.

\section{Markov Models, Graphs, Matrices, and Signal Models}
\label{mc}

The construction of the time model, the space model, and the generic
nearest neighbor (GNN) model in Sections~\ref{ztrafo}, \ref{Ctrafo},
and
\ref{genshiftmodel}, respectively, consisted of two essential steps:
\begin{enumerate}
\item The introduction of (time/space) marks $t_n$, the abstract shift
operator $q$, and its operation on the $t_n$.
\item Extension to linear combinations of marks on the one hand to
form the signal module, and to linear combinations of multiple shift operators
on the other hand to form the filter algebra.
\end{enumerate}
The second step is a common mathematical procedure and the effect, of
course, is that we obtain a {\em linear} signal model. The question
remains to identify the mathematical nature of the first step, and we
will give a partial answer by relating this step, under certain
assumptions, to the definition of a discrete Markov chain. Afterwards,
we will discuss the general relationship between the concepts of
finite-state Markov chains, weighted graphs, signal models, and square
matrices.

\subsection{Discrete Markov Chains}\label{dmc}

We start with the definition of a discrete Markov chain, following
\cite{Romanovsky:70}.  

\mypar{Definition} We assume a system $S$ of a discrete number of
mutually exclusive states, abstractly denoted by $t_n$, $n\in
I\subseteq \Z$. At an initial instant of time, $k = 0$, the uncertainty
of the state is described by a random variable $X_0$, whose
distribution $(a_{0,n})_{n\in I}$ describes the probability of $S$
being in state $t_n$, $n\in I$. The system is now observed at regular
time instances $k\in\N$ and the state distribution evolves according
to a fixed stochastic rule. Namely, if at time $k$, the system is
in state $t_n$, then the probability that at time $k+1$ the system is
in state $t_m$ is $p_{m,n}$, independent of the states at previous
time instances $0,\dots,k-1$:
$$
p_{m,n} = \prob(X_{k+1}=t_m|X_k=t_n).
$$ 
For simplicity, we assume that for any $n$ there are only finitely
many $p_{m,n}\neq 0$.  Under these conditions the sequence
$X_0,X_1,\dots$ is called a {\em Markov chain with discrete state
space and of discrete time}. Further, we require that $p_{m,n}$ is
independent of $k$, which means that the chain is {\em homogeneous}.
We always assume discrete time and homogeneity and just say {\em
discrete-time Markov chain}.  If the set of states $t_n$, i.e,
$I\subseteq Z$ is finite, we call it  a {\em finite Markov chain}.

A discrete Markov chain is completely described by its initial 
distribution and its  transition probability matrix
$$
Q = [p_{m,n}]_{m,n\in I}.
$$
If the distribution of $X_k$ is the row vector $a_k = (a_{k,n})_{n\in I}$, 
then the distribution of $X_{k+1}$ is $a_{k+1}=a_kQ$.  

The matrix $Q$ is {\em stochastic}, i.e., all entries are non-negative,
all column sums are equal to 1, and no row contains only zeros.

\mypar{Markov chains and signal models} We relate a signal model as
constructed in Sections~\ref{ztrafo}, \ref{Ctrafo}, and \ref{genshiftmodel}
to a discrete Markov chain by relating the (time/space) marks to the 
states and the shift matrix to the  transition probability matrix $Q$.
%
\begin{table}[ht]\centering
\ra{1.2}
\begin{tabular}{@{}ll@{}}\toprule
signal model & discrete Markov chain \\ \midrule
(time/space) marks $t_n$ & states $t_n$ \\
matrix $\phi(q)$ for shift operator q & probability transition matrix Q\\
\bottomrule
\end{tabular}
\end{table}

For example, in the infinite discrete-time model
(Figure~\ref{timeshift}), $\phi(q)$ is an infinite matrix with 1's on
the lower diagonal and zeros elsewhere, and is thus stochastic, i.e., a
candidate $Q$. The states are the time marks $t_n$. Thus, the discrete
time model is a Markov chain 
 with
the following intuitive interpretation: if the discrete time is at
time $t_n$ (state $t_n$), then, independently of the past, it is with
probability 1 in the next observation at time $t_{n+1}$.  

Similarly, the matrix $\phi(q)$ corresponding to the space model
(Figure~\ref{spaceshift}) is stochastic, but this time the transition
from a location $t_n$ in space is with probability 1/2 to $t_{n-1}$ or
$t_{n+1}$, respectively. This also motivates why we defined the shift
with the scaling 1/2 (even though omitting the scaling would have led
to an equivalent, modified version of Chebyshev polynomials), namely,
to obtain a stochastic matrix.

In the generic next neighbor shift (Figure~\ref{genshift}), the
situation is slightly different. To interpret the model as a Markov
chain, it is necessary, that $a_n,b_n,c_n\geq 0$. In this case, we can
then normalize by dividing the $n$th column of $\phi(q)$ by
$a_n+b_n+c_n$ to make the matrix stochastic.

We have seen that the realization may modify a model, as in the case
of the infinite space model given by the $C$-transform, see
Figure~\ref{spacegraph}. For the $W$-transform, negative numbers occur
in $\phi(q)$, which destroys the connection to Markov chains in the
above sense. In the other cases, after possible normalization, we
obtain again a discrete Markov chain.

Similar observations hold for {\em finite} signal models, which
correspond to finite Markov chains (finite number of states).  For
example, among the shifts corresponding to the 16 DTTs, negative
numbers occur for the last row and the last column of
Table~\ref{chebmods}, as can be seen from \eqref{T1diag} and
Table~\ref{betas}. The other 9 DTTs correspond to signal models that
can be interpreted as Markov chains.

\mypar{Conclusion} The above discussion is somewhat philosophical, but
the connection is still rigorous in a mathematical sense. Note that a
discrete Markov chain does not imply linearity, which means a
non-linear signal model could possibly be built from it.

Also we want to note that the above Markov chain interpretation is
different from a stochastic process or random field.  In a random
field, random variables describe uncertainty in the {\em signal} and
they take values in $\R$ or $\C$. In the above interpretation, the
random variables explain the uncertainty in the {\em location} (in time or
space) and they take values $t_n$. In other words, random fields model
the signal values $s_n$, whereas in the discussion in this section Markov chains model the signal
indices $n$. Under certain assumptions to be discussed in
Section~\ref{sec:gmrf}, random fields are equivalent to signal models.

\subsection{Markov Models, Graphs, Matrices, and Polynomial Algebras}

Above we discussed the relationship between signal models and discrete
Markov chains. Besides this relation, we repeatedly visualized signal
models using graphs (using Definition~\ref{visdef}) or by the shift
matrix $\phi(x)$, i.e., the coordinate matrix for the respective
shift.  Since $x$ generates $\alg$, it contains most if not all
information about the signal model.

We investigate in this section the exact connection between these
four concepts. We focus on the case of a finite shift-invariant signal
model provided by a module $\C[x]/p(x)$. The four concepts are:
\begin{itemize}
\item a square matrix $A\in\C^{n\times n}$;
\item a weighted graph with $n$ vertices;
\item a finite Markov chain with $n$ states (see Section~\ref{dmc});
\item a shift-invariant signal model $(\alg,\md,\Phi)$ with
$\md = \C[x]/p(x)$.
\end{itemize}

The relationship is displayed in Figure~\ref{allrel} and discussed
next. Every solid arrow in Figure~\ref{allrel} signifies that one can
translate the source concept into the target concept. The dashed
arrows mean that this translation depends on a
condition, which is then given next to this arrow.
As central concept, we use the square matrix $A$.

\newcommand{\twolines}[2]{\begin{tabular}{cc}#1\\#2\end{tabular}}
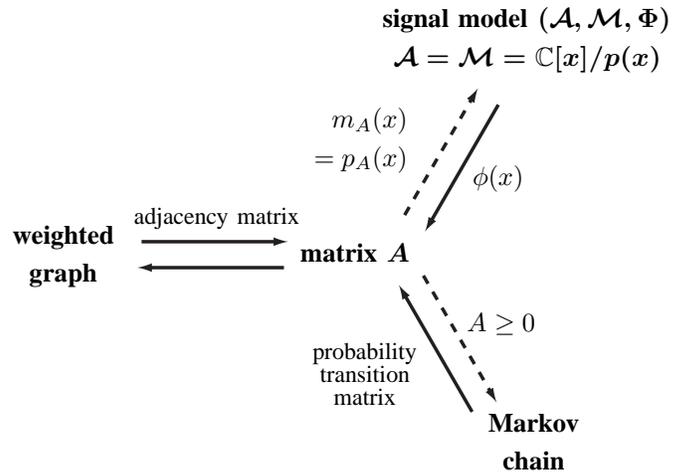
\begin{figure}\centering
\isdraft{\renewcommand{\baselinestretch}{1}\small}{}
\begin{picture}(250,170)
\put(127,73){\makebox(0,0){\includegraphics[scale=0.7]{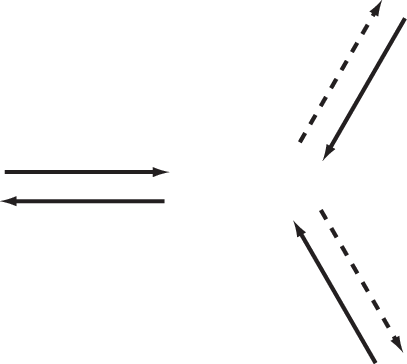}}}
\put(5,71){\makebox(0,0)[l]{\twolines{\bf weighted}{\bf graph}}}
\put(120,72){\makebox(0,0)[l]{\bf matrix \boldmath$A$}}
\put(185,0){\makebox(0,0)[l]{\twolines{\bf Markov}{\bf chain}}}
\put(145,153){\makebox(0,0)[l]{\twolines%
  {\bf signal model \boldmath$(\alg,\md,\Phi)$}
  {\bf \boldmath$\alg=\md=\C[x]/p(x)$}}}
\put(120,115){%
  \makebox(0,0)[l]{$\begin{array}{rr}m_A(x)\\=p_A(x)\end{array}$}}
\put(183,45){\makebox(0,0)[l]{$A\geq 0$}}
\put(115,25){\makebox(0,0)[l]{%
  \renewcommand{\arraystretch}{0.7}
  \begin{tabular}{cc}\small probability\\
    \small transition\\ \small matrix\end{tabular}}}
\put(57,85){\makebox(0,0)[l]{\small adjacency matrix}}
\put(185,100){\makebox(0,0)[l]{$\phi(x)$}}
\end{picture}
\vspace*{5pt}
\caption{The connection between square matrices $A$, weighted graphs, 
regular signal models, and finite Markov chains.\label{allrel}}
\end{figure}

\mypar{Matrix and weighted graph}
Every weighted graph uniquely determines an adjacency matrix. Conversely,
every matrix $A$ can be viewed as the adjacency matrix of a weighted 
graph.

\mypar{Matrix and Markov chain} As discussed in the previous Section~\ref{dmc},
every Markov chain uniquely determines a probability transition
matrix. Conversely, a given matrix $A$ can be viewed as a probability
transition matrix only if $A\geq 0$ (all entries non-negative) and no
row is equal to zero. In this case the columns can be scaled to have sum~1 to yield a stochastic matrix.

\mypar{Matrix and regular signal model} Given a regular signal model
with $\md = \alg = \C[x]/p(x)$, $\md$ uniquely determines the shift
matrix $A = \phi(x)$. The converse is the subject of the following lemma.

\begin{lemma}\label{matsm}
Let $A\in\C^{n\times n}$. There exists a polynomial $p(x)$ of degree $n$
and a basis of $\C[x]/p(x)$ such that $\phi(x) = A$ if and only if the
minimal polynomial $m_A(x)$ and the characteristic polynomial $p_A(x)$
of $A$ coincide:
$$
m_A(x) = p_A(x).
$$
In this case, $p(x) = p_A(x)$.
\end{lemma}
\begin{proof}
Let $A\in\C^{n\times n}$ and let $m_A(x) = p_A(x)$. In this case $A$
is similar to the companion matrix $C$ of $p_A(x)=x^n+\sum_{0\leq
i<n}\beta_ix^i$ (see \cite[p.~187, Theorem~4]{Gantmacher:59}), which
is shown in \eqref{companion}. In other words, $A = TCT^{-1}$ for a
suitable invertible matrix $T$. Defining $p = p_A$ and $b =
(1,x,\dots,x^{n-1})$ as basis in $\md = \C[x]/p(x)$, we have $\phi(x)
= C$. A base change in $\md$ with $T$ yields the result.

For the converse, let $A\in\C^{n\times n}$ and assume that $m_A(x)
\neq p_A(x)$ and that $A = \phi(x)$ for some module $\md = \C[x]/p(x)$ and
basis $b$. Let $J$ be the Jordan normal form of $A$, $J = TAT^{-1}$.
If $\phi'$ is the representation afforded by $\md$ with basis $b'=Tb$,
then $J = \phi'(x)$. Since $m_A\neq p_A$,
$A$ has eigenspace of at least dimension 2, say, for eigenvalue
$\lambda$. Locating these in $J$, there are $p_k,p_\ell\in b'$, $p_k\neq
p_\ell$, such that
\begin{eqnarray*}
&& xp_k = \lambda p_k\text{ and }xp_\ell = \lambda p_\ell\\
&\Leftrightarrow & (x-\lambda)p_k\equiv(x-\lambda)p_\ell\equiv 0
  \text{ mod }p\\
&\Leftrightarrow & (x-\lambda)p_k = (x-\lambda)p_\ell = p\\
&\Leftrightarrow & p_k = p_\ell
\end{eqnarray*}
which is a contradiction.

Finally, if $\alg = \C[x]/p(x)$ and $\phi(x) = A$, then, since $\phi$
is a homomorphism of algebras, $0 = \phi(p(x)) = p(\phi(x)) = p(A)$.
Thus $m_A|p$ and $m_A = p_A$  yields $p = p_A$.
\end{proof}

For example, Lemma~\ref{altfin} shows that for $a=1$ and an even size
$n$, the matrix \eqref{aspacemat} has the minimal polynomial
$T_{n/2+1}-T_{n/2-1}$ of degree $n/2+1<n$. Thus, we cannot realize the
matrix as a $\phi(x)$. And indeed, in the alternative space model, it
is realized as $\phi((x^{-1}+x)/2)$, where $\phi$ is the finite time
model, which is regular. However, this means we can still create a
signal model that has \eqref{aspacemat} as a shift matrix, but the
algebra is {\em smaller} than the module, i.e., the module or the model
is not regular.

\mypar{Matrix and non-regular signal model} The previous discussion
opens a new question: namely, given a matrix $A$, under which
condition is it possible to realize $A = \phi(q(x))$, where $\phi$ is
the regular representation of a suitably chosen regular module $\md =
\C[x]/p(x)$ with basis $b$. This relaxed requirement (when compared to
Lemma~\ref{matsm}) allows now for non-regular modules, i.e., those
with an algebra {\em smaller} than the module; namely, the algebra is
now a subalgebra of $\C[x]/p(x)$ generated by a suitable element
$q(x)$.

To answer this question, we first have to understand, under which
condition a polynomial $q(x)\in\alg$ generates a proper subalgebra of
$\alg = \C[x]/p(x)$. We only consider the case of a separable $p$.

\begin{lemma}\label{subalgebra}
Let $p(x)$ be a separable polynomial with zeros $\alpha =
(\alpha_0,\dots,\alpha_{n-1})$, $\alg=\C[x]/p(x)$, and let
$q(x)\in\alg$.  Then 
 $q(x)$ generates the algebra~$\alg$,
 if and only if $q$, viewed as a mapping on
$\alpha$, is injective. In other words, if $q(\alpha_i)\neq
q(\alpha_j)$ for $i\neq j$.
\end{lemma}
\begin{proof}
The key to the proof is to apply the CRT to $\alg$ and investigate
the subalgebra generated by $q$ in the decomposed domain.
The CRT yields
$$
\alg = \C[x]/p(x)\cong\bigdirsum_{0\leq k<n}\C[x]/(x-\alpha_k)
$$ 
(now viewed as an isomorphism of algebras). The subalgebra $\langle
q(x)\rangle$ generated by $q(x)$ is equal to $\alg$ if and only if
$x\in\langle q(x)\rangle$, i.e., if and only if there is a polynomial
$r(x)$ such that $r(q(x))\equiv x\text{ mod }p(x)$.  In the decomposed
domain, this condition translates into the existence of a polynomial
$r(x)$ that maps the list $(q(\alpha_0),\dots,q(\alpha_{n-1})$ onto
the list $\alpha = (\alpha_0,\dots,\alpha_{n-1})$. This is possible if
and only if $q$ is injective on $\alpha$.
\end{proof}

Back to the previous example, from Lemma~\ref{altfin}, we have $\alg =
\C[x]/(x^n-1)$, and $q(x) = (x^{-1} + x)/2 = (x^{n-1}+x)/2$.  Since
$q(x)$ maps conjugate zeros of $x^n-1$ to the same value (namely to
their real part):
$$
q(\omega_n^k) = q(\omega_n^{n-k}) = \cos(2k\pi/n),
$$
$q$ generates a proper subalgebra.  This mapping to
the same value corresponds precisely to the effect of collapsing
frequency responses discussed in Section~\ref{altdtttransform}, which
can now be stated more precisely.  Namely, if a regular signal model
$\alg = \md = \C[x]/p(x)$ is given, where $p$ is separable with list
of zeros $\alpha$, and if the algebra of filters is reduced to the
subalgebra generated by $q(x)$, then the frequency responses collapse
precisely as determined by the mapping
$$
(\alpha_0,\dots,\alpha_{n-1})\mapsto
  (q(\alpha_0),\dots,q(\alpha_{n-1}).
$$

Before we extend Lemma~\ref{matsm}, we need another auxiliary lemma.
\begin{lemma}\label{genrep}
Let $A\in\C^{n\times n}$ with minimal polynomial $m_A(x)$.
Then
$$
\begin{array}{rcl}
\phi:\ \C[x]/m_A(x) & \rightarrow & \C^{n\times n},\\
x & \mapsto & A.
\end{array}
$$
is a representation of $\alg$, i.e., a homomorphism of algebras.
\end{lemma}
\begin{proof}
Follows from $m_A(A) = 0$.
\end{proof}
This lemma provides the algebra $\alg$ for any matrix $A$ selected as
shift matrix. The remaining question is under which conditions there exists
a module of the form $\md = \C[x]/p(x)$ that affords the
representation in Lemma~\ref{genrep}?  This is answered in the next
lemma, again only considering a separable $p$.

\begin{lemma}\label{diagsm}
Let $A\in\C^{n\times n}$. There exists a module $\md = \C[x]/p(x)$
with separable $p$ and with regular representation $\phi$ (w.r.t.~a
suitably chosen basis $b$) such that $\phi(q(x)) = A$, if and only if
$A$ is diagonalizable.

In other words, every diagonalizable matrix can become a shift matrix,
where the shift generates a subalgebra $\alg = \langle
q(x)\rangle<\C[x]/p(x)$. The module affording the representation 
is $\md = \C[x]/p(x)$.

In this case
$$
\alg\cong\C[x]/m_A(x),
$$
where $m_A$ is the minimal polynomial of $A$.
\end{lemma}
\begin{proof}
First assume a regular module $\md=\C[x]/p(x)$ is given with separable
$p$ and basis $b$ and afforded representation $\phi$.  Then $\phi(x)$
and thus $\phi(q(x))$ is diagonalized by any Fourier transform for
$\md$ (Theorem~\ref{diagprop}).

For the converse assume $A\in\C^{n\times n}$ is diagonalizable with
list of eigenvalues $\beta=(\beta_0,\dots,\beta_{n-1})$, which may
contain duplicates. Consider an arbitrary regular module $\C[x]/p(x)$
with separable $p$, basis $b$, and afforded representation $\phi$. Let
$\alpha=(\alpha_0,\dots,\alpha_{n-1})$ be the list of zeros of $p$.
Since $\alpha$ has no duplicates, there is a polynomial $q$ of degree
at most $n-1$ with $q(\alpha_k)=\beta_k$. Thus $\phi(q(x)) =
q(\phi(x))$ is similar to $A$: $B\phi(q(x))B^{-1} = A$ with a suitable
invertible matrix $B$. A base change in $\md$ yields the result.

The second assertion follows from Lemma~\ref{genrep}.
\end{proof}
The proof is constructive and shows the large degree of freedom in
choosing the module: any $\C[x]/p(x)$ with separable polynomial $p$.

\section{Gauss-Markov Random Fields and Signal Models}
\label{randomfield}

Signal processing usually adopts either a deterministic point of view
with linear transforms and filtering as the basic processing tools or
a stochastic perspective where signals are modeled as stochastic
processes or random fields. We have argued elsewhere~\cite{Moura:98}
that there is an intimate connection between these two seemingly
disparate approaches. This connection establishes that linear
transforms like the discrete Fourier transform, the discrete cosine
transform, or the discrete sine transform are the Karhunen-Lo\`eve
transforms~(KLT) associated with a discrete time (or causal) or
discrete space (or noncausal) random field. In this section we show
how this connection is articulated by the algebraic theory of signal
processing: how the signal model in Definition~\ref{sigmoddef} relates
to Gauss-Markov random fields~(GMRF) and how Fourier transforms, as
the generic diagonalizing transforms in the algebraic approach, relate
to the KLT of suitably defined random fields.  We start by introducing
Gauss-Markov random fields and KLTs in Section~\ref{sec:gmrf}. Then we
consider two examples, again modeling time and space, that show
similarities and differences between the algebraic and the stochastic
signal model. In Section~\ref{sec:gmrfandsignalmodel}, we extend this
connection to generic results connecting the stochastic and the
algebraic approaches. We focus on \emph{finite}, real-valued signal
models.

\subsection{Gauss-Markov Random Fields}\label{sec:gmrf}

\mypar{Random fields} Consider the~$n$ random variables
$s_0,\dots,s_{n-1}$ that follow the difference equation
\begin{multline}\label{gmrf}
s_k = a_{k,k-m}s_{k-m} + \dots + a_{k,k}s_k + \isdraft{}{\\}
\dots + a_{k,k+m}s_{k+m} + \nu_k,
\end{multline}
for $0\leq k<n$. We call $\{s_k\}_{0\leq k<n}$ a \emph{noncausal} or
\emph{acausal} random field, or random field for short. The
$a_{k,\ell}\in\R$ are the field parameters and the $\nu_k$ are random
variables representing the error or noise in the representation. The
noise $\nu_k$ is assumed here to be zero mean Gauss, which, because of
the linearity in~\eqref{gmrf}, makes $\{s_k\}_{0\leq k<n}$ a Gauss
random field. The noise and signal covariances will be specified
below. Each signal value~$s_k$ depends on at most $m$ right and left
neighbors up to the uncertainty in $\nu_k$.  If, for all $k$,
$a_{k,k+1}=\dots=a_{k,k+m}=0$, and the noise is white, then the random
field is \emph{causal} and usually called a {\em process} instead of a
field.  If in \eqref{gmrf}, the $a_{k,\ell}$ do not depend on $k$,
then the field is called {\em homogeneous}, or {\em stationary} if it
is a process.  At the boundaries, indices outside the signal index
scope $0\leq k < n-1$ occur in \eqref{gmrf}, which poses the need
for~$m$ left and~$m$ right boundary conditions that express
$s_{-1},\dots,s_{-m}$ and $s_n,\dots,s_{n+m-1}$ as linear combinations
of $s_0,\dots,s_{n-1}$. We assume that the boundary conditions~(\bc)
are already implicit in \eqref{gmrf}, i.e., the values $a_{k,\ell}$
have been adjusted accordingly.

After the \bc's are fixed, the~$n$ equations~\eqref{gmrf} can be
combined into vector form. Namely, the~$s_k$ and~$\nu_k$ are collected
into random vectors $\coord{s}$, $\coorda{\nu}$, respectively, to get
\begin{gather}
\coord{s} = A\coord{s} + \coorda{\nu},\quad
A = [a_{k,\ell}]_{0\leq k,\ell<n},\quad\text{or}\nonumber\\
(\one_n-A)\coord{s} = \coorda{\nu}.\label{gmrfvec}
\end{gather}
This is the initial representation of a random field we will work with.

\mypar{GMRF: MMSE estimate} We now discuss the covariance structure
of~$\coorda{\nu}$, i.e., its covariance matrix~$\Sigma_\coorda{\nu}$,
to make the field~$\coord{s}$ a GMRF and to derive from~\eqref{gmrf}
the minimum mean square error (MMSE) representation of the GMRF. We
distinguish several cases. For simplicity we assume that $\one_n-A$
has full rank, i.e., does not have the eigenvalue 0.

Case~1: $\one_n-A$, i.e., $A$, is symmetric and 
$\one_n-A$ is positive definite. 
We assume that the $\coorda{\nu}$ is correlated noise with covariance
matrix $\Sigma_\coorda{\nu}$ given by
\begin{equation*}
\Sigma_\coorda{\nu} =  \sigma^2(\one_n-A).
\end{equation*}
Then, it is well known, e.g., \cite{Moura:92}, that~\eqref{gmrf},
or~\eqref{gmrfvec}, is the MMSE representation of an $m$th-order GMRF
and that the covariance~$\Sigma_\coord{s}$ of the field~$\coord{s}$,
and the field and noise cross
covariance~$\Sigma_{\coord{s},\coorda{\nu}}$ are given by
\begin{equation}\label{mmsesym}
\begin{array}{rcl}
\Sigma_\coord{s} & = & \sigma^2(\one_n-A)^{-1},\\
\Sigma_{\coord{s},\coorda{\nu}} & = & \sigma^2\one_n.
\end{array}
\end{equation}
The cross-covariance~$\Sigma_{\coord{s},\coorda{\nu}}$ being diagonal
reflects the orthogonality between the field~$\coord{s}$ and the
noise~$\coorda{\nu}$ in the MMSE representation.

Case~2: $\one_n-A$, i.e., $A$, is symmetric, and $\one_n-A$ has
negative eigenvalues. We now assume that the noise is white with
covariance
\begin{equation*}
\Sigma_\coorda{\nu} =  \sigma^2\one_n.
\end{equation*}
The MMSE representation of the field follows from~\eqref{gmrfvec} by
multiplying both sides by $\one_n-A^T=\one_n-A$. The GMRF MMSE
representation hence becomes
\begin{equation}
(\one_n-A)^2\coord{s} = (\one_n-A)\coorda{\nu}.
\label{gmrfvec-2}
\end{equation}
The GMRF covariance is now
\begin{equation}\label{mmsesym1}
\Sigma_\coord{s}  =  \sigma^2(\one_n-A)^{-2}.
\end{equation}

Case 3: $\one_n-A$, i.e., $A$, is not symmetric. In this case, we assume
the noise in~\eqref{gmrfvec} to be white with covariance
\begin{equation}
\Sigma_\coorda{\nu}  =  \sigma^2\one_n.
\end{equation}
The MMSE representation follows from~\eqref{gmrfvec} by multiplying
both sides with $\one_n-A^T$. As a result we get the signal covariance
to be
\begin{equation}\label{mmsenotsym}
\Sigma_\coord{s} = \sigma^2(\one_n-A)^{-1}(\one_n-A^T)^{-1}.\\
\end{equation}

\mypar{KLT} The Karhunen-Lo\`eve transform (KLT) is defined in 
statistics as the linear transform that diagonalizes the covariance of
a random vector. For an $m$-th order GMRF, 
 the KLT is a unitary matrix~$F$ such that
\begin{equation}\label{klt}
F\Sigma_\coord{s}F^T = F\Sigma_{\coord{s}}F^{-1}\text{ is diagonal,}
\end{equation}
where $\Sigma_\coord{s}$ is given by~\eqref{mmsesym} or~\eqref{mmsesym1}
for a symmetric $A$, and by \eqref{mmsenotsym} for a non-symmetric $A$.

The transformed random vector $F\coord{s}$ has a diagonal covariance
matrix, i.e., the KLT decomposes, via a base change, a signal into its
statistically independent components.

We adopt the terminology in~\cite{Moura:92} and refer to the inverse
covariance matrix as the \emph{potential} matrix; it follows that the
KLT diagonalizes the potential matrix. As suggested in
\cite{Moura:92}, we will mostly work with the potential matrix.

\subsection{Examples: Discrete Time and Space}\label{dists}

To understand the similarities and differences between an algebraic
signal model and a corresponding stochastic model, or random field, we
consider two examples: discrete time and discrete space. The general 
connection is presented afterwards.

The correspondence between algebraic and stochastic modeling is
established in an intuitive way by relating the matrix~$A$
in~\eqref{gmrfvec} to the shift, or shift matrix, in the algebraic
model.

\mypar{Discrete space} We construct a random field that expresses the
spatial structure given by the space shift (Figure~\ref{spaceshift})
or the visualization of the space model in
Figure~\ref{abstractspacegraph}.  It is given by the first-order
homogeneous GMRF
\begin{equation}\label{spacegmrf}
s_k = a(s_{k-1} + s_{k+1}) + \nu_k,\quad k\in\Z.
\end{equation}
The noise~$\{\nu_k\}$ has the appropriate covariance structure as
discussed above.  The parameter~$a$ could be chosen as~$1/2$ (as in
the space shift), but we will see that the exact choice is not
crucial.

To construct a finite-length GMRF for the index scope $0\leq k <n$, we
need to introduce \bc's that express $s_{-1}$ and $s_n$ as linear
combinations of the $s_k$, $0\leq k<n$.  Unfortunately, there are no
clear guidelines as to which \bc's are natural choices. In the
literature, these choices are related to the \bc's used in solving
partial differential equations \cite{Moura:92,Strang:99}.  In the
algebraic finite space model, we chose the \bc's to yield a monomial
signal extension, but for the GMRF in \eqref{spacegmrf}, we don't have
a notion of signal extension beyond $s_{-1}$ and $s_n$, since there is
no underlying algebraic structure.

We now relate~\eqref{spacegmrf} to the finite space model to be able
to draw from the algebraic theory.  We choose as an example a pair of
\bc's that is associated to a finite space model, namely $s_{-1} =
s_0$ and $s_n = s_{n-1}$, which leads to
\begin{equation}
\label{eq:amatrix}
A = a\cdot
\left[
\begin{array}{ccccccc}
1 & 1 & & & & \\
1 & 0 & 1 \\
0 & 1 & 0 & \cdot \\
 & & 1 & \cdot & 1\\
 & & & \cdot & 0 & 1 \\
 & & & & 1 & 1
\end{array}
\right].
\end{equation} 
Clearly, $A$, and thus $\one_n-A$, in \eqref{gmrfvec} is symmetric, so
we can be in either Case~1 or Case~2 above.  To find out which case,
we first determine when $\one_n-A$ is positive definite. To answer
this question, we apply our algebraic theory. Namely, from
Table~\ref{chebmods} and using~\eqref{T1diag}, we know that $\one_n-A
= \phi(1-2ax)$, where $\phi$ is the representation afforded by the
regular module $\md=\C[x]/(x-1)U_{n-1}$ with $V$-basis.  The zeros of
$(x-1)U_{n-1}$ are given by $\cos(k\pi/n)$, $0\leq k<n$
(Table~\ref{4cheb} in Appendix~\ref{chebs}).  Using
\eqref{diagpropDTT}, the eigenvalues of $\one_n-A$ are hence given by
the evaluations of $1-2ax$ at these zeros:
$$
(1 - 2a\cos(k\pi/n)),\quad 0\leq k<n.
$$ 
It follows that the matrix~$\one_n-A$ is positive definite (Case~1)if and
only if $-1/2\leq a <1/2$. By~\eqref{mmsesym}, the covariance of the
field~$\coord{s}$ is $\Sigma_\coord{s} = (\one_n-A)^{-1}$ in this
case.  The KLT in this case is also provided by the algebraic theory
and is unique up to a unitary diagonal matrix~$D$. It is given by the
orthogonal $\DCTt{2}_n$.

If $\one_n-A$ has negative eigenvalues (Case~2), i.e., $|a|>1/2$, then
the covariance of~$\coord{s}$ is provided by~\eqref{mmsesym1} as
$\Sigma_\coord{s}=(\one_n-A)^{-2}$. Thus, the orthogonal $\DCTt{2}$ is
also a KLT in this case, but may not be unique (up to a diagonal),
since $(\one_n-A)^2$ may have duplicate eigenvalues. The cases in
which this occurs can be computed explicitly by again using
\eqref{diagpropDTT}, which provides these eigenvalues as the
evaluation of $(1-2ax)^2$ at $\cos(k\pi/n)$, $0\leq k<n$.

We can ask what other GMRFs have the~$\DCTt{2}$ as the KLT. To
address this question, we consider {\em all} matrices diagonalized by
$\DCTt{2}$---these are exactly the matrices $A=\phi(h)$, where
$h\in\alg=\C[x]/(x-1)U_{n-1}$. In particular, for a fixed $m < n$, let
$h = 2a_1T_1 + 2a_2T_2 + \dots + 2a_mT_m$. Then $A = \phi(h)$ is
naturally expressed in terms of $\phi(T_k)$, where $T_k$ are the
$k$-fold space shift operators, and so $A$ is an $m$-banded matrix
(only the $m$ upper and lower diagonals are non-zero) besides the
entries due to the signal extension. The matrix $\one_n-A$ is positive
definite, if and only if $1-h(x)$ takes only positive values at
$\cos(k\pi/n)$, $0\leq k<n$. Under these conditions, $h$ determines a
homogeneous $m$th-order GMRF with monomial \bc's $s_{-i-1} = s_{i}$
and $s_{n+i} = s_{n-1-i}$, $0\leq i<m$, and the KLT is the $\DCTt{2}$
as desired.

In this example, up to now, we considered only the matrix~$A$
in~\eqref{eq:amatrix}. We derived the conditions on~$a$ under which
the $\DCTt{2}$ is the KLT for a homogeneous symmetric $m$th-order GMRF
with monomial \bc's $s_{-i-1} = s_{i}$ and $s_{n+i} = s_{n-1-i}$,
$0\leq i<m$. We can adapt this discussion to all 16 DTTs, as long as
we make sure that the corresponding~$A$ is symmetric. In our algebraic
theory, this leads to the {\em symmetric} signal models
(Definition~\ref{symsm}) associated to the orthogonal DTTs (see
Section~\ref{orthodtts}). As explained in that section, some of the 16
symmetric models may not be strictly homogeneous, i.e.,
\eqref{spacegmrf} may need to be adjusted for $k=0,1$ or
$k=n-2,n-1$.

In summary, we obtain the following theorem, which completes the
result from \cite{Moura:98}.
\begin{theorem}[DTTs as KLTs]\label{dttsklts}
Let $1\leq m<n$. Every $\DTT_n$ is a KLT for the 
homogeneous $m$th-order GMRF
$$
s_k = a_1(s_{k-1}+s_{k+1}) +\dots + a_{m}(s_{k-m}+s_{k+m})+\nu_k,
$$
with monomial \bc's given by the signal extension of the algebraic 
signal model associated to the respective DTT.
\end{theorem}

In the alternative finite space model (Section~\ref{altrealfin}), we
identified also the periodic boundary conditions $s_{-1} = s_{n-1}$
and $s_n = s_0$ as possible choice for a monomial signal
extension. The matrix $A$ then becomes
\begin{equation}\label{agmrf}
A =
a\cdot\left[
\begin{array}{ccccccc}
0 & 1 & & & & 1 \\
1 & 0 & 1 \\
0 & 1 & 0 & \cdot \\
 & & 1 & \cdot & 1\\
 & & & \cdot & 0 & 1 \\
1 & & & & 1 & 0
\end{array}
\right],
\end{equation}
which is also symmetric. If we choose $\md = \C[x]/(x^n-1)$ with basis
$(1,x,\dots,x^{n-1})$ and $\phi$ is the representation of $\alg=\md$
afforded by $\md$, then $\one_n-A = \phi(1-a(x^{-1}+x))$. The zeros of
$x^n-1$ are the $n$th roots of unity, and hence the eigenvalues of 
$\one_n-A$ are the evaluations of $1-a(x^{-1}+x)$ at these zeros:
$$
(1-2a\cos(2k\pi/n)),\quad 0\leq k\leq n/2.
$$ 
It follows, similar to above, that $\one_n-A$ is positive definite
if and only if $-1/2\leq a<1/2$ for odd $n$, and $-1/2<a<1/2$ for even
$n$.  Also, similar to above, we can extend this discussion to higher
order GMRFs by setting $A =
\phi(1-a_1(x^{-1}+x^1)-\dots-a_m(x^{-m}+x^m))$, $m\leq n/2$. The
algebraic theory now also establishes the set of all KLTs for the GMRF
in this case as the set of matrices
$$
X\cdot\RDFT'_n,
$$
where $\RDFT'_n$ is the orthogonal RDFT, and $X$ is any unitary
x-shaped matrix of the form \eqref{xshape}.

\begin{theorem}[DFT and real DFTs as KLTs]\label{dftsklts}
Let $1\leq m\leq n/2$. The orthogonal DFT, RDFT, and DHT are 
KLTs for the homogeneous $m$th-order GMRF
$$
s_k = a_1(s_{k-1}+s_{k+1}) +\dots +a_{m}(s_{k-m}+s_{k+m})+\nu_k,
$$
with periodic \bc's.
\end{theorem}
As mentioned before, from a strict computational point of view
the choice should be the RDFT in this case.

In summary, the concepts of ``algebraic space model'' and associated
``homogeneous symmetric GMRF'' are (essentially) equivalent. As a
consequence, every orthogonal Fourier transform for the former is a
KLT for the latter and vice-versa.

For the algebraic \emph{time} model, or \emph{causal} GMRF, the situation is
different as we explore next.

\mypar{Discrete time} Again, we consider first an example, before
dealing with the general case in the next subsection. We proceed as in the
case of discrete space above.  The structure of the time-shift
(Figure~\ref{timeshift}) is described by the first-order {\em causal}
stationary GMRF (or stochastic process)
\begin{equation}\label{timegmrf}
s_k = as_{k-1} + \nu_k,\quad k\in\Z.
\end{equation}
The parameter~$a$ is left undetermined at this point. The
noise~$\nu_k$ is white.

To construct a GMRF for the finite index scope $0\leq k<n$, we need to
define \bc's for $s_{-1}$ and $s_n$.  We choose the periodic \bc
$s_{-1}=s_n$ identified by the algebraic theory and get as~$A$
in~\eqref{gmrfvec} a scaled cyclic shift matrix:
\begin{equation}
\label{eq:amatrix-cyclic}
A = a\cdot
\left[
\begin{array}{cccc}
 &&& 1 \\
1 \\
 & \ddots \\
 && 1 
\end{array}
\right].
\end{equation}
Since~$A$, and thus $\one_n-A$, is not symmetric, we are in
Case~3. The covariance of the signal~$\coord{s}$ is given
by~\eqref{mmsenotsym} as $\Sigma_\coord{s} =
(\one_n-A)^{-1}(\one_n-A^T)^{-1}$, which is guaranteed to have no
negative eigenvalues, regardless of the value of~$a$. Now, however, in
contradistinction to the finite space model of the previous example,
there is a difference between the algebraic model and the associated
GMRF. Namely, between the Fourier transform (for the algebraic model),
which diagonalizes $A$, and the KLT (for the GMRF), which diagonalizes
$\Sigma_\coord{s}$, or the potential matrix $(\one_n-A^T)(\one_n-A)$.
In general, these diagonalization problems are distinct.

However, with the particular~$A$ in~\eqref{eq:amatrix-cyclic}, we
still have a close relationship. Since~$A$
in~\eqref{eq:amatrix-cyclic} is orthogonal,
\begin{equation}\label{ts}
\Sigma_\coord{s}^{-1}=(\one_n-A)(\one_n-A^T) = (1+a^2)\one_n - A',
\end{equation}
where $A'$ is the matrix in~\eqref{agmrf}. 
Since $A'=A+1/a^{n-2}A^{n-1}$, every matrix that diagonalizes $A$ also 
diagonalizes $A'$, and hence $\Sigma_\coord{s}$. In other words,
every unitary Fourier transform for the finite
algebraic time model (with periodic \bc's) is a KLT for its GMRF
counterpart. The converse does not hold in general. For example,
the (orthogonal) RDFT is a KLT, but not a Fourier transform.

As an aside, \eqref{ts} shows that every KLT for the
GMRF~\eqref{timegmrf} is a KLT for the GMRF~\eqref{spacegmrf}, if, for
both, periodic \bc's are chosen. In other words, the causal
GMRF~\eqref{timegmrf} and the noncausal GMRF~\eqref{spacegmrf} with
periodic boundary conditions are essentially equivalent.

The above examples for discrete \emph{space} and discrete \emph{time}
convey the essential relationship between the algebraic signal model
with shift matrix~$A$ and the GMRF with matrix~$A$
in~\eqref{gmrfvec}. Next, we consider the general case.

\subsection{GMRFs and Signal Models}
\label{sec:gmrfandsignalmodel}

We now consider the relation between the statistical model and the
algebraic approach in the general case. The connection is established
by relating the matrix~$A$ in~\eqref{gmrfvec} to the shift matrix in
the algebraic model, and is provided by Lemmas~\ref{matsm}
and~\ref{diagsm}. The diagonalization property of
the KLT in~\eqref{klt} will be related to the diagonalization property of
the Fourier transform in Theorem~\ref{diagprop}.

\mypar{Equivalence in the symmetric case} We identify
under which conditions GMRFs and signal models are equivalent.

\begin{theorem}\label{gmrfequiv}
The following two concepts of signal models are equivalent: A GMRF
with symmetric matrix~$A$ in~\eqref{gmrfvec} and a symmetric algebraic
signal model $(\alg,\md,\Phi)$ with shift matrix $\phi(x) =
A$. Further:
\begin{itemize}
\item If $\one_n-A$ is positive definite, the KLT for the GMRF is an
orthogonal Fourier transform for the signal model, and vice-versa.
\item If $\one_n-A$ has negative eigenvalues, then every
orthogonal Fourier transform is a KLT for the GMRF, but, in general,
the converse does not hold. Namely, if $\md = \R[x]/p(x)$, then the
converse holds if and only if the mapping $x\mapsto x^2$ is injective
on the zeros of $p$.
\end{itemize}
\end{theorem}
\begin{proof}
The proof is straightforward. The main point is what we noted above
that diagonalizing $A$ is equivalent (and accomplished by the same
matrices) as diagonalizing $\sigma^2(\one_n-A)^{-1}$ in \eqref{klt}.

Let a GMRF with symmetric matrix~$A$ be given, and let $\one_n-A$ be
positive definite. Then, by \eqref{mmsesym}, a given KLT $F$
diagonalizes $\Sigma_\coord{s} = (\one_n-A)^{-1}$, or, equivalently,
$A$.  Since $A$ is symmetric it is diagonalizable and we can apply
Lemma~\ref{diagsm} to obtain a signal model with $\md=\R[x]/p(x)$,
$\alg=\langle q(x)\rangle<\R[x]/p(x)$, and $\phi(q(x)) = A$. Since $F$
diagonalizes $A$, it has to be an orthogonal Fourier transform
for this signal model.

Conversely, let $(\alg,\md,\Phi)$ with $\md = \R[x]/p(x)$ be a real,
symmetric signal model, and let $\alg<\R[x]/p(x)$ be generated by
$q(x)$. Set $A = \phi(q(x))$. Then there is an $a>0$ such that
$\one_n-aA$ is positive definite, and $aA$ defines a GMRF via
\eqref{gmrfvec} with covariance $\Sigma_\coord{s} =
(\one_n-aA)^{-1}$.  Further, the set of all Fourier transforms is the
set of all matrices diagonalizing $A$ or $aA$. Since $aA$ is symmetric,
it has an orthogonal Fourier transform, which is thus a KLT for the
GMRF defined by $aA$.

To prove the final assertion, we note that, if $\one_n-aA$ has negative
eigenvalues, the covariance of the GMRF~$\coord{s}$  defined by~$A$ is given by
\eqref{mmsesym1} as $\Sigma_\coord{s}=(\one_n-A)^{-2}$. Thus, the KLT
is characterized by the diagonalization of $(\one_n-A)^{-2}$. It follows that
every orthogonal Fourier transform for the algebraic signal model
(which diagonalizes $A$) is a KLT, but not vice-versa. From
Lemma~\ref{subalgebra} it follows that equivalence holds if and only
if $x^2$ is injective on the zeros of $p$.
\end{proof}

Theorem~\ref{gmrfequiv} can be applied to port concepts from the
algebraic theory to the theory of GMRFs. 

\begin{itemize}
\item In Section~\ref{dists} we applied the algebraic theory to
identify the ``good'' \bc's for an $m$th-order homogeneous GMRF and
obtained the 16 DTTs as associated KLTs. Further, every DTT was
identified as a KLT for GMRFs of order $1\leq m < n$ with respect to
these \bc's. Finally, these $m$th-order GMRF could be conveniently
written in terms of
$$
A = \phi\bigl(\sum_{1\leq k\leq m}h_kT_k\bigr).
$$
\item We obtain a classification of all possible (variant) 
symmetric 1st-order
GMRFs, which correspond precisely to the symmetric GNN models
(Section~\ref{orthgnn}). In particular, this guarantees the existence
of fast, $O(n\log^2(n))$, algorithms for all KLTs for these GMRFs
(Section~\ref{fastgnn}). Further, as in the previous item, since the
KLT diagonalizes the shift matrix $A = \phi(x)$, it diagonalizes all
matrices $\phi(q(x))$ in the generated algebra, thus a KLT for a
1st-order GMRF is automatically also a KLT for all higher order GMRFs
with suitable \bc's.

\item In Section~\ref{mc}, we established the connection between
signal models, graphs, and Markov chains. This connection now extends
to GMRFs.  For example, a GMRF with symmetric matrix~$A$ in
\eqref{gmrfvec} is equivalent to an undirected weighted graph. More
interestingly, we obtain that for $A\geq 0$ a GMRF, which models the
signal values as random variables, is equivalent to a Markov chain,
which models the signal {\em indices} as random variables.
\end{itemize}

\mypar{Connection in the non-symmetric case} Theorem~\ref{gmrfequiv}
cannot be extended to the general case of a non-symmetric matrix $A$,
since diagonalization of $A$ (as done by the Fourier transform for the
signal model for which $A$ is a shift matrix) is different from the
diagonalization of the potential matrix $(\one_n-A)(\one_n-A^T)$ (as
done by the KLT for the GMRF defined by $A$; w.l.o.g.~we omitted
$\sigma^2$). Also, $A$ is not necessary diagonalizable, whereas the
symmetric matrix $(\one_n-A)(\one_n-A^T)$ always is. An example for
this situation was the time model in Section~\ref{dists}, but in that
case, there was still a connection between the Fourier transform and
the KLT. We establish a general condition under which this connection
holds.

First, we remind the reader of the following known property
\cite[p.~272]{Gantmacher:59}.

\begin{lemma}\label{normalpoly}
Let $A\in\R^{n\times n}$. Then $A^T$ is a polynomial in $A$,
$$
A^T = q(A),
$$
if and only if $A$ is normal, i.e., $AA^T = A^TA$.
\end{lemma}

Note that normal matrices are precisely those matrices that can be
diagonalized by a unitary matrix \cite[p.~273]{Gantmacher:59}.  In
particular, in the real case, (scaled) orthogonal and symmetric
matrices are normal.

Now we can extend Theorem~\ref{gmrfequiv}.
\begin{theorem}\label{gmrfequiv1}
Let $A\in\R^{n\times n}$ normal, i.e., $A^T = q(A)$.  Further, let
$(\alg,\md,\Phi)$ be the signal model for which $A$ is the shift
matrix, and which is obtained from Lemma~\ref{diagsm}. Then, this
model has a unitary Fourier transform, and every such unitary Fourier
transform is a KLT for the GMRF defined by $A$ via
\eqref{gmrfvec}. The converse holds if and only if the polynomial
$$
(1-x)(1-q(x))
$$ 
generates $\alg$, or equivalently (Lemma~\ref{subalgebra}), if
$(1-x)(1-q(x))$ is an injective mapping on the zeros of $p$.
\end{theorem}

\begin{proof}
Let a real, normal matrix $A$ be given, which implies $A^T = q(A)$ for
a suitable polynomial $q$.  Further, let $(\alg,\md,\Phi)$, $\md =
\R[x]/p(x)$, be a signal model for which $A$ is the shift matrix,
obtained by using Lemma~\ref{diagsm}. Note that this lemma is
applicable, since $A$ as a normal matrix is diagonalizable.  Since $A$
is normal, the signal model has a unitary Fourier transform
$\four$. Since $\four$ diagonalizes $A$, it also diagonalizes
$(\one_n-A)(\one_n-q(A))$ as a polynomial in $A$. Thus,
$\four$ is a KLT for the GMRF defined by $A$.  

Conversely, the diagonalization of $(\one_n-A)(\one_n-q(A))$ implies
the diagonalization of $A$ only if the polynomial $(1-x)(1-q(x))$
generates $\alg$. Using Lemma~\ref{subalgebra}, this is the case 
if this polynomial is injective on the zeros of $p$.
\end{proof}

Theorem~\ref{gmrfequiv1} tells us that, for normal $A$, there may be
more KLTs for the GMRF than (unitary) Fourier transforms for the
associated signal model generated by $A$. This implies that a KLT may
be {\em cheaper} in terms of arithmetic cost than any of these Fourier
transforms.  This was precisely the case for the time model with
periodic \bc's in Section~\ref{dists}. The algebraic signal model
requires as Fourier transform the DFT, whereas for the associated
GMRF, the RDFT or DHT was sufficient.  Indeed, the polynomial $q$ in
this case is $q(x) = x^{n-1}$, and the mapping $(1-x)(1-x^{n-1})$ maps
conjugate roots of unity to the same value and is hence not injective.

We restricted the discussion of the relationship between GMRFs and
signal models in our definition \eqref{gmrf} to the finite case,
similar to Section~\ref{mc}. As a consequence, we  did not discuss the
asymptotic equivalence of a GMRF or a signal model to its infinite
counterpart.  However, in the case of a 1st-order GMRF corresponding
to the GNN model, the method in \cite{Yemini:79} provides a general
recipe for asymptotic analysis based on orthogonal polynomials, thus,
fitting into our framework. As a special case, \cite{Yemini:79}
considers the DCT, type 2. The analysis shows, in our language, the
role of the signal extension associated to this transform. In
particular, it seems the monomiality of the signal extension is
crucial for convergence. However, with Theorems~\ref{dttsklts} and
\ref{dftsklts} in mind, it appears that the appropriateness of the
chosen \bc's should have a stronger impact on the transform's
performance than its asymptotic properties.

\section{Conclusions}\label{conclusions}

We briefly summarize the main contributions of the paper and then
briefly discuss the further development of the algebraic theory of
signal processing that will be the subject of future publications.

\mypar{The algebraic foundation of signal processing} The first main
goal of the paper is to establish the algebraic structure and
generalize the fundamental concepts of signal processing. Towards
this, we determined the following:
\begin{itemize}
\item We introduced the \emph{signal model} as a triple
$(\alg,\md,\Phi)$ of a filter algebra, a signal module, and a
bijective linear map. We explained how the \emph{representation
theory} of algebras provides the main ingredients for signal
processing including the notions of filtering, spectrum, frequency
response, and Fourier transform.
\item We recognized the fundamental role of the \emph{shift} as a
suitably chosen generator of the filter algebra and provided a general
recipe to derive infinite and finite signal models from the shift.  We
used this procedure to derive many different signal models for time
and space, infinite and finite, complex and real, and showed that in
each case all important signal processing concepts are well-defined
and available.
\item In particular, by doing so, we generalized the $z$-transform,
which is a special case of a signal model, and introduced the
$C$-transform as the equivalent of the $z$-transform for space models.
\item We identified the equivalence between shift-invariant systems
and commutative algebras.
\end{itemize}

\mypar{Discrete time and space and trigonometric transforms} As the
second main contribution of this paper and a first application of the
algebraic theory of signal processing, we explained how to derive
infinite and finite signal models for time and space, and, in doing
so, we discovered practically all existing trigonometric transforms
and completed this class with a few new transform not introduced
before. We showed that each trigonometric transform is the Fourier
transform for either a time or a space model. The most important
results include the following:
\begin{itemize}
\item We showed that finite and infinite signal processing are
instantiations of the same general theory, and explained how to derive
finite signal models from their infinite counterparts.
\item The need for boundary conditions in the finite case proved to be
central to our theory. In particular, the paper makes the role of
boundary conditions and their relation to signal extensions and signal
models completely transparent and explains the importance of choosing
a simple, that is, monomial, signal extension.
\item We showed that time and space signal processing are
instantiations of the same theory by deriving infinite and
finite \emph{space} models in complete analogy, albeit with noted
important differences, to their time counterparts. As a major
insight, we showed that the 16 DCTs and DSTs are, in a rigorous sense,
the space analogue of the DFT.
\item We extended the idea of space models to generic next neighbor
(GNN) models and showed how their realization naturally connects to
the theory of orthogonal polynomials.
\end{itemize}

\mypar{Algebraic signal models and GMRFs}
Finally, as our third main contribution, we established the
equivalence (under certain conditions) between shift-invariant signal
models, graphs, and Markov chains, and, perhaps most importantly,
Gauss-Markov random fields. The connection between these concepts
makes it possible to look at signal models from different perspectives
(an important example is the visualization of a signal model that we
defined) and to understand the inherent limitations of the models and 
thus the signal processing they provide. Further, we showed the 
similarities and differences between our signal models and stochastic 
Gauss-Markov models. An important insight was the connection between 
the Fourier transform and the Karhunen-Lo\`eve transform.

\mypar{Evolution of the algebraic theory of signal processing} This
paper shows that signal processing is algebraic in nature. In other
words, the goal is not to impose a mathematical theory on existing
signal processing theories, but rather to expose the mathematical
structure that is the essence or foundation of signal processing.  In
other words, the paper shows that many apparently distinct concepts
are instantiations of the same concepts, for example such as the
trigonometric transforms as a special examples of Fourier transforms.
 
The algebraic theory of signal processing can be used to derive new
results in signal processing and fast algorithms for computing linear
transforms. Already in this paper, besides the classification of all
existing trigonometric transforms, we introduced several new
transforms. In a future paper~\cite{Pueschel:05} we use the algebraic
theory of signal processing to classify and discover fast algorithms.
We have started this algebraic theory of fast algorithms already in
\cite{Egner:01,Pueschel:03a}.  Besides that, we have already used the
algebraic theory to derive new signal models and new signal
processing schemes. An example is a signal model, and its associated
Fourier transform, for non-separable signal processing on a
2-D triangular grid \cite{Pueschel:04a,Pueschel:04b}.  We are confident
that the further development and the application of the algebraic
theory of signal processing will lead to many other new applications.


\bibliographystyle{IEEE}
\bibliography{paper}

\begin{thebibliography}{10}

\bibitem{Nussbaumer:82}
H.~J. Nussbaumer,
\newblock {\em Fast {F}ourier Transformation and Convolution Algorithms},
\newblock Springer, 2nd edition, 1982.

\bibitem{Gray:88}
Robert~M. Gray,
\newblock {\em Probability, Random Processes, and Ergodic Properties},
\newblock Springer-Verlag, New York, 1988.

\bibitem{Pueschel:04a}
M.~P{\"u}schel and M.~R{\"o}tteler,
\newblock ``{The Discrete Triangle Transform},''
\newblock in {\em Proc.~ICASSP}, 2004.

\bibitem{Pueschel:05a}
M.~P{\"u}schel and M.~R{\"o}tteler,
\newblock ``{Fourier Transform for the Spatial Quincunx Lattice},''
\newblock in {\em Proc.~ICIP}, 2005.

\bibitem{Nicholson:71}
P.J. Nicholson,
\newblock ``Algebraic theory of finite {F}ourier transforms,''
\newblock {\em Journal of Computer and System Sciences}, vol. 5, pp. 524--547,
  1971.

\bibitem{Winograd:78}
S.~Winograd,
\newblock ``On computing the discrete {F}ourier transform,''
\newblock {\em Mathematics of Computation}, vol. 32, pp. 175--199, 1978.

\bibitem{Auslander:84}
L.~Auslander, E.~Feig, and S.~Winograd,
\newblock ``Abelian semi-simple algebras and algorithms for the discrete
  {F}ourier transform,''
\newblock {\em Advances in Applied Mathematics}, vol. 5, pp. 31--55, 1984.

\bibitem{Beth:84}
Th. Beth,
\newblock {\em {V}erfahren der {S}chnellen {F}ouriertransformation [Methods for
  the Fast {F}ourier Transform]},
\newblock Teubner, 1984.

\bibitem{VanLoan:92}
C.~Van~Loan,
\newblock {\em Computational Framework of the Fast {F}ourier Transform},
\newblock Siam, 1992.

\bibitem{Tolimieri:97}
R.~Tolimieri, M.~An, and C.~Lu,
\newblock {\em Algorithms for Discrete Fourier Transforms and Convolution},
\newblock Springer, 2nd edition, 1997.

\bibitem{Tolimieri:97a}
R.~Tolimieri, M.~An, and C.~Lu,
\newblock {\em Mathematics of Multidimensional Fourier Transform Algorithms},
\newblock Springer, 2nd edition, 1997.

\bibitem{Pueschel:03a}
M.~P{\"u}schel and J.~M.~F. Moura,
\newblock ``The algebraic approach to the discrete cosine and sine transforms
  and their fast algorithms,''
\newblock {\em SIAM Journal of Computing}, vol. 32, no. 5, pp. 1280--1316,
  2003.

\bibitem{Pueschel:03}
M.~P{\"u}schel,
\newblock ``Cooley-{T}ukey {FFT} like algorithms for the {DCT},''
\newblock in {\em Proc.~ICASSP}, 2003, vol.~2, pp. 501--504.

\bibitem{Pueschel:05}
M.~P{\"u}schel and J.~M.~F. Moura,
\newblock ``Algebraic signal processing theory: 1-d cooley-tukey type
  algorithms, part {I},'' In preparation.

\bibitem{Pueschel:04b}
M.~P{\"u}schel and M.~R{\"o}tteler,
\newblock ``Cooley-{T}ukey {FFT} like fast algorithms for the discrete triangle
  transform,''
\newblock in {\em Proc.~11th IEEE DSP Workshop}, 2004.

\bibitem{Kalman:69}
Kalman~R. E., Falb~P. L., and Arbib~M. A.,
\newblock {\em Topics in Mathematical System Theory},
\newblock McGraw-Hill, 1969.

\bibitem{Basile:69}
G.~Basile and G.~Marro,
\newblock ``Controlled and conditioned invariant subspaces in linear system
  theory,''
\newblock {\em J.~Optimization Theory Applications}, vol. 3, pp. 306--315,
  1969.

\bibitem{Wonham:70}
W.~M. Wonham and A.~S. Morse,
\newblock ``Decoupling and pole assignment in linear multivariable systems: A
  geometric approach,''
\newblock {\em SIAM J.~Control and Optimization}, vol. 8, pp. 1--18, 1970.

\bibitem{Willems:71}
J.~C. Willems and S.~K. Mitter,
\newblock ``Controllability, obersvability, pole allocation, and state
  reconstruction,''
\newblock {\em IEEE Transactions on Automatic Control}, vol. 16, pp. 582--595,
  1971.

\bibitem{Johnston:73}
Robert deB. Johnston,
\newblock {\em Linear Systems Over various Rings},
\newblock Ph.D. thesis, Massachusetts Institute of technology, Cambridge, MA,
  1973.

\bibitem{Fuhrman:76}
Paul~A. Fuhrman,
\newblock ``Algebraic system theory: An analyst's point of view,''
\newblock {\em Journal of the Franklin Institute}, vol. 301.

\bibitem{Fuhrman:96}
Paul~A. Fuhrman,
\newblock {\em A Polynomial Approach to Linear Algebra},
\newblock Springer Verlag, New York, 1996.

\bibitem{Cooley:65}
J.~W. Cooley and J.~W. Tukey,
\newblock ``An algorithm for the machine calculation of complex {F}ourier
  series,''
\newblock {\em Math. of Computation}, vol. 19, pp. 297--301, 1965.

\bibitem{Heideman:85}
M.~T. Heideman, D.~H. Johnson, and C.~S. Burrus,
\newblock ``{G}auss and the {H}istory of the {F}ast {F}ourier {T}ransform,''
\newblock {\em Archive for History of Exact Sciences}, vol. 34, pp. 265--277,
  1985.

\bibitem{Apple:70}
G.~G. Apple and P.~A. Wintz,
\newblock ``Calculation of {F}ourier transforms on finite abelian groups,''
\newblock {\em {IEEE Trans.~on Information Theory}}, vol. IT-16, pp. 233--234,
  1970.

\bibitem{Cairns:71}
T.~W. Cairns,
\newblock ``On the fast {F}ourier transform on finite abelian groups,''
\newblock {\em IEEE Trans.~on Computers}, vol. C-21, pp. 569--571, 1971.

\bibitem{Winograd:79}
S.~Winograd,
\newblock ``On the multiplicative complexity of the discrete {F}ourier
  transform,''
\newblock {\em Advances in Mathematics}, vol. 32, pp. 83--117, 1979.

\bibitem{Winograd:80}
S.~Winograd,
\newblock {\em Arithmetic Complexity of Computation},
\newblock Siam, 1980.

\bibitem{Auslander:84a}
L.~Auslander, E.~Feig, and S.~Winograd,
\newblock ``The multiplicative complexity of the discrete {F}ourier
  transform,''
\newblock {\em Advances in Applied Mathematics}, vol. 5, pp. 87--109, 1984.

\bibitem{Edwards:67}
R.~E. Edwards,
\newblock {\em {F}ourier Series: A Modern Introduction}, vol.~I,
\newblock Holt, Rinehart and Winston, 1967.

\bibitem{Edwards:67a}
R.~E. Edwards,
\newblock {\em {F}ourier Series: A Modern Introduction}, vol.~II,
\newblock Holt, Rinehart and Winston, 1967.

\bibitem{Rudin:62}
W.~Rudin,
\newblock {\em Fourier Analysis on Groups},
\newblock Number~12 in Interscience tracts in pure and applied mathematics.
  Interscience, New York, 1962.

\bibitem{Karpovsky:77}
M.~G. Karpovsky,
\newblock ``Fast {F}ourier transforms on finite non-abelian groups,''
\newblock {\em IEEE Trans.~on Computers}, vol. C-26, pp. 1028--1030, 1977.

\bibitem{Beth:87}
Th. Beth,
\newblock ``{On the computational complexity of the general discrete {F}ourier
  transform},''
\newblock {\em Theoretical Computer Science}, vol. 51, pp. 331--339, 1987.

\bibitem{Clausen:88}
M.~Clausen,
\newblock {\em {B}eitr{\"a}ge zum {E}ntwurf schneller
  {S}pektraltransformationen ({H}abilitationsschrift)},
\newblock Univ. Karlsruhe, 1988.

\bibitem{Clausen:89}
M.~Clausen,
\newblock ``Fast generalized {F}ourier transforms,''
\newblock {\em Theoretical Computer Science}, vol. 67, pp. 55--63, 1989.

\bibitem{Clausen:93}
M.~Clausen and U.~Baum,
\newblock {\em Fast Fourier Transforms},
\newblock BI-Wiss.-Verl., 1993.

\bibitem{Maslen:95}
D.~Maslen and D.~Rockmore,
\newblock ``Generalized {FFT}s -- a survey of some recent results,''
\newblock in {\em Proceedings of IMACS Workshop in Groups and Computation},
  1995, vol.~28, pp. 182--238.

\bibitem{Rockmore:90}
D.~Rockmore,
\newblock ``Fast {F}ourier analysis for abelian group extensions,''
\newblock {\em Advances in Applied Mathematics}, vol. 11, pp. 164--204, 1990.

\bibitem{Diaconis:90}
P.~Diaconis and D.~Rockmore,
\newblock ``Efficient computation of the {F}ourier transform on finite
  groups,''
\newblock {\em Amer. Math. Soc.}, vol. 3(2), pp. 297--332, 1990.

\bibitem{Maslen:00}
D.~Maslen and D.~Rockmore,
\newblock ``Double coset decompositions and computational harmonic analysis on
  groups,''
\newblock {\em Journal of Fourier Analysis and Applications}, vol. 6, no. 4,
  2000.

\bibitem{Diaconis:89}
P.~Diaconis,
\newblock ``A generalization of spectral analysis with applications to ranked
  data,''
\newblock {\em Annals of Statistics}, vol. 17, pp. 949--979, 1989.

\bibitem{Diaconis:88}
P.~Diaconis,
\newblock {\em Group Representations in Probability and Statistics},
\newblock Lecture Notes---Monograph Series. IMS, 1988.

\bibitem{Driscoll:94}
J.~R. Driscoll and D.~M. Healy~Jr.,
\newblock ``Computing {F}ourier transforms and convolutions on the 2-sphere,''
\newblock {\em Advances in Applied Mathematics}, vol. 15, pp. 203--250, 1994.

\bibitem{Foote:00}
R.~Foote, G.~Mirchandi, D.~Rockmore, D.~Healy, and T.~Olson,
\newblock ``A wreath product approach to signal and image processing: Part
  i---multiresolution analysis,''
\newblock {\em IEEE Trans.~on Signal Processing}, vol. 48, no. 1, pp. 102--132,
  2000.

\bibitem{Mirchandini:00}
G.~Mirchandi, R.~Foote, D.~Rockmore, D.~Healy, and T.~Olson,
\newblock ``A wreath product approach to signal and image processing---part ii:
  Multiresolution analysis,''
\newblock {\em IEEE Trans.~on Signal Processing}, vol. 48, no. 3, pp. 749--767,
  2000.

\bibitem{Resnikoff:98}
Howard~L. Resnikoff and Raymond~O. Wells,
\newblock {\em Wavelet Analysis},
\newblock Springer, 1998.

\bibitem{Shokrollahi:01}
A.~Shokrollahi, B.~Hassibi, B.~M. Hochwald, and W.~Sweldens,
\newblock ``Representation theory for high-rate multiple-antenna code design,''
\newblock {\em IEEE Trans.~on Information Theory}, vol. 47, no. 6, pp.
  2335--2367, 2001.

\bibitem{Minkwitz:93}
T.~Minkwitz,
\newblock {\em {A}lgorithmensynthese f{\"u}r lineare {S}ysteme mit
  {S}ymmetrie},
\newblock Ph.D. thesis, Universit{\"a}t Karlsruhe, Informatik, 1993.

\bibitem{Minkwitz:95}
T.~Minkwitz,
\newblock ``Algorithms explained by symmetry,''
\newblock {\em Lecture Notes on Computer Science}, vol. 900, pp. 157--167,
  1995.

\bibitem{Egner:01}
S.~Egner and M.~P{\"u}schel,
\newblock ``Automatic generation of fast discrete signal transforms,''
\newblock {\em IEEE Trans.~on Signal Processing}, vol. 49, no. 9, pp.
  1992--2002, 2001.

\bibitem{Egner:04}
S.~Egner and M.~P{\"u}schel,
\newblock ``Symmetry-based matrix factorization,''
\newblock {\em Journal of Symbolic Computation, special issue on "Computer
  Algebra and Signal Processing"}, vol. 37, no. 2, pp. 157--186, 2004.

\bibitem{Pueschel:02}
M.~P{\"u}schel,
\newblock ``Decomposing monomial representations of solvable groups,''
\newblock {\em Journal of Symbolic Computation}, vol. 34, no. 6, pp. 561--596,
  2002.

\bibitem{Steidl:91}
G.~Steidl and M.~Tasche,
\newblock ``A polynomial approach to fast algorithms for discrete
  {F}ourier-cosine and {F}ourier-sine transforms,''
\newblock {\em Mathematics of Computation}, vol. 56, no. 193, pp. 281--296,
  1991.

\bibitem{Moura:98}
J.~M.~F. Moura and M.~G.~S. Bruno,
\newblock ``{DCT/DST} and {G}auss-{M}arkov fields: Conditions for
  equivalence,''
\newblock {\em IEEE Trans.~on Signal Processing}, vol. 46, no. 9, pp.
  2571--2574, 1998.

\bibitem{Strang:99}
G.~Strang,
\newblock ``The discrete cosine transform,''
\newblock {\em SIAM Review}, vol. 41, no. 1, pp. 135--147, 1999.

\bibitem{Jacobson:74}
N.~Jacobson,
\newblock {\em Basic Algebra I},
\newblock W.~H.~Freeman and Co., 1974.

\bibitem{James:93}
G.~James and M.~Liebeck,
\newblock {\em Representations and Characters of Groups},
\newblock Cambridge Univ. Pr., 1993.

\bibitem{Curtis:62}
W.~C. Curtis and I.~Reiner,
\newblock {\em Representation Theory of Finite Groups},
\newblock Interscience, 1962.

\bibitem{Kashin:89}
B.~S. Kashin and A.~A. Saakyan,
\newblock {\em Orthogonal Series},
\newblock American Mathematical Society, 1989.

\bibitem{Sandberg:98}
Irwin~W. Sandberg,
\newblock ``A representation theorem for linear systems,''
\newblock {\em IEEE Trans.~on Circuits and Systems---1: Fundamental Theory and
  Applications}, vol. 45, no. 5, pp. 578--580, 1998.

\bibitem{Cox:97}
D.~Cox, J.~Little, and D.~O'Shea,
\newblock {\em Ideals, Varieties, and Algorithms},
\newblock Springer, 1997.

\bibitem{Driscoll:97}
J.~R. Driscoll, D.~M. Healy~Jr., and D.~Rockmore,
\newblock ``Fast discrete polynomial transforms with applications to data
  analysis for distance transitive graphs,''
\newblock {\em SIAM Journal Computation}, vol. 26, pp. 1066--1099, 1997.

\bibitem{Potts:98}
D.~Potts, G.~Steidl, and M.~Tasche,
\newblock ``Fast algorithms for discrete polynomial transforms,''
\newblock {\em Mathematics of Computation}, vol. 67, no. 224, pp. 1577--1590,
  1998.

\bibitem{Nussbaumer:79}
Henri~J. Nussbaumer and Philippe Quandalle,
\newblock ``Fast computation of discrete {F}ourier transforms using polynomial
  transforms,''
\newblock {\em IEEE Trans.~on Acoustics, Speech, and Signal Processing}, vol.
  ASSP-27, no. 2, pp. 169--181, 1979.

\bibitem{Kailath:97}
Th. Kailath and V.~Olshevsky,
\newblock ``Displacement structure approach to polynomial {V}andermonde and
  related matrices,''
\newblock {\em Linear Algebra and Applications}, vol. 261, pp. 49--90, 1997.

\bibitem{Bongiovanni:76}
G.~Bongiovanni, P.~Corsini, and G.~Frosini,
\newblock ``One-dimensional and two-dimensional generalized discrete {F}ourier
  transform,''
\newblock {\em IEEE Trans.~on Acoustics, Speech, and Signal Processing}, vol.
  ASSP-24, no. 2, pp. 97--99, 1976.

\bibitem{Martucci:94}
S.~A. Martucci,
\newblock ``Symmetric convolution and the discrete sine and cosine
  transforms,''
\newblock {\em IEEE Trans.~on Signal Processing}, vol. 42, no. 5, pp.
  1038--1051, 1994.

\bibitem{Britanak:99}
V.~Britanak and K.~R. Rao,
\newblock ``The fast generalized discrete {F}ourier transforms: A unified
  approach to the discrete sinusoidal transforms computation,''
\newblock {\em Signal Processing}, vol. 79, pp. 135--150, 1999.

\bibitem{Pan:01}
V.~Y. Pan,
\newblock {\em Structured Matrices and Polynomials},
\newblock Birkh{\"a}user Springer, 2001.

\bibitem{Beth:83}
Th. Beth, W.~Fumy, and W.~M{\"u}hlfeld,
\newblock ``Zur algebraischen diskreten {F}ourier-{T}ransformation [on the
  algebraic discrete {F}ourier transform],''
\newblock {\em Archiv der Mathematik}, vol. 6, no. 3, pp. 238--244, 1983.

\bibitem{Hong:94}
J.~Hong, M.~Vetterli, and P.~Duhamel,
\newblock ``Basefield transforms with the convolution property,''
\newblock {\em Proceedings of the IEEE}, vol. 82, no. 3, pp. 400--412, 1994.

\bibitem{Beth:89}
Th. Beth,
\newblock ``Generating fast {H}artley transforms,''
\newblock in {\em Proceedings URSI-ISSSE}, 1989, pp. 688--692.

\bibitem{Bracewell:83}
R.~N. Bracewell,
\newblock ``Discrete {H}artley transform,''
\newblock {\em J. Optical Society America}, vol. 73, no. 12, pp. 1832--1835,
  1983.

\bibitem{Wang:81}
Z.~Wang,
\newblock ``Harmonic analysis with a real frequency function. i. aperiodic
  case,''
\newblock {\em Appl. Math. Comput.}, vol. 9, pp. 53--73, 1981.

\bibitem{Wang:81a}
Z.~Wang,
\newblock ``Harmonic analysis with a real frequency function. i. periodic and
  bounded cases,''
\newblock {\em Appl. Math. Comput.}, vol. 9, pp. 153--163, 1981.

\bibitem{Wang:81b}
Z.~Wang,
\newblock ``Harmonic analysis with a real frequency function. i. data
  sequence,''
\newblock {\em Appl. Math. Comput.}, vol. 9, pp. 245--255, 1981.

\bibitem{Wang:85}
Z.~Wang and B.~R. Hunt,
\newblock ``The discrete {W} transform,''
\newblock {\em Applied Mathematics and Computation}, vol. 16, pp. 19--48, 1985.

\bibitem{Hartley:42}
R.~V.~L. Hartley,
\newblock ``A more symmetrical {F}ourier analysis applied to transmission
  problems,''
\newblock {\em Proc. IRE}, vol. 30, pp. 144--150, 1942.

\bibitem{Moura:92}
J.~M.~F. Moura and Nikhil Balram,
\newblock ``Recursive structure of noncausal {G}auss {M}arkov random fields,''
\newblock {\em IEEE Trans.~Information Theory}, vol. 38, no. 2, pp. 334--354,
  March 1992.

\bibitem{Alexitis:61}
G.~Alexitis,
\newblock {\em Convergence Problems of Orthogonal Series},
\newblock Pergamon Press, 1961.

\bibitem{Ahmed:74}
N.~Ahmed, T.~Natarajan, and K.~R. Rao,
\newblock ``Discrete cosine transform,''
\newblock {\em {IEEE Trans.~on Computers}}, vol. C-23, pp. 90--93, 1974.

\bibitem{Kailath:96}
Th. Kailath and V.~Olshevsky,
\newblock ``Displacement structure approach to discrete trigonometric transform
  based preconditioners of {G}.~{S}trang and {T}.~{C}han type,''
\newblock {\em Calcolo}, vol. 33, pp. 191--208, 1996.

\bibitem{Sanchez:95}
V.~S{\'a}nchez, P.~Garc{\'{\i}}a, A.~M. Peinado, J.~C. Segura, and A.~J. Rubio,
\newblock ``Diagonalizing properties of the discrete cosine transforms,''
\newblock {\em IEEE Trans.~on Signal Processing}, vol. 43, no. 11, pp.
  2631--2641, 1995.

\bibitem{Szegoe:67}
G.~Szeg{\"o},
\newblock {\em Orthogonal Polynomials},
\newblock Amer.~Math.~Soc.~Colloq.~Publ., 3rd edition, 1967.

\bibitem{Yemini:79}
Y.~Yemini and J.~Pearl,
\newblock ``Asymptotic properties of discrete unitary transforms,''
\newblock {\em IEEE Trans.~on Pattern Analysis and Machine Intelligence}, vol.
  PAMI-1, no. 4, pp. 366--371, 1979.

\bibitem{Chihara:78}
T.~S. Chihara,
\newblock {\em An Introduction to Orthogonal Polynomials},
\newblock Gordon and Breach, 1978.

\bibitem{Gantmacher:59}
F.~R. Gantmacher,
\newblock {\em Matrix Theory}, vol.~I,
\newblock Chelsea, 1959.

\bibitem{Wang:82}
Z.~Wang,
\newblock ``A fast algorithm for the discrete sine transform implemented by the
  fast cosine transform,''
\newblock {\em IEEE Trans.~on Acoustics, Speech, and Signal Processing}, vol.
  ASSP-30, no. 5, pp. 814--815, 1982.

\bibitem{Rayes:98}
M.~O. Rayes, V.~Trevisan, and P.~S. Wang,
\newblock ``Factorization of {C}hebyshev polynomials,''
\newblock Tech. {R}ep. ICM-199802-0001, Kent State University, 1998.

\bibitem{Mandyam:96}
G.~Mandyam and N.~Ahmed,
\newblock ``The discrete {L}aguerre transform: derivation and applications,''
\newblock {\em IEEE Trans.~on Signal Processing}, vol. 44, no. 12, pp.
  2925--2931, 1996.

\bibitem{Martens:90}
J.-B. Martens,
\newblock ``The {H}ermite transform---theory,''
\newblock {\em IEEE Trans.~on Acoustics, Speech, and Signal Processing}, vol.
  38, no. 9, pp. 1595--1605, 1990.

\bibitem{Aburdene:94}
M.~F. Aburdene and J.~E. Dorband,
\newblock ``Unification of {L}egendre, {L}aguerre, {H}ermite, and binomial
  discrete transforms using {P}ascal's matrix,''
\newblock {\em {Multidimensional Systems and Signal Processing}}, vol. 5, no.
  3, pp. 301--305, 1994.

\bibitem{Haddad:88}
R.~A. Haddad and A.~N. Akansu,
\newblock ``A new orthogonal transform for signal coding,''
\newblock {\em IEEE Trans.~on Acoustics, Speech, and Signal Processing}, vol.
  36, no. 9, pp. 1404--1411, 1988.

\bibitem{Mukundan:01}
R.~Mukundan, S.~H. Ong, and P.~A. Lee,
\newblock ``Image analysis by tchebichef moments,''
\newblock {\em IEEE Trans.~on Image Processing}, vol. 10, no. 9, pp.
  1357--1363, 2001.

\bibitem{Romanovsky:70}
V.~I. Romanovsky,
\newblock {\em Discrete Markov Chains},
\newblock Wolters-Noordhoff, 1970.

\bibitem{Rivlin:74}
T.~J. Rivlin,
\newblock {\em The {C}hebyshev Polynomials},
\newblock Wiley Interscience, 1974.

\end{thebibliography}

\appendices

\section{Algebraic Background}\label{algdefs}

\mypar{Algebraic definitions}
We provide here the formal definitions of the most important algebraic
concepts used in this paper.

\begin{definition}[Algebra]\label{algebradef}
A $\C$-algebra $\alg$ is a ring that is at the same time a $\C$-vector
space, such that the addition in the ring and the addition in the
vector space coincide. In addition, for $\alpha\in\C$ and $g,h\in\alg$,
$$
\alpha(gh) = (\alpha g)h = g(\alpha h)
$$
has to hold.
\end{definition}

\begin{definition}[Module]\label{moduledef}
Let $\alg$ be a $\C$-algebra. A (left) $\alg$-module is a $\C$-vector space
$\md$ that permits an operation
$$
\alg\times\md\rightarrow\md,\quad
(a, m)\mapsto am,
$$
which satisfies, for $a, b, 1\in\alg$ and $m, n\in\md$,
\begin{eqnarray*}
a(m+n) & = & am + an\\
(a+b)m & = & am + bm\\
(ab)m & = & a(bm) \\
1m & = & m.
\end{eqnarray*}
\end{definition}

\begin{definition}[Homomorphism of algebras]\label{alghomdef}
Let $\alg, {\cal B}$ be $\C$-algebras. A homomorphism is a mapping
$\phi: \alg\rightarrow{\cal B}$ that satisfies, for $a,b\in\alg$,
$\alpha\in\C$,
\begin{eqnarray*}
\phi(a+b) & = & \phi(a)+\phi(b)\\
\phi(ab) & = & \phi(a)\phi(b)\\
\phi(\alpha a) & = & \alpha\phi(a).
\end{eqnarray*}
\end{definition}

\begin{definition}[Homomorphism of modules]\label{hommoddef}
Let $\md, {\cal N}$ be $\alg$-modules. A homomorphism is a mapping
$\phi: \md\rightarrow{\cal N}$ that satisfies, for $a\in\alg$,
$m, n\in\md$,
\begin{eqnarray*}
\phi(m+n) & = & \phi(m)+\phi(n)\\
\phi(am) & = & a\phi(m).
\end{eqnarray*}
\end{definition}

\mypar{Vector spaces} We assume the reader is familiar with standard
linear algebra and only define the notion of direct sum that we use in
this paper.

\begin{definition}[Direct sum of vector spaces]\label{directsumV}
Let $V$ be a vector space, and let $U,W\leq V$ be subvector spaces
with $V = U + W$ and $U\cap W = \{0\}$. Then $V= U+W = U\dirsum W$ is
called the {\em inner} direct sum of $U$ and $W$. Every element $v\in
V$ can now be uniquely represented as $v = u+w$, $u\in U, w\in W$.  We
now consider the {\em outer} direct sum.  Represent $v$ as the pair
$(u,w)$ and define addition and scalar multiplication
componentwise. We denote the set of these pairs also as $U\dirsum W$
and call it the {\em outer} direct sum of $U$ and $W$. So the outer
direct sum of two vector spaces is their Cartesian product with
componentwise addition and scalar multiplication.  Every inner direct
sum corresponds to an outer direct sum, but the latter is more
general, since it can be applied to any pair of vector spaces not
contained in a common larger vector space.
\end{definition}

\mypar{Chinese remainder theorem} The most commonly used form of the
Chinese remainder theorem (CRT) is for the ring of integers. It states that
if $n = pq$, $\gcd(p,q) = 1$, then
\begin{equation}\label{crtint}
\Z/n\Z \cong \Z/p\Z \times \Z/q\Z.
\end{equation}
Here, $\Z/n\Z$ denotes the ring of integers $\{0,\dots,n-1\}$ with
addition and multiplication modulo $n$, and $\times$ denotes the
Cartesian product with elementwise operation.  The isomorphism in
\eqref{crtint} is given by, for $k\in\Z/n\Z$,
$$
\phi:\ k\mapsto (k\text{ mod }p, k\text{ mod }q).
$$
In words, the CRT states that ``computing (addition and
multiplication) modulo $n$ is equivalent to computing in parallel
modulo $p$ and modulo $q$.''

The CRT also holds for the ring, or algebra, of polynomials $\C[x]$.
In this case $\C[x]/p(x)\C[x]$ denotes the ring, or algebra, of
polynomials of degree less than $n$ with addition and multiplication
modulo $p(x)$. These {\em polynomial algebras} are discussed in
detail in Section~\ref{polyalgs}. We write for short
$$
\C[x]/p(x) = \C[x]/p(x)\C[x].
$$

\begin{theorem}[Chinese remainder theorem for polynomials]\label{crt}
Let $p(x)\in\C[x]$, and let $p(x) = r(x)s(x)$ with $\gcd(r(x),s(x)) =
1$. Then
$$
\C[x]/p(x)\cong \C[x]/r(x)\dirsum\C[x]/s(x).
$$
This isomorphism is given by the mapping
$$
q(x)\mapsto(q(x)\text{ mod }r(x), q(x)\text{ mod }s(x)).
$$
\end{theorem}
Note that, for polynomial algebras, we write $\dirsum$ instead of
$\times$ in \eqref{crtint}, since they carry the additional structure
of a vector space, and thus the Cartesian product is equivalent to the
outer direct sum in Definition~\ref{directsumV}.

The Chinese remainder theorem can be generalized to the algebra of
polynomials in several variables $\C[\overline{x}]=\C[x_1,\dots,x_k]$,
but the theory is in general more involved. In this case the ``factor
algebras'' have the form
$$
\C[\overline{x}]/(p_1(\overline{x})\C[\overline{x}]+\dots+
  p_\ell(\overline{x})\C[\overline{x}]),
$$
which will write for short as
$$
\C[\overline{x}]/\langle p_1(\overline{x}),\dots,
  p_\ell(\overline{x})\rangle.
$$

\mypar{Graphs} We introduce some basic notions for graphs that we use
in this paper.

\begin{definition}[Graph]\label{graph}
A weighted, directed graph ${\cal G}$ is a triple $(V, E, w)$, where
$V$ is the set of {\em vertices}, $E\subset V\times V$ the set of
directed {\em edges}, given by ordered pairs, and $w:\ E\rightarrow\C$
a {\em weight function} that assigns to each edge a complex number.

Each graph is uniquely described by its square adjacency matrix
$A_{\cal G}$ defined as follows. $A_{\cal G}$ is a complex $|V|\times
|V|$ matrix, in which rows and columns are indexed with $V$. At
position $(v,w)\in V\times V$, the matrix has the entry $\omega(v,w)$,
if $(v,w)\in E$, and 0 else.
\end{definition}

\section{Module Property of $\ell^p$}\label{modproplp}

The following theorem shows that $\ell^p(\Z)$, $1\leq p\leq\infty$, is
an $\ell^1(\Z)$-module. An analogous statement holds when $\Z$ is
replaced by $\N$. We provide the proof for completeness.

\begin{theorem}\label{lpmod}
Let $1\leq p\leq \infty$. Then $\ell^p(\Z)$ is an $\ell^1(\Z)$-module
with the operation being convolution of sequences.
\end{theorem}
\begin{proof}
Before we start the proof, we remind the reader of H{\"o}lder's
inequality, which for sequences takes the following form.
Let $s\in\ell^p(\Z)$ and $t\in\ell^q(\Z)$, where $1/p+1/q=1$. Then
$$
||s\cdot t||_1\leq ||s||_p ||t||_q,
$$
where ``$\cdot$'' is the pointwise product.  We begin the proof of
the theorem and denote convolution by ``$\star$.'' Let
$s\in\ell^p(\Z)$ and $h\in\ell^1(\Z)$. It is clear that the
convolution $t=h\star s$ exists and that the result is in
$\ell^\infty(\Z)$, since $\ell^p(\Z)\subseteq\ell^\infty(\Z)$ and $h$ is
BIBO stable.  To prove the Theorem, what we need to show is that $t =
h\star s$ is also in $\ell^p(\Z)$.  Using H{\"o}lder's inequality in
the third step, we get for the $n$th output
\begin{eqnarray*}
|t_n| & = & \bigl|\sum_{k\in\Z} h_k s_{n-k}\bigr| \\
& \leq & \sum_{k\in\Z}|s_{n-k}||h_k|^{1/p}\cdot |h_k|^{1-1/p} \\
& \leq & \bigl( \sum_{k\in\Z}|s_{n-k}|^p|h_k|\bigr)^{1/p}
  \bigl(\sum_{k\in\Z}|h_k|\bigr)^{1-1/p}.
\end{eqnarray*}
Raising to the $p$-th power on each side and summing over $\Z$ yields
\begin{eqnarray*}
\sum_{n\in\Z}|t_n|^p & \leq & 
  \sum_{n\in\Z}\biggl( 
    \bigl( \sum_{k\in\Z}|s_{n-k}|^p|h_k|\bigr)
    \bigl(\sum_{k\in\Z}|h_k|\bigr)^{p-1}
  \biggl) \\
& = & \bigl(\sum_{k\in\Z}|h_k|\bigr)^{p-1}\cdot 
  \bigl( \sum_{n\in\Z}\sum_{k\in\Z}|s_{n-k}|^p|h_k|\bigr)\\
& = &  \bigl(\sum_{k\in\Z}|h_k|\bigr)^{p-1}\cdot 
  \sum_{k\in\Z}|h_k|\cdot \sum_{n\in\Z}|s_{n-k}|^p\\
& = & \bigl(\sum_{k\in\Z}|h_k|\bigr)^p\cdot 
  \sum_{n\in\Z}|s_{n}|^p
\end{eqnarray*}
which yields the desired result. In the second step we pulled out the
last factor in the sum since it does not depend on $n$. In the third
step we exchanged the order of summation in the second sum (which is
absolute convergent).
\end{proof}

\section{Chebyshev Polynomials}\label{chebs}

\begin{table*}\centering
\caption{Four series of Chebyshev polynomials. The range for the zeros
is $0\leq k < n$. In the trigonometric closed form $\cos\theta = x$
and in the power form $(u+u^{-1})/2 = x$.\label{4cheb}}
\ra{1.5}
$
\begin{array}{@{}lllll@{\,}l@{\,}lll@{}}\toprule
\text{polynomial} & n = 0,1 & \text{closed form} & \text{power form} &
  \multicolumn{3}{l}{\text{symmetry}} & \text{zeros} & \text{weight }
  w(x)\\ \midrule
T_n & 1, x & \cos(n\theta) & \frac{u^n + u^{-n}}{2} & 
  T_{-n} & = & T_n & \cos\frac{(k+\frac 12)\pi}{n} & (1 - x^2)^{-1/2}\\ 
U_n & 1, 2x & \frac{\sin(n+1)\theta}{\sin\theta} &
  \frac{u^{n+1} - u^{-(n+1)}}{u - u^{-1}} & U_{-n} & = & -U_{n-2} &
  \cos\frac{(k+1)\pi}{n+1} & (1 - x^2)^{1/2}\\ 
V_n & 1, 2x-1 & \frac{\cos(n+\fh)\theta}{\cos \fh\theta} & 
  \frac{u^{n+1/2} + u^{-(n+1/2)}}{u^{1/2} + u^{-1/2}} & 
  V_{-n} & = & V_{n-1} & \cos\frac{(k+\frac 12)\pi}{n+\frac 12} & 
  (1+x)^{1/2}(1-x)^{-1/2}\\ 
W_n & 1, 2x+1 & \frac{\sin(n+\fh)\theta}{\sin \fh\theta} & 
  \frac{u^{n+1/2} - u^{-(n+1/2)}}{u^{1/2} - u^{-1/2}} & 
  W_{-n} & = & -W_{n-1} & \cos\frac{(k+1)\pi}{n+\frac 12} & 
  (1+x)^{-1/2}(1-x)^{1/2} \\
\bottomrule
\end{array}$
\end{table*}

Chebyshev polynomials, and the more general class of orthogonal
polynomials, have many interesting properties and play an important
role in different areas of mathematics, including statistics,
approximation theory, and graph theory. An excellent introduction to
the theory of orthogonal polynomials can be found in the books of
Chihara, Szeg\"o, and Rivlin
\cite{Chihara:78,Szegoe:67,Rivlin:74}. In this section we give the
main properties of Chebyshev polynomials that we will use in this
paper.

We call every sequence $C = (C_n\mid n\in\Z)$ of polynomials that
satisfies the three-term recurrence
\begin{equation}\label{gencheb}
C_{n+1}(x) = 2xC_{n}(x) - C_{n-1}(x)
\end{equation}
a sequence of {\em Chebyshev polynomials} ($C$ stands for Chebyshev).
Using \eqref{gencheb}, the sequence $C$ is uniquely determined by the
initial polynomials $C_0, C_1$.  The most important---and commonly
known---are the Chebyshev polynomials of the {\em first kind}, denoted
by $C_n = T_n$ and determined by $T_0 = 1$ and $T_1 = x$. We provide a
few examples:
$$
\begin{array}{cccccc}
T_{-2} & T_{-1} & T_0 & T_1 & T_2 & T_3 \\ \hline
2x^2-1 & x & 1 & x & 2x^2-1 & 4x^3-3x \\
\end{array}
$$
For $x\in [-1,1]$, $T_n$ can be written in
closed (or parameterized) form as
\begin{equation}\label{trigform}
T_n = \cos n\theta,\quad\cos\theta = x.
\end{equation}
The closed form exhibits the {\em symmetry property},
\begin{equation}\label{Tsym}
T_{-n} = T_n,
\end{equation}
and can be used to readily derive the zeros of $T_n$, and
to show that $T = (T_n\mid n\in\Z)$ is orthogonal on $(-1,1)$ with
respect to the weight function $w(x) = (1-x^2)^{-1/2}$, i.e., 
$$
\int_{-1}^1 T_n(x)T_m(x)w(x)dx = 0,\quad\text{for }n\neq m.
$$
We will occasionally use another parameterization of $T_n$,
which we call power form, given by
\begin{equation}\label{powerform}
T_n = \frac{u^n + u^{-n}}{2},\quad
\frac{u+u^{-1}}{2} = x.
\end{equation}
By substituting $u = e^{j\theta}$ we obtain \eqref{trigform}.

In this paper, we also consider the Chebyshev polynomials of the
second, third, and fourth kind, denoted by $U_n, V_n, W_n$,
respectively, that arise from different initial polynomials $C_0,
C_1$. Each of these sequences exhibits a symmetry property similar to
\eqref{Tsym} and is orthogonal on the interval $(-1,1)$ with respect
to a weight $w(x)$. Furthermore, $U_n, V_n, W_n$ can be written in
closed form. These properties are summarized in Table~\ref{4cheb}.

In addition, we will need the following properties that are shared by
all sequences of Chebyshev polynomials including $T,U,V,W$ (see
\cite{Chihara:78}).
\begin{lemma}\label{chebprop}
Let $C = (C_n\mid n\in\Z)$ be a sequence of Chebyshev polynomials. Then
the following holds:
\begin{enumerate}
\renewcommand{\labelenumi}{\roman{enumi})} 
\item The sequence $C$ is determined by any two successive polynomials
  $C_n,C_{n+1}$.
\item $\deg(C_0) = 0, \deg(C_1) = 1 \Rightarrow \deg(C_n) = n, \text{
  for } n\geq 0$.
\item $C_n = C_1\cdot U_{n-1} - C_0\cdot U_{n-2}$.
\item $T_k\cdot C_n = (C_{n+k} + C_{n-k})/2$.
\end{enumerate}
\end{lemma}

In this paper, we consider only Chebyshev polynomials
$C$ that satisfy $C_0 = 1$, and $\deg(C_1) = 1$.

The following properties of the four kinds of Chebyshev polynomials
introduced above are a direct consequence of their relationship to
cosine and sine function and trigonometric identities. The proof is
straightforward by induction and is omitted.

\begin{lemma}\label{cheb1}
\begin{enumerate}
\renewcommand{\labelenumi}{\roman{enumi})} 
\item The leading coefficient of $T_n$ is $2^{n-1}$; the leading
  coefficient of $U_n, V_n, W_n$ is $2^n$.
\item $T_n(1) = 1, U_n(1) = n+1, V_n(1) = 1, W_n(1) = 2n+1$
\item $ T_n(-1) = (-1)^n, U_n(-1) = (-1)^n(n+1), 
V_n(-1) = (-1)^n(2n+1), W_n(-1) = (-1)^n$
\end{enumerate}
\end{lemma}

\section{Direct Derivation of Orthogonal DTTs}\label{directorth}

In Section~\ref{orthodtts} we derived and explained the orthogonal
versions of 4 of the 16 DTTs by using the Christoffel-Darboux formula
(Theorem~\ref{cd}). To do the same for the other 12 DTTs, we need to
derive variants of \eqref{CD}. We use the $\DCTt{1}$ as an example;
the other DTTs are handled similarly.

The $\DCTt{1}$ has the associated module $\C[x]/(T_n-T_{n-2})$ and
base vectors $T_k$. The zeros of $T_n -T_{n-2} = 2(x^2-1)U_{n-2}$ are
$\alpha_k = \cos k\pi/(n-1)$, $0\leq k < n$.  Since the goal is to
make the top case ($x\neq y$) in \eqref{CD} vanish for the zeros of
$T_n - T_{n-2}$, it seems natural to substitute $P_n$ in \eqref{CD} by
$T_n - T_{n-2}$ to obtain the expression
\isdraft{%
\begin{equation*}
A(x,y) =\\
\frac{T_{n-1}(y)(T_n(x)-T_{n-2}(x)) -
T_{n-1}(x)(T_n(y)-T_{n-2}(y))}{x-y},
\end{equation*}
}{%
\begin{multline*}
A(x,y) =\\
\frac{T_{n-1}(y)(T_n(x)-T_{n-2}(x)) -
T_{n-1}(x)(T_n(y)-T_{n-2}(y))}{x-y},
\end{multline*}
}
which vanishes for any choice $x = \alpha_i, y = \alpha_j$, $i\neq j$.
Manipulation, including the use of the recurrence \eqref{gencheb} for
$T_n$, yields
\begin{gather*}
A(x,y) = 2B(x,y) - 2T_{n-1}(x)T_{n-1}(y),\text{ with}\\
B(x,y) = \frac{T_{n-1}(y)T_n(x) - T_n(y)T_{n-1}(x)}{x-y},
\end{gather*}
or
\begin{equation}\label{CDadjust}
B(x,y) = A(x,y)/2 + T_{n-1}(x)T_{n-1}(y).
\end{equation}
$B(x,y)$ is the original expression in the top case of
\eqref{CD}. Equation~\eqref{CDadjust} explains how to adjust \eqref{CD}
to yield orthogonality of the transform, namely, by subtracting
$c_nT_{n-1}(x)T_{n-1}(y)$ on both sides to get
\isdraft{%
\begin{equation*}
\sum_{0\leq k < n}\mu_k^{-1}T_k(x)T_k(y) - c_nT_{n-1}(x)T_{n-1}(y) \\
=
\begin{cases}
c_n/2\cdot A(x,y), & x\neq y,\\
c_n\cdot q(x), & x = y,
\end{cases}
\end{equation*}
}{%
\begin{multline*}
\sum_{0\leq k < n}\mu_k^{-1}T_k(x)T_k(y) - c_nT_{n-1}(x)T_{n-1}(y) \\
=
\begin{cases}
c_n/2\cdot A(x,y), & x\neq y,\\
c_n\cdot q(x), & x = y,
\end{cases}
\end{multline*}
}
with
$$
q(x) = T_{n-1}(x)T_n'(x) - T_n(x)T_{n-1}'(x) - T_{n-1}(x)T_{n-1}(x).
$$
To obtain the actual numbers, we compute $\mu_0 = \pi$, $\mu_k =
\pi/2$, $k > 0$. Since $\mu_k^{-1}$ occurs in all summands, we can
drop $\pi$. From Lemma~\ref{cheb1}, i) (Appendix\ref{chebs}), $\beta_n
= 2^{n-1}$, and thus $c_n = 1$, for $n > 1$. It remains to evaluate
$q(x)$ at $\alpha_k = \cos k\pi/(n-1)$, using $T_n' = nU_{n-1}$.  To
obtain $q(\alpha_k)$ we use for $k\neq 0, n-1$ the closed forms of
$T_n, U_n$ (Table~\ref{4cheb}), and Lemma~\ref{cheb1}, ii) else. We
get
$$
q(\alpha_k) = 
\begin{cases}
n-1 & k\neq 0,n-1,\\
2(n-1) & k = 0,n-1.
\end{cases}
$$
Similar to \eqref{diags}, we set
$$
\begin{array}{rcl}
D & = & \diag_{0\leq k<n}(q(\alpha_k)^{-1}) \\
& = & \diag(\frac{1}{2(n-1)}, \frac{1}{n-1},\dots,
  \frac{1}{n-1},\frac{1}{2(n-1)}) \\
E & = & \diag(\mu_k^{-1}) \\
 & = & \diag(1, 2, \dots, 2, 1)
\end{array}
$$ 
and $\sqrt{D}\cdot\pDCTt{1}\cdot\sqrt{E}$ is orthogonal
(Table~\ref{dttorth}).

\end{document}